%% file: thesis7.tex
\documentclass{ucbthesis}
\usepackage[backend=bibtex]{biblatex}
\usepackage{amsmath}
\usepackage{amsfonts}
\usepackage{amssymb}
\usepackage{graphicx}
\usepackage{mathdots}
\usepackage{rotating}
\usepackage{color}
\usepackage{appendix}

\bibliography{references}

\begin{document}


\title{Quantum Bayesian networks with application to games displaying Parrondo's paradox}
\author{Michael Pejic}
\degreesemester{Fall}
\degreeyear{2014}
\degree{Doctor of Philosophy}
\chair{Professor F. Alberto Gr\"{u}nbaum}
\othermembers{Professor Nicolai Reshetikhin\\Professor Umesh Vazirani}
\numberofmembers{3}
\field{Mathematics}
\campus{Berkeley}


\maketitle
\copyrightpage

\include{abstract}

\begin{frontmatter}

\begin{dedication}
\null\vfil
\begin{center}
To my parents.
\end{center}
\vfil\null
\end{dedication}


\tableofcontents
\listoffigures

\begin{acknowledgements}
Michael Pejic acknowledges support from the Applied Math. Sciences subprogram of the Office of Energy Research, US Department of Energy, under
Contract DE-AC03-76SF00098, and from AFOSR grant FA95501210087 through a subcontract to Carnegie Mellon University.
\end{acknowledgements}

\end{frontmatter}

\pagestyle{headings}

\include{chapter1ver2}

\part{Quantum Bayesian networks}
\include{chapter2ver3}
\include{chapter3ver2}
\include{chapter4}
\include{chapter5ver2}
\include{chapter6}
\part{Parrondo's paradox and a Parrondo-like paradox}
\include{chapter7}
\include{chapter8}
\include{chapter9}
\printbibliography
\begin{appendices}
\include{appendix1ver2}

\include{appendix2ver3}
\include{appendix3ver2}

\end{appendices}

\end{document}

%% file: abstract.tex

\begin{abstract}
Bayesian networks and their accompanying graphical models are widely used for prediction and analysis across many disciplines. We will reformulate these in terms of linear maps. This reformulation will suggest a natural extension, which we will show is equivalent to standard textbook quantum mechanics. Therefore, this extension will be termed \textit{quantum}. However, the term \textit{quantum} should not be taken to imply this extension is necessarily only of utility in situations traditionally thought of as in the domain of quantum mechanics. In principle, it may be employed in any modeling situation, say forecasting the weather or the stock market--it is up to experiment to determine if this extension is useful in practice. Even restricting to the domain of quantum mechanics, with this new formulation the advantages of Bayesian networks can be maintained for models incorporating quantum and mixed classical-quantum behavior. The use of these will be illustrated by various basic examples.

Parrondo's paradox refers to the situation where two, multi-round games with a fixed winning criteria, both with probability greater than one-half for one player to win, are combined. Using a possibly biased coin to determine the rule to employ for each round, paradoxically, the previously losing player now wins the combined game with probability greater than one-half. Using the extended Bayesian networks, we will formulate and analyze classical observed, classical hidden, and quantum versions of a game that displays this paradox, finding bounds for the discrepancy from naive expectations for the occurrence of the paradox. A quantum paradox inspired by Parrondo's paradox will also be analyzed. We will prove a bound for the discrepancy from naive expectations for this paradox as well. Games involving quantum walks that achieve this bound will be presented.
\end{abstract}

%% file: chapter1ver2.tex
\chapter{Introduction}
\subsubsection*{Outline of the work}
Bayesian networks and graphical models are useful for classical systems because they are much more intuitive than a list of conditional dependencies. It is also sometimes useful to introduce additional, hypothesized nodes to break complicated dependencies into simpler, potentially universal modules. Usually these are treated just as observable nodes which are always hidden; however, this imposes constraints that are metaphysical in origin and raises difficulties of interpretation. We will give an alternate approach using linear maps on measures, with additional constructions to those generally utilized in graphical models, that resolves those issues. While, in certain situations, this introduces additional maps, not previously available, these new maps are not only in and of themselves of limited interest, but also introduce undesired complications. However, what is extremely fruitful is simply the conceptual leap. Thinking in terms of linear maps on spaces of measures immediately raises the question of looking at linear maps on other spaces. This approach leads to a natural extension, which we will prove is equivalent to standard, textbook quantum mechanics (which is infamous for its apparently unmotivated and incomprehensible formulation). Therefore, this extension will be termed \textit{quantum}. To avoid being swept away by a flood of details, propositions of a more general nature, together with their proofs, needed to show the sensibility and consistency of the extension and its equivalence to quantum mechanics are placed in appendices.

However, the term \textit{quantum} should not be taken to imply this extension is necessarily only of utility in situations traditionally thought of as in the domain of quantum mechanics. In principle, it may be employed in any modeling situation, say forecasting weather or stock prices--it is up to experiment to determine if this extension is useful in practice. In particular, there is no reason for $ \hslash $ to necessarily enter into these models if they are outside the realm of physics. Even restricting to the traditional domain of quantum mechanics, with this new formulation the advantages of Bayesian networks can be maintained for models incorporating quantum and mixed classical-quantum behavior. The use of these will be illustrated by various examples. In particular, we will show that some of the supposed hallmarks of quantum mechanics, no-cloning and teleportation, apply for classical hidden systems as well. 

In the second part, we will utilize these extended Bayesian networks in the study of various games displaying Parrondo's paradox--the phenomenon of two games each winning for one player with probability greater than one-half, yet their convex combination (in a sense to be specified) paradoxically winning for the previously losing player with probability greater than one-half. We will prove bounds for the discrepancy from naive expectations for classical versions of a game; those bounds will then be shown to be broken by a quantum analogue of the game.~\cite{grunbaumpejic}

A quantum paradox inspired by Parrondo's paradox will also be analyzed. We will prove a bound for the discrepancy from naive expectations for this paradox. Games involving quantum walks that achieve this bound will be presented.
\subsubsection*{Philosophical interlude--rejection of metaphysics}
\begin{quote}
For a man will attain unto nothing more perfect than to be found to be most learned in the ignorance which is distinctly his. The more he knows that he is unknowing, the more learned he will be.--Nicholas of Cusa~\cite{nicholas} 
\end{quote}
\begin{quote}
Natural scientists may adopt whatever attitude they please; they are still under the domination of philosophy.--F. Engels ~\cite{engels} 
\end{quote}
Physicists generally pretend to ignore philosophy, but since physics (unlike mathematics) deals with knowledge of external reality, it is not possible to avoid epistemology. A commonly expressed (although often unacknowledged) philosophy is that of naive realism: While other sciences have to deal with human-created concepts, physics is a ``god's-eye view" of reality as it really is. This is presented pictorially in \textit{Figure 1.1}. However, there are several problems with this: (\textit{i}) it does not agree with what physicists actually do in practice~\cite{cartwright}~\cite{cartwright2}; (\textit{ii}) it needlessly encumbers physics with many questions that are metaphysical in nature--the direction of time, the reason for existence of nature, the collapse of the wavefunction, and so on; and, most importantly for our work in this dissertation, (\textit{iii}) it makes it seem as if \textit{quantum} is a kind of physics with no connection to modeling.
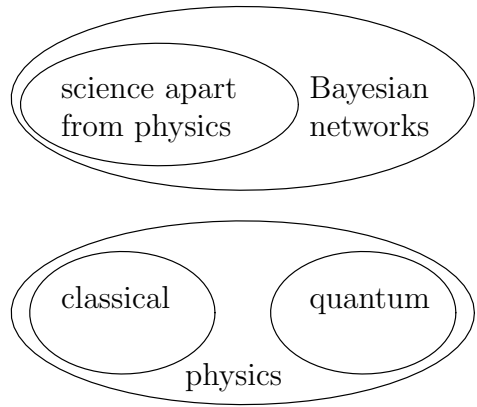
\begin{figure}[p]\label{fig:tradphilosophy}
\setlength{\unitlength}{1 in}
\begin{center}
\begin{picture}(5.5,2.3)
\put(1,-.7){\includegraphics[scale=.6]{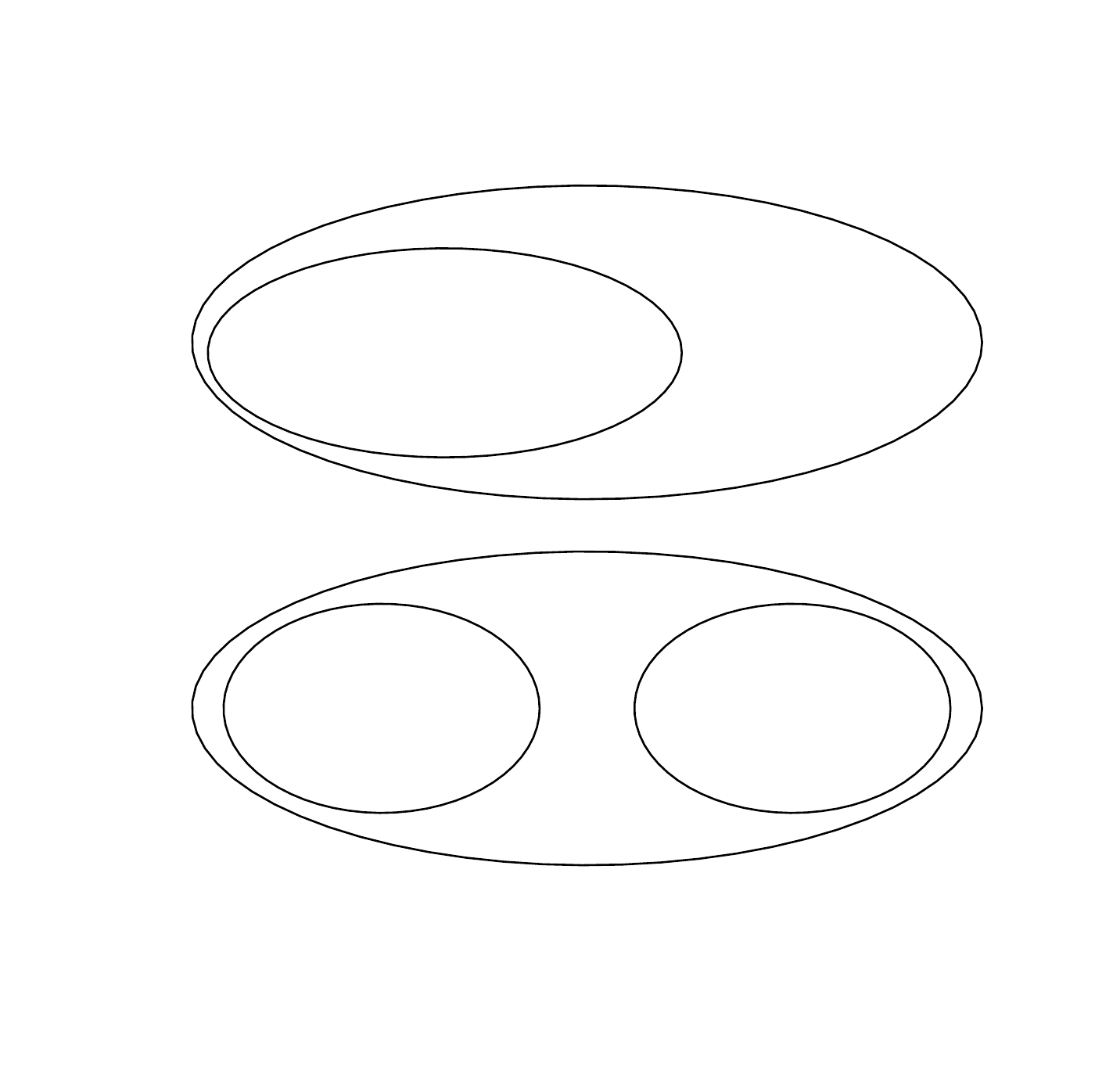}}
\put(3.15,1.6){Bayesian}
\put(3.15,1.4){networks}
\put(2.5,.1){physics}
\put(1.85,.5){classical}
\put(3.15,.5){quantum}
\put(1.85,1.6){science apart}
\put(1.85,1.4){from physics}
\end{picture}
\end{center}
\caption{Traditional conception of the relation of physics to Bayesian networks.}
\end{figure}

The philosophy we employ in this work is one with a long-standing pedigree~\cite{nicholas2}: We know nothing about reality as it really is; hence, external truths are necessarily tentative and model-dependent. This is presented pictorially in \textit{Figure 1.2}. This credo was widely disseminated by Pragmatists in the late nineteenth century who were influenced by the methodology of science~\cite{vaihinger}~\cite{nietzsche}~\cite{poincare}~\cite{james1}~\cite{james2}, and it has continued to be advocated by philosophers of science in the century since~\cite{lee}~\cite{jeans}~\cite{popper}~\cite{hesse}~\cite{rescher}~\cite{lackoff}~\cite{altmann}~\cite{baggott}. The benefit of this philosophy is that not only does it resolve the first two questions raised above, but, most pertinent to this present work, it shows \textit{quantum} is actually a sort of model. Quantum mechanics is then the overlap of physics with this class of models.
\begin{figure}[p]\label{fig:pragphilosophy}
\setlength{\unitlength}{1 in}
\begin{center}
\begin{picture}(5.5,2.3)
\put(.1,-.7){\includegraphics[scale=.9]{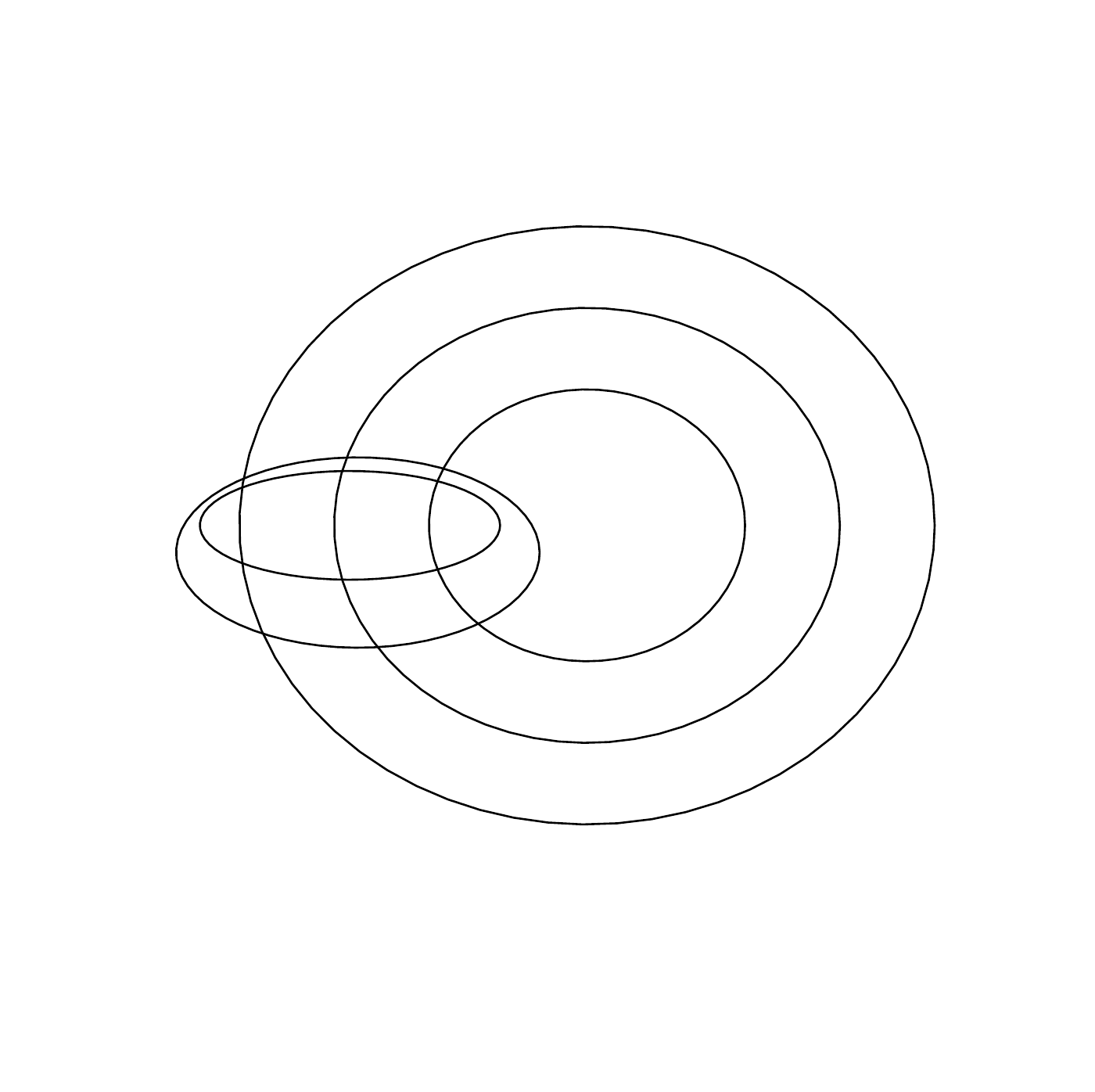}}
\put(3.15,2.2){Bayesian networks}
\put(1.45,1.8){physics}
\put(1.45,1.5){science}
\put(2.55,1.5){traditional}
\put(2.4,1.1){non-traditional}
\put(3.3,1.8){classical}
\put(2.75,.7){quantum}
\put(3.4,.4){novel?}
\end{picture}
\end{center}
\caption{Proposed relation of physics to Bayesian networks.}
\end{figure}
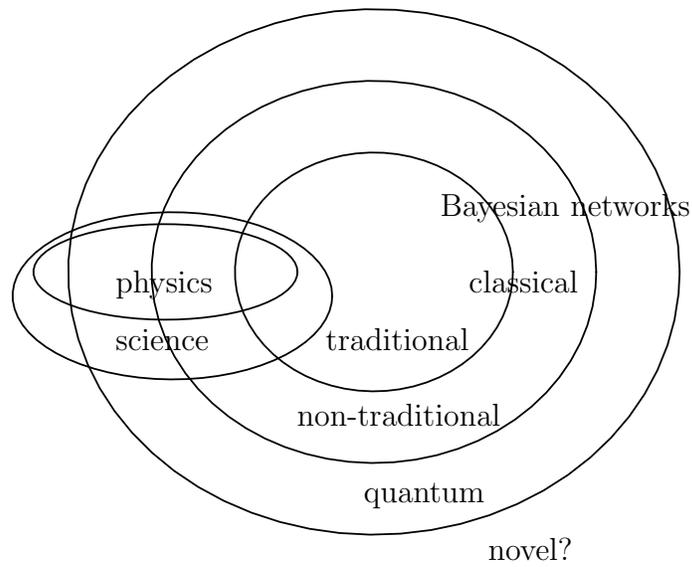

There is no reason to believe reality is doing calculations at all similar to the ones we employ in our hypothesized models\footnote{Human mathematics is an abstraction from our experiences as embodied beings~\cite{lackoff}--perhaps reality uses a form of mathematics that cannot be conceived of in this manner.} or even that it is doing calculations at all. In particular there is the issue of contextuality--we want to employ potentially universal modules in our models since only they have predictive ability in novel situations, but perhaps reality is fundamentally contextual\footnote{Gravity--a long-range force that cannot be screened--is a clue that this may indeed be the case. Isolated subsystems are only possible since gravity is so weak it can be generally ignored.}. 
\subsubsection*{Constraints imposed on our models}
In keeping with the expressed philosophy, no metaphysical constraints arising from claims to know what is ``really going on" will be placed upon the mathematical operations and constructs that can be employed in the models. Rather, there are only three rules that will be enforced. Firstly, the quantities calculated by the mathematical models must be interpretable as probabilities; in particular, they must be positive\footnote{To aid readability, \textit{positive} is used instead of \textit{nonnegative} throughout. Wherever strict positivity is required, the word \textit{strict} will be added.}. Secondly, the mathematical models must be composed of linear maps (which we will show is the weakening of a principle already in wide use, if not always acknowledged). Thirdly, the mathematical models must be composed of potentially universal modules (in a manner that we will precisely define). Of course, the particular model employed in a particular situation may fail to be universal when actually employed in a different context; the point is that this failure should be as a result of experiment and not be preordained as result of our choice of mathematics employed in modeling.

These constraints are extremely restrictive. For maps to be linear, they must clearly live on linear spaces. Furthermore, the constraints also impose strong restrictions on the linear spaces these maps live on. Thus far, we are only aware of three classes of linear spaces that meet the imposed restrictions: \textit{(i)} certain subspaces of measures; \textit{(ii)} density matrices on complex Hilbert spaces ; and \textit{(iii)} tensor products of these. The former gives what is traditionally thought of as classical behavior and we will prove the latter two give behavior that has traditionally been taken the domain of quantum mechanics. However, since we do not start from quantum mechanics, but instead only from the above principles, models involving maps on density matrices may be found to be of utility in situations not traditionally thought of as related to quantum mechanics.

%% file: chapter2ver3.tex
\chapter{Bayesian networks-graphical models}\label{ch:bayesiannetworks}
\section{Graphical models as the form in which information is presented}\label{sec:form}
There are many ways to present joint probability in terms of other quantities and mathematical constructs. Graphical models~\cite{jensen} are a useful way to sort through what otherwise can seem hopelessly complicated in the usual notation. For instance, suppose we are given a probability space $ (\Omega,\mathcal{E},\pi) $, with $ \Omega $ a set, $ \mathcal{E} $ a $ \sigma $-algebra \footnote{A $ \sigma $\textit{-algebra} is a collection of subsets of a set $ \mathcal{X} $, including both $ \varnothing $ and $ \mathcal{X} $, that is closed under relative complementation and countable unions.} of subsets of $ \Omega $, and $ \pi:\mathcal{E}\to \mathbb{R} $ a probability measure. Furthermore suppose we have two generalized random variables, $ X:\Omega\to\mathcal{X} $ and $ Y:\Omega\to\mathcal{Y} $ for sets $ \mathcal{X} $ and $ \mathcal{Y} $. Then, for any $ A\in X(\mathcal{E}) $ and $ B\in Y(\mathcal{E}) $, the joint probability Prob$(X\in A\text{ and } Y\in B)$ is $\pi\,(X^{-1}(A)\cap Y^{-1}(B))$. The graphical model corresponding to presenting the joint probability in this manner--namely by giving ($ \Omega $, $ \mathcal{E} $, $ \pi $, $ \mathcal{X} $, $ X $, $ \mathcal{Y} $, $ Y $)--is
\begin{align}\label{eq:theistic}
\setlength{\unitlength}{1 in}
\begin{picture}(3,1)
\put(1,.2){\circle{.2}}
\put(2,.2){\circle{.2}}
\put(1.5,.8){\circle{.2}}
\put(.75,.2){$ X $}
\put(1.25,.8){$ \Omega $}
\put(2.15,.2){$Y$}
\put(.9,.2){\line(1,0){.19}}
\put(1,.109){\line(0,1){.185}}
\put(1.9,.2){\line(1,0){.19}}
\put(2,.109){\line(0,1){.185}}
\put(1.42,.75){\vector(-3,-4){.36}}
\put(1.435,.735){\line(-3,-4){.33}}
\put(1.58,.75){\vector(3,-4){.36}}
\put(1.565,.735){\line(3,-4){.33}}
\end{picture}
\end{align}
where the double arrows stand for deterministic causation.

The resulting joint probability determines a probability space $ (\mathcal{X}\times\mathcal{Y},\mathcal{F},\rho) $ with the probability measure $ \rho $ given on rectangular subsets by $ \rho(A\times B)=\pi\,(X^{-1}(A)\cup Y^{-1}(B)) $, which can be readily extended to a probability measure for the $ \sigma $-algebra $ \mathcal{F}=(\mathcal{X}\times\mathcal{Y})(\mathcal{E}) $. The graphical model corresponding to presenting the joint probability in this manner--namely by giving ($ \mathcal{X}\times\mathcal{Y} $, $ \mathcal{F} $, $ \rho $)--is
\begin{align}\label{eq:stuffhappens}
\setlength{\unitlength}{1 in}
\begin{picture}(3,.5)
\put(1,.2){\circle{.2}}
\put(2,.2){\circle{.2}}
\put(.75,.2){$ X $}
\put(2.15,.2){$Y$}
\put(.9,.2){\line(1,0){1.19}}
\put(1,.109){\line(0,1){.185}}
\put(2,.109){\line(0,1){.185}}
\end{picture}
\end{align}

For each $ B\in Y(\mathcal{E}) $, the probability for \textit{X} determines a probability space $ (\mathcal{X},\mathcal{G},\mu_{B}) $ with $ \sigma $-algebra $ \mathcal{G}=X(\mathcal{E}) $ and with probability measure $ \mu_{B} $ given by $ \mu_{B}(A)=\rho(A\times B)=\pi(X^{-1}(A)\cup Y^{-1}(B))=\text{Prob}\,(X\in A\text{ and }Y\in B) $. The marginal probability for \textit{X} is then given by $ \mu=\mu_{\mathcal{Y}} $.  Let $  \mathcal{G}\times\mathcal{Y} $ signify the $ \sigma $-algebra of all rectangular sets of the form $ A\times\mathcal{Y}  $ for $ A\in\mathcal{G} $. Since $ \mu $ is a finite measure (hence, $ \sigma $-finite\footnote{A measure $ \mu $ on set $\mathcal{X} $ is $ \sigma $\textit{-finite} if there are a countable collection of $ \mu $-measurable subsets $ \lbrace B_{j}\rbrace$ such that $ \bigcup_{j}B_{j}=\mathcal{X} $ with each $ \mu(B_{j}) $ finite.}) and $ \mu_{B} $ is absolutely continuous\footnote{For measure spaces $ (\mathcal{X},\mathcal{E},mu) $ and $ (\mathcal{X},\mathcal{E},nu) $, $ \nu $ is \textit{absolutely continuous} with respect to $ \mu $ if $ \mu(A)=0 $ implies $ \nu(A)=0 $ for all subsets $ A\in \mathcal{E} $.} with respect to $ \mu $, by the Radon-Nikod\'{y}m theorem~\cite{roydenradonnikodym}, the measures $ \lbrace\mu_{B}\rbrace $ determine the conditional probability (shown in two common notations) $ \tau(B\vert \mathcal{G}\times\mathcal{Y})(x)$ or $\tau(
B\vert x) $ as the Radon-Nikod\'{y}m derivative $ \dfrac{d\mu_{B}}{d\mu}(x) $.

The Radon-Nikod\'{y}m derivative $ \dfrac{d\mu_{B}}{d\mu}(x) $ is actually an equivalence class of functions\footnote{By abuse of nomenclature it is by convention termed a ``function"--we will maintain the quotes for this latter case since we need to maintain the distinction.}. Let $ L^{0}(\mathcal{X};\mu) $ denote the space of equivalence classes of $ \mu $-measurable functions\footnote{For a given measure space $ (\mathcal{X},\mathcal{E},\mu) $, let $ \mathcal{E}_{0} $ be the $ \sigma $-algebra given by the completion of $ \mathcal{E} $ with respect to $ \mu $ (see~\cite{roydencompletemeasure}). The extension of $ \mu $ to $ \mathcal{E}_{0} $ will still be denoted $ \mu $. Then we will say $ \mu $ is a $ \mathcal{F} $\textit{-measure} if $ \mathcal{F} $ is a sub-$ \sigma $-algebra of $ \mathcal{E}_{0} $. An element of $ \mathcal{E}_{0} $ will be termed $ \mu $\textit{-measurable}. A function $ f:\mathcal{X}\to\mathbb{R}\cup\lbrace\infty, -\infty,\text{undefined}\rbrace $ is $ \mu $\textit{-measurable} if it takes the later three values only on a $ \mu $-null subset and is elsewhere measurable with respect to $\mathcal{E}_{0} $ in the usual sense.} that differ on $ \mu $-null subsets. As usual, for $ p\in[0,\infty) $, let $ L^{p}(\mathcal{X};\mu) $ be the space of ``functions" $ f $ in $  L^{0}(\mathcal{X};\mu) $ such that $ \int_{\mathcal{X}}\vert f\vert^{p}\,d\mu $ is finite (the value of the integral is independent of the representative function in the equivalence class). For $ p\in[1,\infty) $, these are Banach spaces for the usual norm
\begin{equation}
\Vert f\Vert=\sqrt[p]{\int_{\mathcal{X}}\vert f\vert^{p}\,d\mu }
\end{equation}
Furthermore, let $ L^{\infty}(\mathcal{X};\mu) $ be the space of ``functions" $ f $ in $  L^{0}(\mathcal{X};\mu) $ that are essentially bounded\footnote{Bounded on all but a $ \mu $-null subset} in magnitude. This is a Banach space with the norm $ \Vert f\Vert $ being the least essential bound on $ \vert f\vert $. Then $ \dfrac{d\mu_{B}}{d\mu}=\tau(
B\vert x) $ is actually a ``function" in $ L^{1}(\mathcal{X};\mu)\cap L^{\infty}(\mathcal{X};\mu) $. This space is clearly isometrically isomorphic to $ L^{1}(\mathcal{X}\times \mathcal{Y};\rho\vert_{\mathcal{G}\times\mathcal{Y}})\cap L^{\infty}(\mathcal{X}\times \mathcal{Y};\rho\vert_{\mathcal{G}\times\mathcal{Y}}) $, which is emphasized by the notation $ \tau(B\vert \mathcal{G}\times\mathcal{Y})(x)$.

Let $ \mathcal{H} $ be the $ \sigma $-algebra $ Y(\mathcal{E}) $. For any disjoint, countable collection $ \lbrace B_{j}\rbrace\subset \mathcal{H} $, 
\begin{equation}
\mu_{\bigcup_{j}B_{j}}=\rho(\cdot\times \bigcup_{j}B_{j})=\sum_{j}\rho(\cdot\times B_{j})=\sum_{j}\mu_{B_{j}}
\end{equation}
with convergence in norm. Since $ \left\Vert\dfrac{d\mu_{B}}{d\mu}\right\Vert_{L^{1}(\mathcal{X};\mu)}=\Vert\mu_{B} \Vert$, 
\begin{align}
\tau\left(\left.\bigcup_{j}B_{j}\right\vert\cdot\right)=\sum_{j}\tau\left(B_{j}\vert\cdot\right)
\end{align}
with convergence in $ L^{1}(\mathcal{X};\mu) $-norm. Hence, the conditional probability $ \tau(\cdot\vert\cdot) $ is a $ L^{1}(\mathcal{X};\mu) $-valued vector measure\footnote{A \textit{vector measure} is a countably-additive set function with values in a Banach space where the convergence for the countably-additivity is in norm.} on $ \mathcal{Y} $.

Then the joint probability Prob$(X\in A\text{ and } Y\in B)$ is given by
\begin{equation}
\int_{x\in A}\tau(B\vert x)\,d\mu(x)
\end{equation}
The directed graphical model corresponding to presenting the joint probability in this manner--namely by giving ($ \mathcal{X} $, $ \mathcal{G} $, $ \mathcal{Y} $, $ \mathcal{H} $,$ \mu $, $ \tau(\cdot\vert\cdot) $)--is
\begin{align}\label{eq:cause}
\setlength{\unitlength}{1 in}
\begin{picture}(3,.5)
\put(1,.2){\circle{.2}}
\put(2,.2){\circle{.2}}
\put(.75,.2){$ X $}
\put(2.15,.2){$Y$}
\put(.9,.2){\line(1,0){.19}}
\put(1,.109){\line(0,1){.185}}
\put(1.9,.2){\line(1,0){.19}}
\put(2,.109){\line(0,1){.185}}
\put(1.09,.2){\vector(1,0){.82}}
\end{picture}
\end{align}

Let $ \nu $ be the marginal probability for \textit{Y}, $ \nu(B)=\rho(\mathcal{X}\times B) $. Then $ \tau(\cdot\vert\cdot) $ is absolutely continuous with respect to $ \nu $ in the sense that $ \tau(B\vert\cdot) $ is the zero ``function" for every \textit{B} such that $ \nu(B)=0$. However, unless we are in the common case where the $ \sigma $-algebra on $\mathcal{X}$ is generated by a countable collection of atoms\footnote{A set in a set algebra is an \textit{atom} if it is indivisible in the set algebra.}, $ L^{1}(\mathcal{X};\mu) $ does not have the Radon-Nikod\'{y}m property\footnote{A Banach space \textsf{B} has the \textit{Radon-Nikod\'{y}m property} if, for any \textsf{B}-valued vector measure $ \nu $ on a set $ \mathcal{X} $ which is absolutely continuous with respect to some $ \sigma $-finite measure $ \mu $ on $ \mathcal{X} $, there is a Bochner integrable, \textsf{B}-valued ``function", $ \frac{d\nu}{d\mu}$, such that $ \nu(A)=\int_{A}\frac{d\nu}{d\mu}\,d\mu $ for any $ \nu $-measurable \textit{A}.}~\cite{ryanl1notradon} (the example given above demonstrates this); hence, there is in general no $f\in L^{1}(\mathcal{Y};\nu;L^{1}(\mathcal{X};\mu)) $ (which by Fubini's theorem~\cite{roydenfubini} is the same as $ L^{1}(\mathcal{X}\times\mathcal{Y};\mu\times\nu) $ for the product measure\footnote{Unfortunately, by convention the tensor product of measures is called the \textit{product} measure and written using $ \times $ instead of the more appropriate $ \otimes $ (however, see~\cite{semadeniproduct} for a use of the latter notation).} $ \mu\times\nu $) such that the joint probability Prob$(X\in A\text{ and } Y\in B)$ is given by $ \int_{(x,y)\in A\times B}f(x,y)\,d(\mu\times\nu)(x,y) $.

Lastly, defining $ \zeta(\cdot\vert\cdot) $ symmetrically to $ \tau(\cdot\vert\cdot) $, the directed graphical model corresponding to presenting the joint probability
\begin{equation}
\text{ Prob}(X\in A\text{ and } Y\in B)=\int_{y\in B}\zeta(A\vert y)\,d\nu(y)
\end{equation}
by giving the marginal probability $ \nu $ and the conditional probability $ \zeta(\cdot\vert\cdot) $ is
\begin{align}\label{eq:causereverse}
\setlength{\unitlength}{1 in}
\begin{picture}(3,.5)
\put(1,.2){\circle{.2}}
\put(2,.2){\circle{.2}}
\put(.75,.2){$ X $}
\put(2.15,.2){$Y$}
\put(.9,.2){\line(1,0){.19}}
\put(1,.109){\line(0,1){.185}}
\put(1.9,.2){\line(1,0){.19}}
\put(2,.109){\line(0,1){.185}}
\put(1.91,.2){\vector(-1,0){.82}}
\end{picture}
\end{align}

As an example, consider calculating the conditional probability 
\begin{equation}
\text{Prob}\, (Y\in B\vert X\in A)=\dfrac{\text{Prob} (Y\in B\text{ and } X\in A)}{\text{Prob} ( X\in A)}
\end{equation}
(for $ \text{Prob} ( X\in A)\neq 0 $--otherwise the joint probability does not determine the conditional probability) using the information presented in the manner corresponding to each of the four graphical models (the filled circle indicates which node is being conditioned on):
\begin{align}
\setlength{\unitlength}{1 in}
\begin{picture}(6,1)
\put(1,.2){\circle*{.2}}
\put(2,.2){\circle{.2}}
\put(1.5,.8){\circle{.2}}
\put(.35,.2){$ X\in A $}
\put(1.25,.8){$ \Omega $}
\put(2.15,.2){$Y$}
\put(1.9,.2){\line(1,0){.19}}
\put(2,.109){\line(0,1){.185}}
\put(1.42,.75){\vector(-3,-4){.36}}
\put(1.435,.735){\line(-3,-4){.33}}
\put(1.58,.75){\vector(3,-4){.36}}
\put(1.565,.735){\line(3,-4){.33}}
\put(3,.4){$\dfrac{\pi\left(X^{-1}(A)\cap Y^{-1}(B)\right)}{\pi\left(X^{-1}(A)\right)}$}
\end{picture}
\end{align}
\begin{align}
\setlength{\unitlength}{1 in}
\begin{picture}(6,.5)
\put(1,.2){\circle*{.2}}
\put(2,.2){\circle{.2}}
\put(.35,.2){$ X\in A $}
\put(2.15,.2){$Y$}
\put(1.09,.2){\line(1,0){1}}
\put(1,.109){\line(0,1){.185}}
\put(2,.109){\line(0,1){.185}}
\put(3,.2){$ \dfrac{\rho\left(A\times B\right)}{\rho\left(A\times\mathcal{Y}\right)} $}
\end{picture}
\end{align}
\begin{align}
\setlength{\unitlength}{1 in}
\begin{picture}(6,.5)
\put(1,.2){\circle*{.2}}
\put(2,.2){\circle{.2}}
\put(.35,.2){$ X\in A $}
\put(2.15,.2){$Y$}
\put(1.9,.2){\line(1,0){.19}}
\put(2,.109){\line(0,1){.185}}
\put(1.09,.2){\vector(1,0){.82}}
\put(3,.2){$\dfrac{\int_{x\in A}\tau(B\vert x)\,d\mu(x)}{\mu(A)}$}
\end{picture}
\end{align}
\begin{align}
\setlength{\unitlength}{1 in}
\begin{picture}(6,.5)
\put(1,.2){\circle*{.2}}
\put(2,.2){\circle{.2}}
\put(.35,.2){$ X\in A $}
\put(2.15,.2){$Y$}
\put(1.9,.2){\line(1,0){.19}}
\put(2,.109){\line(0,1){.185}}
\put(1.91,.2){\vector(-1,0){.82}}
\put(3,.2){$\dfrac{\int_{y\in B}\zeta(A\vert y)\,d\nu(y)}{\int_{y\in \mathcal{Y}}\zeta(A\vert y)\,d\nu(y)}$}
\end{picture}
\end{align}
For instance, Bayes' theorem is simply the calculation corresponding to the presentation of information by the last graphical model. 

A similar situation holds for any finite number of random variables~\cite{rao}, with a graphical model with deterministic causation emanating from a fundamental, hidden probability space, a graphical model with a clique of all the nodes for the random variables, and various directed models with probabilistic causation. For the common case where the various $ \sigma $-algebras are generated by finitely many atoms, the various measures become vectors, the conditional probabilities become stochastic matrices\footnote{A matrix is \textit{stochastic} if all entries are either positive or zero and all column sums are one.} or tensors, and the integrations become sums.  
\section{Transition probability functions}\label{sec:transition}
Depending on how the information for the calculation of the joint probability is presented, we may imagine different ways of varying it. For (\ref{eq:theistic}), it is most natural to imagine independently varying the maps \textit{X} and \textit{Y}. For (\ref{eq:stuffhappens}), there is nothing to independently vary other than the joint probability itself. For (\ref{eq:cause}), we would like to imagine varying the marginal probability $ \mu $ and the conditional probability $ \tau(\cdot\vert\cdot) $ independently. This is a problem because the space $ \tau(\cdot\vert\cdot) $ lives in--the $ L^{1}(\mathcal{X};\mu) $-valued vector measures--depends on $ \mu $. This problem will exist even in the commonly occurring case where the $ \sigma $-algebra $ \mathcal{G} $ is generated by a countable collection of atoms if $ \mu $ is zero on some atom (other than the empty set), since then the conditional probability when conditioning on that atom is not well-defined.

One solution to this problem, following~\cite{rao}, is provided by introducing the following notion:
\paragraph{Definition 2.2.1} For $ \sigma $-algebras $ \mathcal{G} $ on $ \mathcal{X} $ and $ \mathcal{H} $ on $ \mathcal{Y} $, a function $ \tau(\cdot\vert\cdot):\mathcal{H}\times\mathcal{X}\to\mathbb{R} $ is a \textit{transition probability function} if: \textit{(i)} for each $ x\in\mathcal{X} $, $ \tau(\cdot\vert x)$ is a probability measure on $ \mathcal{Y} $ with event $ \sigma $-algebra $ \mathcal{H} $; and \textit{(ii)} for each $ B\in \mathcal{H}$, $ \tau(B\vert \cdot)$ is a bounded, $ \mathcal{G} $-measurable function on $ \mathcal{X} $.
\medskip\\
By taking $ \tau(\cdot\vert\cdot) $ to be a transition probability function rather than a conditional probability, it is now possible to vary $ \mu $ and $ \tau(\cdot\vert\cdot) $ independently.

Following~\cite{rao}, one interpretation of the transition probability function is a function on $ \mathcal{X} $ with values in the probability measures on $ \mathcal{Y} $; naively, one is tempted to imagine $ \tau(\cdot\vert x) $ is the probability given $ X=x $, although this is meaningless unless the marginal probability $ \mu $ has a strictly positive mass atom at \textit{x}. Note the function is not generally Bochner integrable\footnote{For any Banach space \textsf{B}, a \textsf{B}-valued function on a set $ \mathcal{X} $ with a measure $ \mu $ is \textit{Bochner integrable} if there is a sequence of simple functions (functions taking only finitely many values with each value achieved on a set with finite $ \mu $-measure) converging to it, both pointwise in \textsf{B}-norm almost everywhere and in $ L^{1}(\mathcal{X};\mu;\mathsf{B}) $-norm.} unless the $ \sigma $-algebra $ \mathcal{H} $ on $\mathcal{Y}$ is generated by a countable collection of atoms. For example, if $ \mathcal{X}=\mathcal{Y} $ and the random variables \textit{X} and \textit{Y} are the same, then $ \tau(B\vert x)=1_{B}(x)=\delta_{x}(B) $, where the second interpretation as $ \tau(\cdot\vert x)=\delta_{x} $ is not Bochner integrable unless the $ \sigma $-algebra on $\mathcal{X}$ is generated by a countable collection of atoms.

Also note that the interpretation as a function-valued measure is potentially problematic since the result is not generally a vector measure because convergence for countable additivity is generally pointwise rather than in norm (using the supremum norm which is natural for functions); for instance, consider the preceding example. By the Dominated Convergence theorem, however, this pointwise convergence is adequate for integration.~\cite{roydendominated}  

We now raise the question of whether it is always possible to upgrade a conditional probability to a transition probability function. Under rather broad conditions, this is indeed possible. Firstly, to make things precise we have the following definition:
\paragraph{Definition 2.2.2} A transition probability function $ \tau(\cdot\vert\cdot) $ is a \textit{lift} of a conditional probability $ \nu(\cdot\vert\cdot) $ if $ \tau(B\vert\cdot)\in \nu(B\vert\cdot) $ for every subset $ B $ in the $ \sigma $-algebra.
\medskip\\
Then, by \textbf{A3.7}, for any conditional probability there is a lift to a transition probability function (not unique in general) given either of the following sufficient conditions (which are generally met in practice): (\textit{i}) the $ \sigma $-algebra $ \mathcal{H} $ on $ \mathcal{Y} $ is generated by countably many atoms; (\textit{ii}) $ \mathcal{Y} $ is an uncountable, complete, separable, metric space with $ \mathcal{H} $ the Borel $ \sigma $-algebra; or (\textit{iii}) the $ \sigma $-algebra $ \mathcal{G} $ on $ \mathcal{X} $ is generated by countably many atoms. See \textbf{A3.10} for additional sufficient conditions.
\subsection*{Removing metaphysical constraints}\label{subsec:removemeta}
Imagining one can vary $ \mu $ and keep the transition probability function $ \tau(\cdot\vert\cdot) $ fixed, there is an intuitive interpretation of $ \tau(\cdot\vert\cdot) $ as an idealized conditional probability\footnote{As noted in~\cite{rao}.}, with $ \tau(B\vert x) $ being the probability to observe $ B $ given the event $ \lbrace x\rbrace $, even if the latter has probability zero or is not even in the event $ \sigma $-algebra (although in this latter case it is constant within any atom). This leads to a metaphysical notion of the actual existence of a variable taking an actual value with probability reflecting our ignorance of its value. This may be of value for nodes in a graphical model which are observable; however, it is common to add hypothesized, hidden nodes to a directed graphical model in order to (hopefully) break it up into smaller, more manageable pieces~\cite{jensenhm}. For these, there is no justification to necessarily limit oneself to transition probability functions. Furthermore, the probabilities and conditional probabilities involving the hidden nodes lack any meaning in either the Bayesian or frequentist interpretations--the word \textit{probability} then only means positive and norm-one.   

For a fixed choice of $ \tau(\cdot\vert\cdot) $, there is a convex linear map \textit{L} from probability measures on $ \mathcal{X} $ to probability measures on $ \mathcal{Y} $ given by 
\begin{equation}\label{eq:transitioninducedmap}
(L\mu)(B)=\int_{x\in\mathcal{X}}\tau(B\vert x)\,d\mu(x)
\end{equation}
The map \textit{L} extends to a linear map on more general measures and signed measures. Hence, another solution (among many others) to the above posed problem is therefore to take the independently varied objects to be the marginal probability $ \mu $ and the linear map \textit{L}. Not all linear maps that take probability measures to probability measures are necessarily induced by some transition probability function as in (\ref{eq:transitioninducedmap}) (however, they are all induced by some \textit{pseudo-transition ``function"}--see \textbf{B2.7}, \textbf{B2.8}, and \textbf{B2.9}), so this generally introduces additional maps. These additional maps, involving operations such as Lebesgue decomposition~\cite{roydenlebesgue}, are not in and of themselves of great interest; in fact the raise rather undesired complications (see \S\ref{sec:optionIandII}). Furthermore, in the common case where the $ \sigma $-algebras are generated by countably many atoms, any linear map is induced by some transition probability function, so there are no additional maps. However, what is fruitful is the conceptual shift; as will be explored in the following chapter, once we are thinking in terms of linear maps, we are immediately drawn to consider the question of looking at linear maps between spaces other than spaces of measures.
\section{Graphical models as constraints}\label{sec:constraints}
In addition to showing the form in which information is presented, graphical models can also show constraints on the information in a far simpler form than the usual notation. For instance, consider the Markov chain with three random variables $ X,Y,Z $. For the graphical model
\begin{align}
\setlength{\unitlength}{1 in}
\begin{picture}(3,1.1)
\put(1,.2){\circle{.2}}
\put(2,.2){\circle{.2}}
\put(1.5,.8){\circle{.2}}
\put(1.5,.2){\circle{.2}}
\put(.75,.2){$ X $}
\put(1.25,.8){$ \Omega $}
\put(1.65,.2){$Y$}
\put(2.15,.2){$Z$}
\put(.9,.2){\line(1,0){.19}}
\put(1,.109){\line(0,1){.185}}
\put(1.4,.2){\line(1,0){.19}}
\put(1.5,.109){\line(0,1){.185}}
\put(1.9,.2){\line(1,0){.19}}
\put(2,.109){\line(0,1){.185}}
\put(1.42,.75){\vector(-3,-4){.36}}
\put(1.435,.735){\line(-3,-4){.33}}
\put(1.49,.7){\vector(0,-1){.4}}
\put(1.51,.7){\line(0,-1){.35}}
\put(1.58,.75){\vector(3,-4){.36}}
\put(1.565,.735){\line(3,-4){.33}}
\end{picture}
\end{align}
with the joint probability Prob$(X\in A\text{ and } Y\in B\text{ and } Z\in C)$ given by $\pi\,(X^{-1}(A)\cap Y^{-1}(B)\cap Z^{-1}(C))$, it is necessary to explicitly add the constraint that for every $ C\in Z(\mathcal{E}) $, the conditional probability $ \tau(C\vert x,y)=\frac{d\mu_{C}}{d\mu} $, for $ \mu_{C}(A\times B)=\pi\,(X^{-1}(A)\cap Y^{-1}(B)\cap Z^{-1}(C)) $ and $ \mu=\mu_{\mathcal{Z}} $, is independent of \textit{x} (in the almost-everywhere, probabilistic sense). If, as above, we imagine varying the maps \textit{X} and \textit{Y}, it is not at all clear how to do this while maintaining the constraint.

Similarly, for the graphical model
\begin{align}
\setlength{\unitlength}{1 in}
\begin{picture}(3,1.1)
\put(1,.2){\circle{.2}}
\put(2,.2){\circle{.2}}
\put(1.5,.8){\circle{.2}}
\put(.75,.2){\textit{X}}
\put(1.25,.8){\textit{Y}}
\put(2.15,.2){\textit{Z}}
\put(.9,.2){\line(1,0){.19}}
\put(1,.109){\line(0,1){.185}}
\put(1.4,.8){\line(1,0){.19}}
\put(1.5,.709){\line(0,1){.185}}
\put(1.9,.2){\line(1,0){.19}}
\put(2,.109){\line(0,1){.185}}
\put(1.435,.735){\line(-3,-4){.36}}
\put(1.565,.735){\line(3,-4){.36}}
\put(1.09,.2){\line(1,0){.8}}
\end{picture}
\end{align}
with the joint probability Prob$(X\in A\text{ and } Y\in B\text{ and } Z\in C)$ given by $ \rho(A\times B\times C)$, it is necessary to explicitly add the constraint that for every $ C\in \mathcal{I}=Z(\mathcal{E}) $, the conditional probability $ \tau(C\vert x,y)=\frac{d\mu_{C}}{d\mu} $, for $ \mu_{C}(A\times B)=\rho(A\times B\times C) $ and $ \mu=\mu_{\mathcal{Z}} $, is independent of \textit{x} (in the almost-everywhere, probabilistic sense). If, as above, we imagine varying the joint probability $ \rho $, it is not at all clear how to do this while maintaining the constraint.

However, consider the directed graphical model:
\begin{align}
\setlength{\unitlength}{1 in}
\begin{picture}(4,.6)
\put(1,.2){\circle{.2}}
\put(2,.2){\circle{.2}}
\put(3,.2){\circle{.2}}
\put(.75,.2){\textit{X}}
\put(2,.35){\textit{Y}}
\put(3.15,.2){\textit{Z}}
\put(.9,.2){\line(1,0){.19}}
\put(1,.109){\line(0,1){.185}}
\put(1.9,.2){\line(1,0){.19}}
\put(2,.109){\line(0,1){.185}}
\put(2.9,.2){\line(1,0){.19}}
\put(3,.109){\line(0,1){.185}}
\put(1.09,.2){\vector(1,0){.82}}
\put(2.09,.2){\vector(1,0){.82}}
\end{picture}
\end{align}
which corresponds to presenting the information to calculate the joint probability as $ \phi $, $ \eta(\cdot\vert\cdot) $, and $ \tau(\cdot\vert\cdot) $ where
\begin{equation}\label{eq:jointchain}
\text{Prob} (X\in A,Y\in B\text{, and }Z\in C )=\int_{(x,y)\in A\times B}\tau(C\vert y)\,d\mu(x,y)=\int_{y\in B}\tau(C\vert y)\,d\xi(y)
\end{equation}
for the marginal probabilities $ \mu $ on $ \mathcal{X}\times\mathcal{Y} $ and $ \xi $ on $ \mathcal{Y} $ given by
\begin{equation}
\mu(A\times B)=\int_{x\in A}\eta(B\vert x)\,d\phi(x), \xi(B)=\mu(\mathcal{X}\times B)=\int_{x\in \mathcal{X}}\eta(B\vert x)\,d\phi(x)
\end{equation}
The restriction is that the conditional probability $ \tau(\cdot\vert\cdot)$ is independent of \textit{x} (in the almost-everywhere, probabilistic sense). It is not necessary to give this constraint explicitly since it is indicated by the graphical model through the lack of an arrow directly from \textit{X} to \textit{Z}. If, as above, we take $ \eta(\cdot\vert\cdot) $ and $ \tau(\cdot\vert\cdot) $ as transition probability functions instead of conditional probabilities, it is now easy to see how to vary $ \phi $, $ \eta(\cdot\vert\cdot) $, and $ \tau(\cdot\vert\cdot) $ while maintaining the constraint--namely, by only allowing $ \tau(\cdot\vert\cdot) $ that are independent of \textit{x}.

Note if the wrong directed model is chosen, the constraint can be masked. For instance, for the graphical model
\begin{align}\label{eq:jointchainreverse}
\setlength{\unitlength}{1 in}
\begin{picture}(3,1.1)
\put(1,.2){\circle{.2}}
\put(2,.2){\circle{.2}}
\put(1.5,.8){\circle{.2}}
\put(.75,.2){\textit{X}}
\put(1.25,.8){\textit{Y}}
\put(2.15,.2){\textit{Z}}
\put(.9,.2){\line(1,0){.19}}
\put(1,.109){\line(0,1){.185}}
\put(1.4,.8){\line(1,0){.19}}
\put(1.5,.709){\line(0,1){.185}}
\put(1.9,.2){\line(1,0){.19}}
\put(2,.109){\line(0,1){.185}}
\put(1.435,.735){\vector(-3,-4){.36}}
\put(1.93,.27){\vector(-3,4){.36}}
\put(1.9,.2){\vector(-1,0){.8}}
\end{picture}
\end{align}
which corresponds to presenting the information to calculate the joint probability as $ \nu $, $ \theta(\cdot\vert\cdot) $, and $ \zeta(\cdot\vert\cdot) $ where
\begin{equation}\label{eq:jointreversechain}
\text{Prob} (X\in A,Y\in B\text{, and }Z\in C )=\int_{(y,z)\in B\times C}\zeta(A\vert y,z)\,d\kappa(y,z)
\end{equation}
for the marginal probability $ \kappa $ on $ \mathcal{Y}\times\mathcal{Z} $ given by
\begin{equation}
\kappa(B\times C)=\int_{z\in C\times }\theta(B\vert z)\,d\nu(z)
\end{equation}
Once again, it is necessary to explicitly add the constraint that for every $ C\in \mathcal{I} $, the conditional probability $ \tau(C\vert x,y)=\frac{d\mu_{C}}{d\mu} $, for $ \mu_{C}(A\times B)=\int_{(y,z)\in B\times C}\zeta(A\vert y,z)\,d\kappa(y,z) $ and $ \mu=\mu_{\mathcal{Z}} $, is independent of \textit{x} (in the almost-everywhere, probabilistic sense).

This can be readily generalized to more complicated graphical models--any graph that is not simply a clique\footnote{A \textit{clique} is a group of nodes that are all connected to one another.} of all the nodes implies constraints on the allowed joint probabilities. In this manner the various dependencies are displayed in a far more intuitive manner than through a long list of opaque constraints. Of course, it is always possible to impose additional constraints explicitly.
\section{Directed graphical models as tensor networks}\label{sec:tensornetworks}
Tensor networks are a commonly employed, diagrammatic device for contracting tensors and vectors. For the common case where the various $\sigma $-algebras are generated by finitely many atoms, a directed graphical model together with all its information corresponds to a tensor network or, if conditioning is present, the ratio of tensor networks. If the conditioning is only on nodes without parents, the denominator is necessarily one, so these can also be considered tensor networks. Each node in the graphical model with either no children or only one child becomes one node in the tensor network. For nodes in the graphical model with multiple children, it is best to replace them with two nodes, one of which takes in all the inputs  and has a single connection to the other, which is a \textit{copying} or \textit{diagonal} node that is zero unless all its connections are the same, when it has the value one, which branches out to all the outputs.

For example, consider the graphical model:
\begin{align}
\setlength{\unitlength}{1 in}
\begin{picture}(4,1.1)
\put(1,.2){\circle{.2}}
\put(1,.8){\circle{.2}}
\put(2,.5){\circle{.2}}
\put(3,.2){\circle{.2}}
\put(3,.8){\circle*{.2}}
\put(.73,.2){\textit{X}}
\put(.73,.8){\textit{Y}}
\put(2,.25){\textit{V}}
\put(3.15,.2){\textit{Z}}
\put(3.15,.8){\textit{W}}
\put(.9,.2){\line(1,0){.19}}
\put(1,.109){\line(0,1){.185}}
\put(.9,.8){\line(1,0){.19}}
\put(1,.709){\line(0,1){.185}}
\put(2.9,.2){\line(1,0){.19}}
\put(3,.109){\line(0,1){.185}}
\put(1.09,.22){\vector(3,1){.82}}
\put(1.09,.78){\vector(3,-1){.82}}
\put(2.09,.52){\vector(3,1){.82}}
\put(2.09,.48){\vector(3,-1){.82}}
\end{picture}
\end{align}
with corresponding information Prob$ (X=x) $, Prob$ (Y=y) $, Prob$(V=v\vert X=x,Y=y)  $, Prob$(Z=z\vert V=v) $, and Prob$(W=w\vert V=v) $, so the conditional probability Prob$ (X=x,Y=y,Z=z\vert W=w) $ is 
\begin{equation}
\dfrac{\begin{array}{c}\sum_{v\in\mathcal{V}}\left( \text{Prob}(X=x)\,\text{Prob}(Y=y)\,\text{Prob}(V=v\vert X=x,Y=y)\right.\\\left. \text{Prob}(Z=z\vert V=v)\,\text{Prob}(W=w\vert V=v)\right)\end{array}}{\begin{array}{c}\sum_{x'\in\mathcal{X},y'\in\mathcal{Y},v'\in\mathcal{V}}\left( \text{Prob}(X=x')\,\text{Prob}(Y=y')\,\text{Prob}(V=v'\vert X=x',Y=y')\right.\\\left. \text{Prob}(W=w\vert V=v')\right)\end{array}} 
\end{equation}
The corresponding tensor network for the numerator is:
\begin{align}
\setlength{\unitlength}{1 in}
\begin{picture}(4,1.1)
\put(2,.5){\circle*{.1}}
\put(2.3,.5){\circle*{.1}}
\put(2.7,.3){\circle*{.1}}
\put(2.7,.7){\circle*{.1}}
\put(.95,.2){\textit{x}}
\put(.95,.8){\textit{y}}
\put(1.95,.3){A}
\put(2.6,.1){C}
\put(2.8,.6){B}
\put(2.05,.62){Copy}
\put(3,.1){\textit{z}}
\put(3,.8){\textit{w}}
\put(1.1,.2){\line(3,1){.9}}
\put(1.1,.8){\line(3,-1){.9}}
\put(2.04,.5){\line(1,0){.21}}
\put(2.34,.52){\line(2,1){.65}}
\put(2.34,.48){\line(2,-1){.65}}
\end{picture}
\end{align}
where $ A_{xyv}= \text{Prob}(X=x)\,\text{Prob}(Y=y)\,\text{Prob}(V=v\vert X=x,Y=y)$, which is equal to $\text{Prob}(V=v, X=x,Y=y)$, the Copy tensor is zero unless all its subscript are equal, in which case it has value one, $ B_{vw}=\text{Prob}(W=w\vert V=v) $ and $ C_{vz}=\text{Prob}(Z=z\vert V=v) $. The value for the tensor network is then
\begin{equation}
\sum_{v,v',v''\in\mathcal{V}}A_{xyv}\text{Copy}_{vv'v''}B_{v'w}C_{v''z}=\sum_{v\in\mathcal{V}}A_{xyv}B_{vw}C_{vz}
\end{equation}
which equals the numerator. For the denominator, the tensor network in this case is simply
\begin{align}
\setlength{\unitlength}{1 in}
\begin{picture}(1,.5)
\put(.2,.2){\circle*{.1}}
\put(.24,.2){\line(1,0){.6}}
\put(.85,.15){\textit{w}}
\put(0,.2){\textit{D}}
\end{picture}
\end{align}
where
\begin{equation}
D_{w}=\sum_{x\in\mathcal{X},y\in\mathcal{Y},v\in\mathcal{V}} \text{Prob}(X=x)\,\text{Prob}(Y=y)\,\text{Prob}(V=v\vert X=x,Y=y)\text{Prob}(W=w\vert V=v)
\end{equation}
which is equal to $ \text{Prob}(W=w)$.

As this example illustrates, the advantages of the Bayesian network over the tensor network are that: \textit{(i)} it is possible to show which nodes are being observed, marginalized, or conditioned on; and \textit{(ii)} the nodes in the Bayesian network have a more intuitive interpretation. On the other hand, the tensor network does highlight the importance of copying for there to be multiple child nodes, which will be important later for incorporating quantum nodes (see \S\ref{subsec:commentsonoptions}, \S\ref{sec:nocopy}, and \S\ref{subsec:splitter}).
\section{The Copy map and restriction maps}\label{sec:copymap}
Going along with the linear maps on measures induced by transition probability function (see (\ref{eq:transitioninducedmap})), we have the following additional useful linear maps for the evaluation of the joint probability for a directed graphical model. For a set $ \mathcal{X} $ with $ \sigma $-algebra $ \mathcal{E} $, there is a Copy map from $\mathcal{E}$-measures on $ \mathcal{X} $ to $\mathcal{F}$-measures on $ \mathcal{X}\times\mathcal{X}  $, where $ \mathcal{F} $ is the $ \sigma $-algebra generated by the rectangular sets $ \mathcal{E}\times\mathcal{E}  $. It is given by, for any set $ A\in\mathcal{F} $ and $\mathcal{E}$-measure $ \mu $, $ \text{Copy}(\mu)(A)=\mu\left( \lbrace x\in\mathcal{X}\vert (x,x)\in A\rbrace\right) $. It is induced by the transition probability function $ \tau(\cdot\vert\cdot) $ given by 
\begin{equation}
\tau(A\vert x)=\begin{cases} 1&\text{if }(x,x)\in A\\0&\text{otherwise}\end{cases}
\end{equation} 
This can clearly be generalized for creating any finite number of copies.

For each $ A\in\mathcal{E} $, there is a restriction map, which is an idempotent, sending $\mathcal{E}$-measures on $ \mathcal{X} $ to $\mathcal{E}$-measures\footnote{We adopt the convention that a function (or ``function") before a measure, $ f\mu $, is the signed measure $ f\mu(A)=\int_{A} f\,d\mu$.} on $ \mathcal{X} $, $ \mu\to 1_{A}\mu=\mu(A\cap\cdot) $. It is induced by the transition probability function $ \tau(\cdot\vert\cdot) $ given by $ \tau(A\vert x)=1_{A}(x) $.
\section{Determinism}\label{sec:determinism}
An additional restriction that can be placed on the directed graphical models is one of \textit{determinism}, which we will interpret to mean that each node with a parent is associated to a transition probability function (rather than merely a conditional probability) and that all transition probability functions take values only zero or one. In that way, all uncertainty is due to the input probability measures coming from parentless nodes, which can be interpreted as arising from our uncertainty on initial values which are imagined to actually have specific, unknown values. For rather broad conditions (specified in the following), it is always possible to replace a directed graphical model with one that is deterministic and gives identical results for the joint probability. For each transition probability function, we add an auxiliary measure space, namely $ [0,1] $ with the standard topology, the Borel $ \sigma $-algebra, and Lebesgue measure as its marginal probability. We then constrict a deterministic transition probability function involving this additional space such that when the auxiliary space is marginalized, we recover the original transition probability function. Graphically, this means making the replacement for each node with a parent
\begin{align}\label{eq:determinism}
\setlength{\unitlength}{1 in}
\begin{picture}(5,1.1)
\put(1,.4){\circle{.2}}
\put(4,.4){\circle{.2}}
\put(4,.8){\circle{.2}}
\put(.9,.4){\line(1,0){.19}}
\put(1,.309){\line(0,1){.185}}
\put(3.9,.4){\line(1,0){.19}}
\put(4,.309){\line(0,1){.185}}
\put(.55,.1){\vector(4,3){.36}}
\put(.55,.7){\vector(4,-3){.36}}
\put(1.07,.47){\vector(4,3){.36}}
\put(1.07,.33){\vector(4,-3){.36}}
\put(3.55,.1){\vector(4,3){.36}}
\put(3.55,.7){\vector(4,-3){.36}}
\put(4.07,.47){\vector(4,3){.36}}
\put(4.07,.33){\vector(4,-3){.36}}
\put(4.01,.7){\vector(0,-1){.2}}
\put(3.99,.7){\line(0,-1){.17}}
\put(3.55,.12){\line(4,3){.32}}
\put(3.55,.72){\line(4,-3){.32}}
\put(.9,.1){\textit{Y}}
\put(3.9,.1){\textit{Y}}
\put(2.5,.4){$ \Longrightarrow $}
\end{picture}
\end{align}
where the double arrows are used to indicate the associated transition probability function is deterministic.

By \textbf{A3.8}, sufficient conditions for this are given by similar sufficient conditions (which are generally met in practice) for the existence of a lift in \S\ref{sec:transition}, so either: (\textit{i}) the $ \sigma $-algebra $ \mathcal{H} $ on $ \mathcal{Y} $ is generated by countably many atoms; or (\textit{ii}) $ \mathcal{Y} $ is an uncountable, complete, separable, metric space with $ \mathcal{H} $ the Borel $ \sigma $-algebra. Then for any transition probability function $ \tau(\cdot\vert\cdot):\mathcal{H}\times\mathcal{X}\to[0,1] $ there is a deterministic transition probability function $ \xi(\cdot\vert\cdot):\mathcal{F}\times\mathcal{X}\times[0,1]\to\lbrace 0,1\rbrace $ such that  
\[\tau(B\vert x)=\int_{s\in[0,1]}\xi(B\vert x,s)\;d\lambda(s)\]
where $ \lambda $ is Lebesgue measure on $ \mathbb{R} $. By \textbf{A3.9}, deterministic transition probability functions are induced by point transformations, which is another common interpretation of determinism.

Of course, just because we can restrict to working with deterministic models does not mean we must or that we should--there are many arbitrary choices made in the construction of $ \xi(\cdot\vert\cdot) $ and it is more straightforward in general to simply work directly with $ \tau(\cdot\vert\cdot) $. Hence, we will discard the restriction of determinism. (Similarly, in \S\ref{sec:textbookrules}, we will show commonly employed restrictions for quantum models, such as unitarity, are equally without merit.) 

%% file: chapter3ver2.tex
\chapter{Hidden classic and quantum nodes}\label{ch:hidden}
\section{Principles of linearity and potential universality}\label{sec:linearity}
The traditional approach to using Bayesian networks with directed graphical models gives a marginal probability measure for each input\footnote{All nodes without parents.} node and a transition probability function for each of the remaining nodes. Then the following principle~\cite{lorenzoprinciple} is utilized, which is so reasonable it almost always goes unmentioned, but is just implicitly assumed:
\paragraph{Principle 3.1.1--Measurement independence} The input marginal probabilities can be varied independently of each other and the transition probability functions.\footnote{Also known as the \textit{free choice} or \textit{free will} principle.}
\medskip\\
As an example of both the utility and reasonableness of this principle, consider a box with a horn and a switch. Suppose we know the horn buzzes with 0.9 probability if the switch is in the left position, and buzzes with 0.3 probability if the switch is in the right position. The graphical model for this is:
\begin{align}
\setlength{\unitlength}{1 in}
\begin{picture}(3,.7)
\put(1,.2){\circle*{.2}}
\put(2,.2){\circle{.2}}
\put(.75,.4){switch}
\put(1.8,.4){horn}
\put(2.4,.1){box}
\put(1.9,.2){\line(1,0){.19}}
\put(2,.109){\line(0,1){.185}}
\put(1.09,.2){\vector(1,0){.82}}
\put(.5,0){\dashbox{.1}(2.3,.6)}
\end{picture}
\end{align}
Now suppose we have a balanced stick that falls with equal probability to the left or the right when we release it. Now suppose we arrange it so it activates the switch when it falls. The graphical model now is:
\begin{align}
\setlength{\unitlength}{1 in}
\begin{picture}(3,.7)
\put(.5,.2){\circle{.2}}
\put(1.5,.2){\circle{.2}}
\put(2.5,.2){\circle{.2}}
\put(.3,.4){stick}
\put(1.25,.4){switch}
\put(2.3,.4){horn}
\put(2.9,.1){box}
\put(2.4,.2){\line(1,0){.19}}
\put(2.5,.109){\line(0,1){.185}}
\put(.4,.2){\line(1,0){.19}}
\put(.5,.109){\line(0,1){.185}}
\put(1.4,.2){\line(1,0){.19}}
\put(1.5,.109){\line(0,1){.185}}
\put(.59,.2){\vector(1,0){.82}}
\put(.59,.18){\line(1,0){.78}}
\put(1.59,.2){\vector(1,0){.82}}
\put(1,0){\dashbox{.1}(2.3,.6)}
\end{picture}
\end{align}
The initial probability measure here is 0.5 to fall to the right and 0.5 to fall to the left. The action of the stick on the switch is deterministic, as is indicated by the double arrow. Once the switch has been activated, which we observe, the incoming probability measure will be either 1 for the switch to be right, 0 to be left or 0 for the switch to be right, 1 to be left, so we can reuse the transition probability function from above. Then, marginalizing the observations of which way the stick fell and the resulting position of the switch, there is a $ 0.5\cdot 0.9+0.5\cdot 0.3=0.6 $ marginal probability the horn will sound.

Now consider placing a screen that blocks our view of which way the stick falls, but still allows us to hear the horn if it sounds. The graphical model is
\begin{align}
\setlength{\unitlength}{1 in}
\begin{picture}(3,.7)
\put(.5,.2){\circle{.2}}
\put(1.5,.2){\circle{.2}}
\put(2.5,.2){\circle{.2}}
\put(.3,.4){stick}
\put(1.25,.4){switch}
\put(2.3,.4){horn}
\put(2.9,.1){box}
\put(2.4,.2){\line(1,0){.19}}
\put(2.5,.109){\line(0,1){.185}}
\put(.59,.2){\vector(1,0){.82}}
\put(.59,.18){\line(1,0){.78}}
\put(1.59,.2){\vector(1,0){.82}}
\put(1,0){\dashbox{.1}(2.3,.6)}
\end{picture}
\end{align}
If we wish to predict the marginal probability the horn will sound, we have a problem. We are given the behavior of the box for incoming probability measures either: 1 for the switch to be right, 0 to be left; or 0 for the switch to be right, 1 to be left--but not for 0.5 for the switch to be right, 0.5 to be left. This is where we must appeal to the preceding principle, which gives that there is still $ 0.5\cdot 0.9+0.5\cdot 0.3=0.6 $ probability the horn will sound. 

From the metaphysical viewpoint this principle makes perfect sense. If there are actually existing variables that take actual values (such as the position of the switch), observed outcomes depend only these and not on probability measures, which are only a reflection of our ignorance. As has already been commented on, this justification fails for hidden nodes, which are hypothetical constructs we introduce. Without this metaphysical backing, the principle is actually far stronger than what is required in that it assumes the existence of transition probability functions.

If the principle is assumed to hold, then, from (\ref{eq:transitioninducedmap}), the calculation of the joint probability reduces to some combination of composition and tensor product of of linear maps, involving both those induced by the given transition probability functions, restriction maps, and possibly the Copy map (the last two of which are also induced by certain transition probability functions--see \S\ref{sec:copymap}). By \textbf{A3.2} and \textbf{A3.3}, this calculation is well-defined. For instance, the calculation of the joint probability for (\ref{eq:jointchainreverse}), with $ \theta(\cdot\vert\cdot) $ and $ \zeta(\cdot\vert\cdot) $ as transition probability functions, can be given as\footnote{We adopt the usual mathematical convention of maps acting on the left. The opposite convention of maps acting on the right is also commonly employed in the literature for the classical observed case~\cite{norris}.}\footnote{We will follow the convention that the tensor product of vector spaces consists of all finite linear combinations (the algebraic tensor product) except if both are Hilbert spaces, in which case it is the Hilbert space given by the completion using the standard induced inner-product. The tensor product $ K\otimes L $ of linear maps $ L:\mathsf{A}\to\mathsf{C} $ and $ K:\mathsf{B}\to\mathsf{D} $ is, for closed, linear spaces $ \mathsf{E}\subset\mathsf{A}\otimes \mathsf{B} $ and $ \mathsf{F}\subset\mathsf{C}\otimes \mathsf{F} $, the set of all linear maps $ M:\mathsf{E}\to\mathsf{F} $ which, when restricted to $ \mathsf{A}\otimes \mathsf{B} $, agree with $ K\otimes L $. If this set consists of a single map, $ K\otimes L $ is termed \textit{well-defined}.}
\begin{equation}
(R_{A}\circ K\circ((R_{B}\circ L)\otimes I)\circ\text{Copy }\circ R_{C}\nu)(\mathcal{X})
\end{equation}
for \textit{L} the map induced by $ \theta(\cdot\vert\cdot) $, \textit{K} the map induced by $ \zeta(\cdot\vert\cdot) $, \textit{I} the identity map on measures, and $ R_{A} $, $ R_{B} $, $ R_{C} $ restriction maps. By introducing an initializing map, $ L_{i} $, on the trivial measure space\footnote{A measure on a set $ \mathcal{X} $ is \textit{trivial} if the only subsets in its $ \sigma $-algebra are $ \lbrace\varnothing, \mathcal{X}\rbrace $.}, which is isomorphic to $ \mathbb{R} $, with constant value $ \mu $ and a terminal map, $ L_{t} $, which evaluates the measure on $ \mathcal{X} $ (hence, is a map to the trivial measure space), this can be written purely in terms of maps:
\begin{equation}
L_{t}\circ R_{A}\circ K\circ((R_{B}\circ L)\otimes I)\circ\text{Copy }\circ R_{C}\circ L_{i}
\end{equation}
Later, we will introduce hidden nodes of a special form to account for initializing or terminating maps (see \S\ref{subsec:nodesandterminators}). 
  
Therefore, the joint probability is a multilinear functional on the input marginal probabilities. Furthermore, by a convergence theorem for sequences of measures~\cite{roydengeneralconvergence}, it is not just linear for finite linear combinations in each marginal measure, but absolutely convergent countable linear combinations as well. Hence, for any one input node, if $ \Phi $ is the functional (with all the marginal probability measures for other nodes fixed) and $ \left\langle \mu_{j}\right\rangle  $ is a sequence of marginal probability measures for that node with $ \sum_{j}\Vert \mu_{j}\Vert $ finite, then $ \Phi\left(\sum_{j} \mu_{j}\right)=\sum_{j}\Phi \mu_{j}$. It is obvious that it would still be possible to make the calculation for the above example with the box based simply on this linearity property, which is weaker than the above principle.

As we have already mentioned, we will consider hidden nodes associated to maps on spaces other than the space of measures. Therefore, we introduce the following weaker and more general version of the above principle:
\paragraph{Principle 3.1.2--Linearity} The maps for a Bayesian network are linear and bounded (hence, continuous). 
\medskip\\

Moreover, we want the maps for graph fragments to be universal in the sense that the modules (such as the box in the preceding example) can be used to make predictions in novel situations. While this may fail in practice, this should be as a result of experiment and not be preordained by the mathematical models employed. We insist, therefore, on potential universality in the sense that any possible tensor product (not just those for a particular network) of the linear maps employed should always be well-defined. 
\paragraph{Principle 3.1.3--Potential universality} The space of linear maps employed must be such that any tensor product of maps in the space is well-defined.
\medskip\\
The repercussions of the latter two principles will be studied in the following.
\section{Options I and II}\label{sec:optionIandII}
\subsection*{A problem arising from potential universality}
Generalizing from maps induced by transition probability functions to more general linear maps, composition is not an issue. However, the principle of potential universality does not hold in general for maps on measures on specified sets for specified $ \sigma $-algebras of events. One problem is the lack of uniqueness. Suppose one has the identity map \textit{I} on Borel\footnote{The \textit{Borel} $ \sigma $-algebra on a topological space is that generated by the open subsets.} measures on the interval $ [0,1] $ with the usual topology. Define $ I\otimes I $ to be the set of all bounded, linear maps \textit{L} on the Borel measures on $ [0,1]\times[0,1] $, with the usual product topology, such that, restricted to product measures, $ \mu\times\nu $, $ L(\mu\times\nu)=(I\mu)\times(I\nu)=\mu\times\nu $. One obvious member of $ I\otimes I $ is simply the identity map on the Borel measures on $ [0,1]\times[0,1] $. This is the only weak*-continuous\footnote{Using the Riesz theorem~\cite{roydenriesz} which states that Radon (defined in the following) measures on a compact set are dual to the continuous functions on that set. A measure on a topological space is \textit{inner regular} if the measure of any set is approximated by the measure of compact sets it contains. It is \textit{outer regular} if the measure of any set is approximated by the open sets that contain it. A measure is \textit{Radon} if it is Borel and inner regular. If the space is compact and Hausdorff, and the measure is finite, then it is also necessarily outer regular. If the space is compact, Hausdorff, metric, and separable, then Borel measures are necessarily Radon~\cite{roydenborel}.} map in the set. Another map in the set is given by $ K:\rho\to \bigvee (\rho\parallel (\mu\times\nu)) $, where the supremum is taken over all product, finite, Borel measures and $ \rho\parallel (\mu\times\nu) $ is the part of $ \rho $ absolutely continuous with respect to $ \mu\times \nu $ using Lebesgue decomposition~\cite{roydenlebesgue}. This is a well-defined map by \textbf{B2.1} that differs from the identity map. For instance, let $ \rho $ be the diagonal Lebesgue measure,
\begin{equation}
\rho(A)=\lambda\left(\lbrace a\in[0,1]\vert (a,a)\in A\rbrace\right)
\end{equation} 
Then the identity map sends $ \rho $ to itself, whereas $ K\rho=0 $ (see \textbf{B1.8} for details).

Another possible problem is the lack of existence. Not every linear map can be extended. For example, the space of sequences with limit zero, $ c_{0} $, is a norm-closed\footnote{Using the supremum norm.}, weak*-dense\footnote{Using the duality of $ \ell_{1} $ and $ \ell_{\infty} $.} subset of the bounded sequences, $ \ell_{\infty} $. However, there is no extension of the identity map $ c_{0}\to c_{0} $ to a projection $ \ell_{\infty}\to c_{0} $.~\cite{narici}~\cite{wilanskynoextension} However, it is not clear whether there is a similar problem with extending the tensor product of linear maps.
\subsection*{Resolving the problem--two options}
There are various ways to resolve this dilemma. One is to revert to only considering linear maps induced by transition probability functions, which works by \textbf{A3.3}. We will not pursue this approach since it has no ready generalization to linear maps on spaces other than those of measures. Instead, we will consider two approaches which do. The first, which will be termed option \textbf{I}, is to limit the space of measures so as to eliminate measures such as $ \rho $ above. The alternative, which will be termed option \textbf{II}, is to impose additional structure on the sets and then to limit the space of maps, so as to eliminate maps such as \textit{K} above. 

To implement option \textbf{I}, a basic property that we want for our subsets of measures is defined by:
\paragraph*{Definition 3.2.1} A subset \textit{A} of measures is \textit{absolutely-continuous-complete} if, for any $ \mu $ in \textit{A}, all measures absolutely continuous with respect to $ \mu $ are also in \textit{A}.
\medskip\\
By the Radon-Nikod\'{y}m theorem and the density of simple functions\footnote{A \textit{simple function} takes on only finitely many values.} in $ L^{1} $-spaces, a subset of finite measures \textit{A} will have this property if, given any measure $ \mu\in A $, all the restrictions $ \mu\vert_{E}  $ over $ \mu $-measurable subsets $ E $ are also in $ A $. Hence, this property is the minimal requirement to show that a map is indeed positive. Then from \textbf{A1.2}, \textbf{A1.3}, \textbf{B1.3}, \textbf{B1.4}, and \textbf{B2.5}, the necessary and sufficient condition to implement option \textbf{I} is that the tensor product of the subsets of measures for each set is norm-dense\footnote{Using the total-variation norm.} in the subset of measures for the product set.

A sufficient method, termed option \textbf{I'}, to insure this is met is for each set to have an associated \textit{base measure}. The base measure need not be finite, but it must be $ \sigma $-finite. Then the associated subset of measures is the set of all finite measures absolutely continuous with respect to the base measure, which by the Radon-Nikod\'{y}m theorem is equivalent to the space of $ L^{1} $-functions with respect to the base measure. The base measure for the direct product of sets must be the product of the base measures for the individual sets. The sufficiency of this prescription is given by \textbf{B3.1}.

Another sufficient method is to instead use the space of atomic measures\footnote{A measure is \textit{atomic} if there is a union of countably many atoms in the $ \sigma $-algebra such that the complement of the union has measure zero.}. If there are uncountably many atoms in the $ \sigma $-algebra, this is distinct from option \textbf{I'}; otherwise, the counting measure that assigns one to each atom is a base measure. The sufficiency of this prescription is given by \textbf{B6.1} and \textbf{B6.2}. It is also possible to combine these two sufficient approaches, say by using the atomic, $ L^{1}(\mathcal{X};\mu) $-valued vector $ \mathcal{E} $-measures on $ \mathcal{Y} $. This is sufficient by \textbf{B3.1}, \textbf{B6.1}, and \textbf{B6.2}.

The implementation of option \textbf{II} is more straightforward. Each set has a topological structure that makes it a compact, Hausdorff space. By Tychonoff's theorem~\cite{munkrescompactdirectproduct}~\cite{roydencompactdirectproduct}, the direct product of compact spaces with the product topology is necessarily compact. All the $ \sigma $-algebras are required to be the Borel $ \sigma $-algebra. The linear maps are restricted to those that are weak* continuous; in other words, those maps that are the adjoints to linear maps on continuous functions going in the opposite direction. In practice, rather that working with the adjoint maps, one works with the linear maps in the opposite direction. The propositions \textbf{A1.3}, \textbf{C1.1}, and \textbf{C2.4} gives the sufficiency of this prescription.    
\subsection*{Comments on the two options}\label{subsec:commentsonoptions}
For option \textbf{I'}, the need for base measures is not generally a troublesome issue. For $ \sigma $-algebras generated by a countable number of atomic subsets\footnote{A subset in a $ \sigma $-algebra is \textit{atomic} if it is indivisible in the $ \sigma $-algebra.}, the counting measure that assigns one to each atom is a base measure for any finite measure. For classical physics, with configuration space $ \mathcal{X} $ and phase-space given by the cotangent bundle $ T^{*}\mathcal{X} $, the symplectic phase-space volume-form\footnote{As a measure, locally $ \Omega $ is simply Lebesgue measure with respect to any canonical choice of local position and momentum coordinates.} $ \Omega $ provides a natural base measure, since by Heisenberg's uncertainty principle, not even all measures absolutely-continuous with respect to $ \Omega $ are accessible, let alone more singular measures.

Note that allowing the base measure to be a $ \sigma $-finite measure is only for purposes of convenience in allowing the commonly employed Lebesgue measure on unbounded subsets of $ \mathbb{R}^{n} $, as the following theorem shows:
\paragraph{Theorem 3.2.2}Given any $ \sigma $-finite measure $ \mu $ on a set $ \mathcal{X} $, there is a finite measure $ \nu $ such that $ L^{1}(\mathcal{X};\mu) $ is isometric to $ L^{1}(\mathcal{X};\nu) $, where the isomorphism is a pointwise scaling.
\paragraph*{Proof} If $ \mu $ is finite, there is nothing to show, so assume it is not. Since $ \mu $ is $ \sigma $-finite, there is a countable collection of disjoint subsets $ \lbrace B_{j}\rbrace $ of $ \mathcal{X} $ such that $ \mu(B_{j}) $ is finite and nonzero for each $ j\in\lbrace 1,2,\ldots\rbrace $ and $ \bigcup_{j=1}^{\infty}B_{j}=\mathcal{X} $. Then let the finite measure $ \nu $ be given by $ \nu\vert_{B_{j}}=\dfrac{\mu\vert_{B_{j}}}{2^{j}\mu(B_{j})} $. $ \square $
\medskip\\

However, there are two complaints with option \textbf{I'}. One is that for base measures with an infinitely-divisible\footnote{A $ \sigma $-finite measure $ \mu $ for a measure space $ (\mathcal{X},\mathcal{E},\mu) $ is \textit{infinitely-divisible} if for any $ \varepsilon>0 $, there is a countable (finite if $ \mu $ is finite) partition of $ \mathcal{X} $, $ \lbrace B_{j}\rbrace $, with each $ \mu(B_{j})<\varepsilon $.} part (see \textbf{B1.7}), there is no Copy map (by the same argument as in the proof of \textbf{B1.8}); hence, except for this (effectively discrete) case, hidden nodes either have only one child node or are terminated. The second complaint is that passing a continuously variable parameter to a hidden node as a simple number is not permitted (unless that particular value corresponds to an atom in the base measure); instead, one must use a sharply peaked measure. This adds significant complexity for little gain in cases where one is not especially interested in modeling uncertainty in the inputs (see \S\ref{subsec:commentsblackboxoption} and \S\ref{subsec:completingthedefinition} for instances).

For option \textbf{II}, the restriction to topological spaces and Borel $ \sigma $-algebras is also not troublesome, since these are typically used in any case. The limitation of using compact spaces appears severe, but locally compact spaces\footnote{A space is \textit{locally compact} if it can be compactified by the addition of one point, the \textit{point-at-infinity}.~\cite{munkrespointatinfinity}.~\cite{roydenpointatinfinity}} can also be used with the restriction that the maps take continuous functions vanishing at infinity to continuous functions vanishing at infinity, so the adjoint maps on measures do not ``leak away" measure at infinity. 

In addition, for option \textbf{II}, the first complaint above does not occur since the Copy map is adjoint to the map $ \text{Copy}_{*} $ that takes continuous functions on $ \mathcal{X}\times\mathcal{X} $ to continuous functions on $ \mathcal{X} $ by $ (\text{Copy}_{*}f)(x)=f(x,x) $ (which can obviously be generalized to make any finite number of copies). The second complaint does not occur either since the evaluation map is well-defined for continuous functions. However, there is now the opposite problem in that we wish to calculate probabilities on sets, so we need maps on characteristic functions, not just continuous ones. One solution is to extend each map to one from bounded, Borel measurable functions to bounded, Borel measurable functions; by \textbf{C2.9} this can always be done in a unique manner. Another solution is to use the results on continuous functions to get the result for characteristic functions of open sets as in the proof of the Riesz theorem~\cite{roydenriesz}; then outer regularity gives the result on any characteristic function of a Borel set.

Note that for option \textbf{I'}, the considered linear map $ L: L^{1}(\mathcal{X};\mu)\to L^{1}(\mathcal{Y};\nu) $ is always induced by a conditional probability $ \xi(\cdot\vert\cdot) $ which is given by, for $ \nu $-measurable sets \textit{B}, $ \xi(B\vert\cdot)=L^{*}1_{B} $ (which is of course actually an equivalence class of functions that agree almost everywhere with respect to $ \mu $) with the adjoint map $ L^{*}: L^{\infty}(\mathcal{Y};\nu)\to L^{\infty}(\mathcal{X};\mu) $:
\begin{align}
\int_{y\in B}(Lf)(y)\,d\nu(y)&=\int_{y\in \mathcal{Y}}1_{B}(Lf)(y)\,d\nu(y)=\int_{x\in\mathcal{X}}f(x)(L^{*}1_{B})(x)\,d\mu(x)\\&=\int_{x\in\mathcal{X}}f(x)\xi(B\vert x)\,d\mu(x)\nonumber
\end{align}
for any $ f\in L^{1}(\mathcal{X};\mu) $. By \textbf{A3.7}, if either: (\textit{i}) the $\sigma $-algebra for $ \mathcal{Y} $ is generated by countably many atoms; or (\textit{ii}) $ \mathcal{Y} $ is an uncountable, complete, separable, metric space with the Borel $ \sigma $-algebra--then a transition probability function $ \tau(\cdot\vert\cdot) $ does exist that is a lift of $ \xi(\cdot\vert\cdot) $. Similarly, if we use option \textbf{I} with the measures limited to the atomic measures, the considered linear map $ L:\mathcal{A}(\mathcal{X};\mathcal{E})\to \mathcal{A}(\mathcal{Y};\mathcal{F}) $ is always induced by an object $ \tau(\cdot\vert\cdot) $ which is given by, for sets $ B\in\mathcal{F} $, $ \tau(B\vert x)=(L\delta_{A})(B) $ for $ x $ in the atomic set $ A\in\mathcal{E} $:
\begin{equation}
(L\mu)(B)=\int_{x\in\mathcal{X}}\tau(B\vert x)\,d\mu(x)
\end{equation}
for any atomic measure $ \mu\in \mathcal{A}(\mathcal{X};\mathcal{E}) $. However, in general $ \tau(\cdot\vert\cdot) $ will not be a transition probability function since there is no reason for $ \tau(B\vert \cdot) $ to necessarily be $ \mathcal{E} $-measurable. Also, for option \textbf{II}, if $ \mathcal{Y} $ is metric, by \textbf{C2.8}, for the considered linear map $ L:\mathcal{C}(\mathcal{Y})\to\mathcal{C}(\mathcal{X}) $, the adjoint map $ L^{*}:\mathcal{M}(\mathcal{X})\to\mathcal{M}(\mathcal{Y}) $ is induced by the transition probability function $ \tau(\cdot\vert\cdot) $ given by $  \tau(B\vert x)=(L\delta_{x})(B) $ for any $ x\in\mathcal{X} $ and Borel subset $ B\subset\mathcal{Y} $:
\begin{equation}
(L^{*}\mu)(B)=\int_{x\in\mathcal{X}}\tau(B\vert x)\,d\mu(x)
\end{equation}
for any Radon measure $ \mu\in\mathcal{M}(\mathcal{X}) $. However, despite the existence of transition probability functions (or similar objects) in most encountered cases, it is still more fruitful to consider the linear maps themselves as the primary objects of interest rather than the transition probability functions (or similar objects). As is shown in the following section, the linear maps can be generalized to linear maps on structures other than measures, whereas the transition probability functions or similar objects do not generalize.
\section{Quantum nodes}\label{sec:quantumnodes}
\subsection*{Expanding the space of considered maps}
As an alternative to linear maps on measures, consider linear maps on density matrices\footnote{A \textit{Hilbert space} will be taken to be any complete, sesquilinear inner-product space, without regard to cardinality of dimension or separability. The \textit{density matrices} $ \mathcal{D}(\mathsf{H})^{+} $ will be taken to be the self-adjoint, positive operators on a given Hilbert space \textsf{H}.}\footnote{A linear map from $ \mathcal{D}(\mathsf{H}) $ to $ \mathcal{D}(\mathsf{J}) $ is commonly referred to as a \textit{superoperator} in the literature. We choose not to employ this terminology for the following reasons: \textit{(i)} \textit{linear maps} is already standard mathematical terminology and is in common use in the analogous classical situation, for instance \textit{Markov maps}; \textit{(ii)} \textit{superoperator} seems to imply a map on all bounded operators, $ \mathcal{B}(\mathsf{H}) $, when in general it is not possible to extend the domain of the map beyond the trace-class operators, $ \mathcal{S}_{1}(\mathsf{H}) $; and \textit{(iii)} the use of \textit{super-} risks confusion with the unrelated \textit{supersymmetry} and \textit{superstrings}.} or, more generally, density matrix-valued vector measures. We will show below that, with some restrictions on the maps, this can be made to work consistently with the propositions given in \S\ref{sec:linearity}. We will show in \S\ref{sec:textbookrules} that this gives rise to models that are consistent with the usual, textbook quantum mechanics; therefore, nodes whose linear maps involve density matrices will be termed \textit{quantum}. However, the term \textit{quantum} should not be taken to imply these maps are necessarily only of utility in situations traditionally thought of as in the domain of quantum mechanics--in principle, any hidden node in any forecasting situation, say forecasting weather or stock prices, could be a quantum node. It is up to experiment to determine if these are of utility. Thus far, we are unaware of any structures besides measures and density matrices that have the requisite properties to be employed in modelling. The question of whether or not there are such additional structures will be further explored in \S\ref{sec:additional} below. 

Following our work in the preceding section, for option \textbf{I}, instead of linear maps on subsets of real-valued measures, we have linear maps on subsets of the $ \mathcal{D}(\mathsf{H})^{+} $-valued vector measures. For option \textbf{I'}, we will require the vector measures to be absolutely continuous with respect to a base measure. Since $ \mathcal{D}(\mathsf{H}) $ has the Radon-Nikod\'{y}m property~\cite{uhl}, this is equivalent to having $ \mathcal{D}(\mathsf{H})^{+} $-valued, Bochner-integrable functions. The constraint of potential universality mandates having well-defined tensor product of maps; this is maintained by \textbf{A1.3} and \textbf{B4.1} if the tensor product is bounded.

Similarly to the classical case, another sufficient approach to implementing option \textbf{I} is to take the atomic, $ \mathcal{D}(\mathsf{H})^{+} $-valued vector measures, as is shown by \textbf{B6.1} and \textbf{B6.2}. It is also possible to combine these two sufficient approaches, say by using the atomic, $ L^{1}(\mathcal{X};\mu;\mathcal{D}(\mathsf{H})) $-valued vector $ \mathcal{E} $-measures on $ \mathcal{Y} $. This is sufficient by \textbf{B4.1}, \textbf{B6.1}, and \textbf{B6.2}.  

For option \textbf{II}, instead of maps on real-valued, positive, continuous functions, one has maps on continuous functions that take values in the self-adjoint, positive, compact\footnote{An operator is \textit{compact} if the image of any bounded sequence has a convergent subsequence.} operators on a Hilbert space \textsf{H}, $ \mathcal{K}(\mathsf{H})^{+} $. Self-adjoint, compact operators are used since they are the predual to the self-adjoint trace-class operators.~\cite{pedersenpredual} Once again, potential universality mandates having well-defined tensor products of maps; as before, this is maintained if the tensor product is bounded, now by \textbf{A1.3} and \textbf{C3.1}.   
\subsection*{Problem arising from positivity and potential universality}
Of course, everything is not really that simple. The tensor product of bounded maps may not be bounded. Also, even if both maps are positive, their tensor product need not be. To illustrate these problems, take a separable, infinite-dimensional Hilbert space \textsf{H}, fix some orthonormal basis $ \lbrace\mathbf{e}_{j}\rbrace $, and consider the \textit{transpose map} \textit{T} relative to that basis:
\begin{equation}
T\left(\sum_{j,k}a_{jk}\mathbf{e}_{j}\otimes\mathbf{e}_{k}^{*}\right)=\sum_{j,k}a_{kj}\mathbf{e}_{j}\otimes\mathbf{e}_{k}^{*}
\end{equation}
where $ \mathbf{e}_{k}^{*} $ is the functional $ \langle\cdot,\mathbf{e}_{k} \rangle $. This is well-defined on the space of density-matrices on \textsf{H}, $ \mathcal{D}(\mathsf{H})^{+} $, since, by the spectral theorem for compact operators~\cite{dowson}, any such operator can be written in the form of an infinite matrix with finite rank operators corresponding to truncated matrices converging in trace norm. (Of course, since density-matrices are self-adjoint, \textit{T} could also be termed the \textit{conjugate map} relative to the basis). This map is clearly positive and has operator norm one.

However, consider the tensor product map $ T\otimes I_{\mathcal{M}_{n}} $ acting on density matrices $  \mathcal{D}(\mathsf{H}\otimes\mathbb{C}^{n})^{+} $, where $ I_{\mathcal{M}_{n}} $ is the identity map acting on $ n\times n $-matrices. For the rank-one operator $ \psi\otimes\psi^{*} $ with 
\begin{equation}
\psi=\sum_{j=0}^{n-1}\mathbf{e}_{(n+1)j+1}
\end{equation}  
we have $  (T\otimes I_{\mathcal{M}_{n}})(\psi\otimes\psi^{*})=S $. Truncating \textit{S} to the span of $ \lbrace\mathbf{e}_{1},\ldots,\mathbf{e}_{n^{2}}\rbrace $ (it is zero elsewhere), it is the matrix form of the transpose map acting on $ \mathcal{M}_{n} $ written in vector form using the Vec operation\footnote{Vec takes a $ n\times n $-matrix to a column vector of height $ n^{2} $ by stacking columns.}. Therefore, \textit{S} clearly has eigenvalue one with multiplicity $ \frac{n(n+1)}{2} $ and eigenvalue minus one with multiplicity $ \frac{n(n-1)}{2} $. Hence, $ T\otimes I_{\mathcal{M}_{n}} $ is not positive, and
 \begin{equation}
\Vert T\otimes I_{\mathcal{M}_{n}}\Vert_{\text{op}}\geq\frac{1}{\Vert\psi\Vert^{2}}\left(\frac{n(n+1)}{2}\vert 1\vert+\frac{n(n-1)}{2}\vert -1\vert\right)=\frac{n^{2}}{n}=n
\end{equation}
By~\cite{effros}, $ \Vert T\otimes I_{\mathcal{M}_{n}}\Vert_{\text{op}}=n $ and this example is maximal. This is clearly unbounded as $ n\to \infty $. 
\subsection*{Solution to the problem}
The solution to both the positivity and the boundedness problem is to require complete-positivity for the maps. This has several definitions (see \textbf{B2.6}, \textbf{B5.6}, \textbf{C2.1}) that are equivalent (see \textbf{B5.8}, \textbf{C5.6}); the basic notion is that all tensor products with various identity maps should be positive. Since the composition of positive maps is positive, this immediately implies that complete-positivity is preserved under both composition and tensor products. Furthermore, from \textbf{B2.5}, \textbf{B2.5}, \textbf{C2.4}, and \textbf{C5.3}, the operator norm is a cross-norm\footnote{A norm is a \textit{cross-norm} if $ \Vert \mathsf{a}\otimes \mathsf{b} \Vert\leq \Vert \mathsf{a}\Vert\Vert \mathsf{b} \Vert $. Note the property of being a cross-norm depends on the choice of norms for the individual spaces as well as for the larger space containing the tensor products.} for completely-positive maps, so this resolves the boundedness problem as well. The completely-positive maps clearly form a convex cone within the space of all maps; this cone is closed in the norm topology and in various weaker topologies by \textbf{B5.12}, \textbf{B5.14}, and \textbf{C5.10}; however, unlike the cone of positive maps, in infinite dimensions it has no interior in any of these topologies, which raises issues for approximation in numerical calculation. 
\section{No quantum copying}\label{sec:nocopy}
It is commonly stated that cloning is something that is possible classically, but is impossible in quantum mechanics. This is based on false analogy. The correct situation is that there are two different notions, that of copying and that of cloning, that are being confused. Once these two have been separated, we have the following situation:
\[\begin{array}{rcc}&\hspace{.2 in}\underline{\text{classical}}&\hspace{.2 in}\underline{\text{quantum}}\\\\\underline{\text{copying}}&\hspace{.2 in}\parbox{2 in}{Exists and implementable since linear.}&\hspace{.2 in}\parbox{2 in}{Does \textbf{not} exist.}\\\\\underline{\text{cloning}}&\hspace{.2 in}\parbox{2 in}{Exists but \textbf{not} implementable since neither linear nor the ratio of linear maps.}&\hspace{.2 in}\parbox{2 in}{Exists but \textbf{not} implementable since neither linear nor the ratio of linear maps.}\end{array}\]
The confusion is comparing the upper left and lower right entries instead of correctly going across. Cloning is possible for neither classical nor quantum Bayesian networks (as will be shown below in \S\ref{sec:nocloning}) for exactly the same reason, so it does not differentiate the two. On the other hand, copying is possible classically (except for issues arising from potential universality considered above in \S\ref{subsec:commentsonoptions}), but cannot even be defined as a mathematical operation on density matrices.
\subsection*{Classical copying}
As has already been mentioned (see \S\ref{sec:copymap}), there is a Copy map from measures on a set $ \mathcal{X} $ to measures on $ \mathcal{X}\times\mathcal{X}  $. When $ \mathcal{X} $ is a compact set, this map is weak*-continuous, being the adjoint of the previously discussed map $ (\text{Copy}_{*}f)(x)=f(x,x) $. The Clone map is given by $ \mu\to\mu\times\mu $. For the single atom measure for atom \textit{C}, where for any measurable subset $ A\subset \mathcal{X} $,
\begin{equation}
\delta_{C}(A)=\begin{cases}1&\text{if }A\supset C\\0&\text{otherwise}\end{cases}
\end{equation}
we have
\begin{equation}
\text{Copy }\delta_{C}=\delta_{C}\times\delta_{C} =\text{Clone }\delta_{C}
\end{equation}
This is likely the source of confusion between the Copy and Clone maps for the classical case.

Instead of using this explicit form for Copy, an approach that will prove useful in the quantum case is to start with some basic properties, then find the implications. One property of what is commonly accepted as the notion of a copy is that the probability for both copies to have a specified property is equal to that for each copy to have it, which is equal to that of the original having it, so for any unit-norm measure $ \mu $ on $ \mathcal{X} $ and any $ \mu $-measurable set $ A\subset\mathcal{X} $,
\paragraph{Property C} $\text{Copy } \mu(A\times A )=\text{Copy } \mu(A\times\mathcal{X} )=\text{Copy } \mu(\mathcal{X}\times A )=\mu(A)$
\bigskip\\
Note this property implies $\text{Copy } \mu(A\times (\mathcal{X}\setminus A) ) )=\text{Copy } \mu((\mathcal{X}\setminus A)\times A )=0$. Now given a $ \sigma $-algebra $ \mathcal{E} $ of subsets of $ \mathcal{X} $, let $ \mathcal{F} $ be the $ \sigma $-algebra generated by the rectangular subsets $ \mathcal{E}\times\mathcal{E} $. Then we have the following:
\paragraph{Theorem 3.4.1} Any map $ L $ from unit-norm $ \mathcal{E} $-measures on $ \mathcal{X} $ to unit-norm $ \mathcal{F} $-measures on $ \mathcal{X}\times\mathcal{X}$ obeying property \textbf{C} is linear for convex linear combinations.
\paragraph*{Comment} Since any finite measure can be scaled to have unit-norm, this implies the map can be extended to all finite measures, with the extended map being positively linear. By the generating property of measures among signed measures as a result of Jordan decomposition~\cite{roydenjordan}, this implies the map can further be extended to a linear map on signed measures.
\paragraph*{Proof} Let \textit{L} be such a map and $ \mu $ an unit-norm $ \mathcal{E} $-measure on $ \mathcal{X} $. For any subsets $ A,B\in\mathcal{E} $, by the properties of measures,
\begin{align}
(L\mu)(A\times B )=&(L\mu)((A\cap B)\times(A\cap B) )+(L\mu)((A\cap B)\times(B\setminus(A\cap B)) )\\
&+(L\mu)((A\setminus B)\times B )\nonumber
\end{align}
However, $ (A\cap B)\times(B\setminus(A\cap B))\subset(A\cap B)\times(\mathcal{X}\setminus(A\cap B)) $ and $ (A\setminus B)\times B\subset(\mathcal{X}\setminus B)\times B $, so, by property \textbf{C} and its implication,
\begin{equation}
(L\mu)(A\times B )=(L\mu)((A\cap B)\times(A\cap B) )=\mu(A\cap B)=(L\mu)(B\times A )
\end{equation}

Let $ \rho $ be another unit-norm $ \mathcal{E} $-measure on $ \mathcal{X} $. Then for any $ t\in[0,1] $, $ (1-t)\rho+t\mu $ will be a unit-norm $ \mathcal{E} $-measure on $ \mathcal{X} $. Consider the signed measure on $ \mathcal{X}\times\mathcal{X}  $ given by
\begin{equation}
\nu_{t}=L((1-t)\rho+t\mu)-(1-t)L\rho-tL\mu
\end{equation}
Take any $ A\in \mathcal{E}$. Then, by property \textbf{C}, $ \nu_{t}(A\times A)=0 $. However, by the above symmetry property of \textit{L},
\begin{equation}
\nu_{t}(A\times B)=\frac{1}{2}\left(\nu_{t}(A\times B)+\nu_{t}(B\times A) \right)
\end{equation}
which is equal to
\begin{eqnarray}
&\frac{1}{2}\left(\nu_{t}((A\cup B)\times (A\cup B) )-\nu_{t}((A\setminus B)\times (A\setminus B) )\right.\\&\left. -\nu_{t}((B\setminus A)\times (B\setminus A) )+\nu_{t}((A\cap B)\times (A\cap B) ) \right)\nonumber
\end{eqnarray}
which is zero by the preceding property of $ \nu_{t} $. Since $ A,B $ were arbitrary, $ \nu_{t} $ must be the zero measure. $ \square $
\subsection*{Quantum Copying}
The Clone map taking a density matrix on the Hilbert space \textsf{H} to one on $ \mathsf{H}\otimes \mathsf{H}$ is defined as $ \rho\to\rho\otimes\rho $. How to define a Copy map is not obvious. By analogy to the classical case, it should have the following properties for any unit-trace density matrix $ \rho $ and any projector $ E $:
\paragraph{Property Q} $\text{tr }(E\otimes E)\text{ Copy }\rho =\text{tr }(E\otimes I_{\mathsf{H}})\text{ Copy }\rho=\text{tr }(I_{\mathsf{H}}\otimes E )\text{ Copy }\rho=\text{tr } E\rho$
\medskip\\
Note this implies $\text{tr }(E\otimes (I_{\mathsf{H}}-E))\text{ Copy }\rho=\text{tr }((I_{\mathsf{H}}-E)\otimes E)\text{ Copy }\rho=0$. Then we have the following:
\paragraph{Theorem 3.4.2} Any map \textit{L} from unit-trace density matrices on \textsf{H} to unit-trace density matrices on $ \mathsf{H}\otimes\mathsf{H}$ obeying property \textbf{Q} is linear for convex linear combinations.
\paragraph*{Comment} Since any density matrix can be scaled to have trace one, this implies the map can be extended to all density matrices with the extended map being positively linear. By the generating property of density matrices among signed density matrices as a result of the spectral theorem for compact operators, this implies the map can further be extended to a linear map.
\paragraph*{Proof} Let \textit{L} be such a map and $ \rho $ an unit-trace density matrix on \textsf{H}. For any commuting projectors \textit{E}, \textit{F}, 
\begin{align}
\text{tr }(E\otimes F)(L\rho)=&\text{tr }(EF\otimes EF)(L\rho)\\&+\text{tr }(EF\otimes (F-EF))(L\rho)+\text{tr }((E-EF)\otimes F)(L\rho)\nonumber
\end{align}
However, by positivity and the implication of \textbf{Q},
\begin{equation}
0\leq\text{tr }(EF\otimes (F-EF))(L\rho)\leq \text{tr }(EF\otimes (I_{\mathsf{H}}-EF))(L\rho)=0
\end{equation}
and
\begin{equation}
0\leq\text{tr }((E-EF)\otimes F)(L\rho)\leq \text{tr }((I_{\mathsf{H}}-F)\otimes F)(L\rho)=0
\end{equation}
so, using \textbf{Q},
\begin{equation}
\text{tr }(E\otimes F)(L\rho)=\text{tr }(EF\otimes EF)(L\rho)=\text{tr }EF\rho=\text{tr }(F\otimes E)(L\rho)
\end{equation}

Let $ \tau $ be another unit-trace density matrix on \textsf{H}. Then for any $ t\in[0,1] $, $ (1-t)\rho+t\tau $ will be a unit-trace density matrix on \textsf{H}. Consider the signed density matrix on $  \mathsf{H}\otimes\mathsf{H}  $ given by
\begin{equation}
\nu_{t}=L((1-t)\rho+t\tau)-(1-t)L\rho-tL\tau
\end{equation}
Take any projector \textit{E}. Then, by property \textbf{Q}, $ \text{tr }(E\otimes E)\nu_{t}=0 $. However, by the above symmetry property of \textit{L}, for any commuting projectors \textit{E}, \textit{F},
\begin{equation}
\text{tr }(E\otimes F)\nu_{t}=\frac{1}{2}\left(\text{tr }(E\otimes F)\nu_{t}+\text{tr }(F\otimes E)\nu_{t} \right)
\end{equation}
which is equal to
\begin{align}
\frac{1}{2}&\left(\text{tr }((E+F-EF)\otimes (E+F-EF))\nu_{t}-\text{tr }((E-EF)\otimes (E-EF))\nu_{t}\right.\\&\left.-\text{tr }((F-EF)\otimes (F-EF))\nu_{t}+\text{tr }(EF\otimes EF)\nu_{t}\right)\nonumber
\end{align}
which is zero by the preceding property of $ \nu_{t} $. Since $ E,F $ were arbitrary, $ \nu_{t} $ must be the zero operator. $ \square $
\medskip\\
We then have the following theorem, based on the argument of Wooters and Zurek~\cite{wootersnocopy} that density matrices of rank greater than one can be expressed in more than one way (infinitely many ways actually) to create a contradiction.
\paragraph{Theorem 3.4.3} There is no quantum Copy map for non-trivial\footnote{A \textit{trivial} Hilbert space has dimension one.} \textsf{H}.
\paragraph*{Proof} Take any orthonormal $ \lbrace \mathbf{u},\mathbf{v}\rbrace\subset\mathsf{H} $ with corresponding adjoint operators $ \mathbf{u}^{*}=\langle\cdot,\mathbf{u}\rangle_{\mathsf{H}} $ and $ \mathbf{v}^{*}=\langle\cdot,\mathbf{v}\rangle_{\mathsf{H}} $. Consider $ \rho=\frac{1}{2}\left( \mathbf{u}\otimes\mathbf{u}^{*}+\mathbf{v}\otimes\mathbf{v}^{*}\right) $. By linearity,
\begin{equation}
\text{Copy }\rho=\frac{1}{2}\left( \text{ Copy }(\mathbf{u}\otimes\mathbf{u}^{*})+\text{ Copy }(\mathbf{v}\otimes\mathbf{v}^{*})\right)
\end{equation} 
By property \textbf{Q}, for rank one density matrices Copy must be the same as Clone, so Copy $ \rho $ is uniquely given as
\begin{equation}\label{eq:quantumcopyfirst}
\frac{1}{2}\left( \mathbf{u}\otimes\mathbf{u}\otimes\mathbf{u}^{*}\otimes\mathbf{u}^{*}+\mathbf{v}\otimes\mathbf{v}\otimes\mathbf{v}^{*}\otimes\mathbf{v}^{*}\right)
\end{equation}
However, it is also possible to write $ \rho $ as
\begin{equation}
\frac{1}{4}\left( (\mathbf{u}+\mathbf{v})\otimes(\mathbf{u}+\mathbf{v})^{*}+(\mathbf{u}-\mathbf{v})\otimes(\mathbf{u}-\mathbf{v})^{*}\right)
\end{equation}
Then Copy $ \rho $ is uniquely given as
\begin{align}
\frac{1}{8}&\left( (\mathbf{u}+\mathbf{v})\otimes(\mathbf{u}+\mathbf{v})\otimes(\mathbf{u}+\mathbf{v})^{*}\otimes(\mathbf{u}+\mathbf{v})^{*}\right. \\ &\left.+(\mathbf{u}-\mathbf{v})\otimes(\mathbf{u}-\mathbf{v})\otimes(\mathbf{u}-\mathbf{v})^{*}\otimes(\mathbf{u}-\mathbf{v})^{*}\right)\nonumber
\end{align}
\begin{align}\label{eq:quantumcopysecond}
=\frac{1}{4}&\left(\mathbf{u}\otimes\mathbf{u}\otimes\mathbf{u}^{*}\otimes\mathbf{u}^{*}+\mathbf{u}\otimes\mathbf{u}\otimes\mathbf{v}^{*}\otimes\mathbf{v}^{*}+\mathbf{u}\otimes\mathbf{v}\otimes\mathbf{u}^{*}\otimes\mathbf{v}^{*}\right. \nonumber\\&+\mathbf{u}\otimes\mathbf{v}\otimes\mathbf{v}^{*}\otimes\mathbf{u}^{*}+\mathbf{v}\otimes\mathbf{u}\otimes\mathbf{u}^{*}\otimes\mathbf{v}^{*}+\mathbf{v}\otimes\mathbf{u}\otimes\mathbf{v}^{*}\otimes\mathbf{u}^{*}+\nonumber\\&\left. +\mathbf{v}\otimes\mathbf{v}\otimes\mathbf{u}^{*}\otimes\mathbf{u}^{*}+\mathbf{v}\otimes\mathbf{v}\otimes\mathbf{v}^{*}\otimes\mathbf{v}^{*}\right) 
\end{align}
Clearly, (\ref{eq:quantumcopyfirst}) and (\ref{eq:quantumcopysecond}) are unequal, which is a contradiction. $ \square $
\section{Embedding quantum models into classical ones}\label{sec:embeddingquantum}
\subsection*{A construction for option I' using atomic measures}
For option \textbf{I'}, there is a way to embed quantum behavior into a purely classical, but contextual, model. For any Hilbert space \textsf{H}, let $ \mathbb{S}_{\mathsf{H}} $ be the closed unit ball within \textsf{H}. Let $ \mathbb{S}_{\mathsf{H}}/\sim $ be the quotient set formed from $ \mathbb{S}_{\mathsf{H}} $ by the equivalence relation $ \psi\sim\xi $ if there is a phase\footnote{Elements of $ \mathbb{C} $ with magnitude one.} $ w $ such that $ \psi= w\xi $. Clearly $ \mathbb{S}_{\mathsf{H}}/\sim $ is in one-to-one correspondence to rank-one projectors on \textsf{H}. Let $ \mathcal{E} $ be any $ \sigma $-algebra on $ \mathbb{S}_{\mathsf{H}}/\sim $ such that all points are atoms (such as the Borel $ \sigma $-algebra). For any set $ \mathcal{X} $ and base measure $ \mu $ denote the space of atomic, finite-norm, $ L^{1}(\mathcal{X};\mu) $-valued vector $ \mathcal{E} $-measures on $ \mathbb{S}_{\mathsf{H}}/\sim $ by $ \mathcal{A}\left(\mathbb{S}_{\mathsf{H}}/\sim;\mathcal{E};L^{1}(\mathcal{X};\mu)\right) $. This is a Banach space by \textbf{B6.1}. Given $ \tau\in \mathcal{A}\left(\mathbb{S}_{\mathsf{H}}/\sim;\mathcal{E};L^{1}(\mathcal{X};\mu)\right) $, for any $ \mu $-measurable subset $ B\subset\mathcal{X} $, define $ \tau_{B} $ to be the atomic, signed $ \mathcal{E} $-measure on $ \mathbb{S}_{\mathsf{H}}/\sim $ defined by $ \tau_{B}(A)=\int_{B} \tau(A)\,d\mu $ for any $ A\in \mathcal{E} $. Let $ \sim' $ be the equivalence relation on $ \mathcal{A}\left(\mathbb{S}_{\mathsf{H}}/\sim;\mathcal{E};L^{1}(\mathcal{X};\mu)\right) $ given by $ \tau\sim'\chi $ if, for any $ \mu $-measurable subset $ B\subset\mathcal{X} $,
\begin{equation}
\int_{s\in\mathbb{S}_{\mathsf{H}}/\sim}ss^{*}\,d\tau_{B}(s)=\int_{s\in\mathbb{S}_{\mathsf{H}}/\sim}ss^{*}\,d\chi_{B}(s)
\end{equation} 
Since the equivalence class using $ \sim' $ of zero is a closed, linear subspace, the quotient space $ \mathcal{A}\left(\mathbb{S}_{\mathsf{H}}/\sim;\mathcal{E};L^{1}(\mathcal{X};\mu)\right)/\sim' $ is a Banach space using the standard norm for quotient spaces, $ \Vert [\tau]\Vert=\inf_{\chi\in[\tau]}\Vert\chi\Vert $. Define the positive cone on the quotient space to be those equivalence classes with a positive member, using the obvious notion of positivity on $ \mathcal{A}\left(\mathbb{S}_{\mathsf{H}}/\sim;\mathcal{E};L^{1}(\mathcal{X};\mu)\right) $.  

Then we have the following theorem:
\paragraph{Theorem 3.5.1} There is a positive, linear, isometric isomorphism,
\[L^{1}\left(\mathcal{X};\mu;\mathcal{D}(\mathsf{H})\right) \cong\mathcal{A}\left(\mathbb{S}_{\mathsf{H}}/\sim;\mathcal{E};L^{1}(\mathcal{X};\mu)\right)/\sim'\]
\paragraph*{Proof} If the measure $ \mu $ is trivial, then $ L^{1}\left(\mathcal{X};\mu;\mathcal{D}(\mathsf{H})\right)\cong \mathcal{D}(\mathsf{H})$. Define the map $ \Psi:L^{1}\left(\mathcal{X};\mu;\mathcal{D}(\mathsf{H})\right)\to \mathcal{A}\left(\mathbb{S}_{\mathsf{H}}/\sim;\mathcal{E};L^{1}(\mathcal{X};\mu)\right)/\sim' $ by $ \Psi(\rho) $ being the equivalence class of $ \sum_{j}a_{j}\delta_{[\psi_{j}]} $ for $ \rho=\sum_{j}a_{j}\psi_{j}\psi_{j}^{*} $ with countable collections $ \lbrace\psi_{j}\rbrace\in\mathbb{S}_{\mathsf{H}} $ and $ \lbrace a_{j}\rbrace\subset\mathbb{R} $, which is always possible by the spectral theorem for compact operators.

For more general measures on $ \mathcal{X} $, first start with the observation that, given any $ \rho\in L^{1}\left(\mathcal{X};\mu;\mathcal{D}(\mathsf{H})\right) $, each $ \rho(x) $ lives in the same separable subspace of \textsf{H} for almost every $ x $ with respect to $ \mu $, namely the subspace \textsf{G} that the operator $ \int_{x\in\mathcal{X}}\vert\rho(x)\vert\,d\mu\in\mathcal{D}(\mathsf{H}) $ lives in. Let $ \lbrace\mathbf{e}_{j}\rbrace $ be an orthonormal basis for \textsf{G}. For each $ j\in\lbrace 1,2,\ldots\rbrace $, let $ P_{j} $ be the orthogonal projector onto the span of $ \lbrace \mathbf{e}_{1},\mathbf{e}_{2},\ldots,\mathbf{e}_{j}\rbrace $. Since simple functions (more properly, the ``functions" which are equivalence classes containing simple functions) are norm-dense in $ L^{1}\left(\mathcal{X};\mu;\mathcal{D}(\mathsf{H})\right) $, the sequence $ \langle P_{j}\rho P_{j}\rangle $ is a Cauchy sequence by \textbf{C4.3}; hence, it converges in norm by the completeness of the Banach space $ L^{1}\left(\mathcal{X};\mu;\mathcal{D}(\mathsf{H})\right) $. It is readily seen that the limit point is $ \rho $. Now define the map $ \Psi:L^{1}\left(\mathcal{X};\mu;\mathcal{D}(\mathsf{H})\right)\to \mathcal{A}\left(\mathbb{S}_{\mathsf{H}}/\sim;\mathcal{E};L^{1}(\mathcal{X};\mu)\right)/\sim' $ by first defining $ \Psi(\psi\psi^{*}f) $, for $ \psi\in\mathsf{H} $ and $ f\in L^{1}(\mathcal{X};\mu) $ to be the equivalence class of the vector measure $ f\otimes\delta_{[\psi]} $. Since the linear space $ \mathcal{D}(P_{m}\mathsf{H}) $ is spanned by the $ m^{2} $ operators
\begin{equation}
\lbrace \mathbf{e}_{j}\mathbf{e}_{j}^{*}\rbrace_{j\in\lbrace 1,\ldots,m\rbrace}\cup\lbrace (\mathbf{e}_{j}+\mathbf{e}_{k})(\mathbf{e}_{j}+\mathbf{e}_{k})^{*},(\mathbf{e}_{j}+\imath\mathbf{e}_{k})(\mathbf{e}_{j}+\imath\mathbf{e}_{k})^{*} \rbrace_{j,k\in\lbrace 1,\ldots,m\rbrace, j<k}
\end{equation}  
the map $ \Psi $ can be extended by linearity to all $ \rho $ with the property that $ \rho(x) $ lives on the same finite-dimensional subspace of \textsf{H} for almost every \textit{x} with respect to $ \mu $.

Now take any $ \rho $ with this property which is also a simple function. Then, applying the (finite-dimensional) spectral theorem to each of the finitely many values $ \rho $ takes, $ \Psi $ is readily seen to be a positive isometry. Since simple functions are norm-dense, by \textbf{A1.2}, $ \Psi $ is a positive isometry on all $ \rho $ with the property that $ \rho(x) $ lives on the same finite-dimensional subspace of \textsf{H} for almost every \textit{x} with respect to $ \mu $. However, by an above argument, such $ \rho $ are norm-dense in $ L^{1}\left(\mathcal{X};\mu;\mathcal{D}(\mathsf{H})\right) $. Hence, using \textbf{A1.3} to extend $ \Psi $ to all of $ L^{1}\left(\mathcal{X};\mu;\mathcal{D}(\mathsf{H})\right) $, $ \Psi $ is a positive isometry. It remains to show it is surjective, but that is easily seen, with 
\begin{equation}
\Psi^{-1}\left(\left[\sum_{j}f_{j}\otimes\delta_{[\psi_{j}]}\right]\right)=\sum_{j}f_{j}\psi_{j}\psi_{j}^{*}
\end{equation}
for any $ \sum_{j}f_{j}\otimes\delta_{[\psi_{j}]}\in \mathcal{A}\left(\mathbb{S}_{\mathsf{H}}/\sim;\mathcal{E};L^{1}(\mathcal{X};\mu)\right) $. $ \square $
\paragraph{Comment} Since $ \mathbb{S}_{\mathsf{H}}\times \mathbb{S}_{\mathsf{J}}\ncong \mathbb{S}_{\mathsf{H}\otimes\mathsf{J} }$ if neither \textsf{H} nor \textsf{J} are trivial, this construction is necessarily contextual. Therefore, it does not violate Bells' inequality (see \S\ref{sec:bells} below).
\subsection*{The nonexistence of the corresponding construction using a base measure}
The corresponding construction using a base measure $ \nu $ would be for there to be a positive isometry from $ L^{1}(\mathcal{X};\mu;\mathcal{D}(\mathsf{H})) $ to a quotient space of some $ L^{1}(\mathcal{Y};\nu) $. This is impossible, as the following theorem shows:
\paragraph{Theorem 3.5.2}If the Hilbert space \textsf{H} is non-trivial, there are no: \textbf{(i)} set $ \mathcal{Y} $; \textbf{(ii)} $ \sigma $-finite measure $ \nu $; and \textbf{(iii)} equivalence relation $ \sim $ induced by a closed, linear subspace $\mathsf{B}\subset L^{1}(\mathcal{Y};\nu) $--such that there is a positive, linear isomorphism, $ \Psi:L^{1}\left(\mathcal{X};\mu;\mathcal{D}(\mathsf{H})\right)\to L^{1}(\mathcal{Y};\nu)/\sim$, which is also an isometry on the positive cone.
\paragraph*{Proof} Suppose otherwise. Then there is a $ L^{1}\left(\mathcal{X};\mu;\mathcal{D}(\mathsf{H})\right) $-valued vector measure $ \tau $ on $ \mathcal{Y} $ provided by $ \tau(A)=\Psi^{-1}\left([1_{A}]\right) $ for any $ \nu $-measurable subset $ A\subset \mathcal{Y} $. The spaces $ L^{1}(\mathcal{Y};\nu)^{*}\cong L^{\infty}(\mathcal{Y};\nu)$ by Riesz's theorem~\cite{roydenrieszlp}. The dual to $ L^{1}(\mathcal{Y};\nu)/\sim $ is provided by the \textit{annihilator} $ \mathsf{B}^{\perp} $: the closed, linear subspace of $ L^{\infty}(\mathcal{Y};\nu) $ that annihilates \textsf{B}. (Note, in particular that since $ \Psi $ is an isometry on the positive cone, the constant ``function" $ [1_{\mathcal{Y}}]\in \mathsf{B}^{\perp}$.) Therefore, there is an adjoint map $ \Psi^{*}:\mathsf{B}^{\perp}\to L^{1}\left(\mathcal{X};\mu;\mathcal{D}(\mathsf{H})\right)^{*} $ given by
\begin{equation}
\int_{\mathcal{Y}}f\,\Psi\rho\,d\nu=(\Psi^{*}f)\,\rho
\end{equation}
for any $ f\in \mathsf{B}^{\perp} $ and $ \rho\in L^{1}\left(\mathcal{X};\mu;\mathcal{D}(\mathsf{H})\right) $. By the basic properties of Banach spaces, $ (\Psi^{*})^{-1}=(\Psi^{-1})^{*} $, $ \Psi^{*} $ is positive, and $ \Psi^{*} $ is an isometry on the positive cone. 

Then,
\begin{equation}
\int_{A}\Psi^{*-1}\Phi\,d\nu=\Phi(\tau(A))
\end{equation} 
for any linear functional $ \Phi\in L^{1}\left(\mathcal{X};\mu;\mathcal{D}(\mathsf{H})\right)^{*} $ and $ \nu $-measurable subset $ A\subset \mathcal{Y} $. For $ I_{\mathsf{H}} $ the identity operator, $ \Psi^{*-1}\left(I_{\mathsf{H}}1_{\mathcal{X}}\right) $ is the element of $ \mathsf{B}^{\perp} $ that agrees with the norm when integrated with any positive ``function" in $ L^{1}(\mathcal{Y};\nu)/\sim $; hence, it must be $ 1_{\mathcal{Y}} $. Therefore,
\begin{equation}
\int_{\mathcal{X}}\text{tr }\tau(A)\,d\mu=\int_{A}\Psi^{*-1}\left(I_{\mathsf{H}}1_{\mathcal{X}}\right)\,d\nu=\nu(A)
\end{equation}

However, this gives rise to a contradiction. Fix some subset $ B\subset\mathcal{X} $ with $ 0<\mu(B)<\infty $. Take any unit norm $ \psi\in\mathsf{H} $. Since $ \Psi $ is positive, by definition, the equivalence class $ \Psi(\psi\psi^{*}1_{B}) $ has a positive member, call it $ g_{\psi} $. There must be some $ \psi\neq\xi $ such that $ A=\lbrace g_{\psi}>0\rbrace\cap\lbrace g_{\xi}>0\rbrace $ has strictly positive $ \nu $-measure; otherwise, there would be an uncountable collection $ \lbrace g_{\psi}>0\rbrace $ indexed by unit norm $ \psi\in\mathsf{H} $ of subsets of $ \mathcal{Y} $, each with strictly positive $ \nu $ measure, but whose pairwise intersections all have $ \nu $-measure zero. The existence of such a collection would contradict $ \nu $ being $ \sigma $-finite by \textbf{B1.6}. Since $ \Psi $ is an isometry on the positive cone, 
\begin{equation}
\int_{\mathcal{Y}}g_{\psi}\,d\nu=\mu(B)\,\text{tr }I\psi\psi^{*}=\mu(B)=\mu(B)\,\text{tr }\psi\psi^{*}\psi\psi^{*}=\int_{y\in\mathcal{Y}} g_{\psi}(y)\,d\langle \tau\psi,\psi\rangle(y)
\end{equation} 
Hence, $ \langle \tau\psi,\psi\rangle\leq \text{tr }\tau=\nu $ must be equal to $ \nu $ when restricted to $ \lbrace g_{\psi}>0\rbrace\supset A $. By a similar argument, $ \langle \tau\xi,\xi\rangle $ must must be equal to $ \nu $ when restricted to $ \lbrace g_{\xi}>0\rbrace\supset A $. These conditions are impossible to satisfy. $ \square $
\subsection*{The special case of two-dimensional Hilbert spaces}
It is possible to circumvent the conclusion of the preceding theorem if the positive isomorphism is with a closed, linear subspace of $ L^{1}(\mathcal{Y};\nu)/\sim $. Specialize to $ \mathcal{Y}=\mathcal{X}\times\mathcal{Z} $ and $ \nu=\mu\times\eta $ and let \textsf{C} be the closed, linear subspace of $ L^{1}(\mathcal{X}\times\mathcal{Z};\mu\times\eta)/\sim $. Using the notation of the preceding proof, let $ \sim' $ be the equivalence relation on the annihilator $ \mathsf{B}^{\perp}\subset  L^{\infty}(\mathcal{X}\times\mathcal{Z};\mu\times\eta)$ induced by the annihilator $ \mathsf{C}^{\perp}\subset  L^{\infty}(\mathcal{X}\times\mathcal{Z};\mu\times\eta) $, so $ f\sim'g $ if $ \int_{\mathcal{X}\times\mathcal{Z}}[h]f\,d(\mu\times\eta)=\int_{\mathcal{X}\times\mathcal{Z}}[h]g\,d(\mu\times\eta) $ for all $ [h]\in L^{1}(\mathcal{X}\times\mathcal{Z};\mu\times\eta)/\sim  $. From the proof of the preceding theorem, it is then necessary that $ \Psi^{*-1}(I_{\mathsf{H}}1_{\mathcal{X}})=[1_{\mathcal{X}\times\mathcal{Z}}] $. A sufficient way to accomplish this would be for there to be a positive map (not necessarily linear in $ \psi\psi^{*} $) $ \Lambda: \mathbb{S}_{\mathsf{H}}/\sim_{\text{phase}}\to L^{\infty}(\mathcal{Z};\eta) $, where $ \sim_{\text{phase}} $  is the equivalence relation on $ \mathbb{S}_{\mathsf{H}} $ given above, such that 
\begin{equation}
\Psi^{*-1}(\psi\psi^{*}1_{\mathcal{X}})=\left[1_{\mathcal{X}}\otimes \Lambda(\psi)\right]\text{ for all }\psi\in \mathbb{S}_{\mathsf{H}}/\sim_{\text{phase}}
\end{equation}
and, for almost every $ z\in \mathcal{Z} $ with respect to $ \eta $, the map $ \psi\to \Lambda(\psi)(z) $ is a frame function with weight one\footnote{A function $ f:\mathbb{S}_{\mathsf{H}}/\sim_{\text{phase}}\to\mathbb{R} $ is a \textit{frame function with weight w} if it is zero except for a separable subspace of \textsf{H} and $ \sum_{j}f(\mathbf{e}_{j})=w $ for any orthonormal basis $ \lbrace\mathbf{e}_{j}\rbrace $ of that subspace.}.

For \textsf{H} of dimension two, we then have the following based on a construction by Kochen and Specker~\cite{kochenconstruction}. Let $ \omega $ be the usual measure on the sphere $ \mathbb{S}^{2}\subset\mathbb{R}^{3} $, so, with Cartesian coordinates $ (x_{1},x_{2},x_{3}) $ for $ \mathbb{R}^{3} $,
\begin{equation}
\omega(A)=\int_{(x_{1},x_{2},x_{3})\in A}=\left(x_{1}\,dx_{2}\wedge dx_{3}-x_{2}\,dx_{1}\wedge dx_{3}+x_{3}\,dx_{1}\wedge dx_{2}\right)
\end{equation}
for any Borel subset $ A\subset \mathbb{S}^{2} $. Let $ \sim $ be the equivalence relation on $ L^{1}(\mathcal{X}\times\mathbb{S}^{2};\mu\times\omega) $ given by $ f\sim g $ if $ \int_{B\times H}f\,d(\mu\times\omega)=\int_{B\times H}g\,d(\mu\times\omega) $ for all $ \mu $-measurable subsets $ B\subset \mathcal{X} $ and all hemispheres $ H\subset \mathbb{S}^{2} $ (whether the hemispheres are taken open or closed is irrelevant). Equip the quotient space $ L^{1}(\mathcal{X}\times\mathbb{S}^{2};\mu\times\omega)/\sim $ with a norm in the usual way via $ \Vert[f]\Vert=\inf_{g\in[f]}\Vert g\Vert $. Define the positive cone on $ L^{1}(\mathcal{X}\times\mathbb{S}^{2};\mu\times\omega)/\sim$ by those equivalence classes that contain a positive element of $ L^{1}(\mathcal{X}\times\mathbb{S}^{2};\mu\times\omega) $.
\paragraph{Theorem  3.5.3} There is a positive, linear isometry, $ \Psi:L^{1}\left(\mathcal{X};\mu;\mathcal{D}(\mathsf{H})\right)\to L^{1}(\mathcal{X}\times\mathbb{S}^{2};\mu\times\omega)/\sim$ with an associated map $ \Lambda:\mathbb{S}_{\mathsf{H}}/\sim_{\text{phase}}\to L^{\infty}(\mathbb{S}^{2};\omega) $ with the above properties.
\paragraph{Proof} Take any orthonormal basis $ \lbrace \mathbf{e}_{1},\mathbf{e}_{2}\rbrace $ for $ \mathsf{H}\cong\mathbb{C}^{2} $. Let $ \Psi:L^{1}\left(\mathcal{X};\mu;\mathcal{D}(\mathsf{H})\right)\to L^{1}(\mathcal{X}\times\mathbb{S}^{2};\mu\times\omega)/\sim $ be given by first taking $\Psi $ on elements of the form $ \psi\psi^{*}f $ for $ \psi\in \mathbb{S}_{\mathsf{H}} $ and $ f\in L^{1}(\mathcal{X};\mu) $ to be the equivalence class of the positive ``function"
\begin{align}
&f\cdot\begin{cases}\dfrac{1}{\pi}\mathbf{y}\mathbf{s}&\text{if }\mathbf{y}\mathbf{s}>0\\0&\text{otherwise}\end{cases}\\
\text{for }\mathbf{y}&=\left[\begin{array}{ccc}\vert \psi_{1}\vert^{2}-\vert \psi_{2}\vert^{2}&\Re \,2\psi_{1}\overline{\psi_{2}}&\Im\, 2\psi_{1}\overline{\psi_{2}}\end{array} \right]\text{ and }\mathbf{s}=\left[\begin{array}{c}s_{1}\\s_{2}\\s_{3}\end{array} \right]\mathbb{S}^{2}\in \nonumber
\end{align}
where $ \psi_{j}=\langle\psi,\mathbf{e}_{j}\rangle $. Since \textsf{H} is two-dimensional, $ \mathcal{D}(\mathsf{H}) $ is four-dimensional, with basis $ \lbrace \mathbf{e}_{1}\mathbf{e}_{1}^{*},\mathbf{e}_{2}\mathbf{e}_{2}^{*},(\mathbf{e}_{1}+\mathbf{e}_{2})(\mathbf{e}_{1}+\mathbf{e}_{2})^{*},(\mathbf{e}_{1}+\imath\mathbf{e}_{2})(\mathbf{e}_{1}+\imath\mathbf{e}_{2})^{*}\rbrace $. Therefore, $ \Psi $ can be extended by linearity to all of $ L^{1}\left(\mathcal{X};\mu;\mathcal{D}(\mathsf{H})\right) $. 

For $ \rho $ a simple function, by using the spectral theorem for the each of the finite number of values $ \rho $ attains, the map $ \Psi $ is readily seen to be positive and an isometry. Since simple functions (more properly, the ``functions" containing simple functions) are dense in $ L^{1}\left(\mathcal{X};\mu;\mathcal{D}(\mathsf{H})\right) $, by \textbf{A1.2}, $ \Psi $ is a positive isometry for all of $ L^{1}\left(\mathcal{X};\mu;\mathcal{D}(\mathsf{H})\right). $ 

For this $ \Psi $, an associated $ \Lambda $ does exist. It is given by first defining  \begin{equation}\label{eq:wavefunctiontosphere}
\mathbf{z}^{\text{T}}=\left[\begin{array}{ccc}\vert \xi_{1}\vert^{2}-\vert \xi_{2}\vert^{2}&\Re 2\xi_{1}\overline{\xi_{2}}&\Im 2\xi_{1}\overline{\xi_{2}}\end{array} \right]\Leftrightarrow \xi\propto\sqrt{\frac{1+z_{1}}{2}}\mathbf{e}_{1}+\frac{z_{2}+\imath z_{3}}{\sqrt{2(1+z_{1})}}\mathbf{e}_{2}
\end{equation}
where the proportionality for $ \xi $ is up to an irrelevant phase. Then $ \Lambda(\xi)=1_{H_{z}} $, where $ H_{z} $ is the hemisphere centered at $ z $. This has the required properties since \textbf{(i)} for any orthonormal $ \lbrace\xi,\zeta\rbrace $, $ \Lambda(\xi)+\Lambda(\zeta)=1_{H_{z}} +1_{H_{-z}}=1_{\mathbb{S}^{2}}  $, with equality in the $ L^{\infty}(\mathbb{S}^{2};\omega) $-sense of almost everywhere with respect to $ \omega $ and \textbf{(ii)} for any $ \xi,\psi\in \mathbb{S}_{\mathsf{H}} $ and $ f\in L^{1}(\mathcal{X};\mu) $,
\begin{align}
\int_{\mathcal{X}\times\mathbb{S}^{2}}\left(1_{\mathcal{X}}\otimes\Lambda(\xi)\right)\Psi(\psi\psi^{*}f)\,d(\mu\times\omega)=\frac{1}{2}(1+\mathbf{y}\mathbf{z})\left(\int_{\mathcal{X}}f\,d\mu\right)=\vert\langle\xi,\psi\rangle\vert^{2}\left(\int_{\mathcal{X}}f\,d\mu\right)\hspace{.4 in}\square\nonumber
\end{align}
\medskip\\
However, for Hilbert space \textsf{H} of dimension greater than two, there is no construction of this form because Gleason proved~\cite{gleason} that all frame functions on Hilbert spaces of dimension greater than two are regular\footnote{A frame function is \textit{regular} if it is given by $ \psi\to \langle T\psi,\psi\rangle $ for some trace-class, self-adjoint operator \textit{T}.}. 
\paragraph{Theorem 3.5.4} There are no: \textbf{(i)} set $ \mathcal{Z} $; \textbf{(ii)} $ \sigma $-finite measure $ \eta $; and \textbf{(iii)} equivalence relation $ \sim $ induced by a closed, linear subspace $\mathsf{B}\subset L^{1}(\mathcal{X}\times\mathcal{Z};\mu\times\eta) $--such that there is a positive, linear isometry, $ \Psi:L^{1}\left(\mathcal{X};\mu;\mathcal{D}(\mathsf{H})\right)\to L^{1}(\mathcal{X}\times\mathcal{Z};\mu\times\eta)/\sim$, which has an associated map $ \Lambda: \mathbb{S}_{\mathsf{H}}/\sim_{\text{phase}}\to L^{\infty}(\mathcal{Z};\eta) $ with the properties given above.
\paragraph*{Proof} Suppose otherwise. By Gleason's result, there is a $ T\in L^{1}(\mathcal{Z};\eta;\mathcal{D}(\mathsf{H})^{+})$ such that $ T(z) $ has trace one for almost every $ z\in\mathcal{Z} $ with respect to $ \eta $ and 
\begin{equation}
\int_{\mathcal{X}\times\mathcal{Z}}\left(1_{\mathcal{X}}\otimes\langle T\xi,\xi\rangle\right)\Psi(\psi\psi^{*}1_{B})\,d(\mu\times\eta)=\vert\langle\xi,\psi\rangle\vert^{2}\mu(B)
\end{equation}
for all $ \psi,\xi\in \mathbb{S}_{\mathsf{H}}/\sim_{\text{phase}} $ and $  \mu$-measurable $ B\subset\mathcal{X} $. However, then following the argument of theorem \textbf{3.5.2}, there is a contradiction. Fix some subset $ B\subset\mathcal{X} $ with $ 0<\mu(B)<\infty $. Take any unit norm $ \psi\in\mathsf{H} $. Since $ \Psi $ is positive, by definition, the equivalence class $ \Psi(\psi\psi^{*}[1_{B}]) $ has a positive member, call it $ g_{\psi} $. There must be some $ \psi\neq\xi $ such that $ A=\lbrace g_{\psi}>0\rbrace\cap\lbrace g_{\xi}>0\rbrace $ has strictly positive $ \nu $-measure (Since $ g_{\psi},g_{\xi} $ are ``functions", \textit{A} is actually an equivalence class of sets that differ by sets of $ \nu $-measure zero). Since $ \Psi $ is an isometry on the positive cone, 
\begin{align}
\int_{\mathcal{X}\times\mathcal{Z}}g_{\psi}\,d(\mu\times\eta)&=\mu(B)\,\text{tr }I\psi\psi^{*}=\mu(B)=\mu(B)\,\text{tr }\psi\psi^{*}\psi\psi^{*}\\&=\int_{\mathcal{X}\times\mathcal{Z}}\left(1_{\mathcal{X}}\otimes\langle T\psi,\psi\rangle\right) g_{\psi}\,d(\mu\times\eta)\nonumber
\end{align} 
Hence, $ \langle T\psi,\psi\rangle\leq \text{tr }T=1 $ must be equal to one almost everywhere with respect to $ \eta $ when restricted to $ \lbrace g_{\psi}>0\rbrace\supset A $ (once again, this is actually an equivalence class of sets that differ by sets of $ \nu $-measure zero). By a similar argument, $ \langle T\xi,\xi\rangle\leq \text{tr }T=1 $ must be equal to one almost everywhere with respect to $ \eta $ when restricted to $ \lbrace g_{\xi}>0\rbrace\supset A $ (also actually an equivalence class of sets that differ by sets of $ \nu $-measure zero). These conditions are impossible to satisfy. $ \square $
\subsection*{A construction for option II}
For option \textbf{II}, there is also a way to embed quantum behavior into a purely classical, but contextual, model. For any Hilbert space \textsf{H}, let $ \mathbb{B}_{\mathsf{H}} $ be the closed unit ball within \textsf{H}. Equip $ \mathbb{B}_{\mathsf{H}} $ with the weak topology; denote the resulting space by $  \mathbb{B}_{\mathsf{H}}^{\text{weak}}  $. Since Hilbert spaces are reflexive, with $ \mathsf{H}^{*}\cong\mathsf{H} $ by Riesz's theorem \cite{roydenrieszhilbert}, then this is the same as the weak* topology, so by Alaoglu's theorem~\cite{roydenalaoglu}, $ \mathbb{B}_{\mathsf{H}}^{\text{weak}} $ is a compact space. Furthermore, it is Hausdorff since, given any distinct points $ b_{0},b_{1} $, there are separating weak neighborhoods  $ \mathcal{N}\left(b_{0};b_{1}-b_{0};\frac{1}{2}\Vert b_{0}-b_{1}\Vert^{2}\right) $ and $ \mathcal{N}\left(b_{1};b_{1}-b_{0};\frac{1}{2}\Vert b_{0}-b_{1}\Vert^{2}\right) $.

Then, for any compact, Hausdorff space $ \mathcal{X} $, we have the following theorem:
\paragraph{Theorem 3.5.5} There is a positive, isometric isomorphism between $ \mathcal{C}\left(\mathcal{X};\mathcal{K}(\mathsf{H})\right) $ and a closed, linear subspace of $ \mathcal{C}\left(\mathcal{X}\times \mathbb{B}_{\mathsf{H}}^{\text{weak}}\right) $.
\paragraph*{Proof} Consider the map $ \Psi:\mathcal{C}\left(\mathcal{X};\mathcal{K}(\mathsf{H})\right)\to \mathcal{C}\left(\mathcal{X}\times \mathbb{B}_{\mathsf{H}}^{\text{weak}}\right) $ given by
\begin{equation}
(\Psi\varphi)(x,b)=\langle \varphi(x)b,b\rangle
\end{equation} 
The map $ \Psi $ is clearly both positive and an isometry. From the latter property, its image is a closed, linear subspace. The functions in the image of the map are also clearly continuous in \textit{x}; the following lemma shows that they are continuous in \textit{b}, so, as claimed above, they are indeed continuous functions. $ \square $
\paragraph{Lemma 3.5.6} For any compact operator $ \phi\in \mathcal{K}(\mathsf{H}) $, the map $ \Phi:\mathbb{B}_{\mathsf{H}}^{\text{weak}}\to\mathbb{R} $ given by $ \Phi(b)=\langle \phi b,b\rangle $ is continuous.
\paragraph*{Proof} First take the case of $ \phi $ positive and rank one, so $ \phi=\psi\psi^{*} $ for some $ \psi\in\mathsf{H} $. Then, given any $ b_{0}\in \mathbb{B}_{\mathsf{H}} $ and $ \varepsilon>0 $, there is a weak neighborhood $ \mathcal{N}\left(b_{0};\psi;\sqrt{\vert\langle\psi,b_{0}\rangle\vert^{2}+\varepsilon}-\vert\langle\psi,b_{0}\rangle\vert\right) $ such that for every $ b $ in the neighborhood, using the triangle inequality repeatedly,
\begin{align}
\left\vert\Phi(b)-\Phi(b_{0})\right\vert&=\left\vert\vert\langle\psi,b\rangle\vert^{2}-\vert\langle\psi,b\rangle\vert^{2}\right\vert\leq \left(\left\vert\langle\psi,b\rangle\right\vert+\left\vert\langle\psi,b_{0}\rangle\right\vert\right)\left\vert\langle\psi,b\rangle-\langle\psi,b_{0}\rangle\right\vert  \\&< \left(\left\vert\langle\psi,b\rangle\right\vert+\left\vert\langle\psi,b_{0}\rangle\right\vert\right)\left( \sqrt{\vert\langle\psi,b_{0}\rangle\vert^{2}+\varepsilon}-\vert\langle\psi,b_{0}\rangle\vert\right) \nonumber\\&< \left(2\left\vert\langle\psi,b_{0}\rangle\right\vert+\sqrt{\vert\langle\psi,b_{0}\rangle\vert^{2}+\varepsilon}-\vert\langle\psi,b_{0}\rangle\vert\right)\left( \sqrt{\vert\langle\psi,b_{0}\rangle\vert^{2}+\varepsilon}-\vert\langle\psi,b_{0}\rangle\vert\right)\nonumber
\end{align}
which is equal to $ \varepsilon $.

Since finite rank operators are norm-dense in $ \mathcal{K}(\mathsf{H}) $, for any $ \phi\in  \mathcal{K}(\mathsf{H}) $ and $ \varepsilon>0 $ there is are finite collections $ \lbrace\psi_{j}\rbrace\subset  \mathbb{S}_{\mathsf{H}}$ (for $ \mathbb{S}_{\mathsf{H}} $ the unit sphere in \textsf{H}) and $ \lbrace a_{j}\rbrace\subset\mathbb{R} $ such that $ \left\Vert\phi-\sum_{j=1}^{n}a_{j}\psi_{j}\psi_{j}^{*} \right\Vert_{\text{op}}\leq \frac{1}{2}\varepsilon $. Then, by the triangle equality and the above result, for every $ b $ in the neighborhood
\begin{equation}
\bigcap_{j=1}^{n}\mathcal{N}\left(b_{0};\psi_{j};\sqrt{\vert a_{j}\vert\vert\langle\psi_{j},b_{0}\rangle\vert^{2}+\frac{1}{2n}\varepsilon}-\sqrt{\vert a_{j}\vert}\vert\langle\psi,b_{0}\rangle\vert\right) 
\end{equation}
we have $ \left\vert\Phi(b)-\Phi(b_{0})\right\vert<\varepsilon $. $ \square $
\paragraph{Comment} Since $ \mathbb{B}_{\mathsf{H}}\times \mathbb{B}_{\mathsf{J}}\ncong \mathbb{B}_{\mathsf{H}\otimes\mathsf{J} }$ if neither \textsf{H} nor \textsf{J} are trivial, this construction is necessarily contextual. Therefore, it does not violate Bells' inequality (see \S\ref{sec:bells} below). Also, for \textsf{H} finite-dimensional, it is possible to use $ \mathbb{S}_{\mathsf{H}} $ instead of $ \mathbb{B}_{\mathsf{H}} $ and the norm topology (which is equivalent to the weak topology in finite dimensions). In this case, it is also possible to further reduce the space by the equivalence relation on $ \mathbb{S}_{\mathsf{H}} $ used in the preceding construction for option \textbf{I}. 
\section{Embedding classical models into quantum ones}\label{sec:embedclassical}
For $ \sigma$-algebras generated by a countable collection of atoms, using option \textbf{I'} it is always possible to duplicate classical behavior using quantum nodes; one simply embeds $ L^{1}(\mathcal{X};\mu) $ as the diagonal operators in $ \mathcal{D}(L^{2}(\mathcal{X};\mu))^{+} $. Similarly, using option \textbf{II}, if $ \mathcal{X} $ is a finite set of points with the discrete topology, $ \mathcal{C}(\mathcal{X}) $ can be embedded as the diagonal operators in $ \mathcal{K}(L^{2}(\mathcal{X};$counting measure$))^{+} $.

By \textbf{5.28}, this is the only way to linearly embed $  L^{1}(\mathcal{X};\mu)^{+} $ in $ \mathcal{D}(L^{2}(\mathcal{X};\mu))^{+} $ if we want to take advantage of the one-to-one correspondence between functions $ 1_{B} $ in $ L^{\infty}(\mathcal{X};\mu)\cong L^{1}(\mathcal{X};\mu)^{*}$ and the orthogonal projectors $ P_{B}\in \mathcal{B}(L^{2}(\mathcal{X};\mu)) $ to the subspace of ``functions" zero almost everywhere outside \textit{B} provided by the the map diag* defined in the comment following \textbf{B5.19}. However, if the measure has an infinitely-divisible part, then unless $ f\in L^{1}(\mathcal{X};\mu)^{+} $ is zero on this part, by \textbf{B4.3} there is no corresponding diagonal operator in $\mathcal{D}(L^{2}(\mathcal{X};\mu))^{+}  $. One alternative is the map $ \text{diag}^{-1} $ into equivalence classes of trace-class operators with common diagonals defined in \textbf{B5.17} and used in \S\ref{sec:textbookrules} below.
\section{Classical physics}
\subsection*{Classical mechanics}
Another approach, which is that of classical mechanics, is to not completely duplicate quantum behavior, but instead approximate it. The key is to recognize that $ L^{1}(T^{*}\mathcal{X};\Omega)^{+} $ intersects the Hilbert space $ L^{2}(T^{*}\mathcal{X};\Omega) $. Similarly, $ \mathcal{D}(L^{2}(\mathcal{X};\mu))^{+} $ intersects the Hilbert space\footnote{$ \mathcal{S}_{2} $ denotes the Hilbert-Schmidt operators.} $ \mathcal{S}_{2}(L^{2}(\mathcal{X};\mu)) $, which is isomorphic to $ L^{2}(\mathcal{X}\times\mathcal{X};\mu\times\mu) $. Therefore, if we have an isomorphism $ \Psi $ between $ L^{2}(T^{*}\mathcal{X};\Omega) $  and $ L^{2}(\mathcal{X}\times\mathcal{X};\mu\times\mu) $, we can associate some of the elements of $ L^{1}(T^{*}\mathcal{X};\Omega)^{+} $ with those in $ \mathcal{D}(L^{2}(\mathcal{X};\mu))^{+} $.

For $ \mathcal{X}=\mathbb{R}^{n} $ and $ \mu $ mutually absolutely continuous with respect to Lebesgue measure $ \lambda $, this is indeed possible, with $ f(q,p)=(\Psi \rho)(q,p) $ given by
\begin{equation}
\frac{1}{(\pi\hslash)^{n}}\int_{u\in\mathbb{R}^{n}}\exp\left(-\frac{2\imath p\bullet u}{\hslash}\right)\rho(q+u,q-u)\sqrt{\frac{d\mu}{d\lambda}}(q+u)\sqrt{\frac{d\mu}{d\lambda}}(q+u)\,d\lambda(u)
\end{equation}  
and inverse $ \rho(x,x')=(\Psi^{-1} f)(x,x')=(\pi\hslash)^{n}(\Psi^{*} f)(x,x') $ given by
\begin{equation}
\sqrt{\frac{d\lambda}{d\mu}}(x)\sqrt{\frac{d\lambda}{d\mu}}(x')\int_{p\in\mathbb{R}^{n}}\exp\left(\frac{\imath p\bullet (x-x')}{\hslash}\right)f\left( \frac{x+x'}{2},p\right)\,d\lambda(p)
\end{equation}
Intuitively, $ f\in L^{1}(T^{*}\mathcal{X};\Omega)^{+}  $ will be paired with some $ \rho\in \mathcal{D}(L^{2}(\mathcal{X};\mu))^{+} $ if $ \rho  $ is very close to being diagonal and varies slowly along the diagonal whereas \textit{f} obeys a local version of Heisenberg's uncertainty relation, with sharp features in \textit{q} being spread out in \textit{p} and vice versa. Also, these pairs exist for the ground state of the harmonic oscillators, despite $ \rho $ not being very close to being diagonal. Note that when these pairs exist, as a map from $ \mathcal{D}(L^{2}(\mathcal{X};\mu))^{+} $ to $ L^{1}(T^{*}\mathcal{X};\Omega)^{+} $, $ \Psi $ is actually an isometry.

Two observations arise from this. Firstly, if we believe that a quantum description gives better predictions and that a formulation in terms of classical physics is merely an approximation, then the classical predictions are completely untrustworthy for non-integrable classical systems where \textit{f} rapidly develops fine tendrils under time evolution. Evolving the quantum analogue of the system and then using $ \Psi $ would give very different behavior. Either the fine tendrils would not form or the correlation would break down, with $ f=\Psi\rho $ no longer being positive. Secondly, classical physics is simpler than quantum mechanics only because the time evolution for a Hamiltonian system is so simple, being merely a point transformation on a finite dimensional space, $ T^{*}\mathcal{X}\to T^{*}\mathcal{X} $. Working with full stochastic generality using non-Hamiltonian systems in classical physics, the analogous quantum description is no more complicated.

The preceding is evidently dependent on the existence of Lebesgue measure. Since there is no infinite-dimensional analogue of Lebesgue measure~\cite{oxtoby}, it cannot be extended directly to field theories. Instead, some measure which itself depends on $ \hslash $ must be utilized.
\subsection*{Distinguishable versus indistinguishable particles}
It is usually said that there are no indistinguishable particles classically because it is possible to follow trajectories, which are the time-evolution point transformations $ T^{*}\mathcal{X}\to T^{*}\mathcal{X} $. The above analysis gives a different interpretation that is valid in the more general case where there are no trajectories: if indistinguishability is important in a model, so the density matrix $ \rho $ is taken symmetric (or anti-symmetric) under interchange,
\begin{equation}
\rho(x_{1},x_{2},x'_{1},x'_{2})=\pm\rho(x_{2},x_{1},x'_{1},x'_{2})=\pm\rho(x_{1},x_{2},x'_{2},x'_{1})=\rho(x_{2},x_{1},x'_{2},x'_{1})
\end{equation} 
then it is necessarily far from being diagonal, so there is likely no partner in $ L^{1}(T^{*}\mathcal{X};\Omega)^{+} $ to $ \rho $ and a classical model is invalid.

A similar situation holds for the case of $ \sigma$-algebras generated by a countable collection of atoms. If there is classical indistinguishability (such as dollars in a bank account), the measures assigned to certain atoms by any allowed measure must be the same. No issues are raised by embedding the measure as the diagonal in $ \mathcal{D}(L^{2}(\mathcal{X};\mu))^{+} $. However, trying to recast this as quantum indistinguishability then creates off-diagonal entries without classical interpretation.
\section{Additional structures for Bayesian networks?}\label{sec:additional}
It may be possible that additional mathematical structures besides measures and density matrices could be used as the base and target spaces of linear maps to calculate probabilities using Bayesian networks. To avoid topological complications, only the finite-dimensional case will be sought. Looking through the properties of measures and density matrices that are essential, we have the following properties required for an additional mathematical structure, given by a collection $ \mathcal{A} $ of finite-dimensional, normed vector spaces:  \textit{(i)}for any vector spaces $ \mathsf{A},\mathsf{B}\in\mathcal{A} $, the vector space $ \textsf{A}\otimes\textsf{B} $ is also in $ \mathcal{A} $; \textit{(ii)} each $\textsf{B}\in\mathcal{A} $ has a positive cone, $ \mathsf{B}^{+} $ that is generating\footnote{Any \textbf{x} in each $\mathsf{B}\in\mathcal{A} $ can be written as $ \mathbf{y}-\mathbf{z} $ for some $ \mathbf{y},\mathbf{z}\in\mathsf{B} $.};  \textit{(iii)} each $\mathsf{B}\in\mathcal{A} $ has the quasi-\textit{AL}-property that $ \Vert\mathbf{x}\Vert+\Vert\mathbf{y}\Vert=\Vert\mathbf{x}+\mathbf{y}\Vert $ for any $ \mathbf{x},\mathbf{y}\in\mathsf{B}^{+}  $; and \textit{(iv)} for any $ \mathbf{x}\in\mathsf{A}^{+}  $, $ \mathbf{y}\in\mathsf{B}^{+} $, $ \mathbf{x}\otimes\mathbf{y}$ is in $ (\mathsf{A}\otimes\mathsf{B})^{+} $ with $ \Vert\mathbf{x}\otimes\mathbf{y}\Vert=\Vert\mathbf{x}\Vert\Vert\mathbf{y}\Vert $. In addition, \textit{(v)} it is required that there be some nontrivial (other than the identity) maps between these vector spaces that are simultaneously both completely positive and completely bounded. 

Evidently, for measures $ \mathcal{A} $ is a collection of \textit{n}-dimensional vector spaces for \textit{n} a natural number, $ \lbrace 1,2,3,\ldots\rbrace $. For density matrices it is a collection of \textit{n}-dimensional vector spaces for \textit{n} a square number, $ \lbrace 1,4,9,\ldots\rbrace  $. The question, which is open, is whether there are a collection of vector spaces $ \mathcal{A} $ satisfying the preceding properties other than spaces isomorphic to measures, density matrices, or various tensor products of measures and density matrices. Obviously, this question is of vital interest--if the answer is negative, then we are limited to linear maps on measures, density matrices, and their tensor products in formulating Bayesian networks; if the answer is affirmative, then there is the immediate additional question of whether or not the additional structure is of practical utility in making predictions.   

%% file: chapter4.tex
\chapter{Constructing the networks}
\section{Graphical constructs and rules}\label{sec:constructs}
Recall the graphical model simply gives the form for the needed information and illustrates constraints. The actual needed data is given in an accompanying table. In the following, first the rules for the graphical model are given, then what accompanying data needs to be supplied depending on which option, \textbf{I'} or \textbf{II}, is chosen.  
\subsection*{Nodes and terminators}\label{subsec:nodesandterminators}
Nodes that are observable, with the traditional emphasis on the transition probability function, will be indicated in black. Hidden classical nodes, where the linear map on measures is taken to be fundamental, will be red. Quantum nodes, whose linear maps involve non-trivial spaces of density matrices, will be blue. Arrows share the color of their parent. Observable nodes that are being incorporated into the calculation of probability will be crossed. Those that are being marginalized will be left open. Those that are being conditioned on will be filled. Hidden (classical or quantum) nodes will always be left open. To make it visually clear whether or not a graph is complete and represents the calculation of a number (the probability) or is just a graph fragment with unresolved maps, terminal nodes which are either maps from the trivial space (giving initialization) or to the trivial space (giving the evaluation map) will be indicated by half-circle-half-box shapes, with the half-box being the terminating end.
\begin{align}
\setlength{\unitlength}{1 in}
\begin{picture}(4,.6)
\color{red}
\put(1,.2){\oval(.2,.2)[r]}
\put(.9,.3){\line(1,0){.1}}
\put(.9,.1){\line(1,0){.1}}
\put(.9,.1){\line(0,1){.2}}
\put(1.1,.2){\vector(1,0){.5}}
\color{blue}
\put(3,.2){\oval(.2,.2)[l]}
\put(3,.3){\line(1,0){.1}}
\put(3,.1){\line(1,0){.1}}
\put(3.1,.1){\line(0,1){.2}}
\put(2.4,.2){\vector(1,0){.5}}
\end{picture}
\end{align}
\subsection*{Splitters}\label{subsec:splitter}
As has been pointed out (see \S\ref{sec:tensornetworks}), the usual Bayesian network construction depends heavily on the ability to copy, with each child node receiving a copy. Without this ability (see \S\ref{subsec:commentsonoptions} and \S\ref{sec:nocopy}), an additional graphical structure is required, the \textit{splitter}. This is an oval with one or more arrows coming in and more than one arrow coming out--if only one arrow comes out, the splitter can be dispensed with and all the incoming arrows can be redirected to the following node.
\begin{align}
\setlength{\unitlength}{1 in}
\begin{picture}(2,.8)
\color{red}
\put(1,.2){\oval(.1,.5)}
\put(.45,.1){\vector(1,0){.5}}
\put(.45,.5){\vector(2,-1){.5}}
\put(1.05,.1){\vector(1,0){.5}}
\put(1.05,.2){\vector(2,1){.4}}
\end{picture}
\end{align}
\subsection*{Pince-nez}
There is a certain construction that occurs so often for joining hidden nodes to observable ones that it is replaced by its own structure, the \textit{pince-nez}. As the name suggests, this is two circles, one an observable node, the other a hidden one, joined by a bar.
\begin{align}
\setlength{\unitlength}{1 in}
\begin{picture}(3,.8)
\put(1,.5){\circle{.2}}
\put(1,.4){\line(0,1){.2}}
\put(.9,.5){\line(1,0){.19}}
\put(1.085,.52){\vector(3,1){.2}}
\put(1.5,.2){$ \Longrightarrow $}
\put(3,.3){\circle{.2}}
\put(2.895,.2){\line(0,1){.1}}
\put(2.9,.3){\line(1,0){.19}}
\put(3,.2){\line(0,1){.2}}
\put(3.085,.32){\vector(3,1){.5}}
\put(2.5,.42){\vector(2,-1){.39}}
\put(.65,.34){\vector(3,1){.27}}
\put(-.1,.4){\vector(1,-1){.23}}
\color{blue}
\put(.2,.1){\circle{.2}}
\put(.6,.2){\oval(.1,.5)}
\put(3,.1){\circle{.2}}
\put(2.895,.1){\line(0,1){.1}}
\put(3.085,.08){\vector(3,-1){.5}}
\put(2.5,.12){\vector(1,0){.39}}
\put(.3,.1){\vector(1,0){.25}}
\put(.65,.1){\vector(2,-1){.39}}
\put(-.1,.1){\vector(1,0){.2}}
\end{picture}
\end{align}
The neutral term \textit{pince-nez} is employed since the concept of ``measurement"\footnote{We will always place ``measurement" in quotation marks to emphasize its problematic nature.} applied to hidden nodes is so fraught with metaphysical connotations.
\section{Data for the structures using Option I'}
Using \textbf{option I'}, for each hidden node, the most general data is
\[(\mathcal{I},\mu,\mathsf{H};\mathcal{O},\nu,\mathsf{J};L)\]
where $ \mathcal{I} $ is the input set, $ \mu $ the accompanying base measure, \textsf{H} the input Hilbert space, $ \mathcal{O} $ the output set, $ \nu $ the accompanying base measure, \textsf{J} the output Hilbert space, and \textit{L} a completely-positive, norm-preserving (on the positive cone) map in $ \mathcal{B}(L^{1}(\mathcal{I};\mu;\mathcal{D}(\mathsf{H})),L^{1}(\mathcal{O};$\\$\nu;\mathcal{D}(\mathsf{J}))) $. If either Hilbert spaces are the trivial space $ \mathbb{C} $, it can be left off the list. Similarly, if either base measure is trivial, for instance, if $ \mu(\varnothing)=0 $, $ \mu(\mathcal{I})=1 $ with no other sets measurable, then it and its set can be left out as well.

For each splitter, the most general data is
\[ \left(\mathcal{I},\mu,\mathsf{H};\mathcal{O}_{1},\nu_{1},\mathsf{J}_{1},\ldots,\mathcal{O}_{m},\nu_{m},\mathsf{J}_{m};\Psi\right)\] 
where $ \mathcal{I}\cong\mathcal{O}_{1}\times\cdots\times\mathcal{O}_{m} $, $ \mu\cong\nu_{1}\times\cdots\times\nu_{m} $, and $ \mathsf{H}\cong\mathsf{J}_{1}\otimes\cdots\otimes\mathsf{J}_{m}  $, with $ \Psi $ giving the isomorphisms, which should be trivial except for possible permutations to make sure everything goes to the correct place, with more complicated behavior placed in separate nodes. Once again, trivial entries can be left out.

For each pince-nez, the most general data is
\[(\mathcal{I},\mu,\mathsf{H};\mathcal{R},\tau;\mathcal{O},\nu,\mathsf{J};L)\]
where $ \mathcal{I} $ is the input set, $ \mu $ the accompanying base measure, \textsf{H} the input Hilbert space, $ \mathcal{R} $ the observable set, $ \tau $ the accompanying base measure, $ \mathcal{O} $ the output set, $ \nu $ the accompanying base measure, \textsf{J} the output Hilbert space, and \textit{L} a completely-positive, norm-preserving (on the positive cone) map in $ \mathcal{B}\left(L^{1}(\mathcal{I};\mu;\mathcal{D}(\mathsf{H})),L^{1}(\mathcal{R}\times\mathcal{O};\tau\times\nu;\mathcal{D}(\mathsf{J}))\right) $. Again, trivial entries can be left out (however, if $ \tau $ is trivial, then the pince-nez is actually just a node).
\section{Data for the structures using Option II}
Using \textbf{option II}, for each hidden node, the most general data is
\[(\mathcal{I},\mathsf{H};\mathcal{O},\mathsf{J};L) \]
where $ \mathcal{I} $ is the input compact space, \textsf{H} the input Hilbert space, $ \mathcal{O} $ the output compact space, \textsf{J} the output Hilbert space, and \textit{L} a completely-positive, norm-preserving (on the positive cone) map in $ \mathcal{B}\left(\mathcal{C}(\mathcal{O};\mathcal{K}(\mathsf{J})),\mathcal{C}(\mathcal{I};\mathcal{K}(\mathsf{H}))\right) $. As previously noted, the spaces can be locally-compact if \textit{L} is further restricted. Trivial entries can be left off the list.

For each splitter, the most general data is
\[ \left(\mathcal{I},\mathsf{H};\mathcal{O}_{1},\mathsf{J}_{1},\ldots,\mathcal{O}_{m},\mathsf{J}_{m};\Psi\right)\] 
where $ \mathcal{I}\cong\mathcal{O}_{1}\times\cdots\times\mathcal{O}_{m} $ with the product topologies agreeing as well, and $ \mathsf{H}\cong\mathsf{J}_{1}\otimes\cdots\otimes\mathsf{J}_{m}  $, with $ \Psi $ giving the isomorphisms, which should be trivial except for possible permutations to make sure everything goes to the correct place, with more complicated behavior placed in separate nodes. Once again, trivial entries can be left out.

For each pince-nez, the most general data is
\[(\mathcal{I},\mathsf{H};\mathcal{R};\mathcal{O},\mathsf{J};L)\]
where $ \mathcal{I} $ is the input compact space, \textsf{H} the input Hilbert space, $ \mathcal{R} $ the observable compact space, $ \mathcal{O} $ the output compact space, \textsf{J} the output Hilbert space, and \textit{L} a completely-positive, norm-preserving (on the positive cone) map in $  \mathcal{B}\left(\mathcal{C}(\mathcal{R}\times\mathcal{O};\mathcal{K}(\mathsf{J})),\mathcal{C}(\mathcal{I};\mathcal{K}(\mathsf{H}))\right) $.  As previously noted (see \S\ref{sec:optionIandII}), the spaces can be locally-compact if \textit{L} is further restricted. Again, trivial entries can be left out (however, if $ \mathcal{R} $ is trivial in the sense of being the one-point set, then the pince-nez is actually just a node).
\section{Example-double-slit experiment}
So far, the exposition has been quite abstract. To make things more concrete and demonstrate how to use Bayesian networks with linear maps (both classical and quantum) in practice, consider the following example. There is a black-box with a dial which can be set to any position $ \theta $ on a circle, a switch with two settings, and four lights, with lights one and two forming one pair and lights three and four forming another. After study, it is determined to have the following properties: if the switch is off, periodically one of the second pair of lights flashes while the first pair never flashes, whereas, if the switch is on, one of each pair flashes. After more study, it is determined the behavior of each round is independent and, if the switch is off, with probability $ \cos^{2}\theta $ the third light alone flashes and with probability $ \sin^{2}\theta $ the fourth light alone flashes. If the switch is on, the joint probabilities are as follows:
\begin{equation}
\begin{array}{ll}\text{first on, third on, others off}&\frac{1}{4}\\\text{first on, fourth on, others off}&\frac{1}{4}\\\text{second on, third on, others off}&\frac{1}{4}\\\text{second on, fourth on, others off}&\frac{1}{4} \end{array}
\end{equation}
with the other probabilities being zero. Presenting the information in the manner of the preceding joint probabilities corresponds to the following graphical model:
\begin{align}
\setlength{\unitlength}{1 in}
\begin{picture}(3,1.1)
\put(1,.2){\circle{.2}}
\put(2,.2){\circle{.2}}
\put(1,.8){\circle*{.2}}
\put(2,.8){\circle*{.2}}
\put(.5,1){dial}
\put(1.7,1){switch}
\put(-.5,0){first pair of lights}
\put(2.1,0){second pair of lights}
\put(.9,.2){\line(1,0){1.19}}
\put(1,.109){\line(0,1){.6}}
\put(2,.109){\line(0,1){.6}}
\put(1.065,.74){\line(2,-1){.9}}
\put(1.935,.74){\line(-2,-1){.9}}
\end{picture}
\end{align}

Suppose the box is now opened and is found to be composed of five modules. A cable runs from the first module to the second. Two cables come from the second module, one each running to a pair of seemingly identical modules, each of which has one of the first pair of lights. The switch activates switches on the pair in unison. These modules have cables running to the last, which has the dial as a control. If possible, given the joint probabilities, we should confine ourselves to models which are consistent with the  constraints implied by these observations because we would like to be able to predict what would happen if these modules were rewired or taken out and placed in a different context.    
\section{Quantum model}\label{sec:quantumdouble}
Consider first a quantum model with graphical model:
\begin{align}\label{eq:doubleslitgraphical}
\setlength{\unitlength}{1 in}
\begin{picture}(5,2)
\color{blue}
\put(.4,1){\oval(.2,.2)[r]}
\put(.3,1.1){\line(1,0){.1}}
\put(.3,.9){\line(1,0){.1}}
\put(.3,.9){\line(0,1){.2}}
\put(.5,1){\vector(1,0){.7}}
\put(.5,.75){1}
\put(1.3,1){\circle{.2}}
\put(1.3,.75){2}
\put(1.38,1){\vector(1,0){.42}}
\put(1.85,1){\oval(.1,.3)}
\put(1.85,.65){3}
\put(1.9,1.1){\vector(2,1){1}}
\put(1.9,.9){\vector(2,-1){1}}
\put(2.9,.3){\line(0,1){.1}}
\put(3,.3){\circle{.2}}
\put(3,.05){5}
\put(2.9,1.5){\line(0,1){.1}}
\put(3,1.5){\circle{.2}}
\put(3,1.25){4}
\put(3.09,.3){\vector(2,1){1.32}}
\put(3.08,1.47){\vector(3,-1){1.33}}
\put(4.4,.9){\line(0,1){.1}}
\put(4.51,.9){\oval(.2,.2)[l]}
\put(4.51,1){\line(1,0){.1}}
\put(4.51,.8){\line(1,0){.1}}
\put(4.61,.8){\line(0,1){.2}}
\put(4.5,.6){6}
\color{black}{\put(2.9,.4){\line(0,1){.1}}}
\put(3,.5){\circle{.2}}
\put(2.9,1.6){\line(0,1){.1}}
\put(3,1.7){\circle{.2}}
\put(4.4,1){\line(0,1){.1}}
\put(4.5,1.1){\circle{.2}}
\put(2.85,.75){second light}
\put(3,1.9){first light}
\put(2.9,.5){\line(1,0){.19}}
\put(3,.405){\line(0,1){.19}}
\put(2.9,1.7){\line(1,0){.19}}
\put(3,1.605){\line(0,1){.19}}
\put(3.85,1.5){third and fourth}
\put(4.2,1.3){light}
\put(4.4,1.1){\line(1,0){.19}}
\put(4.5,1.02){\line(0,1){.18}}
\put(2.5,1){\circle*{.2}}
\put(4.1,1){\circle*{.2}}
\put(3.85,1){$ \Theta $}
\put(2.65,1){$ S$}
\put(2.57,.94){\vector(2,-3){.33}}
\put(2.57,1.06){\vector(2,3){.33}}
\put(4.18,1){\vector(1,0){.22}}
\put(-.1,.5){\dashbox{.1}(1,1)}
\put(1.1,.5){\dashbox{.1}(1,1)}
\put(2.8,1.1){\dashbox{.1}(1,1)}
\put(2.8,-.05){\dashbox{.1}(1,1)}
\put(3.82,.5){\dashbox{.1}(1.2,1.3)}
\end{picture}
\end{align}
The blue 1,2,3,4,5,6 are for reference. The information presented in the form indicated by the graphical model is as follows:
\paragraph*{node 1} $ (;\mathbb{C}^{2};L_{1}) $, where $ L_{1} $ is the constant map with value $ \left[\begin{array}{c}\frac{1}{\sqrt{2}}\\\frac{1}{\sqrt{2}}\end{array} \right]\left[\begin{array}{cc}\frac{1}{\sqrt{2}} &\frac{1}{\sqrt{2}}\end{array} \right]=\left[\begin{array}{cc}\frac{1}{2}&\frac{1}{2}\\\frac{1}{2}&\frac{1}{2}\end{array} \right]  $.
\paragraph*{node 2} $ (\mathbb{C}^{2};\mathbb{C}^{4};L_{2}) $, where $ L_{2}(\rho) $ is
\begin{equation}
\left[\begin{array}{ccc}0&\begin{array}{cc}0 &0\end{array} &0\\\begin{array}{c}0 \\0\end{array}&\rho&\begin{array}{c}0 \\0\end{array}\\0&\begin{array}{cc}0 &0\end{array} &0\end{array} \right]
\end{equation}
\paragraph*{splitter 3} $ (\mathbb{C}^{4};\mathbb{C}^{2},\mathbb{C}^{2}) $
\paragraph*{pince-nez 4}
\begin{align}
\left(\begin{array}{c}{}\\{}\\{}\end{array}\left\lbrace\begin{array}{l}\text{switch on,}\\\text{switch off} \end{array}\right\rbrace,\begin{array}{l}\text{counting}\\\text{measure}\end{array},\mathbb{C}^{2};\left\lbrace\begin{array}{l}\text{light one on and switch on,}\\\text{light one on and switch off,}\\\text{light one off and switch on,} \\\text{light one off and switch off,}\end{array}\right\rbrace,\begin{array}{l}\text{counting}\\\text{measure}\end{array};\mathbb{C}^{2};L_{4}\right) 
\end{align}
where $ L_{4}(\tau) $ is
\begin{equation}
\begin{cases}\tau\vert_{\text{switch off}}&\text{if light one off and switch off}\\0& \text{if light one on and switch off}\\\left[\begin{array}{cc}1&0\\0&0\end{array} \right]\tau\vert_{\text{switch on}} \left[\begin{array}{cc}1&0\\0&0\end{array} \right]&\text{if first on and switch on}\\\left[\begin{array}{cc}0&0\\0&1\end{array} \right]\tau\vert_{\text{switch on}} \left[\begin{array}{cc}0&0\\0&1\end{array} \right]&\text{if first off and switch on}\end{cases} 
\end{equation}
\paragraph*{pince-nez 5} Same as for pince-nez 4 except the second light replaces the first light.
\paragraph*{pince-nez 6}
\[ \left([0,2\pi),\begin{array}{l}\text{Lebesgue}\\\text{measure}\end{array},\mathbb{C}^{4};\left\lbrace\begin{array}{l}\text{third and fourth on,}\\\text{third on,}\\\text{fourth on,}\\\text{third and fourth off} \end{array}\right\rbrace,\begin{array}{l}\text{counting}\\\text{measure}\end{array};;L_{6}\right) \]
where $ L_{6}(\tau) $ is
\begin{align}
&\int_{0}^{2\pi}\left[\begin{array}{cccc}1&0&0&0\end{array} \right]\tau(\theta) \left[\begin{array}{c}1\\0\\0\\0\end{array} \right]\,d\theta\text{ if both third and fourth on}\\&\int_{0}^{2\pi}\left[\begin{array}{cccc}0&\frac{1}{\sqrt{2}}e^{\imath \theta}&\frac{1}{\sqrt{2}}e^{-\imath \theta}&0\end{array} \right]\tau(\theta)\left[\begin{array}{c}0\\\frac{1}{\sqrt{2}}e^{-\imath \theta}\\\frac{1}{\sqrt{2}}e^{\imath \theta}\\0\end{array} \right]\,d\theta\text{ if third on}\nonumber\\&\int_{0}^{2\pi}\left[\begin{array}{cccc}0&\frac{1}{\sqrt{2}}e^{\imath \theta}&-\frac{1}{\sqrt{2}}e^{-\imath \theta}&0\end{array} \right]\tau(\theta) \left[\begin{array}{c}0\\\frac{1}{\sqrt{2}}e^{-\imath \theta}\\-\frac{1}{\sqrt{2}}e^{\imath \theta}\\0\end{array} \right]\,d\theta\text{ if fourth on}\nonumber\\&\int_{0}^{2\pi}\left[\begin{array}{cccc}0&0&0&1\end{array} \right]\tau(\theta) \left[\begin{array}{c}0\\0\\0\\1\end{array} \right]\,d\theta\text{ if both third and fourth off}\nonumber
\end{align}
\medskip\\
Using this information, the joint probability, given that the switch is off and the dial is set to $ \theta $, for both the first pair being off, the third light being on, and the fourth light being off, is then given by, for positive, unit-norm $ g\in L^{1}([0,2\pi),\text{Lebesgue}) $,
\begin{align}\label{eq:twoslitcalc1}
\left.L_{6}\left((L_{4}\otimes L_{5})\left( L_{2}\left(L_{1}\right) 1_{\text{switch off}}\right)\vert_{\begin{scriptsize}\begin{array}{l}\text{first pair is off,}\\\text{switch is off}\end{array}\end{scriptsize}}\; g\right)\right\vert_{\begin{scriptsize}\begin{array}{l}\text{third is on,}\\\text{fourth is off}\end{array}\end{scriptsize}} \\=\int_{0}^{2\pi}\left\vert\left[\begin{array}{cccc}0&\frac{1}{\sqrt{2}}e^{\imath \theta'}&\frac{1}{\sqrt{2}}e^{-\imath \theta'}&0\end{array} \right]\left[\begin{array}{c}0\\\frac{1}{\sqrt{2}}\\\frac{1}{\sqrt{2}}\\0\end{array} \right]\right\vert^{2}g(\theta')\,d\theta'= \int_{0}^{2\pi}\cos^{2}\theta'\;g(\theta')\,d\theta'
\end{align}
For \textit{g} sufficiently peaked about $ \theta $, the result is approximately $ \cos^{2}\theta $.

Similarly, the joint probability, given that the switch is on and the dial is set to $ \theta $, for both the first and the third light being on, with the second and fourth being off, is then given by
\begin{align}\label{eq:twoslitcalc2}
\left.L_{6}\left( (L_{4}\otimes L_{5})\left( L_{2}\left(L_{1}\right)1_{\text{switch on}}\right)\vert_{\begin{scriptsize}\begin{array}{l}\text{first is on,}\\\text{second is off}\\\text{switch is on}\end{array}\end{scriptsize}}\; g\right)\right\vert_{\begin{scriptsize}\begin{array}{l}\text{third is on,}\\\text{fourth is off}\end{array}\end{scriptsize}}\\=\int_{0}^{2\pi}\left\vert\left[\begin{array}{cccc}0&\frac{1}{\sqrt{2}}e^{\imath \theta'}&\frac{1}{\sqrt{2}}e^{-\imath \theta'}&0\end{array} \right]\left[\begin{array}{c}0\\\frac{1}{\sqrt{2}}\\0\\0\end{array} \right]\right\vert^{2}\;g(\theta')\,d\theta'= \frac{1}{4}
\end{align}
The other joint probabilities can be calculated similarly and agree with the joint probabilities originally determined for the black box.
\subsection*{Comments on the use of option I'}\label{subsec:commentsblackboxoption}
Note the use of option \textbf{I'} leads to extra complexity concerning the dial position $ \Theta $, forcing the introduction of a function \textit{g} peaked at $ \theta $, although  in this example we are not concerned about modelling the uncertainty of that value. In this case, using option \textbf{II} would have been slightly simpler. Then, instead of using counting measure as the base measure on the discrete sets, the discrete topology would be used to make the discrete sets into compact spaces.
\section{The Bayesian network approach versus the standard textbook approach}
Interpreting the switch as controlling the operation of a position ``measurement" at two slits and the dial as selecting a point on a backing screen for another position ``measurement", there is a well-known standard textbook approach using wavefunctions, projectors, and Bohm's postulate that replicates the outcome. By comparison, the Bayesian network for this particular problem may appear cumbersome; however, that is largely a result of familiarity with the firmer. With the network approach, there are none of the seemingly ad hoc rules for dealing with quantum systems; instead, only the simple conditions of positivity, linearity and potential universality. Also, the graphical model is highly intuitive and guides the writing of the correct expression (\ref{eq:twoslitcalc2}). This is of great importance in dealing with more complicated systems.   

In addition, as will be illustrated by more complicated examples to follow (see \S\ref{sec:quantum}), the network approach is far more flexible. Of particular interest, in some special cases it allows the needed Hilbert spaces to be kept to reasonable sizes in the course of the calculation instead of ballooning exponentially. To be more explicit, for a Hilbert space of dimension \textit{n}, the space of operators has dimension of order $ n^{2} $, and the space of maps on these operators has dimension of order $ n^{4} $, which is why it appears cumbersome. However, if, by utilizing the flexibility of the network approach, one is able to avoid dealing with Hilbert spaces of dimension $ n^{N} $ and operator spaces of dimension $ n^{2N} $, where $ N $ is the number of particles, the potential savings is tremendous (as will be seen in \S\ref{sec:quantum}). 
\section{Classical hidden model}\label{sec:classicaldouble}
For the same black box, now consider a classical hidden model with the same graphical model (\ref{eq:doubleslitgraphical}) (except now the hidden nodes and connecting arrows will be red). For simplicity, functions on discrete spaces indexed by numbers will be given as column vectors. The information presented in the form indicated by the graphical model is as follows:
\paragraph*{node 1} $ \left(;\lbrace 1,2\rbrace,\begin{array}{l}\text{counting}\\\text{measure}\end{array};L_{1}\right) $, where $ L_{1} $ is the constant map with value $ \left[\begin{array}{c}\frac{1}{2} \\\frac{1}{2}\end{array} \right]  $.
\paragraph*{node 2} $ \left(\lbrace 1,2\rbrace,\begin{array}{l}\text{counting}\\\text{measure}\end{array};\lbrace (1,1),(1,2),\ldots,(4,4)\rbrace,\begin{array}{l}\text{counting}\\\text{measure}\end{array};L_{2}\right) $, where $ L_{2}(\rho) $ is
\begin{equation}
\left[\begin{array}{cccccccccccccccc}0&0&0&0&0&0&\rho_{1}&0&0&0&0&0&\rho_{2}&0&0&0\end{array}\right]^{\text{T}}
\end{equation}
\paragraph*{splitter 3} $ (\lbrace (1,1),(1,2),\ldots,(4,4)\rbrace;\lbrace 1,2,3,4\rbrace,\lbrace 1,2,3,4\rbrace) $ with the obvious identification.
\paragraph*{pince-nez 4}
\begin{align}
&\left(\left\lbrace\begin{array}{l}\text{switch on,}\\\text{switch off} \end{array}\right\rbrace\times\lbrace 1,2,3,4\rbrace,\begin{array}{l}\text{counting}\\\text{measure}\end{array};\right.\\&\left. \left\lbrace\begin{array}{l}\text{light one on and switch on,}\\\text{light one on and switch off,}\\\text{light one off and switch on,} \\\text{light one off and switch off,}\end{array}\right\rbrace,\begin{array}{l}\text{counting}\\\text{measure}\end{array};\lbrace 1,2,3,4\rbrace,\begin{array}{l}\text{counting}\\\text{measure}\end{array};L_{4}\right)\nonumber
\end{align}
where $ L_{4}(\tau) $ is
\begin{equation}
\begin{cases}\tau\vert_{\text{switch off}}&\text{if light one off and switch off}\\0& \text{if light one on and switch off}\\ \left[\begin{array}{cccc}0&0&0&0\\1&1&0&0\\0&0&0&0\\0&0&0&0\end{array} \right]\tau\vert_{\text{switch on}}& \text{if first on and switch on}\\\left[\begin{array}{cccc}0&0&0&0\\0&0&0&0\\0&0&0&0\\0&0&1&1\end{array} \right]\tau\vert_{\text{switch on}} &\text{if first off and switch on} \end{cases}
\end{equation}
\paragraph*{pince-nez 5} Same as for pince-nez 4 except the second light replaces the first light.
\paragraph*{pince-nez 6}
\begin{align}
&\left(\lbrace (1,1),(1,2),\ldots,(4,4)\rbrace\times[0,2\pi),\begin{array}{l}\text{counting}\\\text{measure}\end{array}\times\begin{array}{l}\text{Lebesgue}\\\text{measure}\end{array};\right.\\&\left. \left\lbrace\begin{array}{l}\text{third and fourth on,}\\\text{third on,}\\\text{fourth on,}\\\text{third and fourth off} \end{array}\right\rbrace,\begin{array}{l}\text{counting}\\\text{measure}\end{array};;L_{6}\right)\nonumber
\end{align}
where $ L_{6}(\tau) $ is
\begin{equation}
\int_{0}^{2\pi}\left[\begin{array}{cccccccccccccccc}1&1&1&1&1&1&0&0&0&0&0&0&0&0&0&0\end{array} \right]\tau(\theta) \,d\theta
\end{equation}
if both third and fourth lights are on,
\begin{equation}
\int_{0}^{2\pi}\left[\begin{array}{cccccccccccccccc}0&0&0&0&0&0&\cos^{2}\theta&\frac{1}{2}&0&0&0&0&\cos^{2}\theta&\frac{1}{2}&0&0\end{array} \right]\tau(\theta)\,d\theta
\end{equation}
if the third light is on and the fourth is off,
\begin{equation}
\int_{0}^{2\pi}\left[\begin{array}{cccccccccccccccc}0&0&0&0&0&0&\sin^{2}\theta&\frac{1}{2}&0&0&0&0&\sin^{2}\theta&\frac{1}{2}&0&0\end{array} \right]\tau(\theta)\,d\theta
\end{equation}
if the fourth light is on and the third is off, and
\begin{equation}
\int_{0}^{2\pi}\left[\begin{array}{cccccccccccccccc}0&0&0&0&0&0&0&0&1&1&1&1&0&0&1&1\end{array} \right]\tau(\theta)\,d\theta
\end{equation}
if both third and fourth lights are off.
\medskip\\
Using this information, the joint probability, given that the switch is off and the dial is set to $ \theta $, for both the first pair being off, the third light being on, and the fourth light being off, is then given by (\ref{eq:twoslitcalc1}), which is $ \int_{0}^{2\pi}\cos^{2}\theta'\;g(\theta') \,d\theta'\approx \cos^{2}\theta$. Similarly, the joint probability, given that the switch is on and the dial is set to $ \theta $, for both the first and the third light being on, with the second and fourth being off, is then given by (\ref{eq:twoslitcalc2}), which is $ \int_{0}^{2\pi}\frac{1}{4}g(\theta') \,d\theta'=\frac{1}{4}$. The other joint probabilities can be calculated similarly and agree with the joint probabilities originally given for the black box.
\section{What is a quantum system?}
For the two-slit experiment, which only required finite-dimensional linear algebra in the quantum model (apart from the already commented on problem of inputting the dial setting $ \Theta $), a classical model with the same behavior also only required finite-dimensional linear algebra. This is atypical. As we have seen in \S\ref{sec:embeddingquantum}, a classical model that duplicates the behavior of a quantum model is generally far more complicated. Also, these classical models are generally inherently contextual.

A reason why the name ``quantum system" could still be applied to this example is universality: while both models have been constructed to be potentially universal, testing will reveal if they fail in this regard. If the black boxes were opened to reveal a laser, beam-splitters, photon detectors, and so on, from experience we would have a lot of confidence that the modules in the quantum model are universal, whereas we would have very little confidence for the modules in the classic model to have that property. Conversely, if the boxes were opened to reveal regular computer circuits, the situation would be reversed.

%% file: chapter5ver2.tex
\chapter{Relation to textbook quantum mechanics}\label{ch:textbookquantum}
\section{Textbook rules for quantum mechanics}\label{sec:textbookrules}
The quantum Bayesian network structure developed so far has constraints only arising from the requirements of positivity, linearity, and potential universality. It is not obvious that it has any connection necessarily to what is usually thought of as quantum mechanics. Using option \textbf{I'}, that is not the situation, as is shown below.

However, while the following justifies the usage of the name \textit{quantum} for the extended Bayesian networks, it should not be taken as a justification of them--quite the opposite. The Bayesian network approach is predicated on a very reasonable basis. The textbook approach to quantum mechanics is only comfortable due to familiarity; on its own merits its rules are incomprehensible and unmotivated. Hence, the following is better taken as a justification of the textbook approach. In other words, if quantum behavior had first been discovered as forecasting arising from a reasonable extension of Bayesian networks, and the usual list of textbook rules were later discovered picking out a certain subset of networks that were sufficient to model any situation, no one would take those rules as primary.  
\subsection*{Rule one}
Textbook quantum mechanics imposes an additional rule: only networks in the form of chains are permitted and all the Hilbert spaces along a chain must be the same (apart for terminating trivial spaces). (This rule may arise from imagining that we have some sort of universal, initial-value, dynamical system). Mathematically, this rule puts no additional constraints on the network formulation; it is always possible to obey this by working in a sufficiently large Hilbert space and waiting until the end of the calculation to perform the reduced traces arising from terminal nodes. However, doing this in practice is unnecessarily difficult--the whole point of Bayesian networks is to try to find a model made of simple and (hopefully) universal parts where the calculations can be done in a manageable manner. Therefore, similarly our treatment of determinism in \S\ref{sec:determinism}, we will discard this rule.     
\subsection*{Rule two}
Another rule from textbook quantum mechanics is: the linear map for a quantum node can only be of the form $ L\rho=U\rho U^{*} $, where $ U $ is unitary.  As we prove in theorem \textbf{5.1.1} below, this is no limitation at all mathematically if option \textbf{I'} is employed. If option \textbf{II} is used, the theorem still holds if \textsf{H} is separable and if there is some strictly positive\footnote{Every open set gets measure greater than zero.}, finite Radon measure on $ \mathcal{X} $. For any node with data $ (\mathcal{I},\mu,\mathsf{H};\mathcal{O},\nu,\mathsf{J};L) $, the map \textit{L} can be represented as the sequence of three operations, using some auxiliary Hilbert space \textsf{M}: (\textbf{i}) injecting $ \rho\in L^{1}(\mathcal{I};\mu;\mathcal{D}(\mathsf{H})) $ as a density matrix on a Hilbert space $ \mathsf{K}\otimes L^{2}(\mathcal{O};\nu)\otimes\mathsf{J} $ utilizing a partial isometry; (\textbf{ii}) taking a reduced trace of the density matrix over \textsf{K}; (\textbf{iii}) mapping the resulting density matrix on $ L^{2}(\mathcal{O};\nu)\otimes\mathsf{J} $ into $ L^{1}(\mathcal{O};\nu;\mathcal{D}(\mathsf{J})) $. By introducing an auxiliary Hilbert space \textsf{L} with
\begin{equation}\label{eq:auxiliaryhilbert}
L^{2}(\mathcal{X};\mu)\otimes\mathsf{H}\otimes \mathsf{L}=L^{2}(\mathcal{O};\nu)\otimes\mathsf{J}\otimes \mathsf{M}
\end{equation}
and some fixed template density-matrix on \textsf{L}, the partial isometry can be upgraded to an unitary operator. In terms of the graphical model, this means replacing a node with the graph fragment:
\begin{align}
\setlength{\unitlength}{1 in}
\begin{picture}(3,1.2)
\color{blue}{\put(1,.5){\circle{.2}}}
\put(.65,.1){\oval(.2,.2)[r]}
\put(.55,.2){\line(1,0){.1}}
\put(.55,0){\line(1,0){.1}}
\put(.55,0){\line(0,1){.2}}
\put(2,.75){\oval(.2,.2)[l]}
\put(2,.85){\line(1,0){.1}}
\put(2,.65){\line(1,0){.1}}
\put(2.1,.65){\line(0,1){.2}}
\put(1.5,.5){\oval(.1,.3)}
\put(.5,.5){\circle{.2}}
\put(2,.25){\circle{.2}}
\put(.8,.7){unitary}
\put(1.09,.5){\vector(1,0){.36}}
\put(1.56,.6){\vector(3,1){.35}}
\put(1.56,.4){\vector(3,-1){.35}}
\put(2.08,.23){\vector(3,-1){.3}}
\put(0,.65){$\mathcal{I},\mu,\mathsf{H}$}
\put(.85,.05){\textsf{L}}
\put(1.6,.72){\textsf{M}}
\put(2.3,.19){$\mathcal{O},\nu,\mathsf{J}$}
\put(.05,.5){\vector(1,0){.34}}
\put(.71,.17){\vector(1,1){.25}}
\put(.59,.5){\vector(1,0){.31}}
\end{picture}
\end{align}
Of course, just because we can do this does not mean we must, or that we should. To obey this rule, a great deal of extra computation and irrelevant, arbitrary choices must be made to no practical benefit. Therefore, this rule will also be discarded. 
\subsection*{Rule three}
Yet another rule from textbook quantum mechanics is: the linear map for a terminal pince-nez (a destructive ``measurement") can only be of the form $ \int_{B}L\rho\,d\nu=\text{tr }E_{B}\rho  $, where $ \lbrace E_{B}\rbrace_{B\in\mathcal{E}} $ are a complete set of mutually commuting, orthogonal projectors\footnote{$ \lbrace E_{B}\rbrace_{B\in\mathcal{E}} $ is a \textit{complete set of mutually commuting, orthogonal projectors} for a $ \sigma $-algebra $ \mathcal{E} $ of subsets of a set $ \mathcal{X} $ if they are all mutually commuting, orthogonal projectors with $ E_{\varnothing}=0 $, $ E_{\mathcal{X}}=I $, and $ E_{\bigcup_{j}B_{j}}=\sum_{j} E_{B_{j}}$ for all countable, disjoint collections $ \lbrace B_{j}\rbrace\subset \mathcal{E} $, with convergence of the sum in the weak* topology induced on $ \mathcal{B}\left(\mathsf{H}\right) $ by its being dual to the trace-class operators $ \mathcal{S}_{1}\left(\mathsf{H}\right) $. In this case, convergence in the ultrastrong-operator, strong-operator, ultraweak-operator (same as weak*), and weak-operator topologies on $\mathcal{B}\left(\mathsf{H}\right)$ are all equivalent (see \textbf{A2.1}).}. For the other pince-nez, only $ \sigma $-algebras generated by a countable set of atoms are allowed, and the map must be of the form
\begin{equation}
\int_{B}L\rho\,d\nu=\sum_{j}E_{A_{j}}\rho E_{A_{j}}\text{ for }\lbrace A_{j}\rbrace\text{ a partition of }B
\end{equation}
This rule can also be stated in the form that every ``measurement" has an associated operator, which are the coordinates \textit{r} (taking values as coordinates in the discrete case) for $ \mathcal{R} $ acting on $ L^{2}(\mathcal{R};\tau) $ in this formulation. An even stronger formulation requires the result of any ``measurement" to be an eigenvalue of the associated operator~\cite{diraceig}. This implicitly assumes any results accrued to unions of atoms is due to post-``measurement" garbling and not inherent in the ``measurement". Furthermore, if the eigenspace corresponding to any eigenvalue has dimension greater than one, this formulation fails to distinguish among the myriad of possible complete sets of mutually commuting, orthogonal projectors consistent with its prescription, giving rise to the ``three-box paradox"~\cite{aharonov} if there is post-conditioning.  

Again employing option \textbf{I'}, as we prove in theorem \textbf{5.1.1} below, this third textbook rule also imposes no mathematical limitation. Using option \textbf{II}, the theorem still holds with the same limitations given before. For any pince-nez with data $ (\mathcal{I},\mu,\mathsf{H};\mathcal{R},\tau;\mathcal{O},\nu,\mathsf{J};L) $, the map \textit{L} can be represented as the sequence of four operations, using some auxiliary Hilbert space \textsf{M}: (\textbf{i}) injecting $ \rho\in L^{1}(\mathcal{I};\mu;\mathcal{D}(\mathsf{H})) $ as a density matrix on a Hilbert space $  L^{2}(\mathcal{R}\times\mathcal{O};\tau\times\nu)\otimes\mathsf{J}\otimes \mathsf{M}$ utilizing a partial isometry; (\textbf{ii}) making a projective ``measurement" on the $  L^{2}(\mathcal{R};\tau) $ portion; (\textbf{iii}) taking a reduced trace of the density matrix over \textsf{L}; and (\textbf{iv}) mapping the resulting density matrix on $ L^{2}(\mathcal{O};\nu)\otimes\mathsf{J} $ into $ L^{1}(\mathcal{O};\nu;\mathcal{D}(\mathsf{J})) $. By introducing an auxiliary Hilbert space \textsf{L} with
\begin{equation}\label{eq:auxiliaryhilbert}
L^{2}(\mathcal{X};\mu)\otimes\mathsf{H}\otimes \mathsf{L}=L^{2}(\mathcal{R}\times\mathcal{O};\tau\times\nu)\otimes\mathsf{J}\otimes \mathsf{M}
\end{equation}
and some fixed template density-matrix on \textsf{L}, the partial isometry can be upgraded to an unitary operator. In terms of the graphical model, this means replacing a pince-nez with the graph fragment:
\begin{align}
\setlength{\unitlength}{1 in}
\begin{picture}(3,1.3)
\put(2,1.15){\line(0,1){.1}}
\put(2,1.24){\line(1,0){.19}}
\put(2.1,1.145){\line(0,1){.19}}
\put(2.1,1.24){\circle{.2}}
\color{blue}
\put(1,.5){\circle{.2}}
\put(.65,.1){\oval(.2,.2)[r]}
\put(.55,.2){\line(1,0){.1}}
\put(.55,0){\line(1,0){.1}}
\put(.55,0){\line(0,1){.2}}
\put(2.3,.5){\oval(.2,.2)[l]}
\put(2.3,.6){\line(1,0){.1}}
\put(2.3,.4){\line(1,0){.1}}
\put(2.4,.4){\line(0,1){.2}}
\put(1.5,.5){\oval(.1,.3)}
\put(.5,.5){\circle{.2}}
\put(2,.25){\circle{.2}}
\put(2.1,1.04){\oval(.2,.19)[l]}
\put(2.1,1.14){\line(1,0){.1}}
\put(2.1,.95){\line(1,0){.1}}
\put(2.2,.95){\line(0,1){.19}}
\put(2,1.05){\line(0,1){.1}}
\put(.8,.7){unitary}
\put(2.3,1.1){projective}
\put(1.09,.5){\vector(1,0){.36}}
\put(1.56,.5){\vector(1,0){.65}}
\put(1.56,.4){\vector(3,-1){.35}}
\put(1.56,.55){\vector(3,4){.44}}
\put(2.08,.23){\vector(3,-1){.4}}
\put(0,.65){$\mathcal{I},\mu,\mathsf{H}$}
\put(.85,.05){\textsf{L}}
\put(2,.55){\textsf{M}}
\put(2.4,.15){$\mathcal{O},\nu,\mathsf{J}$}
\put(1.25,1.1){$  L^{2}(\mathcal{R};\tau) $}
\put(.05,.5){\vector(1,0){.34}}
\put(.71,.17){\vector(1,1){.25}}
\put(.59,.5){\vector(1,0){.31}}
\end{picture}
\end{align}
Once again, just because we can do this does not mean we must, or that we should. To obey this rule, a great deal of extra computation and irrelevant, arbitrary choices must be made to no benefit. Therefore, this rule will also be discarded.
\subsection*{The main theorem}
The statement of the aforementioned theorem is:
\paragraph{Theorem 5.1.1} Any completely-positive map $ L\in\mathcal{CP}(L^{1}(\mathcal{I};\mu;\mathcal{D}(\mathsf{H})), L^{1}(\mathcal{R}\times\mathcal{O};\tau\times\nu;\mathcal{D}(\mathsf{J}))) $ that is an isometry on the positive cone can be expressed as
\[ L\rho=(\text{diag}_{L^{2}\left(\mathcal{R};\tau\right)}\otimes(\text{diag}_{L^{2}\left(\mathcal{O};\nu\right)}\otimes I_{\mathcal{B}(\mathsf{J})})\otimes\text{tr }_{\mathsf{M}})V((\text{diag}_{L^{2}\left(\mathcal{I};\mu\right)}\otimes I_{\mathcal{B}(\mathsf{H})})^{-1}\rho) V^{*}\]
for some Hilbert space \textsf{M} and some partial isometry $ V\in \mathcal{B}(L^{2}\left(\mathcal{I};\mu\right)\otimes\mathsf{H},L^{2}\left(\mathcal{R}\times\mathcal{O};\tau\times\nu\right)\otimes\mathsf{J}\otimes\mathsf{M}) $. If all the Hilbert spaces $ L^{2}\left(\mathcal{I};\mu\right) $, \textsf{H}, $ L^{2}\left(\mathcal{R};\nu\right) $, $ L^{2}\left(\mathcal{O};\nu\right) $, and \textsf{J} have finite dimension, then \textsf{M} can be taken to be $ \mathsf{J}\otimes\dim L^{2}\left(\mathcal{I};\mu\right)\otimes \mathsf{H}$.
\paragraph*{Proof} Apply \textbf{B5.27}. $ \square $
\paragraph*{Comments} The full proof is quite lengthy, occupying from \textbf{B5.17} to \textbf{B5.27} in the appendix. The following is an outline: First a diagonal extraction map diag is defined in \textbf{B5.17}. Using this, we find that the dual space to density-matrix-valued, $ L^{1} $-``functions" is a von Neumann algebra\footnote{A \textit{Banach algebra} is a Banach space equipped with an associative, distributive product satisfying $ \Vert ab\Vert\leq \Vert a\Vert\Vert b\Vert $. A \textit{*-algebra} is a Banach algebra with an antilinear involution. A \textit{C*-algebra} is a *-algebra where $ \Vert a^{*}a\Vert=\Vert a\Vert^{2} $. A \textit{von Neumann algebra} (also termed a \textit{W*-algebra}) is a \textit{C}*-algebra which, as a Banach space, is a dual space.}. Then one theorem by Stinespring~\cite{stinespring} on completely-positive maps on $ C^{*} $-algebras and another by Sakai~\cite{sakai} on representations of von Neumann algebras complete the proof.

For the inverse map in the statement of the theorem, $ (\text{diag}_{L^{2}\left(\mathcal{I};\mu\right)}\otimes I_{\mathcal{B}(\mathsf{H})})^{-1} $, any particular choice in the preimage can be taken--the final result is the same regardless of which choice is made. Also, as claimed, introducing an auxiliary Hilbert space \textsf{L} satisfying (\ref{eq:auxiliaryhilbert}), the partial isometry can be certainly be upgraded (although not in a unique fashion) to an unitary operator \textit{U} satisfying, for some fixed template density-matrix $ \chi\in\mathcal{D}(\mathsf{L}) $,  
\begin{equation}
V((\text{diag}_{L^{2}\left(\mathcal{I};\mu\right)}\otimes I_{\mathcal{B}(\mathsf{H})})^{-1}\rho) V^{*}=U(((\text{diag}_{L^{2}\left(\mathcal{I};\mu\right)}\otimes I_{\mathcal{B}(\mathsf{H})})^{-1}\rho)\otimes \chi) U^{*}
\end{equation}
for all $ \rho\in L^{1}(\mathcal{I};\mu;\mathcal{D}(\mathsf{H})) $. In addition, the statement of the theorem does indeed indicate a projective ``measurement". Let \textit{B} be any $ \tau $-measurable subset of $ \mathcal{R} $. Then, for any $ \varphi\in \mathcal{D}(L^{2}(\mathcal{R};\tau)) $, $ 1_{B} $ as an element of $ L^{\infty}(\mathcal{R};\tau)\cong L^{1}(\mathcal{R};\tau)^{*} $ acting on $ \text{diag}_{L^{2}\left(\mathcal{R};\tau\right)}\varphi $ is
\begin{equation}
\int_{B}\text{diag}_{L^{2}\left(\mathcal{R};\tau\right)}\varphi\;d\tau=\text{tr }P_{B}\varphi
\end{equation}
where $ P_{B}=\text{diag}_{L^{2}(\mathcal{R};\tau)}^{*}1_{B} $ is the orthogonal projector to the subspace of ``functions" in $ L^{2}(\mathcal{R};\tau) $ that are zero almost everywhere outside \textit{B}.
\section{Weak measurements}
\subsection*{Reproducibility as a criteria}
One motivation for the restriction to projective maps, emphasized by Dirac~\cite{diracrepeat}, is reproducibility: if two successive pince-nez have the same data (and the incoming and outgoing data are compatible), then conditioning on the observation for the first being in some measurable subset $ B\subset \mathcal{R} $, the probability measure for the second observation has the property that on measurable sets that do not intersect $ B $, it gives zero, whereas on measurable sets that contain $ B $, it gives one. The Bayesian network for this set-up is:
\begin{align}
\setlength{\unitlength}{1 in}
\begin{picture}(3,.5)
\put(1,.35){\line(0,1){.1}}
\put(1.1,.44){\circle*{.2}}
\put(2,.35){\line(0,1){.1}}
\put(2,.44){\line(1,0){.19}}
\put(2.1,.345){\line(0,1){.19}}
\put(2.1,.44){\circle{.2}}
\color{blue}
\put(1.1,.24){\circle{.2}}
\put(2.1,.24){\circle{.2}}
\put(1,.24){\line(0,1){.1}}
\put(2,.24){\line(0,1){.1}}
\put(.1,.24){\oval(.2,.2)[r]}
\put(0,.34){\line(1,0){.1}}
\put(0,.14){\line(1,0){.1}}
\put(0,.14){\line(0,1){.2}}
\put(3.1,.24){\oval(.2,.2)[l]}
\put(3.1,.34){\line(1,0){.1}}
\put(3.1,.14){\line(1,0){.1}}
\put(3.2,.14){\line(0,1){.2}}
\put(.2,.24){\vector(1,0){.8}}
\put(1.2,.24){\vector(1,0){.8}}
\put(2.2,.24){\vector(1,0){.8}}
\put(1.1,0){1}
\put(2.1,0){1}
\end{picture}
\end{align}
where the repeated `1's indicate the data for the two pince-nez are the same.

While this criteria looks superficially attractive--as is implied in the saw ``measure twice, cut once"--its flaw is readily apparent. Suppose I were asked to see whether or not a light in another room were off. If I see the light is off, I return and report it being off. If I see it is on, I turn it off, then return and also report it being off. This is repeatable, but does not correspond to what we generally mean by the measurement of the state of the light. While this example may seem obtuse, the point is that mathematically projective maps in quantum mechanics act in this manner--they have a significant effect even if the results are ignored. 
\subsection*{Weak versus strong maps for pince-nez}
Building from the preceding example, if  the incoming and outgoing data for a pince-nez are compatible, then the map \textit{L} is termed \textit{strong} if, when marginalizing over the observation, the result (using option \textbf{I'}), $ \int_{\mathcal{R}}L(\rho)\,d\tau $, is far from $ \rho $ for all but a small subset of $ \rho $'s. Conversely, the map \textit{L} is termed  \textit{weak} if $ \int_{\mathcal{R}}L(\rho)\,d\tau $ is close to $ \rho $ for a large subset of $ \rho $'s. Obviously, there is a continuum of possibilities; the strength could be varied depending on an input parameter. For instance, in the double-slit example previously considered, the switch control could be replaced by a continuous slider that would slowly vary the joint probabilities of the output. 

Now we can easily understand the misconception leading to the repeatability criteria; in the classic case, restriction maps (using diagonal projectors) are both repeatable and the weakest possible, having no effect if their observations are ignored. In the quantum case, however, while projective maps (using orthogonal projectors) are still repeatable, they can also be very strong. 
\subsection*{Information as a criteria}
Another misconception is that the result of any pince-nez map can be duplicated by a projective map followed by a garbling of the observation data\footnote{For instance, Dirac seemed to believe ``measurements" were always inherently projective.~\cite{diracrepeat}}. If this were the case, then, by convexity, the information\footnote{Information in the sense of the Shannon definition~\cite{shannon} for the probability measure on the observation set $ \mathcal{R}$.} using a projective map would always be higher than for any other map in every situation. Consequently, the claim by Kochen and Specker that projective ``measurements" give maximal knowledge~\cite{kochenmaximal} would be correct. However, it is easy to see this is false. For instance, the matrices
\begin{equation}
\phi_{1}=\left[\begin{array}{cc}\frac{2}{3}&0\\0&\frac{1}{3}\end{array} \right],\phi_{2}=\left[\begin{array}{cc}\frac{1}{6}&\frac{1}{5}\\\frac{1}{5}&\frac{1}{3}\end{array} \right], \phi_{3}=\left[\begin{array}{cc}\frac{1}{6}&-\frac{1}{5}\\-\frac{1}{5}&\frac{1}{3}\end{array} \right]
\end{equation}
are all positive, sum to the identity, and do not mutually commute. The map $ L:\mathcal{D}(\mathbb{C}^{2})\to \mathbb{R}^{3} $ given by $ L(\rho) =\left(\text{tr }\phi_{1}\rho,\text{tr }\phi_{2}\rho,\text{tr }\phi_{3}\rho\right)$ is completely-positive and norm-preserving (on the positive cone), yet cannot be duplicated by any garbling of a projective map, $ \sum_{k=1}^{2}g_{jk}\text{tr }E_{k}\rho $, for stochastic matrix $ G=[g_{jk}] $ and commuting, complete, orthogonal projectors $ \lbrace E_{1},E_{2}\rbrace $ on $ \mathbb{C}^{2} $ since the matrices $ \lbrace g_{11}E_{1}+g_{12}E_{2},g_{21}E_{1}+g_{22}E_{2},g_{31}E_{1}+g_{32}E_{2}\rbrace $ necessarily mutually commute.

Also, even if it were possible to duplicate the behavior of a particular pince-nez map by an projective map followed by a garbling of the observation data when the module is used in isolation, if the module is then inserted into a larger Bayesian network where conditioning is taking place, the difference in behavior between the particular pince-nez map and the projective map followed by a garbling may become quite large. In particular, for so-called ``weak-measurements"~\cite{aharonov}, which employ both strong and weak pince-nez maps together with both pre- and post-conditioning, the information can be very large. The typical Bayesian network for one of these has the form:
\begin{align}\label{eq:weakmeasurement}
\setlength{\unitlength}{1 in}
\begin{picture}(3,2)
\put(.1,1.78){\circle*{.2}}
\put(.1,1.24){\circle*{.2}}
\put(.1,.38){\circle*{.2}}
\put(1.1,1.18){\circle{.2}}
\put(0,1.68){\line(0,1){.1}}
\put(0,1.14){\line(0,1){.1}}
\put(0,.28){\line(0,1){.1}}
\put(1,1.08){\line(0,1){.1}}
\put(1,1.18){\line(1,0){.19}}
\put(1.095,1.085){\line(0,1){.19}}
\put(3.1,1.78){\circle*{.2}}
\put(3.1,1.18){\circle*{.2}}
\put(3.1,.38){\circle*{.2}}
\put(3,1.68){\line(0,1){.1}}
\put(3,1.08){\line(0,1){.1}}
\put(3,.28){\line(0,1){.1}}
\color{blue}
\put(.1,1.58){\oval(.2,.2)[r]}
\put(0.01,1.68){\line(1,0){.09}}
\put(0.01,1.48){\line(1,0){.09}}
\put(.01,1.48){\line(0,1){.2}}
\put(0,1.58){\line(0,1){.1}}
\put(.1,1.04){\oval(.2,.2)[r]}
\put(.01,1.14){\line(1,0){.09}}
\put(.01,.94){\line(1,0){.09}}
\put(.01,.94){\line(0,1){.2}}
\put(0,1.04){\line(0,1){.1}}
\put(.1,.6){$\vdots$}
\put(.1,.18){\oval(.2,.2)[r]}
\put(.01,.28){\line(1,0){.09}}
\put(.01,.08){\line(1,0){.09}}
\put(.01,.08){\line(0,1){.2}}
\put(0,.18){\line(0,1){.1}}
\put(1.1,.98){\circle{.2}}
\put(1,.98){\line(0,1){.1}}
\put(2.05,.98){\oval(.1,.3)}
\put(3.1,1.58){\oval(.2,.2)[l]}
\put(3.1,1.68){\line(1,0){.1}}
\put(3.1,1.48){\line(1,0){.1}}
\put(3.2,1.48){\line(0,1){.2}}
\put(3,1.58){\line(0,1){.1}}
\put(3.1,.98){\oval(.2,.2)[l]}
\put(3.1,1.08){\line(1,0){.1}}
\put(3.1,.88){\line(1,0){.1}}
\put(3.2,.88){\line(0,1){.2}}
\put(3,.98){\line(0,1){.1}}
\put(3.1,.6){$\vdots$}
\put(3.1,.18){\oval(.2,.2)[l]}
\put(3.1,.28){\line(1,0){.1}}
\put(3.1,.08){\line(1,0){.1}}
\put(3.2,.08){\line(0,1){.2}}
\put(3,.18){\line(0,1){.1}}
\put(.21,1.04){\vector(1,0){.8}}
\put(1.2,.98){\vector(1,0){.8}}
\put(2.1,.98){\vector(1,0){.9}}
\put(.2,1.56){\vector(2,-1){.8}}
\put(.2,.2){\vector(1,1){.8}}
\put(2.1,.98){\vector(1,0){.9}}
\put(2.1,1.05){\vector(3,2){.9}}
\put(2.1,.91){\vector(3,-2){.9}}
\end{picture}
\end{align}
There is both pre- and post-conditioning; those pince-nez have projective maps. The central pince-nez has a map that uses ``fuzzy" projectors with spread of order $ \sqrt{N} $ for \textit{N} particles. Hence, the relative standard deviation in the values assigned to the observation set is only of order $ \frac{1}{\sqrt{N}} $, so the information is high; furthermore, the center of the peak of the distribution can be in unexpected ranges of the observation set, as has been well-publicized recently~\cite{cho}.
\section{Comparison of Bayesian networks to quantum circuits and tensor networks}
A commonly-used, alternative graphical approach for quantum systems is quantum circuits~\cite{mermin}. For example consider the following quantum circuit:
\begin{align}
\setlength{\unitlength}{1 in}
\begin{picture}(3.5,1.2)
\put(1,.7){\oval(.1,1)}
\put(1.7,.9){\oval(.1,.6)}
\put(2.4,.5){\oval(.1,.6)}
\put(3.1,.7){\oval(.1,1)}
\put(1.05,.7){\line(1,0){.6}}
\put(1.05,1.1){\line(1,0){.6}}
\put(1.05,.3){\line(1,0){1.3}}
\put(1.75,.7){\line(1,0){.6}}
\put(1.75,1.1){\line(1,0){1.3}}
\put(2.45,.7){\line(1,0){.6}}
\put(2.45,.3){\line(1,0){.6}}
\put(1,0){$ \psi_{i} $}
\put(1.65,.4){\textit{U}}
\put(2.35,0){\textit{V}}
\put(3.1,0){$ \psi_{f} $}
\put(.7,.25){\textsf{K}}
\put(.7,.65){\textsf{J}}
\put(.7,1.05){\textsf{H}}
\end{picture}
\end{align}
Then for Hilbert spaces \textsf{H}, \textsf{J}, and \textsf{K}, termed the \textit{quantum channels}, and initial state $ \psi_{i}\in \mathsf{H}\otimes\mathsf{J}\otimes\mathsf{K}$, the final state $ \psi_{f} $ is given by $ (I_{\mathsf{H}}\otimes V)(U\otimes I_{\mathsf{K}})\psi_{i} $ for some unitary $ U: \mathsf{H}\otimes\mathsf{J}\to  \mathsf{H}\otimes\mathsf{J} $ and $ V:\mathsf{J}\otimes\mathsf{K}\to\mathsf{J}\otimes\mathsf{K} $. Quantum circuits are obviously directly related to the usual textbook quantum mechanics working with wavefunctions. Following the first textbook rule from above, there is a single overall Hilbert space, the tensor product of all the Hilbert spaces for each quantum channel, that is used throughout. Only unitary operators can be accommodated, following the second textbook rule above.

It is possible to incorporate projective ``measurements" into the quantum circuit, giving a tensor network, but this requires doubling the graph--clearly not an efficient approach graphically. For instance, suppose the initial state is a product state, $ \psi_{i}=\psi_{1}\otimes\psi_{2}\otimes\psi_{3} $ and that a projective ``measurement" is made on the final state, where the projector is of the form $ I_{\mathsf{H}}\otimes P $. Then the calculation of the probability $\Vert I_{\mathsf{H}}\otimes P\psi_{f}\Vert^{2}= \langle I_{\mathsf{H}}\otimes P\psi_{f},\psi_{f}\rangle $ is represented by the tensor network
\begin{align}
\setlength{\unitlength}{1 in}
\begin{picture}(5,1.2)
\put(.3,1.1){\circle{.1}}
\put(.3,.7){\circle{.1}}
\put(.3,.3){\circle{.1}}
\put(1,.9){\oval(.1,.6)}
\put(1.7,.5){\oval(.1,.6)}
\put(2.4,.5){\oval(.1,.6)}
\put(3.1,.5){\oval(.1,.6)}
\put(3.8,.9){\oval(.1,.6)}
\put(4.5,1.1){\circle{.1}}
\put(4.5,.7){\circle{.1}}
\put(4.5,.3){\circle{.1}}
\put(.35,.7){\line(1,0){.6}}
\put(.35,1.1){\line(1,0){.6}}
\put(.35,.3){\line(1,0){1.3}}
\put(1.05,.7){\line(1,0){.6}}
\put(1.05,1.1){\line(1,0){2.7}}
\put(4.45,.7){\line(-1,0){.6}}
\put(4.45,1.1){\line(-1,0){.6}}
\put(4.45,.3){\line(-1,0){1.3}}
\put(1.75,.3){\line(1,0){.6}}
\put(2.45,.3){\line(1,0){.6}}
\put(3.15,.3){\line(1,0){.6}}
\put(1.75,.7){\line(1,0){.6}}
\put(2.45,.7){\line(1,0){.6}}
\put(3.15,.7){\line(1,0){.6}}
\put(.95,.4){\textit{U}}
\put(1.65,0){\textit{V}}
\put(2.35,0){\textit{P}}
\put(3.05,0){\textit{V}*}
\put(3.75,.4){\textit{U}*}
\put(0,.25){$ \psi_{3} $}
\put(0,.65){$ \psi_{2} $}
\put(0,1.05){$ \psi_{1} $}
\put(4.6,.25){$ \psi_{3}^{*} $}
\put(4.6,.65){$ \psi_{2}^{*} $}
\put(4.6,1.05){$ \psi_{1}^{*} $}
\end{picture}
\end{align}
The tensor network has some of the advantages of the Baysian network in that once it is set up, it is possible to look for computational shortcuts.

On the other hand, the Bayesian network has many advantages. It does not depend on the incomprehensible and unmotivated textbook rules, but instead stands on its own reasonable basis. It is graphically more efficient in not having to duplicate itself to include observations. It is also potentially more efficient in allowing non-unitary nodes and non-projective pince-nez, hence avoiding the need to introduce auxiliary spaces, and in having the splitter construction, so it is not necessary to maintain all the quantum circuits throughout the diagram. The Bayesian network indicates which observable nodes are being marginalized or conditioned on and works seamlessly with the usual observable Bayesian networks that are already in common use, so it easily allows models with information coming from previous observations or random factors (such as coin flips). For instance, a tensor network coming from a quantum circuit for the ``weak measurement" example (\ref{eq:weakmeasurement}) would be far more complicated, with the need for auxiliary spaces, yet would still not be able to indicate the post-conditioning graphically (the pre-conditioning could be incorporated into the initial values for the wavefunction).

%% file: chapter6.tex
\chapter{Further examples}
\section{No-cloning--classical and quantum}\label{sec:nocloning}
No-cloning is taken as a hallmark of quantum mechanics. However, it is a property of Bayesian networks more generally, holding even if all the Hilbert spaces are trivial. The non-linear map Clone sends the density matrix-valued vector measure $ \rho $ to $ \rho\otimes\rho $. This cannot be implemented by any device that can be modelled by a Bayesian network since such a network, no matter how complicated, altogether gives rise to a linear map on $ \rho $ by the principle of linearity.

One may still consider something a cloning device if it does not always clone, but only clones conditioned on a observation; for instance, suppose there is a green light for success and a red light for failure. However, to be a cloning device, it must have a finite probability of success for at least two distinct inputs, $ \mu $ and $ \nu $. By the principle of linearity, then every convex combination $ t\mu+(1-t)\nu $ of the two has a finite probability for success. Now consider the following graphical model:
\begin{align}
\setlength{\unitlength}{1 in}
\begin{picture}(5.5,1.2)
\color{blue}
\put(1,.5){\oval(.2,.2)[r]}
\put(.9,.6){\line(1,0){.1}}
\put(.9,.4){\line(1,0){.1}}
\put(.9,.4){\line(0,1){.2}}
\put(1.1,.5){\vector(1,0){.8}}
\put(2.9,.5){\vector(1,0){.8}}
\color{black}
\put(1.9,.1){\dashbox{.1}(1,1)}
\put(3.7,.1){\dashbox{.1}(1,1)}
\put(2.2,.5){cloning}
\put(2.2,.3){device}
\put(4,.5){testing}
\put(4,.3){device}
\put(2.3,.9){\circle{.2}}
\put(2.2,.9){\line(1,0){.19}}
\put(2.3,.809){\line(0,1){.185}}
\put(2.5,.9){light}
\end{picture}
\end{align}
For the device to be cloning, then conditioning on the light being green, the joint probability for the observations on the testing device must be the same as those for the graphical model with the Clone map
\begin{align}
\setlength{\unitlength}{1 in}
\begin{picture}(5.5,1)
\color{blue}
\put(1,.5){\oval(.2,.2)[r]}
\put(.9,.6){\line(1,0){.1}}
\put(.9,.4){\line(1,0){.1}}
\put(.9,.4){\line(0,1){.2}}
\put(1.1,.5){\vector(1,0){1}}
\put(2.7,.5){\vector(1,0){1}}
\put(2.2,.45){Clone}
\color{black}
\put(3.7,.1){\dashbox{.1}(1,1)}
\put(4,.5){testing}
\put(4,.3){device}
\end{picture}
\end{align}
with the same corresponding data for the testing device for each model. For every possibility of what is put in the dashed boxes and what the corresponding information is, there are some linear maps \textit{L} from $ \mathcal{D}(\mathsf{H}) $-valued vector measures on $ \mathcal{X} $ to $ \mathcal{D}(\mathsf{H}\otimes\mathsf{H}) $-valued vector measures on $ \mathcal{X}\times\mathcal{X} $ and \textit{K}, \textit{M} from the latter space to $ [0,1] $ such that the joint probability (conditioning on the light being green) using the purported cloning device is $ \dfrac{(K\circ L)\rho}{(M\circ L)\rho} $, whereas the joint probability using the mathematical Clone map is $K\text{Clone}(\rho)=K(\rho\times \rho)$. With the above assumption, $ (M\circ L)( t\mu+(1-t)\nu))\neq 0 $; however, then we must have
\begin{equation}
(M\circ L)(t\mu+(1-t)\nu)K((t\mu+(1-t)\nu)\times(t\mu+(1-t)\nu))=(K\circ L)(t\mu+(1-t)\nu)\end{equation}
for all $ t\in[0,1] $, which is impossible. 
\section{Teleportation-classical and quantum}
Teleportation is also generally taken as a hallmark of quantum mechanics. However, as will be shown, teleportation is also possible in classical hidden models. 

A device is a teleportation device if, when modelled by a Bayesian network, for a fixed set $ \mathcal{X} $, base measure $ \mu $, Hilbert space $ \mathsf{H} $, and template density-matrix-valued function $ \sigma\in L^{1}(\mathcal{X}\times \mathcal{X};\mu\times\mu;\mathcal{D}(\mathsf{H}\otimes \mathsf{H})^{+}) $, it takes the product function $ \rho\otimes \sigma \in L^{1}(\mathcal{X}\times \mathcal{X}\times \mathcal{X};\mu\times\mu\times\mu;\mathcal{D}(\mathsf{H}\otimes \mathsf{H}\otimes \mathsf{H})^{+}) $ back to $ \rho $ for any $ \rho\in L^{1}(\mathcal{X};\mu;\mathcal{D}(\mathsf{H})^{+}) $ and where a splitter is used so only Alice gets the $ (\mathcal{X},\mu, \mathsf{H}) $ that $ \rho $ lives on, Bob gets the final $ \rho $, and there are no hidden connections between Alice and Bob, but only classic, observable information. Note that because the first textbook rule that the overall Hilbert space is the same throughout is unnecessary for the Bayesian network formulation, there is no need to have another output from Alice (whose value is irrelevant in the context of being a teleportation device)--this leads to simplification in the required calculations.    

There is a Bayesian network for such a device using quantum nodes (based on, but extending, the calculations of Bennett, Brassard, Cr\'{e}peau, Josza, Peres, and Wootters~\cite{bennetttele}) if $ \mathcal{X}=\lbrace 1,2,\ldots, m\rbrace $, $ \mu $ is the counting measure,  $ \mathsf{H} $ is $ \mathbb{C}^{n} $, and $ \sigma  $ is the density-matrix-valued function (written as a column vector with $ m^{2} $ entries, each a $ n^{2}\times n^{2} $-matrix) $ \dfrac{1}{m}\text{ Vec }I_{m}\otimes\left(\dfrac{1}{n}\text{ Vec }I_{n}\left( \text{ Vec }I_{n}\right)^{\text{T}}\right)  $. Note $ \sigma $ is the Kronecker product of Copy applied to the uniform distribution with the usual maximally-entangled state.
The graphical model is:
\begin{equation}
\setlength{\unitlength}{1 in}
\begin{picture}(5.5,2.4)
\color{blue}
\put(1,1.1){\oval(.2,.2)[r]}
\put(.9,1.2){\line(1,0){.1}}
\put(.9,1){\line(1,0){.1}}
\put(.9,1){\line(0,1){.2}}
\put(1,1.6){\oval(.2,.2)[r]}
\put(.9,1.7){\line(1,0){.1}}
\put(.9,1.5){\line(1,0){.1}}
\put(.9,1.5){\line(0,1){.2}}
\put(1.09,1.12){\vector(4,1){.82}}
\put(1.09,1.58){\vector(4,-1){.82}}
\put(1.95,1.35){\oval(.1,.3)}
\put(2.5,2){\oval(.2,.2)[l]}
\put(2.5,2.1){\line(1,0){.1}}
\put(2.5,1.9){\line(1,0){.1}}
\put(2.6,1.9){\line(0,1){.2}}
\put(2.4,1.9){\line(0,1){.1}}
\put(3.5,.3){\circle{.2}}
\put(2,1.4){\vector(1,1){.4}}
\put(2,1.3){\vector(3,-2){1.43}}
\put(3.57,.36){\vector(2,1){.9}}
\put(1,1.75){1}
\put(1,.85){2}
\put(2.4,1.55){4}
\put(3.2,.25){5}
\put(1.9,1){3}
\color{black}
\put(1.7,0){\dashbox{.1}(2.3,2.3)}
\put(2.2,1.2){\dashbox{.1}(1.6,1)}
\put(2.2,.1){\dashbox{.1}(1.6,1)}
\put(2.5,1.8){\circle{.2}}
\put(2.4,1.8){\line(1,0){.19}}
\put(2.5,1.709){\line(0,1){.185}}
\put(2.4,1.8){\line(0,1){.1}}
\put(3.5,.8){\circle{.2}}
\put(3.4,.8){\line(1,0){.19}}
\put(3.5,.709){\line(0,1){.185}}
\put(2.57,1.74){\vector(1,-1){.87}}
\put(2.55,1.72){\line(1,-1){.84}}
\put(3.5,.71){\vector(0,-1){.3}}
\put(3,1.7){Alice}
\put(2.3,.5){Bob}
\put(4.1,.4){teleportation}
\put(4.2,.2){device}
\end{picture}
\end{equation}
The double arrow indicates Bob's observations are identical to Alice's (perfect communication). The information presented in the form according to the graphical model is:
\paragraph*{node 1} $ \left(;\lbrace 1,2,\ldots, m\rbrace,\begin{array}{l}\text{counting}\\\text{measure}\end{array},\mathbb{C}^{n};L_{1}\right) $ where $ L_{1} $ is the constant map with value $ \rho $.
\paragraph*{node 2} $ \left(;\lbrace 1,2,\ldots, m\rbrace\times\lbrace 1,2,\ldots, m\rbrace ,\begin{array}{l}\text{counting}\\\text{measure}\end{array},\mathbb{C}^{n}\otimes \mathbb{C}^{n};L_{2}\right) $ where $ L_{2} $ is the constant map with value $\sigma $.
\paragraph*{splitter 3} $ \left(\lbrace 1,2,\ldots, m\rbrace^{\times 3},\begin{array}{l}\text{counting}\\\text{measure}\end{array},\mathbb{C}^{n^{3}};\right. $
\[\left.\lbrace 1,2,\ldots, m\rbrace\times\lbrace 1,2,\ldots, m\rbrace ,\begin{array}{l}\text{counting}\\\text{measure}\end{array},\mathbb{C}^{n}\otimes \mathbb{C}^{n},\lbrace 1,2,\ldots, m\rbrace,\begin{array}{l}\text{counting}\\\text{measure}\end{array},\mathbb{C}^{n}\right)\]
\paragraph*{pince-nez 4} $\left(\lbrace 1,2,\ldots, m\rbrace\times\lbrace 1,2,\ldots, m\rbrace ,\begin{array}{l}\text{counting}\\\text{measure}\end{array},\mathbb{C}^{n}\otimes \mathbb{C}^{n};\right. $
\[\left. \lbrace 1,2,\ldots, m\rbrace\times\lbrace 1,2,\ldots, n\rbrace\times\lbrace 1,2,\ldots, n\rbrace,\begin{array}{l}\text{counting}\\\text{measure}\end{array} ;;L_{4}\right)\]
Let $ \Omega $ be the $ n\times n $-matrix with the $ n $th roots of unity along its diagonal, $ S $ be the $ n\times n $-shift-matrix with a one in the upper-right corner and ones on the subdiagonal, and $ Q $ be the $ m\times m $-shift-matrix with a one in the upper-right corner and ones on the subdiagonal. For $ k,l\in \lbrace 1,2,\ldots, n\rbrace $, let $ w_{kl} $ be the column vector $ \left( I_{n}\otimes (\Omega^{l-1}S^{k-1})\right)  \text{ Vec }I_{n} $. Then for $ (j,k,l)\in \lbrace 1,2,\ldots, m\rbrace\times\lbrace 1,2,\ldots, n\rbrace\times\lbrace 1,2,\ldots, n\rbrace $, $ L_{4}(\tau)_{jkl} $ is given by
\begin{equation}
\left( \text{ Vec }Q^{j-1}\right)^{\text{T}}\left((I_{m^{2}}\otimes w_{kl}^{*} )\tau(I_{m^{2}}\otimes w_{kl} )\right)
\end{equation} 
\paragraph*{node 5}$ \left(\lbrace 1,2,\ldots, m\rbrace\times\lbrace 1,2,\ldots, m\rbrace\times\lbrace 1,2,\ldots, n\rbrace\times\lbrace 1,2,\ldots, n\rbrace ,\begin{array}{l}\text{counting}\\\text{measure}\end{array},\mathbb{C}^{n};\right. $
\[\left. \lbrace 1,2,\ldots, m\rbrace,\begin{array}{l}\text{counting}\\\text{measure},\end{array} \mathbb{C}^{n};L_{5}\right)\]
The incoming density-matrix-valued function $ \tau $ will be considered as a $ m\times n\times n $-array, indexed by $ j,k,l $, of column vectors of $ m $ entries each, where each entry is a $ n\times n $-matrix. With this convention, $ L_{5}(\tau) $ is given by
\begin{equation}
\sum_{j=1}^{m}\sum_{k,l=1}^{n}((Q^{j-1})^{\text{T}}\otimes I_{n})\left((I_{m}\otimes(\Omega^{l-1}(S^{k-1})^{\text{T}}))\tau_{jkl}(I_{m}\otimes(S^{k-1}\overline{\Omega}^{l-1})\right)
\end{equation} 
\medskip\\
Then, for any following testing device, the incoming density-matrix-valued function from Bob if Alice observes $ (j,k,l)\in \lbrace 1,2,\ldots, m\rbrace\times\lbrace 1,2,\ldots, n\rbrace\times\lbrace 1,2,\ldots, n\rbrace  $ is
\begin{equation}
L_{5}\left(\left.((L_{4}\otimes I_{\mathcal{B}(L^{1}(\mathcal{X};\mu;\mathsf{H}))}))(L_{1}\otimes L_{2}))\right\vert_{j,k,l}1_{\lbrace (j,k,l)\rbrace}\right)
\end{equation}
This apparently complicated expression is just $ \rho $, regardless of which particular values of $ j,k,l $ were observed by Alice and sent to Bob.
\subsection*{The classical case}
Note the preceding is meaningful for the case of trivial Hilbert space, $ n=1 $, so teleportation is not a quantum phenomenon. Another way to achieve the same output in this purely-classical case is to not have a template shared by Alice and Bob, but rather to have the pince-nez for Alice have simply the identity for its map (so the hidden node is not really hidden, but is actually observable). Then Alice sends her information to Bob, who makes a copy using the information. Finally, Alice and Bob forget what the information was (so their nodes are marginalized). The result is that the incoming probability distribution to Alice is the same as that outgoing from Bob. 

However, this second approach is not teleportation because of the forgetting step. Also, for the classical case of the teleportation given above, an eavesdropper to the information sent from Alice to Bob without access to the template cannot replicate the probability distribution, whereas in the second, non-teleportation approach, the eavesdropper could not only replicate the probability distribution by making his own copy, then forgetting, but by not forgetting, would actually have more information.     
\subsection*{Does a teleportation device really teleport?}
Note the graphical model for the teleportation device is simply a graph fragment since the output from Bob is not terminated. One may hope its behavior is universal, but is always possible that additional testing will reveal it is not. For instance, before the discovery of quantum mechanics, it may have been believed that the classical teleportation discussed above was ``true" teleportation. It is possible that additional mathematical structures besides measures and density matrices (see \S\ref{sec:additional} above) can be used to calculate probabilities in Bayesian networks and that these are found to be useful in practice--then the current belief that quantum telportation is ``true" teleportation will also be shown false.   
\section{Bell's inequality for Bayesian networks without metaphysical limitations}\label{sec:bells}
\subsection*{Introduction}
The standard proofs of Bell's inequality~\cite{bell} make assumptions based on assuming the underlying reality of hidden variables, which in our language is equivalent to the existence of transition probability function. As we explored in \S\ref{sec:transition}, this is a metaphysical notion for hidden nodes. For our Bayesian networks, no such limitation is placed, which enlarges the space of possible maps that can be employed. Although, as we already mentioned in \S\ref{sec:transition}, these additional maps are not necessarily of great interest, we would like to show that Bell's inequality still necessarily holds.
\subsection*{Set-up}
Following Clauser, Horne, Shimony, and Holt~\cite{clauser}, consider the case of two observers, Alice and Bob, who are at distant locations. Each has a box, connected by a long cable to the other, with a dial and two lights, marked zero and one, which light periodically. They record their observations, then come together later to compare notes. They find that the flashes are independent in time and depend on the settings of the dial, $ \alpha $ for Alice and $ \beta $ for Bob, but for each round the lights they each saw were not independent of each other, but had a joint probability distribution given by the four functions of $ \alpha,\beta $: $ \text{Prob}\left((0,0)\vert \alpha,\beta\right) $, $ \text{Prob}\left((0,1)\vert \alpha,\beta\right) $, $ \text{Prob}\left((1,0)\vert \alpha,\beta\right) $, and $ \text{Prob}\left((1,1)\vert \alpha,\beta\right) $.
Presenting the information in this manner corresponds to the graphical model:
\begin{align}
\setlength{\unitlength}{1 in}
\begin{picture}(4,1.5)
\put(2,1.3){\circle{.2}}
\put(1.9,1.3){\line(1,0){.19}}
\put(2,1.209){\line(0,1){.185}}
\put(2.07,1.24){\line(1,-3){.29}}
\put(2.37,.3){\circle{.2}}
\put(2.27,.3){\line(1,0){.19}}
\put(2.37,.209){\line(0,1){.185}}
\put(1.55,1.35){\circle*{.2}}
\put(1.6,.1){\circle*{.2}}
\put(1.62,1.32){\line(6,-1){.29}}
\put(1.6,1.28){\line(5,-6){.74}}
\put(1.67,.15){\line(4,1){.6}}
\put(1.65,.17){\line(1,3){.35}}
\put(2.4,1.4){Alice}
\put(1.45,.5){Bob}
\put(1.3,1.25){$ \alpha $}
\put(1.75,.02){$ \beta $}
\end{picture}
\end{align}
There are the additional no-signalling constraints on the probabilities arising from relativity: 
\begin{align}\label{eq:nosignalling}
\text{Prob}\left((0,0)\vert \alpha,\beta\right)+\text{Prob}\left((0,1)\vert \alpha,\beta\right)\text{ is independent of }\beta\\\text{Prob}\left((1,0)\vert \alpha,\beta\right)+\text{Prob}\left((1,1)\vert \alpha,\beta\right)\text{ is independent of }\beta\nonumber\\\text{Prob}\left((0,0)\vert \alpha,\beta\right)+\text{Prob}\left((1,0)\vert \alpha,\beta\right)\text{ is independent of }\alpha\nonumber\\\text{Prob}\left((0,1)\vert \alpha,\beta\right)+\text{Prob}\left((1,1)\vert \alpha,\beta\right)\text{ is independent of }\alpha\nonumber
\end{align}
If these constraints were not met, then Alice and Bob could use their dial settings to transmit information superluminally. However, relativity places no additional constraint on the bounds for the value of, for $ \alpha,\alpha' $ two settings of Alice's dial and $ \beta,\beta' $ two settings of Bob's dial,
\begin{align}\label{eq:Bellsinequality}
&\text{Prob}\left((0,0)\vert \alpha,\beta\right)+\text{Prob}\left((1,1)\vert \alpha,\beta\right)-\text{Prob}\left((0,1)\vert \alpha,\beta\right)-\text{Prob}\left((1,0)\vert \alpha,\beta\right)\\+&\text{Prob}\left((0,0)\vert \alpha',\beta\right)+\text{Prob}\left((1,1)\vert \alpha',\beta\right)-\text{Prob}\left((0,1)\vert \alpha',\beta\right)-\text{Prob}\left((1,0)\vert \alpha',\beta\right)\nonumber\\+&\text{Prob}\left((0,0)\vert \alpha,\beta'\right)+\text{Prob}\left((1,1)\vert \alpha,\beta'\right)-\text{Prob}\left((0,1)\vert \alpha,\beta'\right)-\text{Prob}\left((1,0)\vert \alpha,\beta'\right)\nonumber\\-& \text{Prob}\left((0,0)\vert \alpha',\beta\right)-\text{Prob}\left((1,1)\vert \alpha',\beta\right)+\text{Prob}\left((0,1)\vert \alpha',\beta\right)+\text{Prob}\left((1,0)\vert \alpha',\beta\right)\nonumber
\end{align}
which can still achieve its bound in magnitude arising from the rules of probability, namely four.
\subsection*{The standard hidden variable approach}
A sufficient way to insure the restrictions from relativity are met is to restrict the possible graphical model to the local model:
\begin{align}
\setlength{\unitlength}{1 in}
\begin{picture}(4,1.7)
\put(.7,.75){\circle{.2}}
\put(.77,.81){\vector(3,1){1.2}}
\put(.8,.75){\vector(4,-1){1.52}}
\put(2,1.3){\circle{.2}}
\put(1.9,1.3){\line(1,0){.19}}
\put(2,1.209){\line(0,1){.185}}
\put(2.37,.3){\circle{.2}}
\put(2.27,.3){\line(1,0){.19}}
\put(2.37,.209){\line(0,1){.185}}
\put(1.55,1.35){\circle*{.2}}
\put(1.6,.1){\circle*{.2}}
\put(1.62,1.32){\vector(1,0){.28}}
\put(1.67,.15){\vector(4,1){.6}}
\put(2.4,1.4){Alice}
\put(1.45,.3){Bob}
\put(1.3,1.25){$ \alpha $}
\put(1.75,.02){$ \beta $}
\end{picture}
\end{align}
Then, if the space of values for the marginalized node is $ \mathcal{X} $ and its probability measure is $ \rho $, the joint probability both Alice and Bob get 1 is
\begin{equation}\label{eq:integralformBell}
\int_{x\in \mathcal{X}}\text{Prob}\left(1\text{ for Alice}\vert \alpha,x\right)\text{Prob}\left(1\text{ for Bob}\vert \beta,x\right)\,d\rho(x)
\end{equation}
with the other joint probabilities given similarly. Using the standard approach for Bell's inequality~\cite{clauser}, it can be shown that for the combination of probabilities in (\ref{eq:Bellsinequality}), the bound on its magnitude has now been reduced to two.
\subsection*{Using classical hidden nodes}
A sufficient way to insure the restrictions from relativity are met using classical hidden nodes is to restrict the possible graphical model to to the local model:
\begin{align}
\setlength{\unitlength}{1 in}
\begin{picture}(3,1.9)
\color{red}
\put(.7,.85){\oval(.2,.2)[r]}
\put(.6,.95){\line(1,0){.1}}
\put(.6,.75){\line(1,0){.1}}
\put(.6,.75){\line(0,1){.2}}
\put(2,1.3){\oval(.2,.2)[l]}
\put(2,1.4){\line(1,0){.1}}
\put(2,1.2){\line(1,0){.1}}
\put(2.1,1.2){\line(0,1){.2}}
\put(2.38,.51){\oval(.2,.2)[l]}
\put(2.38,.61){\line(1,0){.1}}
\put(2.38,.41){\line(1,0){.1}}
\put(2.48,.41){\line(0,1){.2}}
\put(1.15,.85){\oval(.1,.3)}
\put(1.9,1.3){\line(0,1){.1}}
\put(2.27,.4){\line(0,1){.1}}
\put(.78,.85){\vector(1,0){.31}}
\put(1.2,.9){\vector(3,2){.7}}
\put(1.2,.8){\vector(3,-1){1.07}}
\put(.7,.6){1}
\put(1.15,.55){4}
\put(2.15,1.3){2}
\put(2.35,.65){3}
\color{black}
\put(1.35,0){\dashbox{.1}(1.5,.8)}
\put(1.35,.9){\dashbox{.1}(1.5,.8)}
\put(2,1.5){\circle{.2}}
\put(1.9,1.5){\line(1,0){.19}}
\put(2,1.409){\line(0,1){.185}}
\put(1.9,1.4){\line(0,1){.1}}
\put(2.37,.3){\circle{.2}}
\put(2.27,.3){\line(1,0){.19}}
\put(2.37,.209){\line(0,1){.185}}
\put(2.27,.3){\line(0,1){.1}}
\put(1.55,1.55){\circle*{.2}}
\put(1.6,.2){\circle*{.2}}
\put(1.62,1.52){\vector(4,-1){.29}}
\put(1.67,.25){\vector(4,1){.6}}
\put(2.4,1.5){Alice}
\put(1.5,.4){Bob}
\put(1.5,1.35){$ \alpha $}
\put(1.72,.12){$ \beta $}
\end{picture}
\end{align}
The boxes are drawn around the two separate locations to indicate these are potentially universal modules.
\paragraph{Theorem 6.3.1} Employing either option \textbf{I} or \textbf{II}, Bell's inequality still holds.
\paragraph*{Proof} Using option \textbf{II}, by the Riesz theorem~\cite{roydenriesz}, each $  \text{Prob}\left((j,k)\vert \alpha,\beta\right) $ is necessarily of the form 
\begin{equation}
\int_{\mathcal{Y}\times\mathcal{Z}}f_{j}(y;\alpha)g_{k}(z;\alpha)\;d\mu(y,z)
\end{equation}
for some positive, continuous functions $ f_{1},f_{2},g_{1},g_{2} $ with $ f_{1}+f_{2}=1_{\mathcal{Y}\times[0,2\pi)}$, $g_{1}+g_{2}=1_{\mathcal{Z}\times[0,2\pi)} $  and some unit-norm, Radon measure $ \mu $ on $ \mathcal{Y}\times\mathcal{Z} $. This is mathematically the same form as (\ref{eq:integralformBell}), if lacking the interpretation in terms of conditional probabilities and probability measures, so the standard argument~\cite{clauser} still applies with bound two for (\ref{eq:Bellsinequality}).

Using option \textbf{I}, each $  \text{Prob}\left((j,k)\vert \alpha,\beta\right) $ is necessarily of the form 
\begin{equation}
(L_{2}\otimes L_{3})\left(\mu\times\rho\times\nu\right)(\lbrace(j,k)\rbrace)
\end{equation}  
where $ \mu $ is some unit-norm measure on $ [0,2\pi) $ concentrated about $ \alpha $ from a collection of measures $ \mathcal{M}_{5} $, $ \nu $ is some unit-norm measure on $ [0,2\pi) $ concentrated about $ \beta $ from the same collection of measures, $\rho $ is some unit-norm measure on $ \mathcal{Y}\times\mathcal{Z} $ from some collection of measures $ \mathcal{M}_{4} $, $ L_{2} $ is a norm-preserving (on the positive cone) map from measures on $ [0,2\pi)\times\mathcal{Y} $ in some collection of measures $ \mathcal{M}_{2} $ to measures on $ \lbrace 0,1\rbrace $ with set algebra the power set $ \lbrace \varnothing, \lbrace 0\rbrace, \lbrace 1\rbrace, \lbrace 0,1\rbrace \rbrace $, and $ L_{3} $ is a norm-preserving (on the positive cone) map from measures on $ \mathcal{Z}\times [0,2\pi)$ in some collection of measures $ \mathcal{M}_{3} $ to measures on $ \lbrace 0,1\rbrace $ with set algebra the power set $ \lbrace \varnothing, \lbrace 0\rbrace, \lbrace 1\rbrace, \lbrace 0,1\rbrace \rbrace $. From the rules for option \textbf{I}, all the collections of measures are absolutely-continuous-complete and are such that $ L_{2}\otimes L_{3} $ is well-defined. However, by \textbf{A1.2}, \textbf{A1.3}, \textbf{B1.3}, \textbf{B1.4}, and \textbf{B2.5}, this implies the bound can be established just be considering $  \text{Prob}\left((j,k)\vert \alpha,\beta\right) $ of the form
\begin{equation}
(L_{2}\otimes L_{3})\left(\sum_{l}\tau_{l}\times \sigma_{l}\right)(\lbrace(j,k)\rbrace)
\end{equation}  
where $ \sum_{l}\tau_{l}\times \sigma_{l} $ is a unit-norm, finite-tensor-rank measure in $ \mathcal{M}_{2}\otimes\mathcal{M}_{3} $. Using this form for $ \text{Prob}\left((j,k)\vert \alpha,\beta\right) $, the usual bound of two readily follows for (\ref{eq:Bellsinequality}). $ \square $
\subsection*{Comments}
Using the well-known quantum model of Clauser, Horne, Shimony, and Holt~\cite{clauser} (which reuses some of the modules from the double-slit experiment (see \S\ref{sec:quantumdouble}) in a different arrangement--an example of universality), it is possible to violate Bell's inequality and the combination of probabilities in (\ref{eq:Bellsinequality}) can even achieve Tsirelson's bound of $ 2\sqrt{2} $~\cite{tsirelson}. Any classical hidden model duplicating these results must be nonlocal; hence, contextual--certainly the modules we used for the classical hidden model for the double slit experiment (see \S\ref{sec:classicaldouble}) would be of no use.

We then have the question of whether Tsirelson's bound can be broken. This may be possible if there are additional mathematical structures besides measures and density matrices that can be utilized in Bayesian networks while respecting the principles of positivity and potential universality (see \S\ref{sec:additional} above).

%% file: chapter8.tex
\chapter{A Parrondo-like paradox for an one-round game}
\section{Defining the game}\label{sec:definingtheparrondolikegame}
Instead of the multi-round game considered in the preceding chapter, now consider a single round quantum game where the winning criteria is still taken to be that the observation is in some specified set \textit{A}. Instead of having two maps which are combined by a coin flip, the map is now fixed and only the initial states are varied. If they were combined in a convex combination by flipping a coin, there would clearly be no Parrondo's paradox since the probabilities depend linearly on the initial state. However, if the initial states were constrained to be rank one (hence, described by a wavefunction), instead of a convex combination, one may consider the minimizing geodesic joining them. If two initial states both give probability greater than one-half for the first player, Alice, to win, but there is somewhere on the minimizing geodesic where the second player, Bob, has probability greater than one-half to win, we will term that a Parrondo-like paradox. 

Suppose there is a continuous control, say a slider with continuous values from zero to one, which varies the initial state along a geodesic in the space of rank-one density matrices. Using the Bayesian network model, this initial state is input into a fixed pince-nez map and the outcome observed. The graphical model is:
\begin{align}\label{fig:quantumoneround}
\setlength{\unitlength}{.8 in}
\begin{picture}(1,1.1)
\color{blue}
\put(.1,.1){\oval(.2,.2)[r]}
\put(0,.2){\line(1,0){.1}}
\put(0,0){\line(1,0){.1}}
\put(0,0){\line(0,1){.2}}
\put(1,.1){\oval(.2,.2)[l]}
\put(1,.2){\line(1,0){.1}}
\put(1,0){\line(1,0){.1}}
\put(1.1,0){\line(0,1){.2}}
\put(.2,.1){\vector(1,0){.7}}
\put(.89,.1){\line(0,1){.1}}
\color{black}
\put(-.1,.8){\circle*{.2}}
\put(.2,.8){$S\in [0,1]$}
\put(-.05,.72){\vector(1,-4){.13}}
\put(1,.3){\circle{.2}}
\put(.9,.3){\line(1,0){.19}}
\put(1,.209){\line(0,1){.185}}
\put(.89,.2){\line(0,1){.1}}
\end{picture}
\end{align}
Employing option \textbf{I'} (see \S\ref{sec:optionIandII}), the data for the initial node is $ ([0,1],\text{Lebesgue};\mathsf{H}; K) $ for some given Hilbert space \textsf{H}. The map \textit{K} will be determined below in \S\ref{subsec:completingthedefinition}. The data for the pince-nez is $ (\mathsf{H};\mathcal{R},\tau;; L) $ for some given observable set $ \mathcal{R} $ with base measure $ \tau $. By the rules of the game, it is possible to reduce to the case where the $ \sigma $-algebra is $ \lbrace\varnothing, A, \tilde{A},\mathcal{R}\rbrace $ with base measure $ \tau $ assigning both \textit{A} and $ \tilde{A} $ the value one. Suppose \textit{L} is also given.

The probability for the observation to be in set \textit{A} (so Alice wins) is then
\begin{equation}
\int_{A}(L\circ K)\nu\, d\tau =((L\circ K)\nu)\vert_{A}
\end{equation} 
\section{Defining geodesics on the space of wavefunctions}
Let $ \mathbb{S}_{\mathsf{H}}\subset\mathsf{H} $ be the unit ball. For any $\zeta,\eta\in \mathcal{S} $, let $ \sim $ be the equivalence relation $ \zeta\sim \eta $ if $ \zeta =\omega \eta $ for some phase $ \omega\in\mathbb{S}^{1}\subset\mathbb{C} $ and let $ [\cdot] $ denote the equivalence classes. Consider a wavefunction $\psi\in \mathbb{S}_{\mathsf{H}} $. Since only the rank-one density matrix $\psi\psi^{*}=\psi\langle\cdot,\psi\rangle $ is meaningful, $ \psi $ is only defined up to an overall phase, so what should actually be considered is the equivalence class $ [\psi] $ in the quotient space $  \mathbb{S}_{\mathsf{H}}/\sim $. The question is then what is the correct topology and metric structure to place on this quotient space.
\subsection*{Choice one--using the trace norm}
Using the topology and metric structure inherited from the placement of rank-one density matrices within all density matrices, consider the choice of metric dist$ {}_{\text{trace}} $ on $ \mathbb{S}_{\mathsf{H}}/\sim $ given by, for any $ [\xi],[\psi]\in \mathcal{S}/\sim $,
\begin{equation}
\text{dist}_{\text{trace}}([\xi],[\psi])=\text{tr }\left\vert\xi\xi^{*} -\psi\psi^{*}\right\vert=2\sqrt{1-\left\vert \langle\psi,\xi\rangle\right\vert^{2}}
\end{equation}
The last equality holds since span${}_{\mathbb{C}}\lbrace \xi,\psi\rbrace $ is an invariant subspace of the operator
\begin{equation}
\left( \xi\xi^{*} -\psi\psi^{*}\right)^{*}\left( \xi\xi^{*} -\psi\psi^{*}\right)= \left( \xi\xi^{*} -\psi\psi^{*}\right)^{2}
\end{equation}
with eigenvalue $ 1-\left\vert \langle\psi,\xi\rangle\right\vert^{2} $ while the orthogonal subspace is the kernel of the operator.
\subsection*{Choice two--using the round metric on $ \mathbb{S}_{\mathsf{H}} $}
Let dist$ {}_{\text{round}} $ be the round metric on $ \mathbb{S}_{\mathsf{H}} $, which is the standard metric induced by its embedding in $ \mathsf{H} $ equipped with its norm, so, for any $\zeta,\eta\in \mathbb{S}_{\mathsf{H}} $,
\begin{equation}
\text{dist}_{\text{round}}(\zeta,\eta)=\arccos\Re\langle\zeta,\eta\rangle 
\end{equation}
Then let the metric dist$ {}_{\sim\text{round}} $ on $ \mathbb{S}_{\mathsf{H}}/\sim $ be given by the usual prescription for quotient spaces, so, for any $ [\xi],[\psi]\in \mathbb{S}_{\mathsf{H}}/\sim $,
\begin{equation}
\text{dist}_{\sim\text{round}}([\xi],[\psi])=\min_{w\in\mathbb{S}^{1}\subset\mathbb{C}}\text{dist}_{\text{round}}(\xi,w\psi)=\arccos\max_{w\in\mathbb{S}^{1}\subset\mathbb{C}}\Re \langle\psi,w\xi\rangle =\arccos\left\vert \langle\psi,\xi\rangle\right\vert
\end{equation}
Then
\begin{equation}
\text{dist}_{\text{trace}}([\xi],[\psi])=2\sin\text{dist}_{\sim\text{round}}([\xi],[\psi])
\end{equation}
\[\Leftrightarrow\text{dist}_{\sim\text{round}}([\xi],[\psi])=\arcsin \frac{1}{2}\text{dist}_{\text{trace}}([\xi],[\psi])\]
In particular, the metrics are equivalent and give rise to the same topology on $  \mathbb{S}_{\mathsf{H}}/\sim $. 
\subsection*{Geodesics on the quotient space}
Fix any $ \xi,\psi\in \mathbb{S}_{\mathsf{H}} $ with corresponding equivalence classes $ [\xi],[\psi]\in \mathbb{S}_{\mathsf{H}}/\sim $. Let $ \hat{\xi} $ be an element of $ [\xi] $ such that $\text{dist}_{\text{round}}(\hat{\xi},\psi)=\text{dist}_{\sim\text{round}}([\xi],[\psi]) \Leftrightarrow   \langle\psi,\hat{\xi}\rangle=\left\vert \langle\psi,\xi\rangle\right\vert$, so $\hat{\xi}=\dfrac{ \langle\psi,\xi\rangle}{\vert \langle\psi,\xi\rangle\vert}\xi$  if $\langle\psi,\xi\rangle\neq 0  $ while $ \hat{\xi} $ can be any element of $ [\xi] $ if $ \langle\psi,\xi\rangle=0 $. Now take the arc $  \gamma:\left[0,\text{dist}_{\text{round}}(\hat{\xi},\psi)\right]\to\mathbb{S}_{\mathsf{H}} $ of the great circle in $ \mathbb{S}_{\mathsf{H}} $ connecting $ \psi $ and $ \hat{\xi} $ with the standard parametrization,
given by
\begin{equation}\label{eq:geodesicequation}
\gamma(\theta)=\left( \cos \theta\right)\psi+\left( \sin \theta\right) \dfrac{\hat{\xi}-\langle\hat{\xi},\psi\rangle\psi}{\Vert\hat{\xi}-\langle\hat{\xi},\psi\rangle\psi\Vert}
\end{equation}
where $ \Vert\hat{\xi}-\langle\hat{\xi},\psi\rangle\psi\Vert=\sqrt{\left\langle\hat{\xi}-\langle\hat{\xi},\psi\rangle\psi,\hat{\xi}-\langle\hat{\xi},\psi\rangle\psi \right\rangle  }= \sqrt{1-\langle\hat{\xi},\psi\rangle^{2} }$.
\paragraph{Proposition 8.2.1}
The curve $ [\gamma] $ in $  \mathbb{S}_{\mathsf{H}}/\sim $ is an unit speed geodesic with respect to the metric dist$ {}_{\sim\text{round}} $.
\paragraph*{Proof} For any $ \theta\in \left(0,\text{dist}_{\text{round}}(\hat{\xi},\psi)\right) $,
\begin{align}
\text{dist}_{\sim\text{round}}([\gamma(\theta)],[\psi])+\text{dist}_{\sim\text{round}}([\xi],[\gamma(\theta)])&=\text{dist}_{\text{round}}(\gamma(\theta),\psi)+\text{dist}_{\text{round}}(\hat{\xi},\gamma(\theta))\\&=\text{dist}_{\text{round}}(\hat{\xi},\psi)\nonumber\\&=\text{dist}_{\sim\text{round}}([\xi],[\psi])\nonumber
\end{align}
so not only is the curve $ [\gamma] $ a geodesic, but by further partitioning, the metric is seen to be precisely the arclength of the geodesic. Also, for any $ \theta\in \left(0,\text{dist}_{\text{round}}(\hat{\xi},\psi)\right) $,
\begin{equation}
\lim_{\varepsilon\to 0^{+}}\frac{1}{\varepsilon}\,\text{dist}_{\sim\text{round}}([\gamma(\theta+\varepsilon) ],[\gamma(\theta)])=\lim_{\varepsilon\to 0^{+}}\frac{1}{\varepsilon}\,\arccos\left(\cos(\varepsilon) \right)=\lim_{\varepsilon\to 0^{+}}\frac{\vert\varepsilon\vert}{\varepsilon}=1
\end{equation}
so the curve is unit speed. $ \square $
\paragraph{Proposition 8.2.2}
The curve $ [\gamma] $ in $  \mathbb{S}_{\mathsf{H}}/\sim $ is a geodesic with respect to the metric dist$ {}_{\text{trace}} $.
\paragraph*{Proof} Since $2\sin z=2z+\mathcal{O}(z^{3})$ as $ z\to 0 $, the arclength of a curve with respect to the metric dist$ {}_{\text{trace}} $ is simply twice the arclength with respect to the metric dist$ {}_{\sim\text{round}} $. Hence, the two metrics have the same geodesics.$ \square $
\subsection*{Completing the definition of the game.}\label{subsec:completingthedefinition}
We still need to define the initialization map \textit{K} from \S\ref{sec:definingtheparrondolikegame} which uses the slider position to determine a point on the geodesic. From the preceding results on geodesics, \textit{K} is determined by first fixing the endpoints $  [\xi],[\psi]\in \mathbb{S}_{\mathsf{H}}/\sim  $. Let $ \delta=\text{dist}_{\sim\text{round}}([\xi],[\psi]) $. Then, for $ \nu $ a measure on $ [0,1] $ absolutely continuous with respect to Lebesgue measure $ \lambda $,
\begin{equation}
K\frac{d\nu}{d\lambda}=\int_{x\in[0,1]}\gamma(x\delta)\gamma(x\delta)^{*}\;d\nu(x)
\end{equation} 
If $ \nu $ is sufficiently concentrated, then, as desired, the image of \textit{K} is approximately rank-one. The complication of dealing with a concentrated measure $ \nu $ for the slider position is another instance of the previously mentioned problem (see \S\ref{subsec:commentsonoptions} and \S\ref{subsec:commentsblackboxoption}) of inputting parameters encountered when employing option \textbf{I'}--the value of the slider cannot be read in directly. For purposes of simplicity and clarity, in the remainder we will instead simply suppose we can directly select the desired point on the geodesic and work with strictly rank-one density matrices.   
\section{Bounds on the extent of the Parrondo-like paradox}
To quantify the extent of the Parrondo-like paradox, by analogy to \S\ref{ch:parrondo} we have $ P_{A}$, $ P'_{A} $,  as the probability Alice wins with initial wavefunction $ [\xi] $ and $ [\psi] $ respectively. Define $ P_{A}^{\text{geo}} $ as the probability Alice wins with initial wavefunction for a specified point on the geodesic joining $ [\xi] $ and $ [\psi] $. Analogously to definition \textbf{7.5.2}, we have the following:
\paragraph*{Definition 8.3.1} The \textit{quantum allowed region}, denoted $ \mathcal{Q}(A,\mathcal{R},\tau,\mathsf{H}) $, is the set of all $\left(P_{A},P'_{A},P_{A}^{\text{geo}}\right)\in[0,1]^{\times 3} $ that occur--for fixed \textit{A}, $ \mathcal{R}$, $ \tau$, and \textsf{H}--over all allowed pince-nez maps \textit{L}, initial wavefunctions $ [\xi] $ and $ [\psi] $, and points on the geodesic joining them.
\medskip\\
The paradox can occur if $ \mathcal{Q} $ intersects the cube $ \left(\frac{1}{2},1\right]\times \left(\frac{1}{2},1\right]\times \left[0,\frac{1}{2}\right) $. We have the following theorem giving $ \mathcal{Q} $ precisely:
\paragraph{Theorem 8.3.2} If the Hilbert space \textsf{H} is nontrivial, $ \mathcal{Q} $ is the closed region 
\[\max\lbrace 0,P_{A}+P'_{A}-1\rbrace\leq P_{A}^{\text{geo}}\leq \min\lbrace P_{A}+P'_{A},1\rbrace\]
\paragraph{Proof} Since the bounded operators are dual to the trace-class ones, there is some self-adjoint operator $ \eta\in\mathcal{B}(\mathsf{H}) $ such that $ L(\rho)\vert_{A}=\text{tr }\eta\rho $ for every $ \rho\in\mathcal{D}(\mathsf{H})^{+} $ and, by \textbf{B5.15} and \textbf{B5.16}, every $ \rho\in\mathcal{S}_{1}(\mathsf{H}) $. The condition on $ \eta $ is that it is in the order interval $ 0\leq \eta \leq I_{\mathsf{H}} $.  Let the $ 2\times 2 $-, Hermitian matrix $ B=[b_{jk}] $ be given by
\begin{equation}\label{eq:defineB}
\left[\begin{array}{cc}L(\psi\psi^{*})\vert_{A}&L(\psi\hat{\xi}^{*})\vert_{A}\\L(\hat{\xi}\psi^{*})\vert_{A}&L(\xi\xi^{*})\vert_{A}\end{array}\right]=\left[\begin{array}{cc}\langle\eta\psi,\psi\rangle&\langle\eta\psi,\hat{\xi}\rangle\\\langle\eta\hat{\xi},\psi\rangle&\langle\eta\xi,\xi\rangle\end{array}\right]
\end{equation}
Let $ \mathcal{B} $ be the set of all such matrices over $ 0\leq \eta \leq I_{\mathsf{H}} $ and $ \psi,\xi\in\mathbb{S}_{\mathsf{H}} $. 

To get a simpler characterization of $ \mathcal{B} $, let $ \mathcal{C} $ be the union of order intervals of $ 2\times 2 $-, Hermitian matrices given by
\begin{equation}
\bigcup_{\delta\in[0,\frac{\pi}{2}]}\left\lbrace 0\leq C\leq \left[\begin{array}{cc}1&\cos \delta\\\cos\delta&1\end{array}\right]\right\rbrace
\end{equation} 
The claim is that $ \mathcal{B}=\mathcal{C} $. To see this is true, take any $ B\in\mathcal{B} $. Since $ \eta\geq 0 $, for any $ a,b\in\mathcal{C} $,
\begin{equation}
\left[\begin{array}{cc} \overline{a}&\overline{b}\end{array}\right]B\left[\begin{array}{c} a\\b\end{array}\right]=\left[\begin{array}{cc} \overline{a}&\overline{b}\end{array}\right]\left[\begin{array}{cc}\langle\eta\psi,\psi\rangle&\langle\eta\psi,\hat{\xi}\rangle\\\langle\eta\hat{\xi},\psi\rangle&\langle\eta\xi,\xi\rangle\end{array}\right]\left[\begin{array}{c} a\\b\end{array}\right]=\left\langle \eta(a\psi+b\hat{\xi}),a\psi+b\hat{\xi}\right\rangle\geq 0 
\end{equation}
Therefore, $ B\geq 0 $. Similarly,  since $ I_{\mathsf{H}}-\eta\geq 0 $, for $ \cos\delta=\langle\psi,\hat{\xi}\rangle $,
\begin{align}
\left[\begin{array}{cc} \overline{a}&\overline{b}\end{array}\right]\left(\left[\begin{array}{cc}1&\cos \delta\\\cos\delta&1\end{array}\right]-B\right)\left[\begin{array}{c} a\\b\end{array}\right]&=\left[\begin{array}{cc} \overline{a}&\overline{b}\end{array}\right]\left[\begin{array}{cc}\langle(I_{\mathsf{H}}-\eta)\psi,\psi\rangle&\langle(I_{\mathsf{H}}-\eta)\psi,\hat{\xi}\rangle\\\langle(I_{\mathsf{H}}-\eta)\hat{\xi},\psi\rangle&\langle(I_{\mathsf{H}}-\eta)\xi,\xi\rangle\end{array}\right]\left[\begin{array}{c} a\\b\end{array}\right]\\&=\left\langle (I_{\mathsf{H}}-\eta)(a\psi+b\hat{\xi}),a\psi+b\hat{\xi}\right\rangle\nonumber
\end{align}
which is always greater than or equal to zero, so $ B\leq \left[\begin{array}{cc}1&\cos \delta\\\cos\delta&1\end{array}\right] $ for $ \delta=\arccos \langle\psi,\hat{\xi}\rangle\in\left[0,\frac{\pi}{2}\right]$. Hence, $ \mathcal{B}\subset\mathcal{C} $.

Now take any $ C\in\mathcal{C} $, so there is some $ \delta\in \left[0,\frac{\pi}{2}\right] $ such that $ C\leq \left[\begin{array}{cc}1&\cos \delta\\\cos\delta&1\end{array}\right] $. Since \textsf{H} is nontrivial, it has a pair of orthonormal vectors, $ \lbrace\mathbf{e}_{1}, \mathbf{e}_{2}\rbrace$. Take $ \psi=\mathbf{e}_{1} $ and $ \xi=\hat{\xi}=\cos\delta \mathbf{e}_{1}+\sin\delta \mathbf{e}_{2}$, which are both clearly of unit norm. Take the operator $ \eta $ to be zero on the complement of the span of $ \lbrace\mathbf{e}_{1}, \mathbf{e}_{2}\rbrace $. On the span, using $ \lbrace\mathbf{e}_{1}, \mathbf{e}_{2}\rbrace $ as the basis, let $ \eta $ be given by
\begin{equation}
C=\left[\begin{array}{cc}\langle\eta\psi,\psi\rangle&\langle\eta\psi,\hat{\xi}\rangle\\\langle\eta\hat{\xi},\psi\rangle&\langle\eta\xi,\xi\rangle\end{array}\right]=\left[\begin{array}{cc}1&0\\\cos \delta&\sin\delta\end{array}\right]\eta\left[\begin{array}{cc}1&\cos \delta\\0&\sin\delta\end{array}\right]
\end{equation}
\[\Leftrightarrow \eta=\left[\begin{array}{cc}1&0\\\cos \delta&\sin\delta\end{array}\right]^{-1}C\left[\begin{array}{cc}1&\cos \delta\\0&\sin\delta\end{array}\right]^{-1}\]
Then, since $ C\geq 0 $, clearly $ \eta\geq 0 $. Since $ C\leq \left[\begin{array}{cc}1&\cos \delta\\\cos\delta&1\end{array}\right] $, for $ \mathbf{v}=\left[\begin{array}{cc}1&\cos \delta\\0&\sin\delta\end{array}\right]^{-1}\left[\begin{array}{c} a\\b\end{array}\right] $,
\begin{align}
\left[\begin{array}{cc} \overline{a}&\overline{b}\end{array}\right](I_{\mathsf{H}}-\eta)\left[\begin{array}{c} a\\b\end{array}\right]=\mathbf{v}^{*}\left(\left[\begin{array}{cc}1&\cos \delta\\\cos\delta&1\end{array}\right]-C\right)\mathbf{v}\geq 0
\end{align}
Hence, $ \eta\leq I_{\mathsf{H}} $, so $ \mathcal{C}\subset\mathcal{B} $ and $ \mathcal{B}=\mathcal{C} $.

Therefore, using the above expression (\ref{eq:geodesicequation}) for the geodesic and $ \Vert\hat{\xi}-\langle\hat{\xi},\psi\rangle\psi\Vert=\sin \delta $, we wish to extremize
\begin{equation}\label{eq:parrondolikemin}
P_{A}^{\text{geo}}=\frac{1}{\sin^{2}\delta}\left[\begin{array}{cc} \sin (\delta-\theta)&\sin \theta\end{array}\right]B\left[\begin{array}{c}\sin (\delta-\theta)\\\sin \theta\end{array}\right] 
\end{equation}
over all $ 0\leq B\leq \left[\begin{array}{cc}1&\cos \delta\\\cos \delta&1\end{array} \right]$, $\delta\in[0,\frac{\pi}{2}]$, and  $\theta\in[0,\delta] $ for fixed $ P_{A}=b_{11} $ and $ P'_{A}=b_{22} $. Setting the imaginary parts of the off-diagonal entries of \textit{B} to zero keeps \textit{B} in the allowed order interval and does not change the value of $ P_{A}$, $P'_{A}$, or $ P_{A}^{\text{geo}} $, so \textit{B} can be taken real.

For fixed $ \delta $ and $ \theta $, the expression in (\ref{eq:parrondolikemin}) is linear in \textit{B}. Since the allowed \textit{B} form a convex set, the extrema are achieved on the set of extreme points, so \textit{B} can be restricted to either \textit{(i)} being rank one or \textit{(ii)} having $ \left[\begin{array}{cc}1&\cos \delta\\\cos \delta&1\end{array} \right]-B$ be rank one. For case \textit{(i)}, we have two subcases for the choice of either $ + $ or $ - $ in
\begin{equation}
B=\left[\begin{array}{cc}b_{11}&\pm\sqrt{b_{11}b_{22}}\\\pm\sqrt{b_{11}b_{22}}&b_{22}\end{array}\right]
\end{equation} 
Choosing the $ + $, then $ b_{11},b_{22}\in[0,1] $ with
\begin{equation}
\max\lbrace 0,\sqrt{b_{11}b_{22}}-\sqrt{(1-b_{11})(1-b_{22})}\rbrace\leq\cos \delta\leq\min\lbrace 1,\sqrt{b_{11}b_{22}}+\sqrt{(1-b_{11})(1-b_{22})}\rbrace
\end{equation} 
and $ \theta\in[0,\delta] $. Let $ f^{\max}_{1}(b_{11},b_{22}) $ be the maximum of (\ref{eq:parrondolikemin}) over all allowed $ \delta,\theta $ for the given $ b_{11},b_{22} $ and $ f^{\min}_{1}(b_{11},b_{22}) $ be the minimum. Then $ f^{\max}_{1}(b_{11},b_{22})=\min\lbrace b_{11}+b_{22},1\rbrace $, with the maximizing $ \delta_{0},\theta_{0} $ given by $ \delta_{0}=\frac{\pi}{2} $, $ \sin\theta_{0}=\sqrt{\frac{b_{22}}{b_{11}+b_{22}}} $ if $ b_{11}+b_{22}\leq 1 $ and by $ \cos\delta_{0}=\sqrt{b_{11}b_{22}}-\sqrt{(1-b_{11})(1-b_{22})} $, $ \sin\theta_{0}=\sqrt{1-b_{11}} $ if $ b_{11}+b_{22}> 1 $. The minimum bound is $ f^{\min}_{1}(b_{11},b_{22})=\min\lbrace b_{11},b_{22}\rbrace $, with the minimizing $ \delta_{0},\theta_{0} $ given by $ \theta_{0}=0 $ if $ b_{11}\leq b_{22} $ and $  \theta_{0}=\delta_{0} $ if $ b_{11}> b_{22} $, with $ \delta_{0} $ arbitrary. Choosing the $ - $, then $ (b_{11},b_{22})\in[0,1]^{\times 2}\cap\lbrace b_{11}+b_{22}\leq 1\rbrace $ with 
\begin{equation}
0\leq\cos \delta\leq\min\lbrace 1,\sqrt{(1-b_{11})(1-b_{22})}-\sqrt{b_{11}b_{22}}\rbrace
\end{equation}
and $ \theta\in[0,\delta] $. Let $ f^{\max}_{2}(b_{11},b_{22}) $ be the maximum of (\ref{eq:parrondolikemin}) over all allowed $ \delta,\theta $ for the given $ b_{11},b_{22} $ and $ f^{\min}_{2}(b_{11},b_{22}) $ be the minimum. Then $ f^{\max}_{2}(b_{11},b_{22})=\max\lbrace b_{11},b_{22}\rbrace $, with the maximizing $ \delta_{0},\theta_{0} $ given complementary to that for the preceding $ f^{\min}_{1} $. The minimum bound is $ f^{\min}_{2}(b_{11},b_{22})=0 $, with the minimizing $ \delta_{0},\theta_{0} $ given by $ \delta_{0}=\frac{\pi}{2}$, $ \sin\theta_{0}=\sqrt{\frac{b_{11}}{b_{11}+b_{22}}} $.

For case \textit{(ii)}, once again we have two subcases for the choice of either $ + $ or $ - $ in
\begin{equation}
B=\left[\begin{array}{cc}b_{11}&\cos \delta\pm\sqrt{(1-b_{11})(1-b_{22})}\\\cos \delta\pm\sqrt{(1-b_{11})(1-b_{22})}&b_{22}\end{array}\right]
\end{equation}
Choosing the $ + $, then $ (b_{11},b_{22})\in[0,1]^{\times 2}\cap\lbrace b_{11}+b_{22}\geq 1\rbrace$ with
\begin{equation}
 0\leq\cos \delta\leq\min\lbrace 1,\sqrt{b_{11}b_{22}}-\sqrt{(1-b_{11})(1-b_{22})}\rbrace
\end{equation}
and $ \theta\in[0,\delta] $. Let $ f^{\max}_{3}(b_{11},b_{22}) $ be the maximum of (\ref{eq:parrondolikemin}) over all allowed $ \delta,\theta $ for the given $ b_{11},b_{22} $ and $ f^{\min}_{3}(b_{11},b_{22}) $ be the minimum. Then $ f^{\max}_{3}(b_{11},b_{22})=1 $, with the minimizing $ \delta_{0},\theta_{0} $ given by $\delta_{0}=\frac{\pi}{2} $, $ \sin\theta_{0}=\sqrt{\frac{1-b_{11}}{2-b_{11}-b_{22}}} $. The minimum bound is $ f^{\min}_{3}(b_{11},b_{22})=\min\lbrace b_{11},b_{22}\rbrace $ with the minimizing $ \delta_{0},\theta_{0} $ given similarly to that for the preceding $ f^{\min}_{1} $. Choosing the $ - $, then $ b_{11},b_{22}\in[0,1]$ with
\begin{equation}
 \max\lbrace 0,\sqrt{(1-b_{11})(1-b_{22})}-\sqrt{b_{11}b_{22}}\rbrace\leq\cos \delta\leq\min\lbrace 1,\sqrt{(1-b_{11})(1-b_{22})}+\sqrt{b_{11}b_{22}}\rbrace
\end{equation}
and $ \theta\in[0,\delta] $. Let $ f^{\max}_{4}(b_{11},b_{22}) $ be the maximum of (\ref{eq:parrondolikemin}) over all allowed $ \delta,\theta $ for the given $ b_{11},b_{22} $ and $ f^{\min}_{4}(b_{11},b_{22}) $ be the minimum. Then $ f^{\max}_{4}(b_{11},b_{22})=\min\lbrace b_{11},b_{22}\rbrace $ with the maximizing $ \delta_{0},\theta_{0} $ given similarly to that for the preceding $ f^{\min}_{1} $. The minimum bound is $ f^{\min}_{4}(b_{11},b_{22})=\max\lbrace 0,b_{11}+b_{22}-1\rbrace $, with the minimizing $ \delta_{0},\theta_{0} $ given by $ \cos\delta_{0}=\sqrt{1-b_{11}}$\\$\sqrt{1-b_{22}}-\sqrt{b_{11}b_{22}} $, $ \sin\theta_{0}=\sqrt{b_{11}} $ if $ b_{11}+ b_{22}\leq 1 $ and $ \delta_{0}=\frac{\pi}{2} $, $ \sin\theta_{0}=\sqrt{\frac{1-b_{22}}{2-b_{11}-b_{22}}} $ if $ b_{11}+ b_{22}>1 $.

Putting the cases and subcases together, the minimum value of $ P_{A}^{\text{geo}} $ for fixed $ P_{A},P'_{A} $ is given by
\begin{equation}
\left(f^{\min}_{1}\wedge f^{\min}_{2}\wedge f^{\min}_{3}\wedge f^{\min}_{4}\right)( P_{A},P'_{A})=f^{\min}_{4}( P_{A},P'_{A})=\max\lbrace 0,P_{A}+P'_{A}-1\rbrace
\end{equation}
The maximum value of $ P_{A}^{\text{geo}} $ for fixed $ P_{A},P'_{A} $ is given by
\begin{equation}
\left(f^{\max}_{1}\vee f^{\max}_{2}\vee f^{\max}_{3}\vee f^{\max}_{4}\right)( P_{A},P'_{A})=f^{\max}_{1}( P_{A},P'_{A})=\min\lbrace P_{A}+P'_{A},1\rbrace
\end{equation}
With fixed values for $ P_{A}$ and $ P'_{A} $, within each case and subcase $ P_{A}^{\text{geo}} $ is a continuous function of $ \delta $ and $ \theta $, so all intermediate values for $ P_{A}^{\text{geo}} $ are achieved. $ \square $
\medskip\\
Note that, by the result of the preceding theorem, the paradox can only occur for values of $ (P_{A},P'_{A}) $ in the triangle bounded by $ P_{A}>\frac{1}{2} $, $ P'_{A}>\frac{1}{2} $, and $ P_{A}+P'_{A}<\frac{3}{2} $.
\section{Conditions for the occurrence of the Parrondo-like paradox}
The matrix \textit{B} defined in the proof of the preceding theorem (\ref{eq:defineB}) puts restrictions on the occurrence of the paradox. If $ B\leq \frac{1}{2}I_{2} $ then both $ P_{A} $ and $ P'_{A} $ are less than or equal to one-half, so the paradox cannot occur for any choices of $ [\psi] $, $ [\xi] $, or point on the geodesic joining them. Similarly, if the trace of \textit{B} is less than or equal to one, then either $ P_{A} $ or $ P'_{A} $ is less than one-half, so the paradox also cannot occur. From these obvious statements, we then have the following nontrivial result:
\paragraph{Theorem 8.4.1} If, for any particular orthonormal $ \mathbf{e}_{1},\mathbf{e}_{2}\in\mathsf{H} $,
\[ \text{tr }\left[\begin{array}{cc}L(\mathbf{e}_{1}\mathbf{e}_{1}^{*})\vert_{A}&L(\mathbf{e}_{1}\mathbf{e}_{2}^{*})\vert_{A}\\L(\mathbf{e}_{2}\mathbf{e}_{1}^{*})\vert_{A}&L(\mathbf{e}_{2}\mathbf{e}_{2}^{*})\vert_{A}\end{array}\right]\leq 1\]
then, for any normalized $ \psi,\xi\in\text{span}_{\mathbb{C}}\lbrace\mathbf{e}_{1},\mathbf{e}_{2}\rbrace $ and point on the geodesic joining $ [\psi] $ and $ [\xi] $, the paradox cannot occur.
\paragraph*{Proof} Let $ \psi=\left[\begin{array}{cc}\mathbf{e}_{1}&\mathbf{e}_{2}\end{array} \right]\left[\begin{array}{c}a\\b\end{array} \right]  $ and $ \hat{\xi}=\left[\begin{array}{cc}\mathbf{e}_{1}&\mathbf{e}_{2}\end{array} \right]\left[\begin{array}{c}c\\d\end{array} \right]  $ with $ \vert a\vert^{2}+\vert b\vert^{2}=\vert c\vert^{2}+\vert d\vert^{2}=1 $ and $ a\overline{c}+b\overline{d}=\cos \delta $ real and greater than or equal to zero. Let 
\begin{equation}
C=\left[\begin{array}{cc}L(\mathbf{e}_{1}\mathbf{e}_{1}^{*})\vert_{A}&L(\mathbf{e}_{1}\mathbf{e}_{2}^{*})\vert_{A}\\L(\mathbf{e}_{2}\mathbf{e}_{1}^{*})\vert_{A}&L(\mathbf{e}_{2}\mathbf{e}_{2}^{*})\vert_{A}\end{array}\right]
\end{equation}
If $ C\leq \frac{1}{2}I_{2} $, then
\begin{equation}
\text{tr }B=\text{tr }\left[\begin{array}{cc}\overline{a}&\overline{b}\\\overline{c}&\overline{d}\end{array}\right]C\left[\begin{array}{cc}a&c\\b&d\end{array}\right]\leq 1
\end{equation} 
so, by the comment preceding the theorem, the paradox cannot occur. Therefore, the only remaining case is where $ C $, whose eigenvalues are necessarily real, has one eigenvalue, $ \lambda_{1}>\frac{1}{2} $, and one eigenvalue, $ \lambda_{2}<\frac{1}{2} $, with $ \lambda_{1}+ \lambda_{2}\leq 1 $. There are corresponding normalized eigenvectors $ \mathbf{v}_{1} $ and $ \mathbf{v}_{2} $, necessarily orthogonal. Writing $ \left[\begin{array}{c}a\\c\end{array} \right]=f_{1}\mathbf{v}_{1}+f_{2}\mathbf{v}_{2} $ and $ \left[\begin{array}{c}b\\d\end{array} \right]=g_{1}\mathbf{v}_{1}+g_{2}\mathbf{v}_{2} $, the above conditions on $ a,b,c,d $ become the following conditions on $ f_{1},f_{2},g_{1},g_{2} $: $ \vert f_{1}\vert^{2}+\vert f_{2}\vert^{2}=1$, $\vert g_{1}\vert^{2}+\vert g_{2}\vert^{2}=1$, and $ f_{1}\overline{g_{1}}+f_{2}\overline{g_{2}}=\cos \delta$ is a positive real or zero.

For the paradox to occur, it must be that both $\vert f_{1}\vert^{2}\lambda_{1}+\vert f_{2}\vert^{2}\lambda_{2}= b_{11}=P_{A}>\frac{1}{2} $ and $ \vert g_{1}\vert^{2}\lambda_{1}+\vert g_{2}\vert^{2}\lambda_{2}=b_{22}=P'_{A}>\frac{1}{2} $; hence,
\begin{align}
\vert f_{1}\vert>\sqrt{\frac{\frac{1}{2}-\lambda_{2}}{\lambda_{1}-\lambda_{2}}}\text{ and }\vert f_{2}\vert<\sqrt{\frac{\lambda_{1}-\frac{1}{2}}{\lambda_{1}-\lambda_{2}}}&\Rightarrow\vert f_{1}\vert>\sqrt{\frac{\frac{1}{2}-\lambda_{2}}{\lambda_{1}-\frac{1}{2}}}\;\vert f_{2}\vert\\
\vert g_{1}\vert>\sqrt{\frac{\frac{1}{2}-\lambda_{2}}{\lambda_{1}-\lambda_{2}}}\text{ and }\vert g_{2}\vert<\sqrt{\frac{\lambda_{1}-\frac{1}{2}}{\lambda_{1}-\lambda_{2}}}&\Rightarrow\vert g_{1}\vert>\sqrt{\frac{\frac{1}{2}-\lambda_{2}}{\lambda_{1}-\frac{1}{2}}}\;\vert g_{2}\vert
\end{align}
Then, since
\begin{equation}
\vert f_{1}\overline{g_{1}}\vert=\vert f_{1}\vert\vert g_{1}\vert>\frac{\frac{1}{2}-\lambda_{2}}{\lambda_{1}-\frac{1}{2}}\;\vert f_{2}\vert\vert g_{2}\vert>\vert f_{2}\vert\vert g_{2}\vert=\vert f_{2}\overline{g_{2}}\vert
\end{equation}
it must be that $ f_{1}\overline{g_{1}}+f_{2}\overline{g_{2}} $ is actually strictly positive and that $ \Re f_{1}\overline{g_{1}}> 0 $. Furthermore, since the imaginary parts of $ f_{1}\overline{g_{1}} $ and $ f_{2}\overline{g_{2}} $ are equal in magnitude, it must be that the real part of $ f_{1}\overline{g_{1}} $ is greater than $ \frac{\frac{1}{2}-\lambda_{2}}{\lambda_{1}-\frac{1}{2}} $ times the magnitude of the real part of $ f_{2}\overline{g_{2}} $ and, therefore, is greater than $ \frac{\frac{1}{2}-\lambda_{2}}{\lambda_{1}-\frac{1}{2}} $ times the real part of $ f_{2}\overline{g_{2}} $. Hence, rearranging terms,
\begin{equation}
\Re\left(\lambda_{1}f_{1}\overline{g_{1}}+\lambda_{2}f_{2}\overline{g_{2}}\right)>\frac{1}{2}\left(f_{1}\overline{g_{1}}+f_{2}\overline{g_{2}}\right)=\frac{1}{2}\cos \delta
\end{equation}
However, then 
\begin{align}
P_{A}^{\text{geo}}&=\frac{1}{\sin^{2}\delta}\left[\begin{array}{cc} \sin (\delta-\theta)&\sin \theta\end{array}\right]B\left[\begin{array}{c}\sin (\delta-\theta)\\\sin \theta\end{array}\right]\\&=\frac{1}{\sin^{2}\delta}\left(\left(\vert f_{1}\vert^{2}\lambda_{1}+\vert f_{2}\vert^{2}\lambda_{2}\right)\sin^{2} (\delta-\theta)+\left(\vert g_{1}\vert^{2}\lambda_{1}+\vert g_{2}\vert^{2}\lambda_{2}\right)\sin^{2} \theta\right.\nonumber\\&\hspace{.2 in}\left. +2\Re\left(\lambda_{1}f_{1}\overline{g_{1}}+\lambda_{2}f_{2}\overline{g_{2}}\right) \sin (\delta-\theta)\sin\theta\right)\nonumber\\&>\frac{1}{\sin^{2}\delta}\left(\frac{1}{2}\sin^{2} (\delta-\theta)+\frac{1}{2}\sin^{2} \theta+\cos \delta\sin (\delta-\theta)\sin\theta\right)\nonumber\\&=\frac{1}{2\sin^{2}\delta}\left(\sin (\delta-\theta)\left(\sin (\delta-\theta)+\cos \delta\sin\theta\right)\right.\nonumber\\&\hspace{.2 in}\left. +\sin \theta\left(\sin(\delta-(\delta-\theta))+\cos \delta\sin (\delta-\theta)\right)\right)\nonumber\\&=\frac{1}{2\sin^{2}\delta}\left(\sin (\delta-\theta)\sin \delta\cos\theta+\sin \theta\sin \delta\cos (\delta-\theta)\right)\nonumber\\&=\frac{\sin^{2}\delta}{2\sin^{2}\delta}=\frac{1}{2}\nonumber
\end{align}
Therefore, if both $ b_{11}=P_{A}>\frac{1}{2} $ and $b_{22}=P'_{A}>\frac{1}{2} $, then $ P_{A}^{\text{geo}}>\frac{1}{2} $ everywhere on the geodesic. $\square$

%% file: chapter9.tex
\chapter{Quantum walks and the Parrondo-like paradox}
\section{Classical random and classical hidden walks}
\subsection{Definitions of classical random and classical hidden walks}
A classical random walk is a special case of the observable Markov chain earlier discussed, with graphical model given in figure (\ref{fig:classicgame}). It models a walker who is in one of a finite number of internal states and occupies one of a countable number of positions at one time, so the space $ \mathcal{X} $ is either $ \mathcal{J}\times \mathbb{Z}^{+} $ for a walk on the half-line, or $ \mathcal{J}\times \mathbb{Z}$ for a walk on the full line. After each  time-step, the walker is at the same or a neighboring location and its internal state can change as well. The process is random, with the transition probability functions possibly dependent on the internal state as well as on the location. This constrains the transition probability functions further than the constraints already imposed by the Markov conditions, but these further constraints are not indicated in the graphical model. If the transition probability functions are spatially translation invariant, the walk is termed homogeneous.

Now we may consider a classical hidden walk. This is in some ways a special case of the classical hidden-Markov chain earlier discussed (\ref{fig:hiddengame}), while in other ways it is a generalization. The hidden set $ \mathcal{H} $ is required to be either $ \mathcal{J}\times \mathbb{Z}^{+} $ for a walk on the half-line, or $ \mathcal{J}\times \mathbb{Z}$ for a walk on the full line. The base measure for the hidden set is required to be the counting measure, and $ L^{1}\left(\mathcal{H};\text{counting measure}\right)\cong\ell^{1} $, so all integrals can be taken to just be sums. The space $ \ell^{1} $ has a natural basis of sequences that have a single nonzero entry with value one. There is a dual ``basis" of sequences in $ \ell^{\infty} $ of the same sequences, which is not a basis in the norm topology, but is one in the weak* topology. This basis and dual ``basis" can be used to assign matrix elements to any operator in $ \mathcal{B}(\ell^{1}) $. Then any operator, \textit{A}, in $ \mathcal{B}(\ell^{1}) $ is in one-to-one correspondence to a certain sequence of matrices, $ \lbrace A_{n}\rbrace $, which may be considered the truncations of the infinite matrix corresponding to the operator. By the triangle inequality, these sequences converge to the operator in the strong-operator topology:
\begin{align}
\lim_{n\to\infty}\Vert A\mathbf{x}-A_{n}\mathbf{x}\Vert&=\lim_{n\to\infty}\Vert A\mathbf{x}-P_{n}AP_{n}\mathbf{x}\Vert\leq\lim_{n\to\infty}\left(\Vert A\mathbf{x}-P_{n}A\mathbf{x}\Vert+\Vert P_{n}A\mathbf{x}-P_{n}AP_{n}\mathbf{x}\Vert\right)\\&\leq\lim_{n\to\infty}\left(\Vert (A\mathbf{x})-P_{n}(A\mathbf{x})\Vert+\Vert P_{n}A\Vert_{\text{op}}\Vert\mathbf{x}-P_{n}\mathbf{x}\Vert\right)=0\nonumber
\end{align}
for any $ \mathbf{x}\in\ell^{1} $, where $ \lbrace P_{n}\rbrace $ are the usual, diagonal projectors onto the span of the first \textit{n} basis elements.

Conversely, given a sequence of matrices, each of which is the truncation of the following, the condition on the sequence so that it actually corresponds to a bounded operator is that the induced operator norm of all the matrices is bounded. However, the operator norm induced by the $ \ell^{1} $ norm is simply the supremum over all columns of the column sum of the magnitudes of the entries, so it is readily calculated. With this form for the maps, the condition that this is a walk (rather than some other sort of hidden-Markov process) is that matrix entries connecting spatial locations that are not neighboring are all zero.

The generalization from the classical hidden-Markov chain is that the last pince-nez map in the chain is no longer required to be the same as the preceding ones; in particular, all the preceding pince-nez can be taken to be simply nodes, so the graphical model is
\begin{align}\label{fig:hiddenwalk}
\setlength{\unitlength}{.8 in}
\begin{picture}(5,.5)
\color{red}
\put(.1,.1){\oval(.2,.2)[r]}
\put(0,.2){\line(1,0){.1}}
\put(0,0){\line(1,0){.1}}
\put(0,0){\line(0,1){.2}}
\put(1,.1){\circle{.2}}
\put(2,.1){\circle{.2}}
\put(4,.1){\circle{.2}}
\put(5,.1){\oval(.2,.2)[l]}
\put(5,.2){\line(1,0){.1}}
\put(5,0){\line(1,0){.1}}
\put(5.1,0){\line(0,1){.2}}
\put(.2,.1){\vector(1,0){.7}}
\put(1.09,.1){\vector(1,0){.8}}
\put(2.09,.1){\line(1,0){.5}}
\put(4.09,.1){\vector(1,0){.8}}
\put(2.7,.1){$\cdots$}
\put(3,.1){$\cdots$}
\put(3.39,.1){\vector(1,0){.5}}
\put(4.5,.1){\vector(1,0){.4}}
\put(4.89,.1){\line(0,1){.1}}
\put(1,.35){1}
\put(2,.35){2}
\put(4,.35){\textit{n}}
\color{black}
\put(5,.3){\circle{.2}}
\put(5,.45){$ n+1 $}
\put(4.9,.3){\line(1,0){.19}}
\put(5,.209){\line(0,1){.185}}
\put(4.89,.2){\line(0,1){.1}}
\end{picture}
\end{align}
Note there is nothing graphically that distinguishes this from a more general Markov process--the constraints that make it a walk are not represented graphically.
\subsection{Connection to orthogonal polynomials and measures on $ \mathbb{R} $}
Orthogonal polynomials\footnote{The following results on orthogonal polynomials are well known and included for comparison to the results for quantum walks given below. For details, see\cite{szego}.} $ \lbrace p_{j}\rbrace $ result from the Gram-Schmidt algorithm applied to $ \lbrace 1,x,x^{2},\ldots\rbrace $ on the real line with inner-product given with respect to some Borel measure $ \mu $, $ \langle f,g\rangle=\int_{\mathbb{R}}fg\,d\mu $. These polynomials all have the maximal number of real roots, which are all simple, else they would not change signs enough times to be orthogonal. For the same reason, the roots are all within the convex hull of the support of $ \mu $ and they interlace as $ j $ increases by one. By orthogonality, the polynomials necessarily obey a three-term recurrence relation, which may be written in matrix form as
\begin{equation}\label{eq:tridiagonal}
x\left[\begin{array}{ccc} p_{0}&p_{1}&\cdots\end{array} \right]=\left[\begin{array}{ccc} p_{0}&p_{1}&\cdots\end{array} \right]\left[\begin{array}{cccc} b_{0}&c_{1}&&\\a_{0}&b_{1}&c_{2}&\\&a_{1}&\ddots&\ddots\\&&\ddots&\ddots\end{array} \right] 
\end{equation}
If the polynomials are normalized to have value one at $ x=1 $, the tridiagonal, infinite matrix on the right has column sum one for each of its columns. If $ \mu $ is such that all the entries in that matrix are positive, then it is a stochastic matrix and can be used for the map per time step for a classical random or classical hidden walk on the half-line where the internal states are trivial.

Conversely, given such a walk, then there is an infinite, stochastic matrix 
\begin{equation}
\left[\begin{array}{cccc} b_{0}&c_{1}&&\\a_{0}&b_{1}&c_{2}&\\&a_{1}&\ddots&\ddots\\&&\ddots&\ddots\end{array} \right]
\end{equation}  
giving the map per time step. If all the \textit{a}'s and \textit{c}'s are strictly positive, then each $ n\times n $-truncation $ A_{n} $ of the matrix is similar to a Hermitian matrix via
\begin{equation}
\left[\begin{array}{cccc} 1&&&\\&\frac{1}{d_{1}}&&\\&&\ddots&\\&&&\frac{1}{d_{n-1}}\end{array} \right]\left[\begin{array}{ccccc} b_{0}&c_{1}&&&\\a_{0}&b_{1}&c_{2}&&\\&a_{1}&\ddots&\ddots&\\&&\ddots&\ddots&c_{n-1}\\&&&a_{n-2}&b_{n-1}\end{array} \right]\left[\begin{array}{cccc} 1&&&\\&d_{1}&&\\&&\ddots&\\&&&d_{n-1}\end{array} \right]
\end{equation}
where $ d_{j}^{2}=\frac{c_{1}\cdots c_{j}}{a_{0}\cdots a_{j-1}} $. Therefore, for each such truncation, the eigenvalues $ \lbrace x^{(n)}_{1},\ldots,x^{(n)}_{n}\rbrace $ are all real. These eigenvalues are all necessarily less than or equal to one in magnitude since the spectral radius of $ A_{n} $ is less than or equal to its operator norm induced by the $ \ell^{1} $-norm, which is one. By the Courant-Fischer minimax theorem~\cite{demmelcourant} the eigenvalues are all simple and they interlace as \textit{n} increases, so they are actually all less than one in magnitude. Define the polynomials $ \lbrace p_{n}\rbrace $ by
\begin{equation}
p_{n}(x)=\frac{\det(xI_{n}-A_{n})}{\det(I_{n}-A_{n})}=\frac{(x-x^{(n)}_{1})\cdots(x-x^{(n)}_{n})}{(1-x^{(n)}_{1})\cdots(1-x^{(n)}_{n})}
\end{equation}
By adding the first row to the second, the second to the third, and so on, it is easy to see that $ \det(I_{n}-A_{n})=a_{0}a_{1}\cdots a_{n-1} $. Then, by expanding $ \det(xI_{n}-A_{n}) $ by minors along its last column and evaluation at the \textit{n} values $ x\in \lbrace x^{(n)}_{1},\ldots,x^{(n)}_{n}\rbrace $ (which is enough to determine a degree-\textit{n} polynomial), the polynomial $x\,p_{n-1}(x) $ obeys the recurrence relation in (\ref{eq:tridiagonal}). 

Let $ \mu_{1} $ be the single atom measure $ \delta_{b_{0}} $ and, for $ n>1 $, let $ \mu_{n} $ be the atomic measure $ \sum_{j=1}^{n}w^{(n)}_{j}\delta_{x^{(n)}_{j}} $, where 
\begin{equation}
w^{(n)}_{j}=\sum_{l=1}^{k}w^{(k)}_{l}\frac{p_{n}\left(x^{(k)}_{l}\right)}{\left(x^{(k)}_{l}-x^{(n)}_{j}\right)p'\left(x^{(n)}_{j}\right)}
\end{equation}
for any $ k\in \left\lbrace \left\lceil\frac{n}{2}\right\rceil,\ldots,n-1\right\rbrace $ (they all give the same result). Furthermore, the sequence of measures $ \langle \mu_{n}\rangle $ stabilizes for any fixed polynomial in the sense that for any degree-\textit{n} polynomial \textit{q}, $ \int_{\mathbb{R}}q\,d\mu_{k} $ is the same for all $ k\geq \left\lfloor\frac{n}{2}+1\right\rfloor $. Therefore, the $ \mu_{n} $ are indeed measures (and not just signed measures) since 
\begin{equation}
w^{(n)}_{j}=\sum_{l=1}^{n-1}w^{(n-1)}_{l}\left(\frac{p_{n}\left(x^{(n-1)}_{l}\right)}{\left(x^{(n-1)}_{l}-x^{(n)}_{j}\right)p'\left(x^{(n)}_{j}\right)}\right)^{2}
\end{equation}
so all the $ w $'s are positive.

Hence, $ \mu_{n}(\mathbb{R})=\int_{\mathbb{R}}1\,d\mu_{n}=1 $ is the total-variation norm of $ \mu_{n} $ for each $ n $. Since the $ \mu_{n} $ are all Radon measures and the interval $ [-1,1] $ is compact, by the Riesz-Markov theorem~\cite{roydenriesz} and Alaoglu's theorem~\cite{roydenalaoglu}, the sequence $ \langle \mu_{n}\rangle $ has a weak* limit point. Since, by the Weierstrass theorem~\cite{semadeniweierstrass}, polynomials are dense in the supremum norm among continuous functions on the compact interval $  [-1,1] $, by the above stabilizing property of the sequence, the limit point is unique and the entire sequence converges to it in the weak* topology. Let this limit be denoted $ \mu $. Then $\lbrace p_{n}\rbrace $ are the orthogonal polynomials corresponding to the measure $ \mu $ on $ \mathbb{R} $. The measure $ \mu $ is unique among Radon measures since any other measure with this property agrees with $ \mu $ on polynomials, but, as stated above, they are dense in norm among continuous functions on $ [-1,1] $, which separate Radon measures; hence, polynomials separate these measures as well.

Furthermore, one may ask if, starting with $ \nu $ such that the tridiagonal, infinite matrix has all positive entries, then forming the measure $ \mu $ following the procedure outlined, it is necessarily the case that $ \nu\propto\mu $ (there may be a scale factor since $ \mu $ necessarily has total-variation norm one). This is true since $ \mu $ and $ \nu $ agree (up to the scale factor) when integrated with polynomials on $ \mathbb{R} $; however, $ \mu $ is supported on $ [-1,1] $ so its moments (and, hence, those of $ \nu $) are bounded. Therefore, by~\cite{durrettmoment}, the moment problem on $ \mathbb{R} $ has a unique solution in this case.

Finally, there is the question of whether every Radon measure on $ [-1,1] $ corresponds to a classical random walk. The answer is no, as is seen by the Jacobi polynomials~\cite{AMSjacobi}, normalized\footnote{Using Pochhammer's symbol, $ (a)_{0}=1 $ and $ (a)_{n}=a(a+1)\cdots (a+n-1) $.} to have the value one at $ x=1 $, $ \left\lbrace \frac{n!}{(1+\alpha)_{n}}P_{n}^{(\alpha,\beta)}\right\rbrace $. These have measure with support on $ [-1,1] $, given there by the measure absolutely continuous with respect to Lebesgue measure and with Radon-Nikod\'{y}m derivative $ (1+x)^{\beta}(1-x)^{\alpha} $. For the recurrence relation, the entries in the tridiagonal, infinite matrix for $ \lbrace a_{n}\rbrace $ and $ \lbrace c_{n}\rbrace $ are always strictly positive (as they must be for any measure with its support on $ (-\infty,1] $), but the $ \lbrace b_{n}\rbrace $ are given by
\begin{equation}
b_{n}=\frac{(\beta^{2}-\alpha^{2})(2n+\alpha+\beta+1)}{(2n+\alpha+\beta)_{3}}
\end{equation}
which are negative for $ \alpha>\beta $. (However, for the Jacobi polynomials shifted to live on the interval $ [0,1] $ with Radon-Nikod\'{y}m derivative $ x^{\alpha}(1-x)^{\beta} $, $ \left\lbrace \frac{(-1)^{n}n!}{(1+\beta)_{n}}P_{n}^{(\alpha,\beta)}(1-2x)\right\rbrace $, there is an associated walk--see~\cite{grunbaumjacobi}). 
\section{Quantum walks}
\subsection{Definition of a quantum walk}
Similarly to the classical hidden walk, the quantum walk is in some ways a special case of the quantum Markov chain earlier discussed (\ref{fig:quantumBayes}), while in other ways it is a generalization. The Hilbert space \textsf{H} is required to be either $ L^{2}(\mathcal{J}\times \mathbb{Z}^{+};\text{counting measure}) $ for a walk on the half-line, or $ L^{2}(\mathcal{J}\times \mathbb{Z};\text{counting measure})$ for a walk on the full line; both are clearly isometrically isomorphic to $ \ell^{2} $, so all integrals can be taken to just be sums. The space $ \ell^{2} $ has a natural basis of sequences that each have a single nonzero entry with value one. This basis and the inner-product can be used to assign matrix elements to any operator $ \mathcal{B}(\ell^{2}) $. Any operator \textit{A} in $ \mathcal{B}(\ell^{2}) $ is in one-to-one correspondence to a certain sequence of matrices, $ \lbrace A_{n}\rbrace $, which may be considered the truncations of the infinite matrix corresponding to the operator. By the triangle inequality, these sequences converge to the operator in the strong-operator topology:
\begin{align}
\lim_{n\to\infty}\Vert A\psi-A_{n}\psi\Vert&=\lim_{n\to\infty}\Vert A\psi-P_{n}AP_{n}\psi\Vert\leq\lim_{n\to\infty}\left(\Vert A\psi-P_{n}A\psi\Vert+\Vert P_{n}A\psi-P_{n}AP_{n}\psi\Vert\right)\\&\leq\lim_{n\to\infty}\left(\Vert (A\psi)-P_{n}(A\psi)\Vert+\Vert P_{n}A\Vert_{\text{op}}\Vert\psi-P_{n}\psi\Vert\right)=0\nonumber
\end{align}
for any $ \psi\in \ell^{2} $, where $ \lbrace P_{n}\rbrace $ are the orthogonal projectors onto the span of the first \textit{n} basis elements.

Conversely, given a sequence of matrices, each of which is the truncation of the following, the condition on the sequence so that it actually corresponds to a bounded operator is that the induced operator norm of all the matrices is bounded. The operator norm induced by the $ \ell^{2} $ norm is the largest singular value, which, unfortunately, is not generally readily calculated. However, for this sequence to correspond to a partial isometry, it is only necessary to show that for all fixed, finite collections of columns, those columns of the $ \lbrace A_{n}\rbrace $ are mutually orthonormal in the limit as $ n\to \infty $. It is readily shown that this condition implies the induced operator norm of each $ A_{n} $ is less than or equal to one. With this form for the maps, the condition that this is a walk (rather than some other sort of quantum Markov process) is that matrix entries connecting spatial locations that are not neighboring are all zero.

The generalization from the quantum Markov chain is that the last pince-nez in the chain is no longer required to be the same as the preceding. In particular, all the preceding pince-nez can be taken to be simply nodes, so the graphical model is
\begin{align}\label{fig:quantumwalk}
\setlength{\unitlength}{.8 in}
\begin{picture}(5,.8)
\color{blue}
\put(.1,.1){\oval(.2,.2)[r]}
\put(0,.2){\line(1,0){.1}}
\put(0,0){\line(1,0){.1}}
\put(0,0){\line(0,1){.2}}
\put(1,.1){\circle{.2}}
\put(2,.1){\circle{.2}}
\put(4,.1){\circle{.2}}
\put(5,.1){\oval(.2,.2)[l]}
\put(5,.2){\line(1,0){.1}}
\put(5,0){\line(1,0){.1}}
\put(5.1,0){\line(0,1){.2}}
\put(.2,.1){\vector(1,0){.7}}
\put(1.09,.1){\vector(1,0){.8}}
\put(2.09,.1){\line(1,0){.5}}
\put(4.09,.1){\vector(1,0){.8}}
\put(2.7,.1){$\cdots$}
\put(3,.1){$\cdots$}
\put(3.39,.1){\vector(1,0){.5}}
\put(4.5,.1){\vector(1,0){.4}}
\put(4.89,.1){\line(0,1){.1}}
\put(1,.35){1}
\put(2,.35){2}
\put(4,.35){\textit{n}}
\color{black}
\put(5,.3){\circle{.2}}
\put(5,.45){$ n+1 $}
\put(4.9,.3){\line(1,0){.19}}
\put(5,.209){\line(0,1){.185}}
\put(4.89,.2){\line(0,1){.1}}
\end{picture}
\end{align}
Note, as for the classical random walks, there is nothing graphically that distinguishes this from a more general quantum Markov process--the constraints that make it a walk are not represented in the graphical model.
\subsection{Orthogonal trigonometric polynomials and measures on $ \mathbb{S}^{1} $}
Trigonometric orthogonal polynomials\footnote{Only a few basic results that are most applicable to quantum walks of this rich topic are presented here. See~\cite{simon} for details and elaboration.} $ \lbrace q_{j}\rbrace $ result from the Gram-Schmidt algorithm applied to $ \lbrace 1,z,z^{-1},z^{2},z^{-2},\ldots\rbrace $ on the unit circle $ \mathbb{S}^{1} $ within $ \mathbb{C} $ with sesquilinear inner-product given with respect to some Borel measure $ \mu $, $ \langle f,g\rangle=\int_{\mathbb{S}^{1}}f\overline{g}\,d\mu $. To form these, it is useful to start with the monic orthogonal polynomials on the unit circle, the Szeg\"{o} polynomials, $ \left\lbrace s_{n}\right\rbrace  $. For any polynomial, define the \textit{reciprocal} polynomial to be the polynomial with its coefficients conjugated and flipped in order, so, if \textit{p} is a \textit{n}th order polynomial,
\begin{equation}
p^{\text{reciprocal}}(z)=z^{n}\overline{p}\left( \frac{1}{z}\right)=z^{n}\overline{p\left( \frac{1}{\overline{z}}\right)}
\end{equation}
where the overline only over the function means to conjugate its coefficients. On $ \mathbb{S}^{1} $, $ p^{\text{reciprocal}}(z)=z^{n}\overline{p\left( z\right)} $. In particular, if $ z $ is a root of $ p $, then $ \dfrac{1}{\overline{z}} $ is a root of $ p^{\text{reciprocal}} $.

Define the \textit{Verblunsky coefficients} by the values of the Szeg\"{o} polynomials at zero,  $ \alpha_{n}=- \overline{s_{n+1}(0)} $. Then, by orthogonality, we have the following Szeg\"{o} recurrence identities for all $ z\in\mathbb{C} $:
\begin{equation}\label{eq:szego1}
z\,s_{n}(z)=s_{n+1}(z)+\overline{\alpha_{n}}\,s_{n}^{\text{reciprocal}}(z)
\end{equation}
\[\Leftrightarrow s_{n+1}^{\text{reciprocal}}(z)=s_{n}^{\text{reciprocal}}(z)- \alpha_{n}z\,s_{n}(z)\]
and
\begin{equation}\label{eq:szego2}
s_{n}(z)=-\overline{\alpha_{n-1}}\,s_{n}^{\text{reciprocal}}(z)+\left(1-\vert\alpha_{n-1}\vert^{2}\right)z\,s_{n-1}(z)
\end{equation}
\[\Leftrightarrow s_{n}^{\text{reciprocal}}(z)=-\alpha_{n-1}\,s_{n}(z)+\left(1-\vert\alpha_{n-1}\vert^{2}\right)s_{n-1}^{\text{reciprocal}}(z)\]
From the first identity, it follows that 
\begin{equation}\label{eq:szegonorm}
\Vert s_{n}\Vert^{2}=\int_{z\in\mathbb{S}^{1}} \left\vert s_{n}(z)\right\vert^{2}\,d\mu(z)=\left(1-\vert\alpha_{n-1}\vert^{2} \right)\cdots\left(1-\vert\alpha_{0}\vert^{2} \right)\mu(\mathbb{S}^{1})
\end{equation}
By Verblunsky's theorem, the measure $ \mu $, the moments of the measure $ \lbrace m_{j}\rbrace $, and the Verblunsky coefficients $ \lbrace \alpha_{j}\rbrace $ all determine each other. The only condition on the Verblunsky coefficients that they do indeed correspond to some measure is that $ \vert\alpha_{j}\vert\leq 1 $ for all \textit{j}. 

The monic orthogonal trigonometric polynomials are then given by $ q_{0}=1 $, and, for $ j\in\lbrace 1,2,\ldots\rbrace $,
\begin{equation}
q_{j}(z)=z^{-(j-1)}s_{2j-1}(z),q_{-j}(z)=z^{-j}s_{2j}^{\text{reciprocal}}(z)
\end{equation}
The orthogonal trigonometric polynomials necessary satisfy a pentadiagonal recurrence relation, with first $ z\,q_{0}(z)=q_{1}(z)+\overline{\alpha_{0}}q_{1}(z) $, then, using (\ref{eq:szego1}) and (\ref{eq:szego2}) repeatedly,
\begin{align}
 zq_{j}(z)&=z^{-j}z^{2}s_{2j-1}(z)=z^{-j}z\left(s_{2j}(z)+\overline{\alpha_{2j-1}}s_{2j-1}^{\text{reciprocal}}(z) \right) \\&=z^{-j}\left(s_{2j+1}(z)+\overline{\alpha_{2j}}s_{2j}^{\text{reciprocal}}(z)+\overline{\alpha_{2j-1}}z\,s_{2j-1}^{\text{reciprocal}}(z) \right)\nonumber\\&=z^{-j}\left(s_{2j+1}(z)+\overline{\alpha_{2j}}s_{2j}^{\text{reciprocal}}(z)\right.\nonumber\\&\hspace{.2 in}\left. +z\overline{\alpha_{2j-1}}\left(-\alpha_{2j-2}\,s_{2j-1}(z)+\left(1-\vert\alpha_{2j-2}\vert^{2}\right)s_{2j-2}^{\text{reciprocal}} (z)\right)  \right)\nonumber\\&=q_{j+1}(z)+\overline{\alpha_{2j}}q_{-j}(z)-\overline{\alpha_{2j-1}}\alpha_{2j-2}q_{j}(z)+\overline{\alpha_{2j-1}}\left(1-\vert\alpha_{2j-2}\vert^{2}\right)q_{-(j-1)}(z)\nonumber
\end{align}
and 
\begin{align}
zq_{-j}(z)&=z^{-j}z\,s_{2j}^{\text{reciprocal}}(z)=z^{-j}z\left(-\alpha_{2j-1}\,s_{2j}+\left(1-\vert\alpha_{2j-1}\vert^{2}\right)s_{2j-1}^{\text{reciprocal}}  \right)\\&=z^{-j}\left(-\alpha_{2j-1}\left(s_{2j+1}+\overline{\alpha_{2j}} s_{2j}^{\text{reciprocal}}\right)\right. \nonumber\\&\hspace{.2 in}\left.+z\left(1-\vert\alpha_{2j-1}\vert^{2}\right)\left(-\alpha_{2j-2}\,s_{2j-1}+\left(1-\vert\alpha_{2j-2}\vert^{2}\right)s_{2j-2}^{\text{reciprocal}}  \right)  \right)\nonumber\\&=-\alpha_{2j-1}q_{j+1}(z)-\overline{\alpha_{2j}}\alpha_{2j-1}q_{-j}(z)-\alpha_{2j-2}\left(1- \vert\alpha_{2j-1}\vert^{2}\right)q_{j}\nonumber\\&\hspace{.2 in}+\left(1- \vert\alpha_{2j-1}\vert^{2}\right)\left(1- \vert\alpha_{2j-2}\vert^{2}\right)q_{-(j-1)}(z)\nonumber
\end{align}
Following Cantero, Moral, and Vel\'{a}zquez~\cite{cmv1}~\cite{cmv2}, writing this in matrix form as
\begin{equation}
\left[\begin{array}{cccc}zq_{0}(z)&zq_{1}(z)&zq_{-1}(z)&\cdots\end{array} \right]=\left[\begin{array}{cccc}q_{0}(z)&q_{1}(z)&q_{-1}(z)&\cdots\end{array} \right]Z
\end{equation}
gives the CMV-matrix \textit{Z} for the monic orthogonal trigonometric polynomials,
\begin{equation}
Z=\left[\begin{array}{cccccc}\overline{\alpha_{0}}&\overline{\alpha_{1}}\left(1- \vert\alpha_{0}\vert^{2}\right)&\left(1- \vert\alpha_{1}\vert^{2}\right)\left(1- \vert\alpha_{0}\vert^{2}\right)&0&\cdots\\1&-\overline{\alpha_{1}}\alpha_{0}&-\left(1- \vert\alpha_{1}\vert^{2}\right)\alpha_{0}&0&\cdots\\0&\overline{\alpha_{2}}&-\alpha_{1}\overline{\alpha_{2}}&\overline{\alpha_{3}}\left(1- \vert\alpha_{2}\vert^{2}\right)&\cdots\\0&1&-\alpha_{1}&-\overline{\alpha_{3}}\alpha_{2}&\cdots\\0&0&0&\overline{\alpha_{4}}&\cdots\\\vdots&\vdots&\vdots&\vdots&\ddots\end{array} \right]
\end{equation}
The matrix \textit{Z} can be written as the product of two block-diagonal matrices (the columns within each $ 2\times 2 $-block correspond to the identities in  (\ref{eq:szego1}) and (\ref{eq:szego2}))
\begin{equation}
\left[\begin{array}{ccccc}\overline{\alpha_{0}}&1- \vert\alpha_{0}\vert^{2}&&&\\1&-\alpha_{0}&&&\\&&\overline{\alpha_{2}}&1- \vert\alpha_{2}\vert^{2}&\\&&1&-\alpha_{2}&\\&&&&\ddots\end{array} \right] \left[\begin{array}{ccccc}1&&&&\\&\overline{\alpha_{1}}&1- \vert\alpha_{1}\vert^{2}&&\\&1&-\alpha_{1}&&\\&&&\overline{\alpha_{3}}&\cdots\\&&&\vdots&\ddots\end{array} \right]
\end{equation}

Using (\ref{eq:szegonorm}), the CMV-matrix \textit{U} for the orthonormal trigonometric polynomials is then given by
\begin{equation}
\left[\begin{array}{ccccc}\overline{\alpha_{0}}&\rho_{0}&&&\\\rho_{0}&-\alpha_{0}&&&\\&&\overline{\alpha_{2}}&\rho_{2}&\\&&\rho_{2}&-\alpha_{2}&\\&&&&\ddots\end{array} \right] \left[\begin{array}{ccccc}1&&&&\\&\overline{\alpha_{1}}&\rho_{1}&&\\&\rho_{1}&-\alpha_{1}&&\\&&&\overline{\alpha_{3}}&\cdots\\&&&\vdots&\ddots\end{array} \right]
\end{equation}
for $ \rho_{j}=\sqrt{1- \vert\alpha_{j}\vert^{2}} $.
\subsection{Connection to quantum walks--the CMV-matrix}
Consider the quantum walk on the half-line $ \mathbb{Z}^{+}\cup\lbrace 0\rbrace $ with internal state set $ \mathcal{J}=\lbrace\uparrow,\downarrow\rbrace $, so the Hilbert space \textsf{H} is $ \ell^{2}(\mathbb{Z}\times\lbrace\uparrow,\downarrow\rbrace )  $. Then, taking the indices as $ 0\uparrow,0\downarrow,1\uparrow,1\downarrow,2\uparrow,\ldots $, following a proposal of Gr\"unbaum in~\cite{grunbaumCMV}, the CMV-matrix \textit{U} above can be used as the map for a quantum walk. With an abuse of notation, for a quantum walk on the line $ \mathbb{Z} $ with the same internal state set and with indices $ \ldots, -2\downarrow,-1\uparrow,-1\downarrow,0\uparrow,0\downarrow,1\uparrow,1\downarrow,2\uparrow,\ldots $, the matrix (underlines indicate indices corresponding spatial location zero) given by the product:
\begin{equation}
\left[\begin{array}{cccccccc}\ddots&&&&&&&\\&\overline{\alpha_{-2}}&\rho_{-2}&&&&&\\&\rho_{-2}&-\alpha_{-2}&&&&&\\&&&\underline{\overline{\alpha_{0}}}&\underline{\rho_{0}}&&&\\&&&\underline{\rho_{0}}&\underline{-\alpha_{0}}&&&\\&&&&&\overline{\alpha_{2}}&\rho_{2}&\\&&&&&\rho_{2}&-\alpha_{2}&\\&&&&&&&\ddots\end{array} \right] 
\end{equation}
\[\left[\begin{array}{ccccccccc}\ddots&\vdots&&&&&&&\\\cdots&\alpha_{-3}&&&&&&&\\&&&\overline{\alpha_{-1}}&\rho_{-1}&&&&\\&&&\rho_{-1}&\underline{-\alpha_{-1}}&\underline{0}&&&\\&&&&\underline{0}&\underline{\overline{\alpha_{1}}}&\rho_{1}&&\\&&&&&\rho_{1}&-\alpha_{1}&&\\&&&&&&&\overline{\alpha_{3}}&\cdots\\&&&&&&&\vdots&\ddots\end{array} \right]\]
will also be called a CMV-matrix and the $ \lbrace \alpha_{j}\rbrace $ called the Verblunsky coefficients, although there is no longer a connection to the orthogonal trigonometric polynomials.
\subsection{Coined quantum walks}
Once again, consider the quantum walk on the half-line $ \mathbb{Z}^{+}\cup\lbrace 0\rbrace $ with internal state set $ \mathcal{J}=\lbrace\uparrow,\downarrow\rbrace $, so the Hilbert space \textsf{H} is $ \ell^{2}(\mathbb{Z}\times\lbrace\uparrow,\downarrow\rbrace )  $. Then, taking the indices as $ 0\uparrow,0\downarrow,1\uparrow,1\downarrow,2\uparrow,\ldots $, if it is either of the form
\begin{equation}
\left[\begin{array}{ccccc}a_{0}&b_{0}&&&\\c_{0}&d_{0}&&&\\&&a_{2}&b_{2}&\\&&c_{2}&d_{2}&\\&&&&\ddots\end{array} \right]  \left[\begin{array}{ccccc}1&&&&\\&0&1&&\\&1&0&&\\&&&0&\cdots\\&&&\vdots&\ddots\end{array} \right]
\end{equation}
or
\begin{equation}
\left[\begin{array}{ccccc}1&&&&\\&0&1&&\\&1&0&&\\&&&0&\cdots\\&&&\vdots&\ddots\end{array} \right]\left[\begin{array}{ccccc}a_{0}&b_{0}&&&\\c_{0}&d_{0}&&&\\&&a_{2}&b_{2}&\\&&c_{2}&d_{2}&\\&&&&\ddots\end{array} \right] 
\end{equation}
then it is termed a \textit{coined walk} with \textit{coins} $ \left\lbrace\left[\begin{array}{cc}a_{2j}&b_{2j}\\c_{2j}&d_{2j}\end{array} \right] \right\rbrace $, which are unitary matrices. The first form is a CMV-matrix if $ b_{2j}=c_{2j} $ is a positive real or zero and if $ \overline{a_{2j}}=-d_{2j} $; then all the Verblunsky coefficients with odd index are zero and the Verblunsky coefficients with even index are given by $ \alpha_{2j}=-d_{2j} $. The second form is the adjoint of a CMV-matrix if the same conditions hold. The difference between the two forms for a quantum walk is clearly just a matter of transforming the initial state by the unitary matrix
\begin{equation}
\left[\begin{array}{ccccc}1&&&&\\&0&1&&\\&1&0&&\\&&&0&\cdots\\&&&\vdots&\ddots\end{array} \right]
\end{equation} 
Therefore, which form is adopted is largely a matter of convention, except for possible restrictions on the initial state.

If all the coins are the same, the quantum walk is termed a coined walk with \textit{constant coin}. Since the overall phase of the wavefunction is irrelevant, the phase of the determinant of the coin is arbitrary. Choosing it to be $ -1 $, the coin is necessarily of the form $ \left[\begin{array}{cc}\overline{\alpha}&\beta\\\overline{\beta}&-\alpha\end{array} \right] $ for some $ \alpha,\beta\in\mathbb{C} $ with $ \vert\alpha\vert^{2}+\vert\beta\vert^{2}=1 $. Then the condition for the unitary matrix for the walk to be a CMV-matrix (or the adjoint of one) is that $ \beta $ is positive real or zero. Similar terminology may be employed for walks on the full line.
\section{The Parrondo-like paradox for quantum walks}
\subsection{Set-up}
Consider the quantum walk on the line $ \mathbb{Z} $ with internal state set $ \mathcal{J}=\lbrace\uparrow,\downarrow\rbrace $, so the Hilbert space \textsf{H} is $ \ell^{2}(\mathbb{Z}\times\lbrace\uparrow,\downarrow\rbrace )  $. Let $ P_{+} $ be the orthogonal projector onto spatial locations with positive index, $ P_{-} $ be the orthogonal projector onto spatial locations with negative index, and $ P_{0} $ be the orthogonal projector onto spatial location zero, so $ P_{-}+P_{0}+P_{+} $ is the identity. Let $ U $ be the unitary operator that gives time evolution for one time step, $ \rho\to U\rho U^{*} $. Then, after \textit{n} time steps, an observation is made with an observation set $ \mathcal{R} $ with $ \sigma $-algebra $ \lbrace\varnothing, A,\tilde{A},\mathcal{R}\rbrace $ and base measure $ \tau $ given by $ \tau(A)=\tau(\tilde{A})=1 $. The pince-nez map \textit{L} is given by  
\begin{equation}
L\rho=1_{A}\text{tr }\rho P_{+}+1_{\tilde{A}}\text{tr }\rho (P_{0}+P_{-})
\end{equation}
Then, by considering two different initial states, $ \psi\psi^{*} $ and $ \xi\xi^{*} $, and the geodesic joining them, we can analyze the occurrence and extent of the Parrondo-like paradox. 
\subsection{Showing the paradox is impossible for certain classes of quantum walks}
Let $ \eta_{0\uparrow} $ be the wavefunction with one for $ \uparrow $ at location zero and all other amplitudes zero and $ \eta_{0\downarrow} $ be the wavefunction with one for $ \downarrow $ at location zero and all other amplitudes zero. Consider the six following cases for the Verblunsky coefficients determining the CMV-matrix \textit{U}: \textit{(i)} $ \alpha_{j}=\omega^{j}\alpha_{-j} $ for some $ \omega\in\mathbb{S}^{1}\subset\mathbb{C} $ and all $ j\in\mathbb{Z} $; \textit{(ii)} $ \alpha_{2j}=-\omega^{2j}\alpha_{-2j},\alpha_{2j+1}=\omega^{2j+1}\alpha_{-2j-1}  $  for some $ \omega\in\mathbb{S}^{1}\subset\mathbb{C} $ and all $ j\in\mathbb{Z} $; \textit{(iii)} $ \alpha_{j}=\overline{\alpha_{-j}} $ for all $ j\in\mathbb{Z} $; \textit{(iv)} $ \alpha_{j}=-\overline{\alpha_{-j}} $ for all $ j\in\mathbb{Z} $; \textit{(v)} $ \alpha_{2j}=\overline{\alpha_{-2j}},\alpha_{2j+1}=-\overline{\alpha_{-2j-1}}  $ for all $ j\in\mathbb{Z} $; and \textit{(vi)} $ \alpha_{2j}=-\overline{\alpha_{-2j}},\alpha_{2j+1}=\overline{\alpha_{-2j-1}}  $ for all $ j\in\mathbb{Z} $. With the preceding set-up, we have the following result:
\paragraph{Theorem 9.3.1} If the Verblunsky coefficients are in any of the preceding six cases, then for any initial wavefunctions $ \psi,\xi $ in the subspace for spatial location zero, $ \text{span}_{\mathbb{C}}\lbrace\eta_{0\uparrow},\eta_{0\downarrow}\rbrace $, the Parrondo-like paradox cannot occur.
\paragraph*{Proof} Let $ \psi^{(n)}=U^{n}\eta_{0\uparrow} $ and $ \xi^{(n)}=U^{n}\eta_{0\downarrow} $. Then $ \xi^{(n)} $ is related to $ \psi^{(n)} $ by the following, for each $ n\in \lbrace 1,\ldots\rbrace $ and $ j\in\mathbb{Z} $, depending on the case:\\
\begin{equation}
\begin{array}{lll}i)&\xi^{(n)}_{j,\uparrow}=-\omega^{j-n}\overline{\psi^{(n)}_{-j,\downarrow}}&\xi^{(n)}_{j,\downarrow}=\omega^{-j-n}\overline{\psi^{(n)}_{-j,\uparrow}}\\
ii)&\xi^{(n)}_{j,\uparrow}=(-\omega)^{j-n}\overline{\psi^{(n)}_{-j,\downarrow}}&\xi^{(n)}_{j,\downarrow}=(-\omega)^{-j-n}\overline{\psi^{(n)}_{-j,\uparrow}}\\
iii)&\xi^{(n)}_{j,\uparrow}=-\psi^{(n)}_{-j,\downarrow}&\xi^{(n)}_{j,\downarrow}=\psi^{(n)}_{-j,\uparrow}\\
iv)&\xi^{(n)}_{j,\uparrow}=\psi^{(n)}_{-j,\downarrow}&\xi^{(n)}_{j,\downarrow}=\psi^{(n)}_{-j,\uparrow}\\
v)&\xi^{(n)}_{j,\uparrow}=(-1)^{j+n+1}\psi^{(n)}_{-j,\downarrow}&\xi^{(n)}_{j,\downarrow}=(-1)^{j+n}\psi^{(n)}_{-j,\uparrow}\\
vi)&\xi^{(n)}_{j,\uparrow}=(-1)^{j+n}\psi^{(n)}_{-j,\downarrow}&\xi^{(n)}_{j,\downarrow}=(-1)^{j+n}\psi^{(n)}_{-j,\uparrow}\end{array}
\end{equation}
Therefore, in any of the six cases,
\begin{align}
\text{tr }\left[\begin{array}{cc}L(U^{n}\eta_{0\uparrow}\eta_{0\uparrow}^{*}U^{*n})\vert_{A}&L(U^{n}\eta_{0\uparrow}\eta_{0\downarrow}^{*}U^{*n})\vert_{A}\\L(U^{n}\eta_{0\downarrow}\eta_{0\uparrow}^{*}U^{*n})\vert_{A}&L(U^{n}\eta_{0\downarrow}\eta_{0\downarrow}^{*}U^{*n})\vert_{A}\end{array}\right]&=\text{tr }\left[\begin{array}{cc}\text{tr } P_{+} \psi^{(n)}\psi^{(n)*}&\text{tr } P_{+} \psi^{(n)}\xi^{(n)*}\\\text{tr } P_{+} \xi^{(n)}\psi^{(n)*}&\text{tr } P_{+} \xi^{(n)}\xi^{(n)*}\end{array}\right]\\&= \left\langle (P_{+}+P_{-}) \psi^{(n)},\psi^{(n)}\right\rangle\leq 1\nonumber
\end{align} 
Hence, by theorem \textbf{8.4.1}, for any initial wavefunctions $ \psi,\xi $ in the subspace for spatial location zero and any point on the geodesic joining $ [\psi] $ and $ [\xi] $, the paradox cannot occur. $ \square $ 
\medskip\\

In particular, note this rules out the paradox for the case of constant coin walks of the first form with initial wavefunctions $ \psi,\xi $ in the subspace for spatial location zero where the coin is of the form $ \left[\begin{array}{cc}\overline{\alpha}&\sqrt{1-\vert\alpha\vert^{2}}\\\sqrt{1-\vert\alpha\vert^{2}}&-\alpha\end{array} \right] $ for some $ \alpha\in\mathbb{C} $ with $ \vert\alpha\vert^{2}\leq 1 $. Using a limit theorem by Konno~\cite{konno1}~\cite{konno2}, we have the following limiting result for more general constant coin walks (not necessarily in the CMV-matrix form) adopting the second form of coined quantum walks:
\paragraph{Theorem 9.3.2} If a quantum walk has constant coin presented in the second form, then for any initial wavefunctions $ \psi,\xi $ in the subspace for spatial location zero, $ \text{span}_{\mathbb{C}}\lbrace\eta_{0\uparrow},\eta_{0\downarrow}\rbrace $, the Parrondo-like paradox cannot occur in the limit as $ n\to\infty $.
\paragraph*{Proof} Adapting the terminology of Konno to our notation, let the coin be given by $ \left[\begin{array}{cc}b&a\\d&c\end{array} \right]$. The wavefunctions $ \eta_{0\uparrow} $ and $ \eta_{0\downarrow} $ are given by $ \beta=1,\alpha=0 $ and $ \beta=0,\alpha=1 $ respectively. Then, employing Konno's limit theorem, the limit of the sum of probabilities,
\begin{equation}
\lim_{n\to\infty}\left(\left\langle P_{+}U^{n}\tau_{0\uparrow} ,U^{n}\tau_{0\uparrow}\right\rangle+ \left\langle P_{+}U^{n}\tau_{0\downarrow} ,U^{n}\tau_{0\downarrow}\right\rangle\right)
\end{equation}
is given by
\begin{equation}
\int_{0}^{\vert a\vert} \frac{2\sqrt{1-\vert a\vert^{2}}}{\pi(1-x^{2})\sqrt{\vert a\vert^{2}-x^{2}}}dx=1
\end{equation}
Hence, by theorem \textbf{8.4.1}, for any initial wavefunctions $ \psi,\xi $ in the subspace for spatial location zero and any point on the geodesic joining $ [\psi] $ and $ [\xi] $, the paradox cannot occur in the limit as $ n\to \infty $. $ \square $
\subsection{Examples of quantum walks displaying the paradox to the maximal extent}
\paragraph{Example 9.3.3} With the above set-up, take initial wavefunctions $ \psi=\frac{1}{\sqrt{2}}\left(\eta_{0\uparrow}+\eta_{0\downarrow}\right) $ and $ \xi=\frac{1}{\sqrt{2}}\left(\eta_{0\uparrow}-\eta_{0\downarrow}\right) $. Then, halfway on the minimizing geodesic between them, the initial wavefunction is $ \eta_{0\uparrow} $. Take all the Verblunsky coefficients to be zero except for $ \alpha_{-1} $, which has value $ \frac{1}{\sqrt{3}} $. Let \textit{U} be the corresponding CMV-matrix.

Let $ \psi^{(n)}=U^{n}\psi $, $ \xi^{(n)}=U^{n}\xi $, and $ \chi^{(n)}=U^{n}\eta_{0\uparrow} $ for $ n\geq 1 $. Then $ \psi^{(n)} $ has all amplitudes zero except for $ \frac{1}{\sqrt{2}} $ for $ \downarrow $ at location \textit{n}, $ -\frac{1}{\sqrt{6}} $ for $ \downarrow $ at location $ n-1 $, and $ \frac{1}{\sqrt{3}} $ for $ \uparrow $ at location $ -n $; $ \xi^{(n)} $ has all amplitudes zero except for $ -\frac{1}{\sqrt{2}} $ for $ \downarrow $ at location \textit{n}, $ -\frac{1}{\sqrt{6}} $ for $ \downarrow $ at location $ n-1 $, and $ \frac{1}{\sqrt{3}} $ for $ \uparrow $ at location $ -n $; and $ \chi^{(n)} $ has all amplitudes zero except for $ \sqrt{\frac{2}{3}} $ for $ \uparrow $ at location $ -n $ and $ -\frac{1}{\sqrt{3}} $ for $ \downarrow $ at location $ n-1 $. Consequently, $ P_{A,n}=P'_{A,n}=\frac{2}{3} $ for all $ n>1 $, yet $ P^{\text{geo}}_{A,n}=\frac{1}{3}$ for the initial wavefuntion halfway on the minimizing geodesic. By theorem \textbf{8.3.2}, this example is on the boundary of allowed values of $ \left(P_{A,n},P'_{A,n},P^{\text{geo}}_{A,n}\right)$ for the paradox.
\paragraph{Example 9.3.4} Again with the above set-up, let \textit{U} be the matrix for the constant coin walk in the second form with coin $ \frac{1}{\sqrt{2}}\left[\begin{array}{cc}1&1\\-1&1\end{array} \right]$. Take $ \varepsilon>0 $ small. Take $ \sigma_{1}>0 $ sufficiently small relative to $ \varepsilon $ such that the normal distribution with mean $ \varepsilon $ and variance $ \sigma_{1}^{2} $ has neglectfully small measure for $ (-\infty,0) $. Take $ a\in \left(-\frac{\pi}{2},0\right) $ and $ \sigma_{2}>0 $ such that: \textit{(i)} the normal distribution with mean $ a $ and variance $ \sigma_{2}^{2} $ has neglectfully small measure outside $ \left(-\frac{\pi}{2},0\right) $ and \textit{(ii)} we have
\begin{equation}
\int_{k\in\left(-\frac{\pi}{2},0\right)}\frac{-\sin k}{\sqrt{1+\cos^{2}k}}\,d\text{Normal}(a,\sigma_{2})(k)=\frac{1}{3}
\end{equation} 

Let $ \varphi,\zeta:\mathbb{Z}\to \mathbb{C} $ be given by
\begin{align}
\varphi_{j}&=\sqrt[4]{\frac{2\sigma_{1}^{2}}{\pi}}e^{-\sigma_{1}^{2}j^{2}+\imath\left(\frac{\pi}{2}-\varepsilon\right)j }=\frac{1}{\sqrt{2\pi}}\int_{-\infty}^{\infty}\frac{\exp\left(-\frac{\left(k-\left(\frac{\pi}{2}-\varepsilon\right)\right)^{2}}{4\sigma_{1}^{2}}+\imath jk\right)}{\sqrt[4]{2\pi\sigma_{1}^{2}}}dk\\&\approx\frac{1}{\sqrt{2\pi}}\int_{-\pi}^{\pi}\frac{\exp\left(-\frac{\left(k-\left(\frac{\pi}{2}-\varepsilon\right)\right)^{2}}{4\sigma_{1}^{2}}+\imath jk\right)}{\sqrt[4]{2\pi\sigma_{1}^{2}}}dk\nonumber\\
\zeta_{j}&=\sqrt[4]{\frac{2\sigma_{2}^{2}}{\pi}}e^{-\sigma_{2}^{2}j^{2}+\imath aj }=\frac{1}{\sqrt{2\pi}}\int_{-\infty}^{\infty}\frac{\exp\left(-\frac{\left(k-a\right)^{2}}{4\sigma_{2}^{2}}+\imath jk\right)}{\sqrt[4]{2\pi\sigma_{2}^{2}}}dk\\&\approx\frac{1}{\sqrt{2\pi}}\int_{-\pi}^{\pi}\frac{\exp\left(-\frac{\left(k-a\right)^{2}}{4\sigma_{2}^{2}}+\imath jk\right)}{\sqrt[4]{2\pi\sigma_{2}^{2}}}dk\nonumber
\end{align}
Then, by the inversion of Fourier series,
\begin{align}
\sum_{j\in\mathbb{Z}}\vert\varphi_{j}\vert^{2}&=\int_{-\infty}^{\infty}\frac{\exp\left(-\frac{\left(k-\left(\frac{\pi}{2}-\varepsilon\right)\right)^{2}}{2\sigma_{1}^{2}}\right)}{\sqrt{2\pi\sigma_{1}^{2}}}dk=1\\\sum_{j\in\mathbb{Z}}\vert\zeta_{j}\vert^{2}&=\int_{-\infty}^{\infty}\frac{\exp\left(-\frac{\left(k-a\right)^{2}}{2\sigma_{2}^{2}}\right)}{\sqrt{2\pi\sigma_{2}^{2}}}dk=1\nonumber\\
\sum_{j\in\mathbb{Z}}\varphi_{j}\overline{\zeta_{j}}&=\int_{-\infty}^{\infty}\frac{\exp\left(-\frac{\left(k-\left(\frac{\pi}{2}-\varepsilon\right)\right)^{2}}{2\sigma_{1}^{2}}-\frac{\left(k-a\right)^{2}}{4\sigma_{2}^{2}}\right)}{\sqrt[4]{2\pi\sigma_{1}^{2}}\sqrt[4]{2\pi\sigma_{2}^{2}}}dk\approx 0\nonumber
\end{align}
Hence, if the initial wavefunctions are taken to be $ \psi=\frac{1}{2}(\zeta+\varphi)\otimes\left[ \begin{array}{c}\imath\\1\end{array}\right] $ and $ \xi=\frac{1}{2}(\zeta-\varphi)\otimes\left[ \begin{array}{c}\imath\\1\end{array}\right] $, they will be properly normalized and approximately orthogonal. Midway on the geodesic joining them, we have $ \chi=\frac{1}{\sqrt{2}}\zeta \otimes\left[ \begin{array}{c}\imath\\1\end{array}\right]$.

Now suppose for the observations we always lump the internal states for any spatial sites, so the $ \sigma $-algebra for the observations is that generated by only the spatial locations. Let $ \lbrace Q_{B}\rbrace $ be the complete set of mutually commuting, orthogonal projectors for this $ \sigma $-algebra. Define $ \kappa:\mathbb{R}\to[-\pi,\pi] $ by
\begin{equation}
\kappa(x)=\begin{cases}\text{arccos}\frac{x}{\sqrt{1-x^{2}}}&\text{if }x\in[-\frac{1}{\sqrt{2}},\frac{1}{\sqrt{2}}]\\0&\text{if }x\in[\frac{1}{\sqrt{2}},\infty)\\\pi&\text{if }x\in(-\infty,-\frac{1}{\sqrt{2}}]\end{cases}
\end{equation}
Then, for each $ n $, let the complete set of commuting projectors $ \lbrace R^{(n)}_{C}\rbrace $ for the Borel $ \sigma $-algebra on $ [0,\pi] $ be given by $ R^{(n)}_{C}=Q_{(n\cdot\kappa^{-1}(C))\cap\mathbb{Z}} $. By Machida's limit theorem~\cite{machida1}~\cite{machida2}, the weak*-limit of the measures $\mu^{(n)}=\left\langle R^{(n)}U^{n}\psi,U^{n}\psi \right\rangle $ exists and is given by
\begin{align}
\mu^{(\infty)}\approx& \frac{1}{2}\text{Normal}\left(\frac{\pi}{2}-\varepsilon,\sigma^{2}_{2}\right)+\frac{1}{4}\left(1-\frac{\sin k}{\sqrt{1+\cos^{2}k}} \right)\text{Normal}(-a,\sigma_{2})\\&+ \frac{1}{4}\left(1+\frac{\sin k}{\sqrt{1+\cos^{2}k}} \right)\text{Normal}(\pi+a,\sigma_{2})\nonumber
\end{align} 
Similarly, the weak*-limit of the measures $\nu^{(n)}=\left\langle Q^{(n)}U^{n}\xi,U^{n}\xi \right\rangle $ exists and is given by the same expression. For both of these, there is approximately $ \frac{1}{2}+\frac{1}{4}\left(1-\frac{1}{3}\right)=\frac{2}{3} $ probability to have $ k\in\left[0,\frac{\pi}{2}\right) $, which corresponds to spatial locations with positive index $ j $. The weak*-limit of the measures $\tau^{(n)}=\left\langle Q^{(n)}U^{n}\chi,U^{n}\chi \right\rangle $ also exists and is given by
\begin{equation}
\tau^{(\infty)}\approx\frac{1}{2}\left(1-\frac{\sin k}{\sqrt{1+\cos^{2}k}} \right)\text{Normal}(-a,\sigma_{2})+ \frac{1}{2}\left(1+\frac{\sin k}{\sqrt{1+\cos^{2}k}} \right)\text{Normal}(\pi+a,\sigma_{2})
\end{equation}
However, now there is only approximately $ \frac{1}{2}\left(1-\frac{1}{3}\right)=\frac{1}{3} $ probability to have $ k\in\left[0,\frac{\pi}{2}\right) $. These values of two-thirds for $ P_{A} $, two-thirds for $ P'_{A} $, and one-third for $P^{\text{geo}}_{A}$ can be arbitrarily closely approached by taking $ \varepsilon $, $ \sigma_{1} $, and $ \sigma_{2} $ sufficiently small. By theorem \textbf{8.3.2}, this example can approach arbitrarily closely to the boundary of allowed values of $ \left(P_{A},P'_{A},P^{\text{geo}}_{A}\right)$ for the paradox. 
\paragraph{Remark} The results of the preceding example also hold for the commonly employed Hadamard coin $ \frac{1}{\sqrt{2}}\left[\begin{array}{cc}1&1\\1&-1\end{array} \right] $. Then the matrix \textit{U} is the adjoint of a CMV-matrix with all Verblunsky coefficients with even index equal to $ \frac{1}{\sqrt{2}} $. For the Hadamard coin, we take initial wavefunctions to be $ \psi=\frac{1}{2}(\zeta+\varphi)\otimes\left[ \begin{array}{c}\imath\\-1\end{array}\right] $ and $ \xi=\frac{1}{2}(\zeta-\varphi)\otimes\left[ \begin{array}{c}\imath\\-1\end{array}\right] $, so we have $ \chi=\frac{1}{\sqrt{2}}\zeta \otimes\left[ \begin{array}{c}\imath\\-1\end{array}\right]$, where 
\begin{align}
\varphi_{j}&=\sqrt[4]{\frac{2\sigma_{1}^{2}}{\pi}}e^{-\sigma_{1}^{2}j^{2}+\imath\varepsilon j }\\
\zeta_{j}&=\sqrt[4]{\frac{2\sigma_{2}^{2}}{\pi}}e^{-\sigma_{2}^{2}j^{2}-\imath(\frac{\pi}{2}+a)j }
\end{align}
for \textit{a}, $ \varepsilon $, $ \sigma_{1} $, and $ \sigma_{2} $ as in the given example. The coin utilized in the example was chosen to agree with that used by Machida~\cite{machida1}~\cite{machida2}.

%% file: appendix1ver2.tex
\chapter{General propositions}
\section{Banach space propositions}
\paragraph{Notation} Let \textsf{A}, \textsf{B},\ldots denote Banach spaces. 
\paragraph{Proposition A1.1} Maps in $ \mathcal{B}(\mathsf{A},\mathsf{B}) $ are continuous in the weak topologies on \textsf{A} and \textsf{B}.
\paragraph*{Proof} Take any such map \textit{L}. For any weak neighborhood 
\begin{equation}
\mathcal{N}(L\mathsf{a};\phi_{1},\ldots,\phi_{n};\varepsilon)=\left\lbrace \mathbf{b}\in\mathsf{B}\vert\, \vert\phi_{j}(\mathbf{b}-L\mathbf{a})\vert<\varepsilon\text{ for }j\in\lbrace 1,\ldots,n\rbrace\right\rbrace 
\end{equation}
with $ \mathsf{a}\in \mathsf{A} $, $ \phi_{1},\ldots,\phi_{n}\in\mathsf{B}^{*} $, and $ \varepsilon>0 $ we have
\begin{equation}
L\left(\mathcal{N}(\mathsf{a};L^{*}\phi_{1},\ldots,L^{*}\phi_{n};\varepsilon)\right)=\mathcal{N}(L\mathbf{a};\phi_{1},\ldots,\phi_{n};\varepsilon)\hspace{.4 in}\square
\end{equation}
\paragraph{Corollary A1.2} A map in $ L\in \mathcal{B}(\mathsf{A},\mathsf{B}) $ is uniquely determined by its values on a weakly dense subset.
\paragraph{Comment} Starting with a weakly dense subset that is a vector space, its norm closure is necessarily a linear subspace by the properties of Cauchy sequences. Since this subspace is convex, by the separating theorem~\cite{roydenseparating}, it cannot be weakly dense unless it is the entire space; hence, a weakly dense subset is necessarily also norm-dense. This argument is not entirely satisfactory since the separating theorem uses the Hahn-Banach theorem~\cite{roydenhahnbanach}, which depends on the axiom of choice~\cite{roydenaxiomofchoice}. This deficiency will be rectified for the situation of interest by proposition \textbf{B1.3} below.

The following important proposition is assigned as an exercise in~\cite{wilanskyextension}. 
\paragraph{Proposition A1.3} Let \textsf{A} and \textsf{B} be Banach spaces. A subset \textsf{V} of \textsf{A} that is a vector space uniquely determines a map $ L\in \mathcal{B}(\mathsf{A},\mathsf{B}) $ if \textsf{V} is dense in the norm topology on \textsf{A} and the operator norm of \textit{L} restricted to \textsf{V} is bounded. The operator norm of \textit{L} shares the bound to the operator norm of its restriction.
\paragraph{Proof} Since a continuous function is uniquely determined by its values on a weakly dense subset, and since the norm topology is finer than the weak topology, if \textit{L} exists, it is unique by \textbf{A1.2}. For any $ \mathsf{a}\in\mathsf{A} $, define $ L\mathsf{a} $ as the limit of the Cauchy sequence $ \left\langle L\mathsf{a}_{j}\right\rangle_{j=1}^{\infty} $, where $ \left\langle \mathsf{a}_{j}\right\rangle_{j=1}^{\infty} $ is any Cauchy sequence converging to \textbf{a} composed of elements of \textsf{V}. The choice of Cauchy sequence does not matter because, given another Cauchy sequence $ \left\langle \mathsf{b}_{j}\right\rangle_{j=1}^{\infty} $ converging to \textsf{a} composed of elements of \textsf{V}, then $ \left\langle \mathsf{a}_{j}-\mathsf{b}_{j}\right\rangle_{j=1}^{\infty} $ is a Cauchy sequence converging to zero, so $ \left\langle L(\mathsf{a}_{j}-\mathsf{b}_{j})\right\rangle_{j=1}^{\infty} $ is a Cauchy sequence converging to zero. Since the product of a Cauchy sequence with a scalar is a Cauchy sequence and the sum of two Cauchy sequences is a Cauchy sequence, \textit{L} is linear. $ \square $
\section{Hilbert space propositions}\label{sec:hilbert}
\paragraph{Notation}
Let \textsf{H} be a Hilbert space.
\paragraph{Proposition A2.1} Convergence of a countable sum of disjoint ($ P_{j}P_{k}=0 $ if $ j\neq k $), orthogonal projectors, $ \sum_{j}P_{j} $, to an orthogonal projector \textit{P} is equivalent in the following four topologies on $ \mathcal{B}(\mathsf{H}) $: ultrastrong-operator, strong-operator, ultraweak-operator\footnote{Same as the weak* topology.}, and weak-operator. 
\paragraph*{Proof} The weak-operator topology is courser than the other three, so convergence in any of the others implies convergence in it. Suppose convergence occurs in the weak-operator topology, so for any $ \psi,\psi'\in \mathsf{H} $,
\begin{equation}
\lim_{k\to \infty}\left\langle \left(P-\sum_{j=1}^{k}P_{j}\right)\psi,\psi'\right\rangle=0 
\end{equation} 
Then, for any fixed \textit{k},
\begin{equation}
\left\langle P\sum_{j=1}^{k}P_{j}\psi,\psi'\right\rangle=\lim_{m\to \infty}\left\langle \sum_{l=1}^{m}P_{j}\sum_{j=1}^{k}P_{j}\psi,\psi'\right\rangle=\left\langle \sum_{j=1}^{k}P_{j}\psi,\psi'\right\rangle
\end{equation}
Since $ \psi,\psi' $ were arbitrary, $ P\sum_{j=1}^{k}P_{j}=\sum_{j=1}^{k}P_{j} $. Similarly, $ \left(\sum_{j=1}^{k}P_{j}\right)P=\sum_{j=1}^{k}P_{j} $. Therefore, $ \left(P-\sum_{j=1}^{k}P_{j}\right)^{2}=P-\sum_{j=1}^{k}P_{j} $, so $ P-\sum_{j=1}^{k}P_{j} $ is itself an orthogonal projector (it is clearly self-adjoint). Then, by taking $ \psi=\psi' $, this implies 
\begin{equation}
\lim_{k\to \infty}\left\Vert \left(P-\sum_{j=1}^{k}P_{j}\right)\psi\right\Vert=0 
\end{equation}
so convergence necessarily also occurs in the strong-operator topology. Since the sequence of operators $ \left\langle P-\sum_{j=1}^{k}P_{j}\right\rangle_{k=1}^{\infty}  $ is bounded in operator norm (all being projectors), the convergence necessarily occurs in the ultrastrong-operator topology as well. However, the ultrastrong-operator topology is finer than the others, so convergence in it implies convergence in the other three. $ \square $ 
\section{Transition function propositions}
\paragraph{Notation} For the following, sets will be denoted $ \mathcal{X}$, $\mathcal{Y}$,$\ldots $ and $ \sigma $-algebras by $ \mathcal{E} $, $ \mathcal{F} $,$ \ldots $. The spaces of finite, signed-measures on the given set with the given $ \sigma $-algebra will be denoted $ \mathcal{M}(\mathcal{X};\mathcal{E})$, $ \mathcal{M}(\mathcal{Y};\mathcal{F}) $,$ \ldots $. These are Banach spaces using the total variation norm. Note--by convention product measures are written using $ \times $ although they are actually tensor products and should be written using $ \otimes $ (see~\cite{semadeniproduct} for a use of the latter notation).

Following~\cite{rao}, we have the following:
\paragraph{Definition A3.1} For $ \sigma $-algebras $ \mathcal{E} $ on $ \mathcal{X} $ and $ \mathcal{F} $ on $ \mathcal{Y} $, a function $ \tau(\cdot\vert\cdot):\mathcal{F}\times\mathcal{X}\to\mathbb{R} $ is a \textit{transition function} if: \textit{(i)} for each $ x\in\mathcal{X} $, $ \tau(\cdot\vert x)\in \mathcal{M}(\mathcal{Y};\mathcal{F})$; and \textit{(ii)} for each $ B\in \mathcal{F}$, $ \tau(B\vert \cdot)$ is a bounded, $ \mathcal{E} $-measurable function on $ \mathcal{X} $.
\medskip\\
If $ \tau(\cdot\vert\cdot) $ is positive and has the additional property that $ \tau(\mathcal{Y}\vert\cdot)=1_{\mathcal{X}} $, then it is termed a \textit{transition probability function}. The transition functions for specified ($ \mathcal{E} $, $ \mathcal{X} $, $ \mathcal{F} $, $ \mathcal{Y} $) clearly form a vector space. They form a Banach space using the norm
\begin{equation}
\Vert\tau(\cdot\vert\cdot)\Vert=\sup_{x\in\mathcal{X}}\Vert \tau(\cdot\vert x)\Vert_{\text{total variation}}
\end{equation} 
A transition function $ \tau(\cdot\vert\cdot) $ with specified data ($ \mathcal{E} $, $ \mathcal{X} $, $ \mathcal{F} $, $ \mathcal{Y} $) induces a linear map $ L\in\mathcal{B}(\mathcal{M}(\mathcal{X};\mathcal{E}),\mathcal{M}(\mathcal{Y};\mathcal{F}))$ via
\begin{equation}
(L\mu)(B)=\int_{x\in\mathcal{X}}\tau(B\vert x)\,d\mu(x)
\end{equation}
Not every bounded linear map is induced by a transition function; in general, a \textit{pseudo-transition ``function"} is required (see \textbf{B2.7} and \textbf{B2.8}). However, we do have the following:
\paragraph{Proposition A3.2} If $ L\in\mathcal{B}(\mathcal{M}(\mathcal{X};\mathcal{E}),\mathcal{M}(\mathcal{Y};\mathcal{F}))$ and $ K\in\mathcal{B}(\mathcal{M}(\mathcal{Y};\mathcal{F}),\mathcal{M}(\mathcal{Z};\mathcal{G}))$ are both induced by transition functions, then the composition $ K\circ L $ is also induced by a transition function.
\paragraph*{Proof} Let \textit{L} and \textit{K} be such maps, with associated transition functions $ \tau(\cdot\vert\cdot) $ and $ \nu(\cdot\vert\cdot)$. Define $ \omega(\cdot\vert\cdot) $ by
\begin{equation}
\omega(C\vert x)=\int_{y\in\mathcal{Y}}\nu( C\vert y)\,d(\tau(\cdot\vert x))(y)
\end{equation}
for $ C\in\mathcal{G} $ and $ x\in\mathcal{X} $. Then $ \omega(\cdot\vert x)\in \mathcal{M}(\mathcal{Z};\mathcal{G}) $. To see that $ \omega(C\vert \cdot) $ is  $ \mathcal{E} $-measurable, note that, since $ \nu( C\vert \cdot) $ is $ \mathcal{F} $-measurable and bounded, there is a sequence of simple functions $ \left\langle \sum_{k}b_{jk}1_{B_{jk}}\right\rangle  $ converging uniformly to it. By the Dominated Convergence theorem~\cite{roydendominated}, we then have
\begin{equation}
\omega(C\vert x)=\lim_{j\to \infty}\sum_{k}b_{jk}\tau(B_{jk}\vert x)
\end{equation} 
By~\cite{roydenmeasurable}, finite sums of measurable functions are measurable and the pointwise limit of a sequence of measurable functions is measurable, so $ \omega(C\vert \cdot) $ is measurable. Hence, $ \omega(\cdot\vert\cdot) $ is a transition function.

Now take any $ \mu\in \mathcal{M}(\mathcal{X};\mathcal{E}) $ and $ C\in \mathcal{G} $. Then, using the preceding results,
\begin{align}
((K\circ L)(\mu))(C)&=\int_{y\in\mathcal{Y}}\nu( C\vert y)\,d(L\mu)(y)=\lim_{j\to \infty}\sum_{k}b_{jk}(L\mu)(B_{jk})\\&=\lim_{j\to \infty}\sum_{k}b_{jk}\int_{x\in\mathcal{X}}\tau(B_{jk}\vert x)\,d\mu(x)=\int_{x\in\mathcal{X}}\omega(C\vert x)\,d\mu(x)\nonumber
\end{align} 
Hence, $ K\circ L $ is induced by the transition function $ \omega(\cdot\vert\cdot) $. $ \square $
\paragraph{Proposition A3.3} If $ L\in\mathcal{B}(\mathcal{M}(\mathcal{X};\mathcal{E}),\mathcal{M}(\mathcal{Z};\mathcal{G}))$ and $ K\in\mathcal{B}(\mathcal{M}(\mathcal{Y};\mathcal{F}),\mathcal{M}(\mathcal{W};\mathcal{H}))$ are both induced by transition functions, then the tensor product map $ K\otimes L $ is well-defined and also induced by a transition function.
\paragraph*{Proof} Let \textit{L} and \textit{K} be such maps, with associated transition functions $ \tau(\cdot\vert\cdot) $ and $ \nu(\cdot\vert\cdot)$. Define $ \omega(\cdot\vert\cdot) $ by $ \omega(\cdot\vert x,y)= \tau(\cdot\vert x)\times \nu(\cdot\vert y) $ for each $ x\in\mathcal{X} $ and $ y\in\mathcal{Y} $. Then $ \omega(\cdot\vert x,y) $ is clearly in $ \mathcal{M}(\mathcal{X}\times\mathcal{Y};\mathcal{I}) $, where $ \mathcal{I} $ is the $ \sigma $-algebra generated by the rectangular subsets $ \mathcal{R}=\mathcal{E}\times\mathcal{F} $. The rectangular subsets form a semialgebra\footnote{A collection of sets is a \textit{semialgebra} if it is closed under intersection and the complement of any set is a finite union of sets in the collection.} Then the finite union of rectangular sets is an algebra. 

Following Hausdorff~\cite{hausdorffborel}, let $ R_{0} $ be the rectangular subsets $ \mathcal{R} $. For each ordinal $ \alpha $, let $ R_{\alpha} $ be the collection of subsets of $ \mathcal{X}\times\mathcal{Y} $ that are the countable intersection of subsets from the various collections $ R_{\beta} $ for ordinals $ \beta<\alpha $ if $ \alpha $ is even and that are the countable union of subsets from the various collections $ G_{\beta} $ for ordinals $ \beta<\alpha $ if $ \alpha $ is odd (where all limit ordinals--those without a predecessor--taken even). For clarification, using the standard notation~\cite{roydenrdelta}, $ R_{1}=\mathcal{R}_{\sigma} $, $ R_{2}=\mathcal{R}_{\sigma\delta} $, and so on for the finite ordinals.

Fix any ordinal $ \alpha $ and suppose that for all subsets $ B $ in all collections $ R_{\beta} $ for $ \beta<\alpha $ we have the following property: there is an $ \mathcal{I} $-measurable $ \omega(B\vert\cdot) $ such that
\begin{equation}
((L\otimes K)\mu)(B)=\int_{(x,y)\in\mathcal{X}\times\mathcal{Y}}\omega( B\vert x,y)\,d\mu(x,y)
\end{equation}
for any $ \mathcal{I} $-measure $ \mu $ on $ \mathcal{X}\times\mathcal{Y} $. Take any $ C\in R_{\alpha} $. If $ \alpha $ is odd, we have a sequence $ \langle B_{j}\rangle $ of subsets from the various $ R_{\beta} $ with $ \beta<\alpha $ such that $ C=\bigcup_{j}B_{j} $. Let $ \omega(C\vert\cdot)=\bigvee_{j}\omega(B_{j}\vert\cdot) $. Then $ \omega(C\vert\cdot) $ is $ \mathcal{I} $-measurable (see~\cite{roydenmeasurable}) and we can extend $ L\otimes K $ by 
\begin{equation}\label{eq:propertytransfinite}
((L\otimes K)\mu)(C)=\int_{(x,y)\in\mathcal{X}\times\mathcal{Y}}\omega( C\vert x,y)\,d\mu(x,y)
\end{equation}
for any $ \mathcal{I} $-measure $ \mu $ on $ \mathcal{X}\times\mathcal{Y} $; hence, since \textit{C} was arbitrary, $ R_{\alpha} $ has the property. Similarly, if $ \alpha $ is even, we have a sequence $ \langle B_{j}\rangle $ of subsets from the various $ R_{\beta} $ with $ \beta<\alpha $ such that $ C=\bigcap_{j}B_{j} $. Let $ \omega(C\vert\cdot)=\bigwedge_{j}\omega(B_{j}\vert\cdot) $. Then $ \omega(C\vert\cdot) $ is measurable and we can extend $ L\otimes K $ by (\ref{eq:propertytransfinite}); hence, $ R_{\alpha} $ also has the property.

However, $ R_{0} $ has the property--for any $ \mathcal{I} $-measure $ \mu $, $ (L\otimes K)\mu $ is uniquely determined on the rectangular subsets by
\begin{equation}
((L\otimes K)\mu)(A\times B)=\int_{(x,y)\in\mathcal{X}\times\mathcal{Y}}\tau(A\vert x) \nu(B\vert y)\,d\mu(x,y)
\end{equation}
and $ \omega(A\times B\vert x,y)=\tau(A\vert x) \nu(B\vert y) $ is $ \mathcal{I} $-measurable (see~\cite{roydenmeasurable})--so by transfinite induction~\cite{hausdorfftransfinite}~\cite{sierpinskitransfinite} all the $ R_{\alpha} $ have the property. Following Kuratowski~\cite{kuratowskiborel}, the $ \sigma $-algebra $ \mathcal{I} $ is given by the union $ \bigcup_{\alpha}R_{\alpha} $. Therefore, $ \omega(\cdot\vert\cdot) $ is a transition function and induces a well-defined tensor product map $ K\otimes L $. $ \square $ 
\paragraph*{Comment} The preceding proof does not depend on the axiom of choice since the union only needs to be taken up to the ordinal number for the minimal uncountable well-ordered set~\cite{hausdorffborel}~\cite{kuratowskiborel}, whose existence does not depend on the axiom of choice~\cite{munkrestransfinite}.
\paragraph*{Lemma A3.4} For any measure space $ (\mathcal{X},\mathcal{E},\mu) $, any ``function" $ f\in L^{0}(\mathcal{X};\mu) $ contains some $ \mathcal{E} $-measurable function.
\paragraph*{Proof} Take any function $ g\in f $. For each $ x\in\mathcal{X} $ where $ g(x)\in  \lbrace\pm\infty, \text{undefined}\rbrace $, redefine \textit{g} so $ g(x)=0 $. The new \textit{g} is still in \textit{f}. Now construct the sequence of $ \mu $-measurable simple functions $ \langle g_{k}\rangle $ converging pointwise to \textit{g} by
\begin{equation}
g_{0}=0, g_{1}=1_{\lbrace g\geq 1\rbrace}-1_{\lbrace g\leq -1\rbrace}
\end{equation}
and for $ k\in \lbrace 1,2\ldots\rbrace $, 
\begin{equation}
g_{k+1}=2^{k}\cdot 1_{\lbrace g\geq 2^{k}\rbrace}+\sum_{m=1}^{4^{k}-1}\frac{m}{2^{k}}\,1_{\lbrace\frac{m+1}{2^{k}}>g\geq \frac{m}{2^{k}}\rbrace}+\sum_{m=-4^{k}+1}^{-1}\frac{m}{2^{k}}\,1_{\lbrace\frac{m-1}{2^{k}}<g\leq -\frac{m}{2^{k}}\rbrace}-2^{k}\cdot 1_{\lbrace g\leq -2^{k}\rbrace}
\end{equation}
By the definition of the completion $ \mathcal{E}_{0} $ of $ \mathcal{E} $~\cite{roydencompletemeasure}, for each $ \mathcal{E}_{0} $-measurable subset $ B $, there is some $ D\in\mathcal{E} $ such that $ D\supset\sim B $ and $ \mu(D)=\mu(\sim B) $; hence $ C=\sim D $ is in $ \mathcal{E} $, is contained within \textit{B}, and has $ \mu(C)=\mu(B) $. Using this, for $ k\in\lbrace 0,1,2\ldots\rbrace $ there are subsets $ \lbrace C^{k}_{m}\rbrace\in\mathcal{E} $ such that:
\begin{align}
&C^{(k+1)}_{2^{k}}\subset\left\lbrace g\geq 2^{k}\right\rbrace\text{ and } \mu\left(D^{(k+1)}_{2^{k}}\right)=0\text{ for }D^{(k+1)}_{2^{k}}=\left\lbrace g\geq 2^{k}\right\rbrace\setminus C^{(k+1)}_{2^{k}}\\&C^{(k+1)}_{m}\subset\left\lbrace\frac{m+1}{2^{k}}>g\geq \frac{m}{2^{k}}\right\rbrace\text{ and } \mu\left(C^{(k+1)}_{m}\right)=0\nonumber\\&\text{ for }D^{(k+1)}_{m}=\left\lbrace\frac{m+1}{2^{k}}>g\geq \frac{m}{2^{k}}\right\rbrace\setminus C^{(k+1)}_{2^{k}},  m\in \left\lbrace 1,2,\ldots, 4^{k}-1 \right\rbrace\\&C^{(k+1)}_{m}\subset\left\lbrace\frac{m-1}{2^{k}}<g\leq -\frac{m}{2^{k}}\right\rbrace\text{ and } \mu\left(D^{(k+1)}_{m}\right)=0\nonumber\\&\text{ for }D^{(k+1)}_{m}=\left\lbrace\frac{m-1}{2^{k}}<g\leq -\frac{m}{2^{k}}\right\rbrace\setminus C^{(k+1)}_{2^{k}},m\in \left\lbrace -1,-2,\ldots, -4^{k}+1 \right\rbrace\\&C^{(k+1)}_{-2^{k}}\subset\left\lbrace g\leq -2^{k}\right\rbrace\text{ and } \mu\left( D^{(k+1)}_{-2^{k}}\right)=0\text{ for }D^{(k+1)}=\left\lbrace g\leq -2^{k}\right\rbrace\setminus C^{(k+1)}_{-2^{k}}
\end{align}
Hence, there is a sequence of $ \mathcal{E} $-measurable simple functions $ \langle h_{k}\rangle $, where
\begin{equation}
h_{0}=0, h_{1}=1_{C^{(1)}_{1}}-1_{C^{(1)}_{-1}},h_{k+1}=\sum_{m=-4^{k}, m\neq 0}^{4^{k}}\frac{m}{2^{k}}\,1_{C^{(k+1)}_{m}} 
\end{equation}
Then $ \langle h_{k}\rangle $ converges pointwise to a $ \mathcal{E} $-measurable function \textit{h}~\cite{roydenmeasurable}. The function \textit{h} differs from \textit{g} on at most the subset $ \bigcup_{k}\bigcup_{m}D_{m}^{(k)} $, which is the countable union of $ \mu $-null subsets; hence, itself $ \mu $-null. Therefore, \textit{h} is necessarily in \textit{f}. $ \square $ 
\medskip\\
For the following, $ (\mathcal{X},\mathcal{E},\mu) $ is a measure space with $ \mu(\mathcal{X})=1 $  and $ \mathcal{F} $ is a $ \sigma $-algebra on $ \mathcal{Y} $.
\paragraph{Definition 3.5} A \textit{conditional probability} $ \nu(\cdot\vert\cdot) $ is a $ L^{1}(\mathcal{X};\mu)^{+} $-valued, vector $ \mathcal{F} $-measure on $ \mathcal{Y} $ with the property that $ \tau(\mathcal{Y}\vert\cdot)=[1_{\mathcal{X}}] $.   
\paragraph{Definition 3.6} A transition probability function $ \tau(\cdot\vert\cdot) $ is a \textit{lift} of a conditional probability $ \nu(\cdot\vert\cdot) $ if $ \tau(B\vert\cdot)\in \nu(B\vert\cdot) $ for every subset $ B\in \mathcal{F} $.
\paragraph{Proposition A3.7} If either: (\textit{i}) $ \mathcal{F} $ is generated by countably many atoms; (\textit{ii}) $ \mathcal{Y} $ is an uncountable, complete, separable, metric space with $ \mathcal{F} $ the Borel $ \sigma $-algebra; or (\textit{iii}) $ \mathcal{E} $ is generated by countably many atoms--then for any marginal probability $ \mu $ on $ \mathcal{X} $ and conditional probability $ \nu(\cdot\vert\cdot) $, there is some (generally non-unique) lift $ \tau(\cdot\vert\cdot) $.
\paragraph*{Proof--case(\textit{i})} For each atom $ A\in\mathcal{F} $, choose a particular, $ \mathcal{E} $-measurable function $ \hat{\nu}(A\vert\cdot)\in\hat{\nu}(A\vert\cdot) $ (these exist by \textbf{A3.4}). The function $ \hat{\nu}(A\vert\cdot) $ can only be less than zero on a set of $ \mu $-measure zero; hence, the set $ \mathcal{B}_{1} $ where any of these functions is less than zero is still of $ \mu $-measure zero. Furthermore, since the union of all atoms is $ \mathcal{X} $, we have
\begin{equation}
\sum_{\text{atoms }A\in\mathcal{F}}\hat{\nu}(A\vert\cdot)\in \nu(\mathcal{X}\vert\cdot)=[1_{\mathcal{X}}]
\end{equation} 
where the sum on the left-hand side converges pointwise; hence, the function given by the left-hand side has value one almost everywhere with respect to $ \mu $. Let $ \mathcal{B}_{2} $ be the set with $ \mu $-measure zero where it is not equal to one. Let $ \mathcal{B} $ be a set in $ \mathcal{E} $ with $ \mu $-measure zero containing $ \mathcal{B}_{1}\cup\mathcal{B}_{2} $ (such a set necessarily exists by the definition of $ \mu $-measure zero~\cite{roydenmeasurable}). Then, define $ \tau(\cdot\vert\cdot) $ by
\begin{equation}
\tau(B\vert x)=\begin{cases}\sum_{\text{atoms }A\in\mathcal{F}, A\subset B}\hat{\nu}(A\vert x)&\text{if }x\notin\mathcal{B}\\\int_{\mathcal{X}}\nu(B\vert\cdot)\;d\mu&\text{if }x\in\mathcal{B}\end{cases}
\end{equation}
for any $ B\in\mathcal{F} $ and $ x\in\mathcal{X} $. It is readily seen that $ \tau(\cdot\vert\cdot) $ is a lift of $ \nu(\cdot\vert\cdot) $.

\paragraph*{Case(\textit{ii})} The obvious approach is to use the separable and metric properties to form countable, locally finite open covers of decreasing radii using the paracompact property of metric spaces~\cite{roydenpara}~\cite{munkrespara}--from these a sequence of refining partitions can be formed. The $ \sigma $-algebra generated by these partitions is the Borel $ \sigma $-algebra; one may then hope that a procedure similar to the preceding with a choice for each element of each partition would give the lift $ \tau(\cdot\vert\cdot) $. However, while it easily seen this approach succeeds if a lift exists (and we will show one does below), it cannot be used to prove the existence of a lift; the problem is that to employ the Extension theorem~\cite{roydenextension} pointwise to get a measure, unlike the situation above, there are an uncountable number of conditions that must be met. While each fails on a set of measure zero, it is not clear that taken altogether the set where any one of them fails is still only of measure zero. (The existence of a lift would give this, but that is what we are trying to prove!) Instead, we will proceed in a different direction. The plan is to add in an auxiliary measure space, namely $ [0,1] $ with the standard topology, the Borel $ \sigma $-algebra, and Lebesgue measure. Then we will construct a deterministic transition probability function\footnote{One taking only the values zero and one.} (see \textbf{A3.9} for the connection to another common concept of determinism); integrating over the auxiliary space will then provide the desired lift.    

Start with the result due to Kuratowski that $ \mathcal{Y} $ is Borel equivalent\footnote{Two topological spaces $ \mathcal{X} $ and $ \mathcal{Y} $ are \textit{Borel equivalent} if there is a bijection $ \psi:\mathcal{X}\to \mathcal{Y} $ such that both $ \psi $ and $ \psi^{-1} $ take Borel sets to Borel sets (so both $ \psi^{-1} $ and $ \psi $ are Borel measurable).} to the interval $ [0,1] $ with the standard topology~\cite{roydenborelequivalent}; let $ \psi:[0,1]\to\mathcal{Y}  $ be the corresponding bijection. For $ \lambda $ Lebesgue measure on $ [0,1] $, let the $ \sigma $-ideal\footnote{A $ \sigma $\textit{-ideal} is a subcollection of a $ \sigma $-algebra that contains all subsets of member sets within the $ \sigma $-algebra and is closed under countable unions.} $ \mathcal{N} $ be the $ \mu\times \lambda $-null subsets of $ \sigma\left(\mathcal{E}\times\text{Borel}([0,1])\right)$. For each $ \alpha\in[0,1] $, let $  \hat{\nu}(\psi([0,\alpha))\vert \cdot) $ be a particular $ \mathcal{E} $-measurable (not merely $ \mu $-measurable) function from $ \nu(\psi([0,\alpha))\vert \cdot) $ (these exist by \textbf{A3.4}). Define the equivalence class $ A_{\alpha}$ of subsets of $\mathcal{X}\times[0,1]$  by
\begin{align}
\left[\bigcup_{s\in [0,1]}\left\lbrace  x\vert \hat{\nu}(\psi([0,\alpha))\vert x)> s\right\rbrace \times \lbrace s\rbrace\right]=\left[\bigcup_{q\in\mathbb{Q}\times [0,1]}\left\lbrace  x\vert \hat{\nu}(\psi([0,\alpha))\vert x)> q\right\rbrace \times [0,q]\right]
\end{align}
where the equivalence class is taken over sets that differ by subsets in the $ \sigma $-ideal $ \mathcal{N} $. Clearly this is independent of the choice of $ \hat{\nu}(\psi([0,\alpha))\vert \cdot) $. Note $ A_{\alpha}=[\mathcal{X}] $ for $ \alpha>1 $ and  $ A_{\alpha}=\mathcal{N} $ for $ \alpha\leq 0 $.

Then the collection $ \lbrace A_{\alpha}\rbrace $ is a soma\footnote{Following~\cite{roydensoma}, for $ \sigma $-algebra $ \mathcal{E} $ with $ \sigma $-ideal given by null sets $ \mathcal{N} $, the collection of equivalence classes indexed by reals $ \lbrace A_{\alpha}\rbrace $ of $ \mathcal{E}/\mathcal{N} $ is a \textit{soma} if: \textit{(i)} $ A_{\alpha}\leq A_{\beta} $ for $ \alpha\leq \beta $; and \textit{(ii)} $ A_{\alpha}=\bigvee_{j}A_{\beta_{j}}  $ for any countable, increasing sequence $ \langle\beta_{j}\rangle $ converging to $ \alpha $.} By~\cite{roydensomafunction}, there is a $ \sigma\left(\mathcal{E}\times\text{Borel}([0,1])\right)$-measurable function $ \varphi:\mathcal{X}\times[0,1]\to [0,1] $ for which the soma\footnote{The \textit{soma} of the equivalence class $ [f] $ of a real valued function \textit{f} is the collection of equivalence classes of preimages, $ \left[f^{-1}((-\infty,\alpha))\right] $, over $ \alpha\in\mathbb{R} $. } of $ [\varphi] $ is $ A_{\alpha} $. (Note $ \varphi $ is not unique; any particular function in $ [\varphi] $ will do.)

Now define $ \tau(\psi([0,\alpha))\vert\cdot) $ by
\begin{equation}
\tau(\psi([0,\alpha))\vert x)=\text{essential sup} \left\lbrace s\in[0,1]\vert \varphi(x,s)\in[0,\alpha)\right\rbrace
\end{equation}
where the essential supremum is taken over equivalence classes of sets differing by Lebesgue-null sets. By Tonelli's theorem~\cite{roydentonelli} , $ \tau(\psi([0,\alpha))\vert\cdot) $ is an $ \mathcal{E} $-measurable function. Furthermore, it is it $ \nu(\psi([0,\alpha))\vert \cdot) $. To see this, note that by Tonelli's theorem, for almost every $ x\in\mathcal{X} $ with respect to $ \mu $, the sets
\begin{equation}
\left\lbrace  s\in[0,1]\vert \hat{\nu}(\psi([0,\alpha))\vert x)> s\right\rbrace\text{ and }\left\lbrace s\in[0,1]\vert \varphi(x,s)\in[0,\alpha)\right\rbrace
\end{equation}
differ by only a Lebesgue measure zero subset. Taking the essential supremum over each set, we have that $ \hat{\nu}(\psi([0,\alpha))\vert \cdot)=\tau(\psi([0,\alpha))\vert \cdot) $ almost everywhere with respect to $ \mu $.

More generally, for any $ B\in\text{Borel}(\mathcal{Y}) $, define $ \tau( B\vert\cdot) $ by
\begin{equation}
\tau( B\vert x)= \lambda\left(\left\lbrace s\in[0,1]\vert \psi\circ\varphi(x,s)\in B \right\rbrace  \right)
\end{equation}
Since the composition of measurable functions is measurable (readily seen by the definition of measurability), then, by Tonelli's theorem, $ \tau( B\vert \cdot) $ is $ \mathcal{E} $-measurable. For each $ x\in\mathcal{X} $, $ \tau(\cdot\vert x) $ is a Borel measure on $ \mathcal{Y} $ with $ \tau(\mathcal{Y}\vert x)=1 $. Finally, since the collection $ \lbrace[0,\alpha)\rbrace $ over $ \alpha\in\mathbb{R} $ generates the Borel $ \sigma $-algebra on $ [0,1] $, the collection $ \lbrace\psi([0,\alpha))\rbrace $ generates the Borel $ \sigma $-algebra on $ \mathcal{Y} $, so $ \tau( B\vert\cdot)\in \nu( B\vert\cdot) $ for any $ B\in\text{Borel}(\mathcal{Y}) $.
\paragraph*{Case(\textit{iii})} For each atom $ A\in\mathcal{E} $ with $ \mu(A)>0 $, $ \nu(B\vert x) $ is well-defined for any $ x\in A $ and $ B\in\mathcal{F} $, so $ \tau(B\vert x) $ is necessarily $\nu(B\vert x)  $. On the remaining atoms, $ \tau( \cdot\vert\cdot) $  can be set rather arbitrarily; take it to be $ \int_{\mathcal{X}}\nu(B\vert\cdot)\;d\mu $ for concreteness. $ \square $ 
\medskip\\
Examining the proof of the preceding theorem, we have the following interesting result:
\paragraph{Proposition A3.8} If either: (\textit{i}) $ \mathcal{F} $ is generated by countably many atoms; or (\textit{ii}) $ \mathcal{Y} $ is an uncountable, complete, separable, metric space with $ \mathcal{F} $ the Borel $ \sigma $-algebra--then for any transition probability function $ \tau(\cdot\vert\cdot):\mathcal{F}\times\mathcal{X}\to[0,1] $ there is a deterministic transition probability function $ \xi(\cdot\vert\cdot):\mathcal{F}\times\mathcal{X}\times[0,1]\to\lbrace 0,1\rbrace $ such that  
\[\tau(B\vert x)=\int_{s\in[0,1]}\xi(B\vert x,s)\;d\lambda(s)\]
where $ \lambda $ is Lebesgue measure on $ \mathbb{R} $.
\paragraph*{Proof} First take case (\textit{i}). By the definition of being countable, there is an ordering of the atoms $ \lbrace A_{j}\rbrace $ for $ j\in\lbrace 1,2,\ldots\rbrace $ (with the indexing set truncating to $ j\in\lbrace 1,2,\ldots, n\rbrace $ if there are only finitely many atoms). For $ m\in\lbrace 1,2,\ldots\rbrace $ (or $ m\in\lbrace 1,2,\ldots, n\rbrace $), define $ B_{m} $ by $ \bigcup_{j=1}^{m}A_{j} $ and $ C_{m}\subset\mathcal{X}\times[0,1] $ by
\begin{equation}
\bigcup_{s\in [0,1]}\left\lbrace  x\vert \tau(B_{m}\vert x)> s\right\rbrace \times \lbrace s\rbrace=\bigcup_{q\in\mathbb{Q}\times [0,1]}\left\lbrace  x\vert\tau(B_{m}\vert x)> q\right\rbrace \times [0,q]
\end{equation}
Then $ \langle C_{m}\rangle $ is an increasing, nested sequence of subsets in $ \sigma(\mathcal{E}\times[0,1]) $ such that $ \bigcup_{m}C_{m}=\mathcal{X}\times[0,1] $. 

Define the function $ \varphi:\mathcal{X}\times[0,1]\to \mathbb{Z}^{+} $ by
\begin{equation}
\varphi(x,s)=\sum_{j}1_{\mathcal{X}\setminus C_{j}}
\end{equation}
Define $ \xi(\cdot\vert\cdot) $ by
\begin{equation}
\xi\left(B\vert x,s\right)=\begin{cases}1&\text{if }A_{\varphi(x,s)}\subset B\\0&\text{otherwise}\end{cases}
\end{equation}
for any $ B\in\mathcal{F} $, $ x\in\mathcal{X}  $, and $ s\in[0,1] $. Define $ \tau'(\cdot\vert\cdot) $ by
\begin{equation}
\tau'(B\cdot x)=\int_{s\in[0,1]}\xi(B\vert x,s)\;d\lambda(s)
\end{equation}
for any $ B\in\mathcal{F} $ and $ x\in\mathcal{X}  $. It is readily seen that $ \xi(\cdot\vert\cdot) $ is a deterministic transition probability function and $  \tau'(\cdot\vert\cdot) $ is a transition probability function. Furthermore, for any $ m\in\lbrace 1,2,\ldots\rbrace $ (or $ m\in\lbrace 1,2,\ldots, n\rbrace $),
\begin{equation}
\tau'(B_{m}\vert x)=\int_{s\in[0,1]}\xi(B_{m}\vert x,s)\;d\lambda(s)=\lambda\left(C_{m}\cap\lbrace x\rbrace\times[0,1]\right)=\tau(B_{m}\vert x)
\end{equation}
Since the collection $ \lbrace B_{m}\rbrace $ generates $ \mathcal{F} $, $ \tau(\cdot\vert\cdot)=\tau'(\cdot\vert\cdot) $.

Now take case (\textit{ii}). As with the proof of the preceding theorem, start with the proposition that $ \mathcal{Y} $ is Borel equivalent to the interval $ [0,1] $ with the standard topology; let $ \psi:[0,1]\to\mathcal{Y}  $ be the corresponding bijection. Define the $ \sigma(\mathcal{E}\times\text{Borel}([0,1])) $-measurable subsets $ A_{\alpha}$ of $\mathcal{X}\times[0,1]$  by
\begin{align}
A_{\alpha}=\bigcup_{s\in [0,1]}\left\lbrace  x\vert \tau(\psi([0,\alpha))\vert x)> s\right\rbrace \times \lbrace s\rbrace\\=\bigcup_{q\in\mathbb{Q}\times [0,1]}\left\lbrace  x\vert\tau(\psi([0,\alpha))\vert x)> q\right\rbrace \times [0,q]
\end{align}
Note $ A_{\alpha}=\mathcal{X} $ for $ \alpha>1 $ and  $ A_{\alpha}=\varnothing $ for $ \alpha\leq 0 $.

Then, following~\cite{roydenordinatefunction} the collection of ordinate sets $ \lbrace A_{\alpha}\rbrace $ defines a $ \sigma(\mathcal{E}\times\text{Borel}([0,1])) $-measurable function $ \varphi:\mathcal{X}\times[0,1]\to [0,1] $ by 
\begin{equation}
\varphi(x,s)=\inf\lbrace \alpha\in\mathbb{R}\vert (x,s)\in A_{\alpha}\rbrace
\end{equation}
Define $ \xi(\cdot\vert\cdot) $ by
\begin{equation}
\xi\left(B\vert x,s\right)=\begin{cases}1&\text{if }\psi\circ\varphi(x,s)\in B\\0&\text{otherwise}\end{cases}
\end{equation}
for any $ B\in\mathcal{F} $, $ x\in\mathcal{X}  $, and $ s\in[0,1] $. Define $ \tau'(\cdot\vert\cdot) $ by
\begin{equation}
\tau'(B\cdot x)=\int_{s\in[0,1]}\xi(B\vert x,s)\;d\lambda(s)
\end{equation}
for any $ B\in\mathcal{F} $ and $ x\in\mathcal{X}  $. It is readily seen that $ \xi(\cdot\vert\cdot) $ is a deterministic transition probability function and $  \tau'(\cdot\vert\cdot) $ is a transition probability function. Furthermore, for any $ \alpha\in \mathbb{R}$, 
\begin{align}
\tau'(\psi([0,\alpha))\vert x)=\int_{s\in[0,1]}\xi(\psi([0,\alpha))\vert x,s)\;d\lambda(s)\nonumber\\=\lambda\left(A_{\alpha}\cap\lbrace x\rbrace\times[0,1]\right)=\tau(\psi([0,\alpha))\vert x)
\end{align}
Finally, since the collection $ \lbrace[0,\alpha)\rbrace $ over $ \alpha\in\mathbb{R} $ generates the Borel $ \sigma $-algebra on $ [0,1] $, the collection $ \lbrace\psi([0,\alpha))\rbrace $ generates the Borel $ \sigma $-algebra on $ \mathcal{Y} $, so $ \tau( \cdot\vert\cdot)= \tau( \cdot\vert\cdot) $. $ \square $
\paragraph*{Comment} Note the preceding can be readily extended to cover cases where $ \mu $ is the zero measure apart from a sub-$ \sigma $-algebra of $ \mathcal{E} $ generated by countably many atoms or the marginal probability on $ \mathcal{Y} $ formed from $ \nu(\cdot\vert\cdot) $ and $ \mu $ is the zero measure apart from a sub-$ \sigma $-algebra of $ \mathcal{F} $ generated by countably many atoms.
\paragraph{Proposition A3.9} If either: (\textit{i}) $ \mathcal{F} $ is generated by countably many atoms; or (\textit{ii}) $ \mathcal{Y} $ is an uncountable, complete, separable, metric space with $ \mathcal{F} $ the Borel $ \sigma $-algebra--then any deterministic transition probability function $ \tau(\cdot\vert\cdot):\mathcal{F}\times\mathcal{X}\to\lbrace 0,1\rbrace $ is induced by some measurable point transformation $ f:\mathcal{X}\to\mathcal{Y} $, $ \tau(B\vert x)=(1_{B}\circ f) (x) $.
\paragraph{Proof} First take case (\textit{i}). For each $ x\in\mathcal{X} $, take $ f(x) $ to be any value in the atom $ A\in\mathcal{X} $ with the property $ \tau(A\vert x)=1\Leftrightarrow \tau(\mathcal{Y}\setminus A\vert x)=0 $.

Now take case (\textit{ii}). As with the preceding proofs, start with the proposition that $ \mathcal{Y} $ is Borel equivalent to the interval $ [0,1] $ with the standard topology; let $ \psi:[0,1]\to\mathcal{Y}  $ be the corresponding bijection. For each $ x\in\mathcal{X} $, take $ f(x) $ to be defined by
\begin{equation}
f(x)=\psi\left(\sup\lbrace \alpha\in[0,1]\vert \tau(\psi([0,\alpha))\vert x)=0\rbrace\right)
\end{equation} 
Then \textit{f} is measurable since $ \lbrace(\alpha,1]\vert \alpha\in\mathbb{R}\rbrace $ generate the Borel subsets of $ [0,1] $, so $\lbrace\psi((\alpha,1])\vert \alpha\in\mathbb{R}\rbrace $ generate the Borel subsets of $\mathcal{Y} $, and 
\begin{equation}
f^{-1}\left(\psi((\alpha,1])\right)=\lbrace x\vert \psi^{-1}\circ f(x)>\alpha\rbrace=\lbrace x\vert\tau(\psi((\alpha,1])\vert x)=1 \rbrace\in\mathcal{E}\hspace{.4 in}\square
\end{equation}
\paragraph{Proposition A3.10} If the marginal probability $ \mu $ on $ \mathcal{X} $ and conditional probability $ \nu(\cdot\vert\cdot) $ are such that the resulting joint probability is absolutely continuous with respect to the product measure $ \mu\times\eta $ for some $ \mathcal{F} $-measure $ \eta $ on $ \mathcal{Y} $, then there is some (generally non-unique) lift $ \tau(\cdot\vert\cdot) $.
\paragraph*{Proof} The joint probability measure $ \rho $ resulting from $ \mu $ and $ \nu(\cdot\vert\cdot) $ is
\begin{equation}
\rho(A\times B)=\int_{A}\nu(B\vert\cdot);d\mu
\end{equation}
for any rectangular subset $ A\times B\in \mathcal{E}\times\mathcal{F} $. By the Radon-Nikod\'{y}m theorem~\cite{roydenradonnikodym}, there is a ``function" $ g\in L^{1}(\mathcal{X}\times\mathcal{Y};\mu\times\eta) $, the Radon-Nikod\'{y}m derivative $ g=\dfrac{d\rho}{d(\mu\times\eta)} $, such that $ \rho=g\;\mu\times\eta $. Choose any particular $ \sigma(\mathcal{E}\times\mathcal{F}) $-measurable function $ \hat{g}\in g $ (such a function necessarily exists by \textbf{A3.4}). The function $\hat{g} $ is less than zero on a subset of $ \mu\times\eta $-measure zero; by changing $\hat{g} $ to be zero there, the new $\hat{g} $ is still in \textit{g}. By Tonelli's theorem, $ \xi(B\vert\cdot)=\int_{y\in B}\hat{g}(\cdot,y)\;d\eta(y) $ is $ \mathcal{E} $-measurable. Take $ \tau(B\vert x) $ to be $ \xi(B\vert x) $ for all \textit{x} where $ \xi(\mathcal{X}\vert x)= 1 $; on the $ \mathcal{E} $-measurable, $ \mu $-null subset where $ \xi(\mathcal{X}\vert \cdot)\neq 1 $, take $ \tau(B\vert x) $ to be $ \rho(\mathcal{X}\times B) $. $ \square $
\paragraph*{Comment} Instead of the approach of \textbf{A3.7}, one may attempt a simpler approach using the notion of liftings from $L^{\infty}(\mathcal{X};\mu)$ to $\mathcal{E} $-measurable functions on $ \mathcal{X} $.~\cite{raolifting}~\cite{traynor}~\cite{tulcea} 
\paragraph*{Definition A3.11} A map from $ L:L^{\infty}(\mathcal{X};\mu)\to\mu $-measurable functions on $ \mathcal{X} $ is a \textit{linear lifting} if it has the following properties: \textit{(i)} \textit{L} is positive, linear, and continuous (bounded); \textit{(ii)} $ Lf\in f $ for any $ f\in L^{\infty}(\mathcal{X};\mu) $; and \textit{(iii)} $ L[1_{\mathcal{X}}]= 1_{\mathcal{X}}$.
\paragraph*{Definition A3.12} A linear lifting is a \textit{lifting} if it also respects the multiplicative structure\footnote{Then the lifting \textit{K} necessarily takes ``functions" containing characteristic sets to characteristic sets, which reduces the problem of the existence of liftings to that of the existence of set liftings.} of $ L^{\infty}(\mathcal{X};\mu) $: \textit{(iv)} $ K(fg)=Kf\, Kg $ for any $ f,g\in L^{\infty}(\mathcal{X};\mu) $.
\paragraph*{Definition A3.13} A lifting is an $ \mathcal{E} $\textit{-lifting} if it maps into $\mathcal{E} $-measurable functions (rather than simply $ \mu $-measurable ones).
\medskip\\
The existence of $ \mathcal{E} $-liftings is in general an open question, even in simple cases.\footnote{For instance, $ \mathcal{X} =[0,1]$, $ \mathcal{G} $ being the Borel $ \sigma $-algebra, and $ \mu $ being Lebesgue measure.~\cite{tulcea}} Note the Lifting theorem~\cite{raolifting}~\cite{traynor} does give the existence of a lifting $ K:L^{\infty}(\mathcal{X};\mu)\to\mathcal{E}_{0} $-measurable functions on $ \mathcal{X} $. However, the completed $ \sigma $-algebra $ \mathcal{E}_{0} $ depends on the marginal probability $ \mu $, so this is not what we desire. Furthermore, the proof of the Lifting theorem relies on the Axiom of Choice and is nonconstructive~\cite{raolifting}, so it is of little practical use. In addition, as the following example demonstrates, this approach cannot work in general:
\paragraph*{Counter-example A3.14} Let $ \mathcal{X}=\mathcal{Y}=[0,1] $ with the usual topology. Let $ \mathcal{E}=\mathcal{F} $ be the resulting Borel $ \sigma $-algebra. Let $ \lambda $ be Lebesgue measure on $[0,1]$ and the joint probability $ \rho $ be the diagonal measure
\begin{equation}
\rho(E)=\lambda\left(\lbrace x\in [0,1]\vert (x,x)\in E\rbrace\right)
\end{equation}
Then the marginal probability $ \mu=\lambda $ and the conditional probability $ \nu(B\vert\cdot)=\left[1_{B}\right] $. There is an obvious lift given by $ \tau(B\vert\cdot)=1_{B} $. However, this cannot be written as a lifting applied to $ \nu(B\vert\cdot) $; else, since $  \tau(B\vert\cdot)\neq  \tau(B'\vert\cdot) $ for $ B\neq B' $ yet $ B$ and $ B' $ differ by a $ \lambda $-null subset, so $ \left[1_{B}\right]=\left[1_{B'}\right] $, we have the contradiction
\[1_{B}=\tau(B\vert\cdot)=K\left[1_{B}\right]=K\left[1_{B'}\right]=\tau(B'\vert\cdot)=1_{B'}\]
We now claim that there is no lift $ \tau(\cdot\vert\cdot) $ of $ \nu(\cdot\vert\cdot) $ which can be written as $ \tau(B\vert\cdot)=K\nu(B\vert\cdot) $ for all Borel subsets \textit{B} for a lifting map \textit{K}.
\paragraph*{Proof} Suppose there were such a lift coming from a lifting map \textit{K}. Since liftings take characteristic functions to characteristic functions, \textit{K} induces a set lifting $ \kappa $ by $ K[1_{B}]=1_{C}\Leftrightarrow C=\kappa B $. Let $ A_{0}=[0,1] $. Then $ \kappa A_{0} $ differs from $ A_{0} $ by a $ \lambda $-null set. (Actually, $ \kappa[0,1]=[0,1] $, but we are seeking to establish a general pattern.) By definition, there is an open set $ B_{0} $ that covers $ A_{0}\setminus \kappa A_{0} $ with $ \lambda(B_{0})<\frac{1}{4} $ (one-quarter can be replaced by anything strictly between one-half and zero). Now, iteratively, define $ A_{j}=\left[\frac{m_{j}}{2^{j}},\frac{m_{j}+1}{2^{j}}\right] $ to be either the subset $ \left[\frac{m_{j-1}}{2^{j-1}},\frac{2m_{j-1}+1}{2^{j}}\right] $ or $ \left[\frac{2m_{j-1}+1}{2^{j}},\frac{m_{j-1}+1}{2^{j-1}}\right] $ depending on which one has 
\begin{equation}
\sum_{k=0}^{j-1}\lambda\left( A_{j}\cap B_{k}\right)\leq \frac{1}{2}\sum_{k=0}^{j-1}\lambda\left( A_{j-1}\cap B_{k}\right)
\end{equation}
Once we have $ A_{j} $, we take $ B_{j} $ to be an open set that covers $ A_{j}\setminus \kappa A_{j} $ with $ \lambda(B_{j})<\frac{1}{4^{j+1}} $, then continue to $ A_{j+1} $.

Since $ [0,1]  $ is compact and each $ A_{j} $ is closed, the decreasing nested sequence of subsets $ \langle A_{j}\rangle $ has nonempty intersection by the finite intersection property~\cite{roydenfiniteintersection}~\cite{munkresfiniteintersection}. Since the diameters of the $ A_{j} $ tend to zero, this intersection is a single point--call it $ x_{0} $. Then $ x_{0} $ is also in the intersection $ \bigcap_{j}\kappa A_{j} $. Suppose otherwise, so it is not in some $ \kappa A_{j} $; hence, it is in $ B_{j} $. Since $ B_{j} $ is open, then there is some ball, $ \left(x_{0}-\varepsilon,x_{0}+\varepsilon \right)  $, also in $ B_{j} $. Then, for $ k $ sufficiently large so $ \varepsilon> 2^{-k} $, $ A_{k}\subset B_{j}$. However,
\begin{align}
\lambda\left(A_{k}\setminus \bigcup_{m=0}^{k}B_{m}\right)\geq \frac{1}{2^{k}}-\sum_{m=0}^{k}\lambda\left(A_{k}\cup B_{m}\right)\nonumber\\\geq \frac{1}{2^{k}}-\frac{1}{2}\sum_{m=0}^{k-1}\lambda\left(A_{k-1}\cup B_{m}\right)-\frac{1}{4^{k+1}}
\end{align}
Continuing, we eventually arrive at
\begin{equation}
\lambda\left(A_{k}\setminus \bigcup_{m=0}^{k}B_{m}\right)\geq\frac{1}{2^{k}}-\sum_{m=0}^{k}\frac{1}{2^{k-m}4^{m+1}}>\frac{1}{2}\cdot\frac{1}{2^{k}}>0
\end{equation}
which is a contradiction to $ A_{k}\subset B_{j}$. Hence, $ x_{0}\in  \bigcap_{j}\kappa A_{j} $. Note, by the positivity of \textit{K}, $ \langle \kappa A_{j}\rangle $ is also a decreasing nested sequence (with $ \kappa A_{j}\subset A_{j} $) and $ x_{0} $ is the only point in the intersection.

However, this leads to a contradiction. For each $ j $, let $ E_{j}=A_{j}\setminus\lbrace x_{0}\rbrace $, so $ \bigcup_{j}E_{j}=\varnothing $. Then, since $ A_{j}\sim E_{j}\sim \kappa A_{j} $, pointwise
\begin{equation}
\lim_{j\to \infty}K\nu\left(\left.E_{j}\right\vert\cdot\right)=\lim_{j\to \infty}K\left[1_{E_{j}}\right]=\lim_{j\to \infty}1_{\kappa A_{j}}
\end{equation}
which is zero except at $ x_{0} $, where it has the value one. For each $ j $, let $ F_{j}=E_{j}\setminus E_{j+1} $, so $ \lbrace F_{j}\rbrace $ is a disjoint collection. Then, pointwise, by the countable additivity of measures,
\begin{equation}
\sum_{j}K \nu\left(\left.F_{j}\right\vert\cdot\right)=K\nu\left(\left.\bigcup_{j}F_{j}\right\vert\cdot\right)=K\nu\left(\left.[0,1]\setminus\lbrace x_{0}\rbrace\right\vert\cdot\right)=1_{\kappa[0,1]}=1_{[0,1]}
\end{equation}
However, for any $ m $,
\begin{equation}
1_{[0,1]}=K\nu\left(\left.\bigcup_{j}F_{j}\right\vert\cdot\right)=K\nu\left(\left.\bigcup_{j=1}^{m-1}F_{j}\cup E_{m}\right\vert\cdot\right)=\sum_{j=1}^{m}K\nu\left(\left.F_{j}\right\vert\cdot\right)+K\nu\left(\left.E_{m}\right\vert\cdot\right)
\end{equation}
Taking the limit as $ m\to \infty $ gives the right-hand side as 
\begin{equation}
\sum_{j}K\nu\left(\left.F_{j}\right\vert\cdot\right)+\lim_{m\to \infty}K\nu\left(\left.E_{m}\right\vert\cdot\right)=1_{[0,1]}+1_{\lbrace x_{0}\rbrace}
\end{equation}
which is a contradiction. Hence, there is no lift $ \tau(\cdot\vert\cdot) $ arising from a lifting map \textit{K} for the conditional probability $ \nu(\cdot\vert\cdot) $. $ \square $

%% file: appendix2ver3.tex
\chapter{Propositions for option I}
\section{Measures}
\paragraph{Notation} For the following, sets will be denoted $ \mathcal{X}$, $\mathcal{Y}$,$\ldots $ and $ \sigma $-algebras by $ \mathcal{E} $, $ \mathcal{F} $,$ \ldots $. The spaces of finite, signed-measures on the given set with the given $ \sigma $-algebra will be denoted $ \mathcal{M}(\mathcal{X};\mathcal{E})$, $ \mathcal{M}(\mathcal{Y};\mathcal{F}) $,$ \ldots $. These are Banach spaces using the total variation norm. For a given measure space $ (\mathcal{X},\mathcal{E},\mu) $, let $ \mathcal{E}_{0} $ be the $ \sigma $-algebra given by the completion of $ \mathcal{E} $ with respect to $ \mu $ (see~\cite{roydencompletemeasure}). The extension of $ \mu $ to $ \mathcal{E}_{0} $ will still be denoted $ \mu $. Then we will say $ \mu $ is a $ \mathcal{F} $\textit{-measure} if $ \mathcal{F} $ is a sub-$ \sigma $-algebra of $ \mathcal{E}_{0} $. An element of $ \mathcal{E}_{0} $ will be termed $ \mu $\textit{-measurable}.
\paragraph{Definition B1.1} A subset of signed-measures, $ M\subset\mathcal{M}(\mathcal{X};\mathcal{E})  $, is \textit{absolutely-continuous-complete} if, for any $ \mu $ in \textit{M}, all signed-measures in $ mathcal{M}(\mathcal{X};\mathcal{E})$ absolutely continuous with respect to $ \vert\mu\vert $ are also in \textit{M}.
\paragraph{Proposition B1.2} Any absolutely-continuous-complete subset that is a vector space is directed-complete\footnote{A partially-ordered subset is \textit{upward-directed} if, given any two elements, there is a third that is greater than or equal to both. A set is \textit{directed-complete} if any bounded, upward-directed subset has a supremum.}.
\paragraph*{Proof} Let $ M\subset\mathcal{M}(\mathcal{X};\mathcal{E}) $ be such a subset. Given any upward-directed subset $ B\subset M $, bounded by above by some $ \rho\in M $, define $ \bigvee_{\mu\in B }\mu $ by, for any $ E\in\mathcal{E} $,
\begin{equation}
\left( \bigvee_{\mu\in B }\mu\right)(E)=\sup_{\mu\in B }\mu(E)
\end{equation}
This exists since it is bounded from above by $ \rho(E) $. It is readily seen that $ \bigvee_{\mu\in B }\mu $ is greater than or equal to (using the partial ordering) any $ \mu\in B $ and that, given any other $ \nu\in M $ with that property, $\bigvee_{\mu\in B }\mu \leq \nu$. Given any countable collection of disjoint, measurable sets, $ \lbrace E\rbrace_{j=1}^{\infty}\subset \mathcal{E} $, by the upward-directed property and the countable additivity of each $ \mu\in B $,
\begin{equation}
\sum_{j=1}^{n}\left( \bigvee_{\mu\in B }\mu\right)(E_{j})+\left( \bigvee_{\mu\in B }\mu\right)\left(\bigcup_{j=n+1}^{\infty}E_{j}\right)\leq\left( \bigvee_{\mu\in B }\mu\right)\left(\bigcup_{j=1}^{\infty}E_{j}\right)\leq\sum_{j=1}^{\infty}\left( \bigvee_{\mu\in B }\mu\right)(E_{j})
\end{equation}
for any $ n\in\lbrace 1,2,\ldots\rbrace $. Since $ \nu\leq\bigvee_{\mu\in B }\mu\leq\rho $ for any $ \nu\in B $, and both $ \rho $ and $ \nu $ are countably additive, both $ \left( \bigvee_{\mu\in B }\mu\right)\left(\bigcup_{j=n+1}^{\infty}E_{j}\right)\to 0 $ and the tail sum $ \sum_{j=n+1}^{\infty}\left( \bigvee_{\mu\in B }\mu\right)(E_{j})\to 0 $ as $ n\to \infty $, so
\begin{equation}
\sum_{j=1}^{\infty}\left( \bigvee_{\mu\in B }\mu\right)(E_{j})\leq\left( \bigvee_{\mu\in B }\mu\right)\left(\bigcup_{j=1}^{\infty}E_{j}\right)\leq\sum_{j=1}^{\infty}\left( \bigvee_{\mu\in B }\mu\right)(E_{j})
\end{equation}
Therefore, $ \bigvee_{\mu\in B }\mu $ is countably additive, so it is a signed $ \mathcal{E} $-measure. Its total variation norm is bounded by the total variation norm of $ \rho\vee(-\nu) $ for any particular $ \nu\in B $. Since $ \nu\leq\bigvee_{\mu\in B }\mu\leq \rho $, $ \bigvee_{\mu\in B }\mu $ is absolutely continuous with respect to $ \vert\rho\vert+ \vert\nu\vert$, so $ \bigvee_{\mu\in B }\mu \in M$. $ \square $
\paragraph{Comment} The following proposition corrects the defect mentioned after \textbf{A1.2} for the particular case required.
\medskip\\
For the following two propositions, let $ \mathcal{G} $ be the $ \sigma $-algebra generated\footnote{For any collection $ \mathcal{A} $ of subsets of $ \mathcal{X} $, the $ \sigma $-algebra \textit{generated by} $ \mathcal{A} $, $ \sigma(\mathcal{A}) $, is the smallest $ \sigma $-algebra of subsets of $ \mathcal{X} $ containing $ \mathcal{A} $.~\cite{roydengenerate}} by the rectangular subsets $ \mathcal{E}\times\mathcal{F} $, denoted $ \mathcal{G}=\sigma
(\mathcal{E}\times\mathcal{F}) $. Let the absolutely-continuous-complete subsets $ M\subset\mathcal{M}(\mathcal{X};\mathcal{E})  $, $ N\subset\mathcal{M}(\mathcal{Y};\mathcal{F}) $, and  $ Q\subset\mathcal{M}(\mathcal{X}\times\mathcal{Y};\mathcal{G}) $ also be vector spaces.
\paragraph{Proposition B1.3} If $ M\otimes N $ is weakly dense in \textit{Q}, then it is norm-dense in \textit{Q}.
\paragraph*{Proof} Suppose there were some measure $ \mu\in Q^{+} $ not in the norm-closure of $ M\otimes N  $, so there is some $ \epsilon >0 $ such that $ \Vert\mu-\nu\Vert_{\text{total variation}}>\varepsilon $ for all $ \nu\in M\otimes N $. By Hahn-decomposition~\cite{roydenhahn} and the absolutely-continuous-completeness property of \textit{M} and \textit{N}, if $ \nu\in M\otimes N $, then $ \vert\nu\vert\in M\otimes N  $. Let $ \mu\perp\vert\nu\vert  $ be the singular part of $ \mu $ with respect to $ \vert\nu\vert $ using Lebesgue decomposition~\cite{roydenlebesgue}. The set of $ \mathcal{G} $-measures $ \left\lbrace \mu\perp\vert\nu\vert\vert \nu\in M\otimes N\right\rbrace  $ is lower-bound by the zero-measure. It is downward-directed since $ \mu\perp(\vert\nu\vert+\vert\nu'\vert) $ is less than or equal to (in the partial ordering) both $ \mu\perp\vert\nu\vert $ and $ \mu\perp\vert\nu'\vert $ for any $ \nu,\nu'\in M\otimes N $. Hence, by the preceding proposition, the $ \mathcal{G} $-measure $ \tau =\bigwedge_{\nu\in M\otimes N}\mu\perp\vert\nu\vert $ exists. Furthermore, from its definition in the preceding proof, this measure has total-variation norm greater than or equal to $ \varepsilon $. Let $ \rho\!\parallel\!\tau  $ be the absolutely continuous part of $ \rho $ with respect to $ \tau $ using Lebesgue decomposition. Now consider the bounded, linear functional $ \Phi\in \mathcal{M}(\mathcal{X}\times\mathcal{Y};\mathcal{G})^{*} $ given by $ \Phi\rho=(\rho\!\parallel\!\tau)(\mathcal{X}\times\mathcal{Y}) $. This is zero on $  M\otimes N $, yet $ \Phi\mu=\Vert\tau\Vert_{\text{total variation}}\geq \varepsilon $; hence, $ \mu $ is not in the weak-closure of $ M\otimes N $. $ \square $
\paragraph{Proposition B1.4} If $ M\otimes N $ is norm-dense in \textit{Q}, then $ M^{+}\otimes_{\mathbb{R}^{+}}N^{+} $ is norm-dense in $ Q^{+} $.
\paragraph*{Proof} By Hahn-decomposition and the absolutely-continuous-completeness property of \textit{M} and \textit{N}, the measures in $ M\otimes N $ are total-variation norm-dense for the measures $ Q^{+} $. Given any measure $ \mu=\sum_{j}\nu_{j}\otimes\tau_{j}\in M\otimes N  $, it is absolutely continuous with respect to the product measure $ \rho=\left(\sum_{j}\vert\nu_{j}\vert\right) \otimes \left(\sum_{k}\vert\tau_{k}\vert\right)$, which is in $ M\otimes N $ by Hahn-decomposition and the absolutely-continuous-completeness property of \textit{M} and \textit{N}. By the Radon-Nikod\'{y}m theorem~\cite{roydenradonnikodym}, there is a positive ``function" $ \frac{d\mu}{d\rho}\in L^{1}(\mathcal{X}\times\mathcal{Y};\rho) $ such that $ \mu=\frac{d\mu}{d\rho}\rho $. By proposition \textbf{B3.1} below, $ \frac{d\mu}{d\rho} $ can be arbitrarily well approximated in total-variation norm by elements of $ L^{1}\left(\mathcal{X};\sum_{j}\vert\nu_{j}\vert\right)^{+}\otimes_{\mathbb{R}^{+}} L^{1}\left(\mathcal{Y};\sum_{k}\vert\tau_{k}\vert\right)^{+} $. $ \square $
\paragraph{Proposition B1.5} The total-variation norm is a cross-norm for product measures.
\paragraph*{Proof} Let $ \mu\in \mathcal{M}(\mathcal{X};\mathcal{E})$ and $ \nu\in \mathcal{M}(\mathcal{Y};\mathcal{F}) $. By Hahn decomposition, $ \vert \mu\times \nu\vert=\vert \mu\vert\times \vert\nu\vert $, so
\begin{equation}
\Vert\mu\times \nu\Vert=\vert \mu\times \nu\vert(\mathcal{X} \times \mathcal{Y})= \vert \mu\vert(\mathcal{X}) \vert\nu\vert(\mathcal{Y})=\Vert\mu \Vert\Vert \nu\Vert\hspace{.4 in}\square
\end{equation}
\paragraph{Proposition B1.6} If $ \mu $ is a $ \sigma $-finite measure on $ \mathcal{X} $, then there is no uncountable collection $ \lbrace A_{\alpha}\rbrace $ of subsets of $ \mathcal{X} $ with the properties that $ \mu(A_{\alpha})>0 $ for all $ \alpha $ and $ \mu(A_{\alpha}\cap A_{\beta})=0 $ for all $ \alpha \neq \beta$.
\paragraph*{Proof} Suppose otherwise. Since $ \mu $ is $ \sigma $-finite, there is a countable, disjoint collection $ \lbrace B_{j}\rbrace $ of subsets of $ \mathcal{X} $ such that each $ B_{j} $ has finite $ \mu $-measure and their union $ \cup_{j}B_{j} $ is the entire space $ \mathcal{X} $. Now suppose there is some $ \varepsilon>0 $ and \textit{j} such that for infinitely many of the $ \lbrace A_{\alpha}\rbrace $, $ \mu(B_{j}\cap A_{\alpha})>\varepsilon $. That immediately contradicts $ \mu(B_{j}) $ being finite. Therefore, it must be the case that for any $ \varepsilon>0 $ and \textit{j}, only finitely many of the $ \lbrace A_{\alpha}\rbrace $ satisfy $ \mu(B_{j}\cap A_{\alpha})>\varepsilon $. Let $ \lbrace A_{k}\rbrace $ be the countable collection formed by taking the union over all $ j $ and over all $ \varepsilon\in \lbrace 1,2^{-1},2^{-2},\ldots\rbrace $ of such $ \lbrace A_{\alpha}\rbrace $. However, this also leads to a contradiction, since for any $ A_{\beta}\notin\lbrace A_{k}\rbrace $, $ \mu(A_{\beta}\cap B_{j})=0$ for all $ j $, yet $ 0<\mu(A_{\beta})=\sum_{j} \mu(A_{\beta}\cap B_{j})=0 $. $ \square $
\paragraph{Proposition B1.7} If a $ \sigma $-finite measure is infinitely-divisible\footnote{A $ \sigma $-finite measure $ \mu $ for a measure space $ (\mathcal{X},\mathcal{E},\mu) $ is \textit{infinitely-divisible} if for any $ \varepsilon>0 $, there is a countable (finite if $ \mu $ is finite) partition of $ \mathcal{X} $, $ \lbrace B_{j}\rbrace $, with each $ \mu(B_{j})<\varepsilon $.}, then it has no atoms-for-the-measure\footnote{For a measure space $ (\mathcal{X},\mathcal{E},\mu) $, a subset $ A\in\mathcal{E} $ is an \textit{atom-for-the-measure} $ \mu $ if  for any $ B\in \mathcal{E} $ with $ B\in A $, either $ \mu(B)=0 $ or $ \mu(A\setminus B)=0 $.}.
\paragraph*{Proof} Consider the following property: a measure $ \sigma $-finite $ \mu $ for a measure space $ (\mathcal{X},\mathcal{E},\mu) $ has the once-divisible property if for any subset $ A\in\mathcal{E} $ with finite $ \mu(A)>0 $, there are disjoint $ B,C\in\mathcal{E} $ with $ B\cup C=A $ such that both $ \mu(B) $ and $ \mu(C) $ are less than $ \frac{3}{4}\mu(A) $. This property is readily seen to be equivalent to being infinitely-divisible. Furthermore, if there is an atom-for-the-measure, then the once-divisible property does not hold. Hence, it remains to show if the once-divisible property does not hold, there is necessarily an atom for the measure.

Suppose there is a subset $ A\in\mathcal{E} $ with finite $ \mu(A)>0 $ such that there are  no disjoint $ B,C\in\mathcal{E} $ with $ B\cup C=A $ such that both $ \mu(B) $ and $ \mu(C) $ are less than $ \frac{3}{4}\mu(A) $. Let \textit{c} be the infinum over $ b\in\left[\frac{3}{4},1 \right]  $ such that the division is possible for both $ \mu(B) $ and $ \mu(C) $ are less than $ b\cdot\mu(A) $. If $ c=1 $, then \textit{A} is an atom-for-the-measure $ \mu $. If there is a division with $ \mu(B)=c\cdot\mu(A) $, then \textit{B} is an atom-for-the-measure $ \mu $. The only remaining possibility is that for every $ \varepsilon>0 $, there is a division with $  \mu(B)<(c+\varepsilon)\mu(A) $. By taking $ \varepsilon\in \lbrace 2^{0},2^{-1},2^{-2},\ldots\rbrace $ successively, we get a sequence of subsets $ \langle B_{j}\rangle $ with $ c\cdot\mu(A)<\mu(B_{j})<(c+ 2^{-j})\mu(A) $. For every $ j<k $ in $ \mathbb{Z}^{+} $, $ c\cdot\mu(A)<\mu(B_{j}\cap B_{k})<(c+ 2^{-k})\mu(A) $; hence, $ \bigcap_{j=0}^{\infty}B_{j} $ is in $ \mathcal{E} $ with $ \mu\left(\bigcap_{j=0}^{\infty}B_{j} \right)= c\cdot\mu(A) $--a contradiction. Hence, this possibility cannot occur. $ \square $
\paragraph*{Comment} Note that an atom-for-the-measure need not be an atom for the $ \sigma $-algebra. For instance, the atomic measure $ \delta_{\lbrace 0\rbrace} $ on $ \mathbb{R} $ has all subsets containing 0 as atoms-for-the-measure. For separable, metric spaces with the Borel $ \sigma $-algebra, all atoms-for-the-measure are induced by atomic measures in this manner.~\cite{roydenseparablemetricatoms} However, there are other ways atoms-for-the-measure can occur. Consider the measure on $ [0,1] $ assigning zero to all countable subsets and one to their complements. Then all the subsets with countable complement are atoms-for-the-measure.

Take the measure $ \mu $ finite. Then by successive partitioning and an arguments similar to that used in the preceding proof, for any $ \varepsilon>0 $ it is possible, with only finitely many steps, to find disjoint atoms-for-the-measure $ \lbrace A_{j}\rbrace $ such that any atoms-for-the-measure \textit{A} in the restriction of $ \mu $ to $ \mathcal{X}\setminus\bigcup_{j}A_{j} $ have $ \mu(A)<\varepsilon $. By taking $ \varepsilon\in \lbrace 2^{0},2^{-1},2^{-2},\ldots\rbrace $ successively, we get a countable partition of subsets in $ \mathcal{E} $ 
\begin{equation}
\left(\mathcal{X}\setminus\bigcup_{j}A_{j}\right)\cup\bigcup_{j}A_{j}
\end{equation} 
where $ \mu $ restricted to $ \mathcal{X}\setminus\bigcup_{j}A_{j} $ is infinitely-divisible and each $ A_{j} $ is an atom-for-the-measure. This can clearly be extended to $ \sigma $-finite $ \mu $. 
\paragraph{Proposition B1.8} For the finite, Borel measure $ \rho $ on $ [0,1]\times[0,1] $ (with the standard topology) and linear map \textit{K} of \S\ref{sec:optionIandII}, $ K\rho $ is the zero measure.
\paragraph*{Proof} Take any finite Borel measures $ \mu,nu $ on $ [0,1] $ (with the standard topology). By the preceding comment,
\begin{equation}
\mu=\mu_{0}+\sum_{j}a_{j}\delta_{\lbrace x_{j}\rbrace},\;\nu=\nu_{0}+\sum_{j}b_{j}\delta_{\lbrace y_{j}\rbrace}
\end{equation} 
where $ \mu_{0},nu_{0} $ are infinitely-divisible, $ \langle a_{j}\rangle,\langle b_{j}\rangle $ are  summable sequences of strictly positive reals, and $ \lbrace x_{j}\rbrace, \lbrace y_{j}\rbrace $ are countable subsets of $ [0,1] $. Consider the partition of $ [0,1]\times[0,1] $ given by $ A\cup B\cup C $, where $ B=\left((\lbrace x_{j}\rbrace\cup\lbrace y_{j})\times [0,1]\right)\cup\left( [0,1]\times(\lbrace x_{j}\rbrace\cup\lbrace y_{j})\right) $, $ C=\lbrace (x,x)\vert x\in[0,1]\rbrace\setminus B $, and $ A=[0,1]\times[0,1] \setminus\lbrace (x,x)\vert x\in[0,1]\rbrace\setminus B $. Then $ \rho $ is clearly the zero measure on $ A\cup B $ and both $ \mu\times \sum_{j}b_{j}\delta_{\lbrace y_{j}\rbrace} $ and $\left(\sum_{j}a_{j}\delta_{\lbrace x_{j}\rbrace}\right)\times \nu_{0} $ are clearly the zero measure on \textit{C}. Take any $ \varepsilon>0 $. Since $ \mu_{0} $ is infinitely-divisible, we have a finite partition $ \lbrace B_{j}\rbrace $ of $ [0,1]\setminus (\lbrace x_{j}\rbrace\cup\lbrace y_{j})$ where $ \mu_{0}(B_{j}) $ is less than $ \varepsilon $ for every \textit{j}. Then 
\begin{align}
\mu_{0}\times \nu_{0} (C)&\leq \sum_{j}\mu_{0}\times \nu_{0} (B_{j}\times B_{j})=\sum_{j}\mu_{0}(B_{j}) \nu_{0} (B_{j})\nonumber\\\leq \varepsilon\sum_{j}\nu_{0} (B_{j})=\varepsilon\nu_{0}([0,1]\setminus (\lbrace x_{j}\rbrace\cup\lbrace y_{j}))
\end{align}
Since $ \varepsilon $ was arbitrary, it must be that $ \mu_{0}\times \nu_{0} $ is the zero measure on \textit{C} as well. Therefore, $  \rho\perp \mu\times \nu$. Since $ \mu,nu $ were arbitrary, $ K\rho $ is the zero measure. $ \square $
\paragraph{Proposition B1.9} For $ \sigma $-finite, $ \mathcal{E} $-measures $ \mu,nu $ on $ \mathcal{X} $, If $ nu $ is absolutely convergent with respect to $ \mu $ and $ \mu $ is infinitely-divisible, then $ \nu $ is infinitely-divisible. 
\paragraph*{Proof} Suppose otherwise, so by \textbf{B1.7} there is a subset \textit{A} that is an atom-for-the-measure $ \nu $. Since $ \mu $ is infinitely-divisible, it has the once-divisible property (see \textbf{B1.7}), so there are disjoint subsets $ B_{1} $ and $ C_{1} $ in $ \mathcal{E} $ with $ B_{1}\cup C_{1}=A $ such that both $ \mu(B_{1}) $ and $ \mu(C_{1}) $ are less than $ \frac{3}{4}\mu(A) $. Take $ E_{1} $ the one of $ B_{1}, C_{1} $ that is an atom-for-the-measure $ \nu $. Repeat the process with $ E_{1} $ for \textit{A}  and iterate to get a nested, decreasing sequence of subsets in $ \mathcal{E} $, $ \langle E_{j}\rangle $. Then $ \mu\left(\bigcap_{j}E_{j}\right)=\lim_{j}\mu(E_{j})=0 $ whereas $ \mu\left(\bigcap_{j}E_{j}\right)=\lim_{j}\mu(E_{j})=\mu(A) $--a contradiction to $ \nu $ being absolutely continuous. Hence, $ \nu $ has no atoms-for-the-measure and is infinitely-divisible. $ \square $
\section{Maps on subspaces of the space of measures}
Let $ M\subset \mathcal{M}(\mathcal{X};\mathcal{E})$ be any absolutely-continuous-complete, norm-closed subspace.
\paragraph{Proposition B2.1} Any positive linear map $ L\in \mathcal{B}(M,\mathcal{M}(\mathcal{Y};\mathcal{F})) $ has
\begin{equation}
\Vert L\Vert_{\text{op}}=\sup_{\mu\in M^{+},\mu(\mathcal{X})\leq 1}L\mu(\mathcal{Y})
\end{equation}
\paragraph*{Proof} Since \textit{L} is positive, $ \vert L\mu\vert=\vert L(\mu^{+}-\mu^{-})\vert=\vert L(\mu^{+})-L(\mu^{-})\vert\leq  L(\mu^{+})+L(\mu^{-})=L\vert \mu\vert $ using Hahn decomposition, so
\begin{equation}
\sup_{\mu\in M^{+},\mu(\mathcal{X})\leq 1}L\mu(\mathcal{Y})\leq\Vert L\Vert_{\text{op}}=\sup_{\vert\mu\vert(\mathcal{X})\leq 1}\vert L\mu\vert(\mathcal{Y})\leq \sup_{\mu\in M^{+},\mu(\mathcal{X})\leq 1}L\mu(\mathcal{Y})\hspace{.2 in}\square
\end{equation}
\paragraph{Proposition B2.2} The space of maps $\mathcal{B}(M,\mathcal{M}(\mathcal{Y};\mathcal{F}))$ is a vector lattice\footnote{A partially-ordered, vector space \textsf{A} is a \textit{vector lattice} if for any $ \mathsf{a},\mathsf{b}\in\mathsf{A} $, there are elements $ \mathsf{a}\vee\mathsf{b} $ and $ \mathsf{a}\wedge\mathsf{b} $ such that $ \mathsf{a}\vee\mathsf{b} $ is greater than or equal to both \textsf{a} and \textsf{b}, but is less than or equal to any other element with that property and $ \mathsf{a}\wedge\mathsf{b} $ is less than or equal to both \textsf{a} and \textsf{b}, but is greater than or equal to any other element with that property.}.
\paragraph{Proof} For any maps $ K,L\in \mathcal{B}(M,\mathcal{M}(\mathcal{Y};\mathcal{F})) $ define $ K\vee L $ by $ ( K\vee L)(\mu)(F) $ for $ \mathbf{\mu}\in M^{+} $ and $ F\in\mathcal{F} $ being
\begin{equation}\label{eq:supformula}
\sup \left\lbrace \sum_{j=1}^{n}\sum_{k=1}^{m}\max\left\lbrace  (K\mu_{j})(F_{k}),(L\mu_{j})(F_{k})\right\rbrace\left\vert\begin{array}{c}n,m\in\lbrace 1,2,\ldots\rbrace\\\mu_{1},\ldots,\mu_{n}\in M^{+}\\\text{disjoint }F_{1},\ldots,F_{m}\in\mathcal{F}\\\sum_{j=1}^{n}\mu_{j}=\mu,\bigcup_{j=1}^{m}F_{j}=F\end{array}\right. \right\rbrace 
\end{equation}
Clearly, $ K\vee L\geq K $ and $ K\vee L\geq L $. Also, for any $ J\in \mathcal{B}(M,\mathcal{M}(\mathcal{Y};\mathcal{F})) $ satisfying $ J\geq K $ and $ J\geq L $, then $ J\geq K\vee L $. Also, clearly $ (K\vee L)(c\;\cdot )=c(K\vee L)$ for any real scalar $ c>0 $. Since \textit{M} is a Banach lattice\footnote{A positive cone is \textit{normal} if $ 0\leq\mathsf{a}\leq\mathsf{b} $ implies $ \Vert\mathsf{a}\Vert\leq\Vert\mathsf{b}\Vert $.  A \textit{Banach lattice} is a complete, normed vector lattice with a normal cone and with $ \Vert\vert\mathsf{a}\vert\Vert=\Vert\mathsf{a}\Vert $.}, it has the Riesz decomposition property~\cite{peressini} that for any $ \rho,\nu\in M^{+} $, any $ \mu_{1},\ldots,\mu_{n}\in M^{+} $ such that $ \sum_{j=1}^{n}\mu_{j}=\rho+\nu$ can be decomposed into $ \mu_{j}=\rho_{j}+\nu_{j} $ with both $ \rho_{j},\nu_{j}\in M^{+} $ for each \textit{j} and with $ \sum_{j=1}^{n}\rho_{j}=\rho $ and $ \sum_{j=1}^{n}\nu_{j}=\nu $. Also, given any partition of \textit{F} into disjoint, $ \mathcal{F} $-measurable subsets, $ F=\bigcup_{j=1}^{m_{1}}G_{j}=\bigcup_{j=1}^{m_{2}}H_{j} $, there is a refinement of both partitions, $ F=\bigcup_{j=1}^{m_{1}}\bigcup_{k=1}^{m_{2}}G_{j}\cap H_{k} $. Therefore, $ (K\vee L)(\rho+\nu)(F) $ is equal to
\begin{align}
&\sup \left\lbrace \sum_{j=1}^{n}\sum_{k=1}^{m}\max\left\lbrace  (K\mu_{j})(F_{k}),(L\mu_{j})(F_{k})\right\rbrace\left\vert\begin{array}{c}n,m\in\lbrace 1,2,\ldots\rbrace\\\mu_{1},\ldots,\mu_{n}\in M^{+}\\\text{disjoint }F_{1},\ldots,F_{m}\in\mathcal{F}\\\sum_{j=1}^{n}\mu_{j}=\rho+\nu,\bigcup_{j=1}^{m}F_{j}=F\end{array}\right. \right\rbrace\\&=\sup \left\lbrace \begin{array}{l}\sum_{j=1}^{n_{1}}\sum_{k=1}^{m_{1}}\max\left\lbrace \begin{array}{l} (K\rho_{j})(G_{k}),\\(L\rho_{j})(G_{k})\end{array}\right\rbrace\\+\sum_{j=1}^{n_{2}}\sum_{k=1}^{m_{2}}\max\left\lbrace  \begin{array}{l}(K\nu_{j})(H_{k}),\\(L\nu_{j})(H_{k})\end{array}\right\rbrace\end{array}\left\vert\begin{array}{c}n_{1},n_{2},m_{1},m_{2}\in\lbrace 1,2,\ldots\rbrace\\\rho_{1},\ldots,\rho_{n_{1}},\\\nu_{1},\ldots,\nu_{n_{2}}\in M^{+}\\\text{disjoint }G_{1},\ldots,G_{m_{1}}\in\mathcal{F},\\\text{disjoint }H_{1},\ldots,H_{m_{2}}\in\mathcal{F}\\\sum_{j=1}^{n_{1}}\rho_{j}=\rho,\sum_{j=1}^{n_{2}}\nu_{j}=\nu\\\bigcup_{j=1}^{m_{1}}G_{j}=\bigcup_{j=1}^{m_{2}}H_{j}=F\end{array}\right. \right\rbrace\nonumber\\&=(K\vee L)(\rho)(F)+(K\vee L)(\nu)(F)\nonumber
\end{align}
Since the cone of measures is generating for signed measures by Hahn decomposition and since \textit{M} has the absolutely-continuous-complete property, $ K\vee L $ extends to a linear map on all of $ M $.

It remains to show the image of the map $ K\vee L $ is indeed the signed measures. Since the cone of measures is generating, it suffices to show this with any $ \mu\in M^{+} $. Take any countable collection of disjoint, measurable subsets $ \lbrace F_{j}\rbrace_{j=1}^{\infty}\subset\mathcal{F} $. By the preceding argument using Riesz decomposition and refinement of partitions, for any $ n\in \lbrace 1,2,\ldots\rbrace $,
\begin{equation}
\sum_{j=1}^{n}(K\vee L)(\mu)(F_{j})+(K\vee L)(\mu)\left(\bigcup_{j=n+1}^{\infty}F_{j}\right)\leq(K\vee L)(\mu)\left(\bigcup_{j=1}^{\infty}F_{j}\right)\leq\sum_{j=1}^{\infty}(K\vee L)(\mu)(F_{j})
\end{equation}
However, $ K\mu\leq (K\vee L)\mu\leq \vert K\mu\vert+\vert L\mu\vert $ and both $ K\mu $ and $ \vert K\mu\vert+\vert L\mu\vert $ are countably additive, so both the tail of the series, $ \sum_{j=n+1}^{\infty}(K\vee L)(\mu)(F_{j}) $, and $ (K\vee L)(\mu)\left(\bigcup_{j=n+1}^{\infty}F_{j}\right) $ go to zero as $ n\to \infty $. Therefore, $ (K\vee L)\mu $ is countably additive. 

Finally, $ K\leq K\vee L\leq \vert K\vert+ \vert L\vert $, so $ K\vee L $ is absolutely continuous with respect to $ \vert K\vert+ \vert L\vert $. Hence, $ K\vee L \in M$. Therefore, $ \mathcal{B}\left(\mathcal{M}(\mathcal{X};\mathcal{E}), \mathcal{M}(\mathcal{Y};\mathcal{F})\right) $ is a vector lattice. $ \square $
\paragraph{Proposition B2.3} With the operator norm induced by the total variation norms,\\$\mathcal{B}(M,\mathcal{M}(\mathcal{Y};\mathcal{F}))$ has a normal cone.
\paragraph*{Proof} By \textbf{B2.1}, it only necessary to consider elements in the positive cone $ M^{+} $ to calculate the operator norm of any $ L\in \mathcal{B}\left(M, \mathcal{M}(\mathcal{Y};\mathcal{F})\right)^{+} $. Hence, for any $ K\in\mathcal{B}\left(M, \mathcal{M}(\mathcal{Y};\mathcal{F})\right)^{+} $ with $ K\leq L $, $ \Vert K\Vert_{\text{op}}\leq\Vert L\Vert_{\text{op}} $. $ \square $
\paragraph{Proposition B2.4} For any map $ L\in\mathcal{B}\left(M, \mathcal{M}(\mathcal{Y};\mathcal{F})\right) $, $ \Vert \vert L\vert\Vert_{\text{op}}=\Vert L\Vert_{\text{op}} $.
\paragraph*{Proof} For any such \textit{L},
\begin{equation}
\Vert L\Vert_{\text{op}}=\sup_{\mu\in M,\Vert\mu\Vert\leq 1}  \Vert L\mu\Vert 
\end{equation}
which, using Hahn decomposition, is equal to (using $ \mu\perp\nu $ to show they are mutually singular)
\begin{equation}
\Vert L\Vert_{\text{op}}=\sup\left\lbrace  \Vert L(\mu-\nu)\Vert \left\vert\begin{array}{l}\mu,\nu\in M^{+}\\\mu\perp\nu,\Vert\mu-\nu\Vert\leq 1
\end{array}\right.\right\rbrace
\end{equation}
Since $ \mu $ and $ \nu $ are mutually singular,
\begin{equation}
\Vert\mu-\nu\Vert=\Vert\mu\Vert+\Vert\nu\Vert=\Vert\mu+\nu\Vert
\end{equation}
Then, since $ \vert L\mu\vert<\vert L\vert\mu $ and $ \vert L\nu\vert<\vert L\vert\nu $,
\begin{equation}
\Vert L\Vert_{\text{op}}\leq \sup\left\lbrace   \vert L\vert(\mu+\nu)(\mathcal{Y}) \left\vert\begin{array}{l}\mu,\nu\in M^{+}\\\mu\perp\nu,(\mu+\nu)(\mathcal{Y})\leq 1
\end{array}\right.\right\rbrace
\end{equation}
which can only be increased by not requiring $ \mu $ and $ \nu $ to be mutually singular, so, rewriting $ \mu+\nu\to \mu $, $ \Vert L\Vert_{\text{op}} $ is less than or equal to $ \sup_{\mu\in M^{+},\Vert\mu\Vert\leq 1}\vert L\vert(\mu)(\mathcal{Y}) $, which, by \textbf{B2.1}, is equal to $ \Vert \vert L\vert\Vert_{\text{op}} $.

On the other hand, by \textbf{B2.1},
\begin{equation}
\Vert\vert L\vert\Vert_{\text{op}}=\sup_{
\mu\in M^{+},\Vert\mu\Vert\leq 1}\vert L\vert(\mu)(\mathcal{Y}) 
\end{equation}
which, using $ \vert L\vert=L\vee(-L) $ and the form of `$ \vee $' in (\ref{eq:supformula}), is equal to
\begin{equation}\label{eq:banachlatticemeasuremaps}
\sup \left\lbrace \sum_{j=1}^{n}\sum_{k=1}^{m}\left\vert (L\mu_{j})(F_{k})\right\vert\left\vert\begin{array}{c}n,m\in\lbrace 1,2,\ldots\rbrace\\\mu_{1},\ldots,\mu_{n}\in M^{+},\\\left\Vert\sum_{j=1}^{n}\mu_{j}\right\Vert\leq 1,\\\text{ disjoint }F_{1},\ldots,F_{m}\in\mathcal{F},\\\bigcup_{j=1}^{m}F_{j}=\mathcal{Y}\end{array}\right. \right\rbrace
\end{equation}
which is less than or equal to
\begin{equation}
\sup \left\lbrace \sum_{j=1}^{n}\left\Vert L\mu_{j}\right\Vert\left\vert\begin{array}{c}n\in\lbrace 1,2,\ldots\rbrace\\\mu_{1},\ldots,\mu_{n}\in M^{+},\\\left\Vert\sum_{j=1}^{n}\mu_{j}\right\Vert\leq 1\end{array}\right. \right\rbrace
\end{equation}
Since, by the \textit{AL}-space\footnote{A Banach lattice is an \textit{AL-space} if $ \Vert \mathsf{a}+\mathsf{b}\Vert=\Vert \mathsf{a}\Vert+\Vert \mathsf{b}\Vert $ for $ \mathsf{a},\mathsf{b} $ in the positive cone.} property of \textit{M}, $ \left\Vert\sum_{j=1}^{n}\mu_{j}\right\Vert=\sum_{j=1}^{n}\left\Vert\mu_{j}\right\Vert $, this is less than or equal to
\begin{equation}
\sup_{\mu\in M^{+},\left\Vert\mu\right\Vert\leq 1}  \Vert L\mu\Vert\leq \sup_{\mu\in M,\left\Vert\mu\right\Vert\leq 1}  \Vert L\mu\Vert=\Vert L\Vert_{\text{op}}\hspace{.4 in}\square
\end{equation}
\paragraph{Comment} By \textbf{B2.2}, \textbf{B2.3}, and \textbf{B2.4}, $\mathcal{B}(M,\mathcal{M}(\mathcal{Y};\mathcal{F}))$ is a Banach lattice.
\medskip\\
For the following proposition, let $ \mathcal{G}=\sigma(\mathcal{E}\times\mathcal{F}) $ and $ \mathcal{J}=\sigma( \mathcal{H}\times\mathcal{I}) $. Let the absolutely-continuous-complete subsets $ M\subset\mathcal{M}(\mathcal{X};\mathcal{E})  $, $ N\subset\mathcal{M}(\mathcal{Y};\mathcal{F}) $, and  $ Q\subset\mathcal{M}(\mathcal{X}\times\mathcal{Y};\mathcal{G}) $ also be vector spaces.
\paragraph{Proposition B2.5} If $ M\otimes N $ is norm-dense in \textit{Q}, then for any linear maps\\$ L\in\mathcal{B}(M,\mathcal{M}(\mathcal{Z};\mathcal{H})) $ and $ K\in\mathcal{B}(N,\mathcal{M}(\mathcal{W};\mathcal{I})) $, the map $ L\otimes K:Q\to\mathcal{M}(\mathcal{Z}\times\mathcal{W};\mathcal{J}))  $ is well-defined and satisfies $ \vert L\otimes K\vert=\vert L\vert\otimes \vert K\vert $ and $ \Vert L\otimes K\Vert_{\text{op}}=\Vert L\Vert_{\text{op}} \Vert K\Vert_{\text{op}} $.
\paragraph*{Proof} Both
\begin{equation}
L\otimes K=(L^{+}-L^{-})\otimes(K^{+}-K^{-})=L^{+}\otimes K^{+}-L^{+}\otimes K^{-}-L^{-}\otimes K^{+}+L^{-}\otimes K^{-}
\end{equation}
and
\begin{equation}
-L\otimes K=-(L^{+}-L^{-})\otimes(K^{+}-K^{-})=-L^{+}\otimes K^{+}+L^{+}\otimes K^{-}+L^{-}\otimes K^{+}-L^{-}\otimes K^{-}
\end{equation}
are clearly less than or equal to (in the partial ordering)
\begin{equation}
\vert L\vert\otimes \vert K\vert=(L^{+}+L^{-})\otimes(K^{+}+K^{-})=L^{+}\otimes K^{+}+L^{+}\otimes K^{-}+L^{-}\otimes K^{+}+L^{-}\otimes K^{-}
\end{equation}
so $ \vert L\otimes K\vert\leq\vert L\vert\otimes \vert K\vert $. However, by \textbf{B1.4}, \textbf{B1.5}, \textbf{B2.1}, \textbf{B2.4}, and the definition of operator norm, $ \Vert\vert L\vert\otimes \vert K\vert\Vert\leq \Vert L\Vert\Vert K\Vert $. Hence, by \textbf{A1.3}, $  L\otimes K $ is well-defined.

However, given any measures $\nu\in M^{+} $ and $\rho\in N^{+} $ and subsets $ H\in \mathcal{H} $ and $ I\in \mathcal{I} $,
\begin{equation}
\vert L\otimes K\vert(\nu\times\rho)(H\times I)=
\sup \left\lbrace \sum_{j=1}^{n}\left\vert (L\otimes K)\mu_{j}\right\vert(H\times I)\left\vert\begin{array}{c}n\in\lbrace 1,2,\ldots\rbrace\\\mu_{1},\ldots,\mu_{n}\in Q^{+}\\\sum_{j=1}^{n}\mu_{j}=\nu\times\rho\end{array}\right. \right\rbrace
\end{equation}
using a similar argument to that for the proof of \textbf{B2.4}. By Riesz decomposition, this is equal to
\begin{equation}
\sup \left\lbrace \sum_{j=1}^{n}\sum_{k=1}^{m}\left\vert (L\otimes K)(\nu_{j}\times \rho_{k})\right\vert(H\times I)\left\vert\begin{array}{c}n,m\in\lbrace 1,2,\ldots\rbrace\\\nu_{1},\ldots,\nu_{n}\in M^{+},\rho_{1},\ldots,\rho_{m}\in N^{+}\\\sum_{j=1}^{n}\nu_{j}=\nu,\sum_{k=1}^{m}\rho_{k}=\rho\end{array}\right. \right\rbrace
\end{equation}
which, by the triangle inequality, is greater than or equal to $ \left\vert (L\nu)\times (K\rho)\right\vert(H\times I) $. By the property of signed measures that $ \vert\mu\times\mu'\vert=\vert\mu\vert\times\vert\mu'\vert$, this is equal to $ \vert L\nu\vert(H)  \vert K\rho\vert(I) $.

The rectangular subsets $  \mathcal{H}\times \mathcal{I} $ generate the $ \sigma $-algebra $ \mathcal{J} $, so $ \vert L\otimes K\vert(\nu\times\rho)(A)$\\$ \geq \left((\vert L\vert\nu)\times(\vert K\vert\rho)\right)(A) $ for any $ A\in\mathcal{J} $. By the assumption in the proposition and \textbf{A1.3}, $ \vert L\otimes K\vert(\mu)(A)\leq \left(\vert L\vert\otimes\vert K\vert\right)(\mu)(A) $ for any $ \mu\in Q^{+} $. Hence, $ \vert L\otimes K\vert\geq \vert L\vert\otimes\vert K\vert $, so $ \vert L\otimes K\vert= \vert L\vert\otimes\vert K\vert $. 

We already have $ \Vert L\otimes K\Vert_{\text{op}}\leq\Vert L\Vert_{\text{op}}\Vert K\Vert_{\text{op}} $. On the other hand, 
\begin{align}
\Vert \vert L\vert\Vert_{\text{op}} \Vert\vert K\vert\Vert_{\text{op}}=\sup\left\lbrace \vert L\vert(\mu)(\mathcal{Z})\vert K\vert(\nu)(\mathcal{W})\left\vert\begin{array}{l}\mu\in M^{+},\mu(\mathcal{X})\leq 1,\\\nu\in N^{+},\nu(\mathcal{Y})\leq 1\end{array}\right. \right\rbrace\\\leq \sup\left\lbrace (\vert L\vert\otimes\vert K\vert)(\mu)(\mathcal{Z}\times\mathcal{W})\left\vert\begin{array}{l}\mu\in Q^{+}\\\mu(\mathcal{X}\times\mathcal{Y})\leq 1\end{array}\right. \right\rbrace=\Vert \vert L\vert\otimes\vert K\vert\Vert_{\text{op}}\nonumber
\end{align}
so $  \Vert L\otimes K\Vert_{\text{op}}=\Vert \vert L\vert\otimes\vert K\vert\Vert_{\text{op}}=\Vert \vert L\vert\Vert_{\text{op}} \Vert\vert K\vert\Vert_{\text{op}}=\Vert  L\Vert_{\text{op}} \Vert K\Vert_{\text{op}}$. $ \square $
\paragraph{Corollary B2.6} Positive maps $ L\in\mathcal{B}\left(M,\mathcal{M}(\mathcal{Y};\mathcal{F})\right)^{+}$ are completely-positive\footnote{A map \textit{L} is \textit{completely-positive} if $ L\otimes I_{\mathcal{B}(N)} $ is positive for every $ N $ of the form previously given.}.
\paragraph*{Proof} Use \textbf{B1.4}, \textbf{B2.5}, and the positivity of both maps \textit{L} and $ I_{\mathcal{B}(N)}  $. $ \square $
\medskip\\
Analogously to the pseudo-functions which compose the doubly-dual space $ \mathcal{C}(\mathcal{Z})^{**} $ for $ \mathcal{C}(\mathcal{Z}) $ the continuous functions on some compact space $ \mathcal{Z} $ (see~\cite{semadenipseudofunction}), we have the following:
\paragraph{Definition B2.7} A \textit{pseudo-transition ``function"} $ \tau.(\cdot\vert\cdot) $ with data (\textit{M}, $ \mathcal{M}(\mathcal{Y};\mathcal{F}) $) has the properties: \textit{(i)} for each measure $ \mu\in M^{+} $, $ \tau_{\mu}(\cdot\vert\cdot) $ is a $ L^{1}(\mathcal{X};\mu) $-valued vector $ \mathcal{F} $-measure on $ \mathcal{Y} $; \textit{(ii)} for each $ B\in \mathcal{F} $ and $ \mu\in M^{+} $, $ \tau_{\mu}( B\vert\cdot) $ is essentially bounded with respect to $ \mu $ (so it is in $ L^{\infty}(\mathcal{X};\mu) $); and \textit{(iii)} if $ \mu  \in M^{+}$ is absolutely continuous with respect to $ \nu\in M^{+} $, then for any $ B\in \mathcal{F} $, $ \tau_{\mu}(B\vert\cdot)=\tau_{\nu}(B\vert\cdot) $ as elements in $ L^{1}(\mathcal{X};\mu) $.
\paragraph*{Comment} Using \textbf{A3.7}, it is possible in many commonly encountered situations to lift a pseudo-transition ``function" to a pseudo-transition function, but we will not further pursue this. The space of pseudo-transition ``functions" is clearly a vector space. It is a Banach space under the norm
\begin{equation}
\Vert\tau.(\cdot\vert\cdot)\Vert=\sup_{\mu\in M^{+}, \Vert\mu\Vert\leq 1}\Vert\tau_{\mu}(\cdot\vert\cdot)\Vert
\end{equation}
\paragraph{Proposition B2.8} The space of maps $ \mathcal{B}\left(M,\mathcal{M}(\mathcal{Y};\mathcal{F})\right) $ is isometrically isomorphic to the space of pseudo-transition ``functions" with data (\textit{M}, $ \mathcal{M}(\mathcal{Y};\mathcal{F}) $). Furthermore, this isomorphism takes the positive cones in each space to one another.
\paragraph*{Proof} Given such a pseudo-transition ``function" $ \tau.(\cdot\vert\cdot) $, define \textit{L} by 
\begin{equation}
(L\mu)(B)=\int_{x\in\mathcal{X}} \tau_{\vert\mu\vert}(B\vert x)\,d\mu(x)
\end{equation}
for any $ B\in \mathcal{F} $ and $ \mu\in M $. Then $ L\mu $ is indeed a signed measure since it is countably additive because $ \tau $ is a vector measure. Since $ \tau_{\vert a\mu\vert}(\cdot\vert\cdot)=\tau_{\vert \mu\vert}(\cdot\vert\cdot) $ for any $ a\in\mathbb{R}\setminus\lbrace 0\rbrace $, $ L (a\mu)=a\,L\mu $ for any $ a\in\mathbb{R} $. Also, for any $ \nu\in M $,
\begin{align}
(L(\mu+\nu))(B)&=\int_{x\in\mathcal{X}} \tau_{\vert\mu\vert}(B\vert x)\,d\mu(x)+\int_{x\in\mathcal{X}} \tau_{\vert\nu\vert}(B\vert x)\,d\nu(x)\\
&=\int_{x\in\mathcal{X}} \tau_{\vert\mu\vert+\vert\nu\vert}(B\vert x)\,d(\mu+\nu)(x)\\
&=\int_{x\in\mathcal{X}} \tau_{\vert\mu+\nu\vert}(B\vert x)\,d(\mu+\nu)(x)
\end{align} 
Hence, \textit{L} is linear. To see that it is bounded, we have
\begin{equation}
\Vert L\Vert_{\text{op}}=\sup_{\mu\in M, \Vert\mu\Vert\leq 1}\Vert L\mu\Vert=\sup_{\mu\in M, \Vert\mu\Vert\leq 1}\vert L\mu\vert(\mathcal{Y})
\end{equation}
By Hahn decomposition~\cite{roydenhahn}, this is equal to
\begin{equation}
\sup\left\lbrace(L\mu)(A)-(L\mu)(B)\left\vert\begin{array}{l}\mu\in M, \Vert\mu\Vert\leq 1,\\\text{disjoint }A,B\in\mathcal{F}\end{array}\right.\right\rbrace 
\end{equation}
which is equal to
\begin{equation}
\sup\left\lbrace\sum_{j}\left((L\mu)(A_{j})-(L\mu)(B_{j})\right)\left\vert\begin{array}{l}\mu\in M, \Vert\mu\Vert\leq 1,\\\text{finite, disjoint collections }\lbrace A_{j},B_{j}\rbrace\subset\mathcal{F}\end{array}\right.\right\rbrace 
\end{equation}
This is bounded above by
\begin{equation}
\sup\left\lbrace\sum_{j}\left\Vert\tau_{\mu}(A_{j}\vert\cdot)\right\Vert\left\vert\begin{array}{l}\mu\in M^{+}, \Vert\mu\Vert\leq 1,\\\text{finite, disjoint collections }\lbrace A_{j}\rbrace\subset\mathcal{F}\end{array}\right.\right\rbrace=\sup_{\mu\in M^{+}, \Vert\mu\Vert\leq 1}\left\Vert\tau_{\mu}(\cdot\vert\cdot)\right\Vert=\Vert\tau.(\cdot\vert\cdot)\Vert
\end{equation}
Finally, if $ \tau.(\cdot\vert\cdot) $ is positive, \textit{L} is clearly positive.

Now suppose we are given such a map \textit{L}. For any measure $ \mu\in M^{+} $, let $ M_{\mu} $ be the subspace that is absolutely continuous with respect to $ \mu $. By the Radon-Nikod\'{y}m theorem, $ M_{\mu} $ is isometrically isomorphic to $ L^{1}(\mathcal{X};\mu) $. Therefore, the adjoint map $ (L\vert_{M_{\mu}})^{*} $ takes $ \mathcal{M}(\mathcal{Y};\mathcal{F})^{*} $ to $ L^{1}(\mathcal{X};\mu)^{*}\cong L^{\infty}(\mathcal{X};\mu) $, which is a subspace of $ L^{1}(\mathcal{X};\mu) $ since $ \mu $ is finite. For each $ B\in\mathcal{F} $, define $ \tau_{\mu}(B\vert\cdot) $ to be $ (L\vert_{M_{\mu}})^{*}\Phi_{B} $, where $ \Phi_{B}\in \mathcal{M}(\mathcal{Y};\mathcal{F})^{*} $ is the linear functional that evaluates a signed measure on the set \textit{B}. Then $ \tau_{\mu}(\cdot\vert\cdot) $ is a $ L^{1}(\mathcal{X};\mu) $-valued vector measure since the countable additivity of $ \mu $ implies $ \tau_{\mu}(\cdot\vert\cdot) $ is countably additive. If $ \mu $ is absolutely continuous with respect to $ \nu\in M^{+} $, then $ M_{\mu}\subset M_{\nu} $, so $ \tau_{\mu}(B\vert\cdot)=\tau_{\nu}(B\vert\cdot) $ as elements of $ L^{1}(\mathcal{X};\mu) $. Therefore, $ \tau.(\cdot\vert\cdot) $ is a pseudo-transition ``function". To see that $ \tau.(\cdot\vert\cdot) $ is bounded in norm, we have
\begin{align}
\Vert\tau.(\cdot\vert\cdot)\Vert&=\sup_{\mu\in M^{+}, \Vert\mu\Vert\leq 1}\left\Vert\tau_{\mu}(\cdot\vert\cdot)\right\Vert\\
&=\sup\left\lbrace\sum_{j}\left\Vert\tau_{\mu}(A_{j}\vert\cdot)\right\Vert\left\vert\begin{array}{l}\mu\in M^{+}, \Vert\mu\Vert\leq 1,\\\text{finite, disjoint collections }\lbrace A_{j}\rbrace,\subset\mathcal{F}\end{array}\right.\right\rbrace\nonumber\\
&\leq\sup\left\lbrace\sum_{j}\mu(A_{j})\left\Vert\tau_{\mu}(A_{j}\vert\cdot)\right\Vert_{L^{\infty}(\mathcal{X};\mu)}\left\vert\begin{array}{l}\mu\in M^{+}, \Vert\mu\Vert\leq 1,\\\text{finite, disjoint collections }\lbrace A_{j}\rbrace,\subset\mathcal{F}\end{array}\right.\right\rbrace\nonumber
\end{align}
However,
\begin{equation}
\sum_{j}\mu(A_{j})\left\Vert\tau_{\mu}(A_{j}\vert\cdot)\right\Vert_{L^{\infty}(\mathcal{X};\mu)}=\sum_{j}\mu(A_{j})\left\Vert(L\vert_{M_{\mu}})^{*}\Phi_{A_{j}}\right\Vert_{L^{\infty}(\mathcal{X};\mu)}\leq \mu\left(\bigcup_{j}A_{j}\right)\Vert(L\vert_{M_{\mu}})^{*}\Vert_{\text{op}}\leq \Vert L\Vert_{\text{op}}
\end{equation}
so $ \Vert\tau.(\cdot\vert\cdot)\Vert\leq \Vert L\Vert_{\text{op}} $. Lastly, if \textit{L} is positive, so must be $ \tau.(\cdot\vert\cdot) $; otherwise, if there were some $ \mu\in M^{+} $ and $ B\in\mathcal{F} $ such that $ \tau_{\mu}(B\vert\cdot) $ were strictly less than zero on a set $ A\in\mathcal{E} $ with $ \mu(A)>0 $, then 
\begin{equation}
(L(1_{A}\mu))(B)=\int_{A}(L\vert_{M_{\mu}})^{*}\Phi_{B}\,d\mu=\int_{A}\tau_{\mu}(A\vert x)\,d\mu<0
\end{equation}
which would be a contradiction. $ \square $
\medskip\\
It is also possible to define $ \tau.(\cdot\vert\cdot) $ in terms of \textit{L} rather than adjoints of restrictions of \textit{L} using the Radon-Nikod\'{y}m derivative:
\begin{equation}
\tau_{\mu}(B\vert\cdot)=(L\vert_{M_{\mu}})^{*}\Phi_{B}=\frac{d\left(\left((L\vert_{M_{\mu}})^{*}\Phi_{B}\right)\mu\right)}{d\mu}=\frac{d\mu_{B}}{d\mu}
\end{equation}
where $ \mu_{B} $ is the $ \mathcal{E} $-measure on $  \mathcal{X} $ given by $ \mu_{B}(A)=\left(L(1_{A}\mu)\right)(B) $. Using this, we have the following:
\paragraph{Proposition B2.9} The positive map $ L\in\mathcal{B}\left(M,\mathcal{M}(\mathcal{Y};\mathcal{F})\right)^{+} $ is norm-preserving on the positive cone if and only if the associated pseudo-transition "function" $ \tau.(\cdot\vert\cdot) $ satisfies $ \tau_{\mu}(\mathcal{Y}\vert\cdot)=[1_{\mathcal{X}}] $ for every measure $ \mu\in M^{+} $. 
\paragraph*{Proof} Suppose such a positive map \textit{L} is norm-preserving on the positive cone; then, for any measure $ \mu\in M^{+} $ and subset $ A\in\mathcal{E} $, 
\begin{equation}
\left(L(1_{A}\mu)\right)(\mathcal{Y})=(1_{A}\mu)(\mathcal{X})=\mu(A)
\end{equation}
Therefore, using the notation of the preceding comment, $ \mu_{\mathcal{Y}}=\mu $, so $ \tau_{\mu}(\mathcal{Y}\vert\cdot)=\frac{d\mu}{d\mu}=[1_{\mathcal{X}}] $. Conversely, suppose $ \tau.(\cdot\vert\cdot) $ is such that $ \tau_{\mu}(\mathcal{Y}\vert\cdot)=[1_{\mathcal{X}}] $ for every measure $ \mu\in M^{+} $. Then, for any measure $ \mu\in M^{+} $, 
\begin{equation}
(L\mu)(\mathcal{Y})=\int_{x\in\mathcal{X}}\tau(\mathcal{Y}\vert x)\,d\mu(x)=\mu(X)
\end{equation}
so \textit{L} is norm-preserving on the positive cone. $ \square $
\section{$ L^{1} $-spaces}
\paragraph{Notation} In the following, let $ \mathcal{X} $, $ \mathcal{Y} $,$ \ldots $ denote  sets and $ \mu $, $ \nu $,$ \ldots $ denote $ \sigma $-finite measures. Hilbert spaces, denoted \textsf{H}, \textsf{J},$ \ldots $  are complete, sesquilinear inner-product spaces, with no restriction as to their dimension or separability. $ \mathcal{D}(\mathsf{H})$, $\mathcal{D}(\mathsf{J})$,$\ldots  $ denote the spaces of density matrices (trace-class, self-adjoint operators) on the specified Hilbert space. These spaces are Banach spaces employing the trace norm. $ L^{1}(\mathcal{X};\mu) $, $ L^{1}(\mathcal{Y};\nu) $,$\ldots$ denote the space of integrable, real-valued ``functions" on the given sets with respect to the given measures. These spaces are Banach spaces employing the $ L^{1} $-norm. $ L^{1}(\mathcal{X};\mu;\mathcal{D}(\mathsf{H})) $, $ L^{1}(\mathcal{Y};\nu;\mathcal{D}(\mathsf{J})) $,$\ldots$ denote the space of Bochner-integrable, density-matrix-valued ``functions" on the given sets with respect the given measures. These spaces are Banach spaces employing first the trace norm pointwise, then the $ L^{1} $-norm. For $ n\in \lbrace 1,2,\ldots\rbrace $, $ \mathcal{M}_{n} $ is the space of $ n\times n $-matrices.
\paragraph{Comment} The following proposition strengthens the well-known result, which is a special case of a result by Grothendieck~\cite{grothendieck}, that $ L^{1}(\mathcal{X}\times\mathcal{Y}; \mu\times\nu)=L^{1}(\mathcal{X}; \mu)\hat{\otimes}L^{1}(\mathcal{Y}; \nu) $, where $ \hat{\otimes} $ indicates completion in the projective norm\footnote{The \textit{projective norm} on $ \mathsf{A}\otimes\mathsf{B} $ is the norm induced by duality with \text{Bilinear}($ \mathsf{A},\mathsf{B} $), $ \Vert\mathsf{c}\Vert_{\wedge}=\inf \sum_{j}\Vert\mathsf{a}_{j}\Vert\Vert\mathsf{b}_{j}\Vert $, where the infinum is taken over all $ \sum_{j}\mathsf{a}_{j}\otimes\mathsf{b}_{j}\in  \mathsf{A}\otimes\mathsf{B} $ that equal \textsf{c}.}.
\paragraph{Proposition B3.1} The finite-nonnegative-tensor-rank\footnote{Using only positive real scalars.} ``functions" in $ L^{1}(\mathcal{X}\times\mathcal{Y}; \mu\times\nu)^{+} $, with respect to ``functions" in $ L^{1}(\mathcal{X}; \mu)^{+} $ and $ L^{1}(\mathcal{Y}; \nu)^{+} $, are dense in the norm topology.
\paragraph*{Proof} Take any $ f\in L^{1}(\mathcal{X}\times\mathcal{Y}; \mu\times\nu)^{+} $. \textit{f} can be arbitrarily well-approximated in $ L^{1}( \mathcal{X}\times \mathcal{Y};\mu\times \nu) $-norm by simple functions: $ \sum_{j}a_{j}1_{A_{j}} $ for finite collections of positive reals $ \lbrace a_{j}\rbrace $ and finite $ \mu\times\nu $-measure subsets $ \lbrace A_{j}\rbrace $. By the construction of product measures (see~\cite{roydenproductmeasures}), each $ A_{j} $ is covered by some finite collection of disjoint, measurable, rectangular subsets $ \lbrace B_{k}\times C_{k}\rbrace $ with $ \mu\times\nu\left(\left(\bigcup_{k}B_{k}\times C_{k}\right)\setminus A_{j} \right) $ arbitrarily small. Hence, \textit{f} can be arbitrarily well-approximated in $ L^{1}( \mathcal{X}\times \mathcal{Y};\mu\times \nu) $-norm by simple functions: $ \sum_{j}a_{j}1_{B_{j}\times C_{j}}=\sum_{j}a_{j}1_{B_{j}}\otimes 1_{C_{j}} $ for finite collections of positive reals $ \lbrace a_{j}\rbrace $, finite $ \mu$-measure subsets $ \lbrace B_{j}\rbrace$, and finite $ \nu$-measure subsets $ \lbrace C_{j}\rbrace$. $ \square $
\section{Density-matrix-valued $ L^{1} $-spaces}
\paragraph{Proposition B4.1} $ L^{1}(\mathcal{X};\mu;\mathcal{D}(\mathsf{H}))\otimes L^{1}(\mathcal{Y};\nu;\mathcal{D}(\mathsf{J}))$ is trace-norm dense within $ L^{1}(\mathcal{X}\times \mathcal{Y};\mu\times\nu;\mathcal{D}(\mathsf{H}\otimes\mathsf{J})) $.
\paragraph*{Proof} Given operator $ \rho\in\mathcal{D}(\mathsf{H}\otimes\mathsf{J})  $ , by the spectral theorem for compact operators, $ \rho $ can be arbitrarily well approximated in trace-norm by sums of the form $ \sum_{j=1}^{n}a_{j}\mathbf{e}_{j}\otimes\mathbf{e}_{j}^{*} $ for some $ \lbrace a_{j}\rbrace_{j=1}^{n}\subset\mathbb{R} $ and orthonormal collection of vectors $ \lbrace\mathbf{e}_{j}\rbrace_{j=1}^{n}\subset\mathsf{H}\otimes \mathsf{J} $. By definition, each $ \mathbf{e}_{j} $ can be arbitrarily well approximated in $ \mathsf{H}\otimes \mathsf{J} $-norm (so $ \mathbf{e}_{j}\otimes\mathbf{e}_{j}^{*} $ will be arbitrarily well-approximated in trace-norm) by sums of the form $ \sum_{k=1}^{m}b_{k}\mathbf{f}_{k}\otimes\mathbf{g}_{k} $ for some $ b_{k}\in\mathbb{C}$, $ \mathbf{f}_{k}\in\mathsf{H} $, and $ \mathbf{g}_{k}\in\mathsf{J} $. Using polarization,
\begin{equation}
\left( \sum_{k=1}^{m}b_{k}\mathbf{f}_{k}\otimes\mathbf{g}_{k}\right)\otimes\left( \sum_{\ell=1}^{m}b_{\ell}\mathbf{f}_{\ell}\otimes\mathbf{g}_{\ell}\right)^{*} 
\end{equation}  
\[=\sum_{k,\ell=1}^{m}b_{k}\overline{b_{\ell}}\frac{1}{4}\left((\mathbf{f}_{k}+\mathbf{f}_{\ell})\otimes(\mathbf{f}_{k}+\mathbf{f}_{\ell})^{*}-(\mathbf{f}_{k}-\mathbf{f}_{\ell})\otimes(\mathbf{f}_{k}-\mathbf{f}_{\ell})^{*}\right. \]
\[\left. +\imath(\mathbf{f}_{k}+\imath\mathbf{f}_{\ell})\otimes(\mathbf{f}_{k}+\imath\mathbf{f}_{\ell})^{*}-\imath(\mathbf{f}_{k}-\imath\mathbf{f}_{\ell})\otimes(\mathbf{f}_{k}-\imath\mathbf{f}_{\ell})^{*}\right)   \]
\[\otimes\frac{1}{4}\left((\mathbf{g}_{k}+\mathbf{g}_{\ell})\otimes(\mathbf{g}_{k}+\mathbf{g}_{\ell})^{*}-(\mathbf{g}_{k}-\mathbf{g}_{\ell})\otimes(\mathbf{g}_{k}-\mathbf{g}_{\ell})^{*}\right. \]
\[\left. +\imath(\mathbf{g}_{k}+\imath\mathbf{g}_{\ell})\otimes(\mathbf{g}_{k}+\imath\mathbf{g}_{\ell})^{*}-\imath(\mathbf{g}_{k}-\imath\mathbf{g}_{\ell})\otimes(\mathbf{g}_{k}-\imath\mathbf{g}_{\ell})^{*}\right)   \]
\[=\sum_{k=1}^{m}\vert b_{k}\vert^{2}\left(\mathbf{f}_{k}\otimes\mathbf{f}_{k}^{*}\right)\otimes\left(\mathbf{g}_{k}\otimes\mathbf{g}_{k}^{*}\right)\]
\[+\frac{1}{8}\sum_{k<\ell}\left( \Re(b_{k}\overline{b_{\ell}})\left( \left((\mathbf{f}_{k}+\mathbf{f}_{\ell})\otimes(\mathbf{f}_{k}+\mathbf{f}_{\ell})^{*}-(\mathbf{f}_{k}-\mathbf{f}_{\ell})\otimes(\mathbf{f}_{k}-\mathbf{f}_{\ell})^{*}\right) \right.\right.  \]
\[\otimes\left((\mathbf{g}_{k}+\mathbf{g}_{\ell})\otimes(\mathbf{g}_{k}+\mathbf{g}_{\ell})^{*}-(\mathbf{g}_{k}-\mathbf{g}_{\ell})\otimes(\mathbf{g}_{k}-\mathbf{g}_{\ell})^{*}\right)\]
\[ +\left((\mathbf{f}_{k}+\imath\mathbf{f}_{\ell})\otimes(\mathbf{f}_{k}+\imath\mathbf{f}_{\ell})^{*}-(\mathbf{f}_{k}-\imath\mathbf{f}_{\ell})\otimes(\mathbf{f}_{k}-\imath\mathbf{f}_{\ell})^{*}\right)   \]
\[\left.  \otimes\left((\mathbf{g}_{k}+\imath\mathbf{g}_{\ell})\otimes(\mathbf{g}_{k}+\imath\mathbf{g}_{\ell})^{*}-(\mathbf{g}_{k}-\imath\mathbf{g}_{\ell})\otimes(\mathbf{g}_{k}-\imath\mathbf{g}_{\ell})^{*}\right)   \right) \]
\[- \Im(b_{k}\overline{b_{\ell}})\left( \left((\mathbf{f}_{k}+\mathbf{f}_{\ell})\otimes(\mathbf{f}_{k}+\mathbf{f}_{\ell})^{*}-(\mathbf{f}_{k}-\mathbf{f}_{\ell})\otimes(\mathbf{f}_{k}-\mathbf{f}_{\ell})^{*}\right) \right.  \]
\[\otimes\left((\mathbf{g}_{k}+\imath\mathbf{g}_{\ell})\otimes(\mathbf{g}_{k}+\imath\mathbf{g}_{\ell})^{*}-(\mathbf{g}_{k}-\imath\mathbf{g}_{\ell})\otimes(\mathbf{g}_{k}-\imath\mathbf{g}_{\ell})^{*}\right)   \]
\[ +\left((\mathbf{f}_{k}+\imath\mathbf{f}_{\ell})\otimes(\mathbf{f}_{k}+\imath\mathbf{f}_{\ell})^{*}-(\mathbf{f}_{k}-\imath\mathbf{f}_{\ell})\otimes(\mathbf{f}_{k}-\imath\mathbf{f}_{\ell})^{*}\right)   \]
\[\left. \left.  \otimes\left((\mathbf{g}_{k}+\mathbf{g}_{\ell})\otimes(\mathbf{g}_{k}+\mathbf{g}_{\ell})^{*}-(\mathbf{g}_{k}-\mathbf{g}_{\ell})\otimes(\mathbf{g}_{k}-\mathbf{g}_{\ell})^{*}\right) \right)\right) \]
Then, using the definition of Bochner integrable functions and following the argument in the proof of \textbf{B3.1} gives the desired result. $ \square $
\paragraph{Corollary B4.2} The Bochner integrable ``functions"\footnote{The space $ \mathcal{S}_{1} $ are the trace-class operators.} $ L^{1}(\mathcal{X};\mu;\mathcal{S}_{1}(\mathsf{H}))\otimes L^{1}(\mathcal{Y};\nu;\mathcal{S}_{1}(\mathsf{J}))$ are trace-norm dense within $ L^{1}(\mathcal{X}\times \mathcal{Y};\mu\times\nu;\mathcal{S}_{1}(\mathsf{H}\otimes\mathsf{J})) $.
\paragraph*{Proof} Using Cartesian decomposition\footnote{Writing $ \rho=\frac{1}{2}(\rho+\rho*)+\imath\cdot\frac{1}{2\imath}(\rho-\rho^{*}) $ for $ \rho\in \mathcal{S}_{1}(\mathsf{K}) $.}, for any Hilbert space \textsf{K}, $ \mathcal{S}_{1}(\mathsf{K})=\mathcal{D}(\mathsf{K})+\imath \mathcal{D}(\mathsf{K}) $. Then using the preceding proposition gives the desired result. $ \square $
\paragraph{Proposition B4.3} If the $ \sigma $-finite measure $ \mu $ is infinitely-divisible, there is no non-zero, diagonal\footnote{An operator $ \rho\in \mathcal{D}(L^{2}(\mathcal{X};\mu)) $ is \textit{diagonal} if $ \langle \rho 1_{B}, 1_{B'}\rangle=0 $ for every disjoint, $ \mu $-measurable $ B,B' $ with finite $ \mu $-measure.} operator in $ \mathcal{D}(L^{2}(\mathcal{X};\mu)) $.
\paragraph*{Proof} Take $ \rho \in \mathcal{D}(L^{2}(\mathcal{X};\mu)) $ to be diagonal. Take any ``function" $ f\in L^{2}(\mathcal{X};\mu)  $. Take any $ \varepsilon>0 $. Since $ \mu $ is infinitely-divisible, the measure $ \vert f\vert^{2}\mu $ is infinitely-divisible by \textbf{B1.9}, so there is a countable partition $ \lbrace B_{j}\rbrace $ of $ \mathcal{X} $ such that $ \vert f\vert^{2}\mu(B_{j})=\int_{B_{j}}\vert f\vert^{2}\;d\mu<\varepsilon $ for every \textit{j}. Then, since $ \rho  $ is diagonal,
\begin{equation}
\langle\rho f,f\rangle=\sum_{j,k} \langle\rho 1_{B_{j}}f,1_{B_{k}}f\rangle=\sum_{j} \langle\rho 1_{B_{j}}f,1_{B_{j}}f\rangle      
\end{equation}
Using the spectral theorem for compact operators to write $ \rho=\sum_{j}\lambda_{j}\psi_{j}\psi_{j}^{*} $ for $ \langle\lambda_{j}\rangle $ an absolutely-summable sequence, orthonormal $ \lbrace\psi_{j}\rbrace\subset \mathcal{X};\mu)$, and $ \psi_{j}^{*} $ the linear functional $ \langle\cdot, \psi_{j}\rangle $, this becomes
\begin{equation}
\langle\rho f,f\rangle=\sum_{j}\sum_{k} \lambda_{k}\left\vert\langle 1_{B_{j}}f,\psi_{k}\rangle\right\vert^{2}
\end{equation}
Since the expression $ \left\vert\langle 1_{B_{j}}f,\psi_{k}\rangle\right\vert^{2} $ is bounded by the Cauchy-Schwartz inequality and $ \langle\lambda_{j}\rangle $ is absolutely-summable, the sum in \textit{k} is uniformly convergent, so $ \langle\rho f,f\rangle $ is equal to
\begin{equation}
\sum_{k}\sum_{j} \lambda_{k}\left\vert\langle 1_{B_{j}}f,\psi_{k}\rangle\right\vert^{2}
\end{equation}
The ``functions" $ \lbrace 1_{B_{j}}f\rbrace $ are orthogonal, so 
\begin{equation}
\langle\rho f,f\rangle\leq \left(\sum_{k} \lambda_{k}\Vert\psi_{k}\Vert^{2}\right)\cdot\sup_{j}\Vert 1_{B_{j}}f\Vert^{2}=\Vert\rho\Vert_{\text{trace}}\cdot \varepsilon
\end{equation}
Since $  \varepsilon $ was arbitrary, it must be that $ \langle\rho f,f\rangle=0 $. Since \textit{f} was arbitrary, it must be that $ \rho $ is the zero operator. $ \square $
\section{Maps on density-matrix-valued $ L^{1} $-spaces}
\paragraph{Proposition B5.1} Any map $ L\in\mathcal{B}\left(L^{1}(\mathcal{X};\mu;\mathcal{D}(\mathsf{H})),L^{1}(\mathcal{Y};\nu;\mathcal{D}(\mathsf{J}) \right) $ satisfies
\[\Vert L\Vert_{\text{op}}=\sup\left\lbrace\Vert L\rho\Vert\left\vert \begin{array}{l}
\rho\in L^{1}(\mathcal{X};\mu;\mathcal{D}(\mathsf{H})),\Vert\rho\Vert\leq 1,\\\rho\text{ is pointwise almost-every-}\\\text{where rank one}\end{array}\right.\right\rbrace\]
\paragraph*{Proof} By the definition of Bochner integrable functions and by the spectral theorem for compact operators, simple functions with values in the finite-rank, self-adjoint operators are $ L^{1}(\mathcal{X};\mu;\mathcal{D}(\mathsf{H})) $-norm dense. Therefore, $ \Vert L\Vert_{\text{op}} $ is equal to
\begin{equation}
\sup\left\lbrace \left\Vert\sum_{j=1}^{n}\sum_{k=1}^{m}\lambda_{jk}L\left(1_{A_{j}}\mathbf{e}_{j}^{k}\otimes\mathbf{e}_{j}^{k*} \right)\right\Vert\left\vert\begin{array}{l}n,m\in\lbrace 1,2,\ldots\rbrace\text{, disjoint }\\\mu\text{-measurable subsets }\lbrace A_{1},\ldots,A_{n}\rbrace,\\\text{collections of orthonormal}\\\text{elements of }\mathsf{H}\\\lbrace\lbrace\mathbf{e}_{1}^{1},\ldots,\mathbf{e}_{m}^{1}\rbrace,\ldots,\lbrace\mathbf{e}_{1}^{n},\ldots,\mathbf{e}_{m}^{n}\rbrace\rbrace,\\ \lbrace\lambda_{11},\ldots,\lambda_{nm}\rbrace\subset\mathbb{R},\\\sum_{j=1}^{n}\sum_{k=1}^{m}\vert\lambda_{jk}\vert\mu(A_{j})\leq 1\end{array}\right.\right\rbrace 
\end{equation}
where $ \mathbf{e}_{j}^{k*} $ is the functional $ \langle\cdot, \mathbf{e}_{j}^{k}\rangle $. Now fix a value of $ j $, the $ A_{1},\ldots,A_{n} $, all the collections of orthonormal vectors, and all the $ \lambda $'s except $ \lambda_{j1},\ldots,\lambda_{jm} $. The set of all elements of of $ L^{1}(\mathcal{Y};\nu;\mathcal{D}(\mathsf{J}) $ given by $ \sum_{j=1}^{n}\sum_{k=1}^{m}\lambda_{jk}L\left(1_{A_{j}}\mathbf{e}_{j}^{k}\otimes\mathbf{e}_{j}^{k*} \right) $ for fixed $ \sum_{k=1}^{m}\vert\lambda_{jk}\vert $ is a finite-dimensional, convex subset; hence, its maximum value of norm necessarily occurs at its extreme points where one $ \lambda_{jk} $ is nonzero whereas $ \lambda_{j1},\ldots,\widehat{\lambda_{jk}},\ldots,\lambda_{jm} $ are all zero. Since the choice of $ j $ was arbitrary, this is true for all $ j $, so the supremum is unchanged by restricting to $ m=1 $. $ \square $
\paragraph{Proposition B5.2} Any positive map $ L\in\mathcal{B}(L^{1}(\mathcal{X};\mu;\mathcal{D}(\mathsf{H})),L^{1}(\mathcal{Y};\nu;\mathcal{D}(\mathsf{J}) )^{+} $ satisfies
\[\Vert L\Vert_{\text{op}}=\sup\left\lbrace\Vert L\rho\Vert\left\vert \begin{array}{l}
\rho\in L^{1}(\mathcal{X};\mu;\mathcal{D}(\mathsf{H}))^{+},\Vert\rho\Vert\leq 1,\\\rho\text{ is pointwise almost-every-}\\\text{where rank one}\end{array}\right.\right\rbrace\]
\paragraph*{Proof} By the proof of the preceding proposition, $ \Vert L\Vert_{\text{op}} $ is equal to
\begin{equation}
\sup\left\lbrace \left\Vert\sum_{j=1}^{n}\lambda_{j}L\left(1_{A_{j}}\mathbf{e}_{j}\otimes\mathbf{e}_{j}^{*} \right)\right\Vert\left\vert\begin{array}{l}n\in\lbrace 1,2,\ldots\rbrace\text{, disjoint}\\\mu\text{-measurable subsets}\lbrace A_{1},\ldots,A_{n}\rbrace,\\\text{unit-norm }\mathbf{e}{1},\ldots,\mathbf{e}_{n}\in\mathsf{H},\\\lbrace\lambda_{1},\ldots,\lambda_{n}\rbrace\subset\mathbb{R},\\\sum_{j=1}^{n}\vert\lambda_{j}\vert\mu(A_{j})\leq 1\end{array}\right.\right\rbrace 
\end{equation}
The supremum can only be reduced or stay the same by restricting to positive $ \lambda_{j} $'s. However, since \textit{L} is positive, by the triangle inequality and the quasi-\textit{AL}-property\footnote{A Banach space has the \textit{quasi-AL-property} if $ \Vert \mathsf{a}+\mathsf{b}\Vert= \Vert \mathsf{a}\Vert+\Vert\mathsf{b}\Vert$ for positive \textsf{a}, \textsf{b}.} of $ L^{1}(\mathcal{Y};\nu;\mathcal{D}(\mathsf{J})) $,
\begin{equation}
\left\Vert\sum_{j=1}^{n}\lambda_{j}L\left(1_{A_{j}}\mathbf{e}_{j}\otimes\mathbf{e}_{j}^{*} \right)\right\Vert\leq \sum_{j=1}^{n}\vert\lambda_{j}\vert\left\Vert L\left(1_{A_{j}}\mathbf{e}_{j}\otimes\mathbf{e}_{j}^{*} \right)\right\Vert=\left\Vert\sum_{j=1}^{n}\vert\lambda_{j}\vert L\left(1_{A_{j}}\mathbf{e}_{j}\otimes\mathbf{e}_{j}^{*} \right)\right\Vert
\end{equation}
so the supremum can also only be increased or stay the same by restricting to positive $ \lambda_{j} $'s. Therefore, it must have the same value. $ \square $ 
\paragraph{Corollary B5.3} The cone of positive maps $\mathcal{B}\left(L^{1}(\mathcal{X};\mu;\mathcal{D}(\mathsf{H})),L^{1}(\mathcal{Y};\nu;\mathcal{D}(\mathsf{J})) \right)^{+}$ is a normal cone for the induced operator norm. 
 \paragraph{Proposition B5.4} If $ L\in\mathcal{B}\left(L^{1}(\mathcal{X};\mu;\mathcal{D}(\mathsf{H})),L^{1}(\mathcal{Y};\nu;\mathcal{D}(\mathsf{J})) \right) $ is completely bounded\footnote{\textit{L} is \textit{completely bounded} if it has finite matrix-norm, $ \Vert L\Vert_{\text{matrix}}=\sup_{n}\left\Vert L\otimes I_{\mathcal{M}_{n}}\right\Vert_{\text{op}}$.}, then, for any space $ \mathcal{Z} $, any measure $ \tau $, and any Hilbert space \textsf{K}, $ \Vert L\otimes I\Vert_{\text{op}}\leq \Vert L\Vert_{\text{matrix}} $ with $ I $ the identity map in $ \mathcal{B}\left(L^{1}(\mathcal{Z};\tau;\mathcal{D}(\mathsf{K}))\right) $.
\paragraph*{Proof} By the definition of operator norm and the definition of the tensor product of maps, $ \Vert L\otimes I\Vert_{\text{op}} $ is equal to
\begin{equation}
\sup\left\lbrace\left\Vert (L\otimes I)\rho\right\Vert\left\vert\begin{array}{l}\text{finite-tensor-rank }\rho\in L^{1}(\mathcal{X}\times\mathcal{Z};\mu\times \tau;\mathcal{D}(\mathsf{H}\otimes\mathsf{K})) \\\text{with }\Vert\rho\Vert\leq 1\end{array}\right.\right\rbrace
\end{equation}
By the proof of \textbf{B5.1}, it is only necessary to take the supremum over simple functions taking values with rank one. By the argument in the proof of \textbf{B3.1}, the sets in the simple functions can be restricted to being rectangular. This eliminates consideration of $ \mathcal{Z} $ and $ \tau $, replacing them with positive, real scalars that can be incorporated into the operators. Finally, rank-one, tensor-rank-\textit{n} operators live on a \textit{n}-dimensional subspace of \textsf{K}, which can be identified with $ \mathbb{C}^{n} $. Making the identification $L^{1}(\text{one point};\text{trivial measure};\mathcal{D}(\mathbb{C}^{n}))\leftrightarrow\mathcal{D}(\mathbb{C}^{n}) $ then gives that $ \Vert L\otimes I\Vert_{\text{op}} $ is equal to $ \sup_{n\leq \dim \mathsf{K}} \Vert L\otimes I_{\mathcal{M}_{n}}\Vert_{\text{op}} $. $ \square $
\paragraph{Proposition B5.5} If the positive map $ L\in\mathcal{B}\left(L^{1}(\mathcal{X};\mu;\mathcal{D}(\mathsf{H})),L^{1}(\mathcal{Y};\nu;\mathcal{D}(\mathsf{J})) \right)^{+} $ is such that, for some space $ \mathcal{Z} $, some measure $ \tau $, and some Hilbert space \textsf{K}, $ L\otimes I $ is positive with $ I $ the identity map in $ \mathcal{B}\left(L^{1}(\mathcal{Z};\tau;\mathcal{D}(\mathsf{K}))\right) $, then $ \Vert L\otimes I\Vert_{\text{op}}=\Vert L\Vert_{\text{op}} $.
\paragraph*{Proof} Starting as with the preceding proof, we arrive at the point where $ \Vert L\otimes I\Vert_{\text{op}} $ is given as
\begin{equation}
\sup\left\lbrace\left\Vert \sum_{j=1}^{n}(L\otimes I)(a_{j}\mathbf{v}_{j}\otimes\mathbf{v}_{j}^{*}\; 1_{A_{j}}\otimes 1_{B_{j}})\right\Vert\left\vert\begin{array}{l}n\in\lbrace 1,2,\ldots\rbrace,\\\text{finite-tensor-rank}\\\lbrace \mathbf{v}_{1},\ldots,\mathbf{v}_{n}\rbrace\in \mathsf{H}\otimes\mathsf{K},\\\lbrace a_{1},\ldots, a_{n}\rbrace\subset\mathbb{R},\\\mu\text{-measurable }\lbrace A_{1},\ldots,A_{n}\rbrace,\\ \tau\text{-measurable }\lbrace B_{1},\ldots,B_{n}\rbrace,\\\text{with }\!\!\!\sum_{j=1}^{n}\vert a_{j}\vert\Vert \mathbf{v}_{j}\Vert^{2}\mu(A_{j})\tau(B_{j})\leq 1\end{array}\right.\right\rbrace
\end{equation}
where $ \mathbf{v}_{j}^{*} $ is the functional $ \langle\cdot, \mathbf{v}_{j}\rangle $. Since $ L\otimes I $ is positive, as in the proof for \textbf{B5.2}, it is possible to restrict to positive $ a_{j} $'s without changing the result. Then $ a_{j} $ can be combined with $ \tau(B_{j}) $ and both incorporated into a change in the norm of $ \mathbf{v}_{j} $. Furthermore, since $ \mathbf{v}_{j}$ is of finite-tensor-rank, it is necessarily in $ \mathsf{H}\otimes\mathsf{L}_{j} $ for some finite-dimensional subspace $ \mathsf{L}_{j}\subset \mathsf{K}$, so it can be written as $ \sum_{k=1}^{\dim\mathsf{L}_{j}}\mathbf{x}_{k}^{j}\otimes\mathbf{e}_{k}^{j} $ with $ \lbrace\mathbf{e}_{k}^{j}\rbrace $ an orthonormal basis for $ \mathsf{L}_{j} $.

Then we have for $ \Vert L\otimes I\Vert_{\text{op}} $,
\begin{equation}
\sup\left\lbrace \sum_{j=1}^{n}\sum_{k=1}^{m}\left\Vert L(\mathbf{x}_{k}^{j}\otimes\mathbf{x}_{k}^{j*}1_{A_{j}})\right\Vert\left\vert\begin{array}{l}n\in\lbrace 1,2,\ldots\rbrace,m\in\lbrace 1,2,\ldots\rbrace, m\leq \dim \mathsf{K},\\\mu\text{-measurable }\lbrace A_{1},\ldots,A_{n}\rbrace,\\\lbrace\mathbf{x}_{j}^{k}\rbrace\subset \mathsf{H}, \sum_{j=1}^{n}\sum_{k=1}^{m}\left\Vert\mathbf{x}_{k}^{j}\right\Vert^{2}\mu(A_{j})\leq 1\end{array}\right.\right\rbrace
\end{equation}
By taking the $ \mathbf{x}$'s to be unit length and introducing new, real scalar variables for their squared norm, then by following the argument in the proof of \textbf{B5.1}, the  supremum is the same if $ m=1 $. Hence, $ \Vert L\otimes I\Vert_{\text{op}} $ is equal to
\begin{equation}
\sup\left\lbrace \sum_{j=1}^{n}\text{tr }\left\Vert L(\mathbf{x}_{j}\otimes\mathbf{x}_{j}^{*}1_{A_{j}})\right\Vert\left\vert\begin{array}{l}n\in\lbrace 1,2,\ldots\rbrace,\\\mu\text{-measurable }\lbrace A_{1},\ldots,A_{n}\rbrace,\\\lbrace\mathbf{x}_{j}\rbrace\subset \mathsf{H}, \sum_{j=1}^{n}\left\Vert\mathbf{x}_{j}\right\Vert^{2}\mu(A_{j})\leq 1\end{array}\right.\right\rbrace
\end{equation} 
which is $ \Vert L\Vert_{\text{op}} $. $ \square $
\paragraph{Corollary B5.6} The completely-positive\footnote{\textit{L} is \textit{completely positive} if $ L\otimes I_{\mathcal{M}_{n}} $ is positive for every $ n\in\lbrace 1,2,\ldots,\rbrace $.} maps
\[ \mathcal{B}\left(L^{1}(\mathcal{X};\mu;\mathcal{D}(\mathsf{H})),L^{1}(\mathcal{Y};\nu;\mathcal{D}(\mathsf{J})) \right)^{\text{cp}}\]
are completely bounded.
\paragraph{Corollary B5.7} The cone of completely-positive maps,
\[\mathcal{B}\left(L^{1}(\mathcal{X};\mu;\mathcal{D}(\mathsf{H})),L^{1}(\mathcal{Y};\nu;\mathcal{D}(\mathsf{J})) \right)^{\text{cp}}\] 
is a normal cone for either the induced operator norm or the matrix norm.
\paragraph{Proposition B5.8} If $ L\in\mathcal{B}\left(L^{1}(\mathcal{X};\mu;\mathcal{D}(\mathsf{H})),L^{1}(\mathcal{Y};\nu;\mathcal{D}(\mathsf{J})) \right) $ is completely positive, then, for any space $ \mathcal{Z} $, any measure $ \tau $, and any Hilbert space \textsf{K}, $  L\otimes I $ is positive, with $ I $ the identity map in $ \mathcal{B}\left(L^{1}(\mathcal{Z};\tau;\mathcal{D}(\mathsf{K}))\right) $. 
\paragraph*{Proof} Since \textit{L} is completely positive, by \textbf{B5.6} it is completely bounded. Hence, $ L\otimes I $ exists by \textbf{B5.4}. Furthermore, by \textbf{B4.1} and \textbf{A1.3}, $ L\otimes I $ is unique, so it is meaningful to speak of it being positive.

Now suppose there were some space $ \mathcal{Z} $, some measure $ \tau $, and some Hilbert space \textsf{K} such that $  L\otimes I $ were not positive. Then there would be some positive $ \rho\in L^{1}(\mathcal{X}\times\mathcal{Z};\mu\times\tau;\mathcal{D}(\mathsf{H}\otimes\mathsf{K}))^{+} $ such that $ (L\otimes I)\rho $ is not positive. Since the cone $  L^{1}(\mathcal{Y}\times\mathcal{Z};\nu\times\tau;\mathcal{D}(\mathsf{J}\otimes\mathsf{K}))^{+} $ is norm-closed and $ L\otimes I $ is continuous, that implies there is a relatively open neighborhood of $ \rho $ in the cone $  L^{1}(\mathcal{X}\times\mathcal{Z};\mu\times\tau;\mathcal{D}(\mathsf{H}\otimes\mathsf{K}))^{+} $ whose image under $ L\otimes I $ does not intersect $  L^{1}(\mathcal{Y}\times\mathcal{Z};\nu\times\tau;\mathcal{D}(\mathsf{J}\otimes\mathsf{K}))^{+} $. 

Now approximating $ \rho $ as in the proof of \textbf{B5.4}, one finds that for this to occur there must be some $ n\in\lbrace 1,2,\ldots\rbrace $ for which $ L\otimes I_{\mathcal{M}_{n}} $ is not positive; however, that is a contradiction. $ \square $
\paragraph{Proposition B5.9} If either $ \dim\mathsf{H} $ or $ \dim\mathsf{J} $ is finite and if a positive map
\[L\in\mathcal{B}\left(L^{1}(\mathcal{X};\mu;\mathcal{D}(\mathsf{H})),L^{1}(\mathcal{Y};\nu;\mathcal{D}(\mathsf{J})) \right)^{+}\]
is such that $ L\otimes I_{\mathcal{M}_{m}} $ is positive for $ m=\min\lbrace\dim\mathsf{H}, \dim\mathsf{J}\rbrace $, then \textit{L} is completely positive.
\paragraph*{Proof} Clearly, since $ L\otimes I_{\mathcal{M}_{m}} $ is positive, so is $ L\otimes I_{\mathcal{M}_{n}} $ for all $ n<m $. Now take $ n>m $. $ L\otimes I_{\mathcal{M}_{n}} $ will be positive if for every $ \rho\in L^{1}(\mathcal{X};\mu;\mathcal{D}(\mathsf{H}\otimes \mathbb{C}^{n}))^{+} $, $ \mathbf{y}\in\mathsf{J}\otimes\mathbb{C}^{n} $, and $ \nu $-measurable $ B\subset \mathcal{Y}$,
\begin{equation}
\int_{B} \langle((L\otimes I_{\mathcal{M}_{n}})\rho)\mathbf{y},\mathbf{y}\rangle_{\mathsf{J}\otimes\mathbb{C}^{n}}\,d\nu\geq 0
\end{equation}
By the definition of Bochner integrable functions and the spectral theorem for compact operators, it is enough to show this for $ \rho $ that are simple functions with value in the rank-one operators. We then have to consider, for any finite collection of vectors $ \lbrace\mathbf{x}_{j}\rbrace\subset\mathsf{H}\otimes\mathbb{C}^{n} $ and $ \mu $-measurable subsets $ \lbrace A_{j}\rbrace $
\begin{equation}\label{eq:finitecompletepositive}
\sum_{j}\int_{B} \langle (L\otimes I_{\mathcal{M}_{n}})(\mathbf{x}_{j}\otimes\mathbf{x}^{*}_{j} \;1_{A_{j}})\mathbf{y},\mathbf{y}\rangle_{\mathsf{J}\otimes\mathbb{C}^{n}}\,d\nu
\end{equation}
where $ \mathbf{x}_{j}^{*} $ is the functional $ \langle\cdot, \mathbf{x}_{j}\rangle $. Writing $ \mathbf{y}=\sum_{k=1}^{n}\mathbf{v}_{k}\otimes\mathbf{e}_{k} $ and $ \mathbf{x}_{j}=\sum_{k=1}^{n}\mathbf{w}_{jk}\otimes\mathbf{e}_{k} $ for $ \lbrace\mathbf{e}_{k}\rbrace $ an orthonormal basis for $ \mathbb{C}^{n} $ gives (\ref{eq:finitecompletepositive}) as
\begin{equation}
\sum_{j}\sum_{k,l=1}^{n}\int_{B} \langle L(\mathbf{w}_{jk}\otimes\mathbf{w}^{*}_{jl}\; 1_{A_{j}})\mathbf{v}_{l},\mathbf{v}_{k}\rangle_{\mathsf{J}}\,d\nu
\end{equation}
However, $ \sum_{l=1}^{n}\mathbf{w}^{*}_{jl}\otimes\mathbf{v}_{l} $ is of rank at most $ m $, so there are $\lbrace \tilde{\mathbf{w}}_{jl}\rbrace $ and $\lbrace \tilde{\mathbf{v}}_{l}\rbrace $ such that
\begin{equation}
\sum_{l=1}^{n}\mathbf{w}^{*}_{jl}\otimes\mathbf{v}_{l}=\sum_{l=1}^{m}\tilde{\mathbf{w}}^{*}_{jl}\otimes\tilde{\mathbf{v}}_{l}
\end{equation}
The condition for $ L\otimes I_{\mathcal{M}_{n}} $ to be positive is then that for every $ \nu $-measurable $ B\subset \mathcal{Y}$, finite collection of $ \mu $-measurable subsets $ \lbrace A_{j}\rbrace $, and finite collections of vectors $ \lbrace\tilde{\mathbf{w}}_{jk} \rbrace\subset \mathsf{H}$ and $ \lbrace\tilde{\mathbf{v}}_{k} \rbrace\subset \mathsf{J}$,
\begin{equation}
\sum_{j}\sum_{k,l=1}^{m}\int_{B} \langle L(\tilde{\mathbf{w}}_{jk}\otimes\tilde{\mathbf{w}}^{*}_{jl}\; 1_{A_{j}})\tilde{\mathbf{v}}_{l},\tilde{\mathbf{v}}_{k}\rangle_{\mathsf{J}}\,d\nu\geq 0
\end{equation}
However, this condition is independent of \textit{n}, as long as it is greater than or equal to \textit{m}. $ \square $ 
\paragraph{Proposition B5.10} The space of completely-bounded maps, \[\mathcal{CB}\left(L^{1}(\mathcal{X};\mu;\mathcal{D}(\mathsf{H})),L^{1}(\mathcal{Y};\nu;\mathcal{D}(\mathsf{J})) \right)\]
is a Banach space with respect to the matrix norm.
\paragraph*{Proof} Let $ \langle L_{j}\rangle$ be a Cauchy sequence in the matrix norm of such maps. Since the matrix-norm is greater than or equal to the operator norm, this is a Cauchy sequence in operator norm, so since the space is a Banach space with respect to the operator norm, it converges to some $ L_{\infty} $ in that norm. It remains to show that $ L_{\infty} $ is completely bounded. For each $ n\in \lbrace 1,2,\ldots\rbrace $, by the triangle inequality, $ \Vert K\otimes I_{\mathcal{M}_{n}}\Vert_{\text{op}}\leq n^{2}\Vert K\Vert_{\text{op}}$ for any linear map $ K \in\mathcal{B}\left(L^{1}(\mathcal{X};\mu;\mathcal{D}(\mathsf{H})),L^{1}(\mathcal{Y};\nu;\mathcal{D}(\mathsf{J})) \right) $; hence, the functional $ K\to \Vert K\otimes I_{\mathcal{M}_{n}}\Vert_{\text{op}} $ is continuous in the operator-norm topology. Since $ \langle L_{j}\rangle $ converges to $ L_{\infty} $ in this topology, it must be that
\begin{equation}
\Vert L_{\infty}\otimes I_{\mathcal{M}_{n}}\Vert_{\text{op}}= \lim_{j\to \infty}\Vert L_{j}\otimes I_{\mathcal{M}_{n}}\Vert_{\text{op}}\leq\lim_{j\to \infty}\Vert L_{j}\Vert_{\text{matrix}} 
\end{equation}
The right-hand limit necessarily exists since $ \langle L_{j}\rangle $ is a Cauchy sequence. Therefore, $ \Vert L_{\infty}\Vert_{\text{matrix}}\leq \lim_{j\to \infty}\Vert L_{j}\Vert_{\text{matrix}} $, so $L_{\infty}$ is completely bounded. $ \square $   
\paragraph{Proposition B5.11} The subset of $\mathcal{B}\left(L^{1}(\mathcal{X};\mu;\mathcal{D}(\mathsf{H})),L^{1}(\mathcal{Y};\nu;\mathcal{D}(\mathsf{J})) \right)$ for which the tensor product with $ I_{\mathcal{M}_{n}} $ is positive for some fixed $ n\in\lbrace 1,2,\ldots\rbrace $ is closed in the weak topology.
\paragraph*{Proof} We will show the complement is open. Take such a map \textit{L} that is not in the subset. By the argument in the proof for \textbf{B5.9}, that implies there are some $ \nu $-measurable $ B\subset \mathcal{Y}$, finite collection of $ \mu $-measurable subsets $ \lbrace A_{j}\rbrace $, finite collections of vectors $ \lbrace\mathbf{w}_{jk} \rbrace\subset \mathsf{H}$ and $ \lbrace\mathbf{v}_{k} \rbrace\subset \mathsf{J}$, and $ \varepsilon>0 $ such that
\begin{equation}
\sum_{j=1}^{m}\sum_{k,l=1}^{n}\int_{B} \langle L(\mathbf{w}_{jk}\otimes\mathbf{w}^{*}_{jl}\; 1_{A_{j}})\mathbf{v}_{l},\mathbf{v}_{k}\rangle_{\mathsf{J}}\,d\nu<-\varepsilon 
\end{equation}
Then, by the triangle inequality, all the maps in the weak neighborhood
\begin{equation}
\bigcap_{j=1}^{m}\bigcap_{k,l=1}^{n}\mathcal{N}\left( L;\mathbf{w}_{jk}\mathbf{w}_{jk}^{*}\,1_{A_{j}};\mathbf{v}_{l}\otimes\mathbf{v}_{l}^{*}\, 1_{B};\frac{\varepsilon}{2n^{4}m}\right) 
\end{equation}
\[+\bigcap_{j=1}^{m}\bigcap_{k=1}^{n}\bigcap_{r<l} \mathcal{N}\left( L;(\mathbf{w}_{jr}\otimes\mathbf{w}_{jl}^{*}+\mathbf{w}_{jl}\otimes\mathbf{w}_{jr}^{*})\,1_{A_{j}};\mathbf{v}_{k}\otimes\mathbf{v}_{k}^{*}\, 1_{B};\frac{\varepsilon}{n^{4}m}\right)\]
\[+\bigcap_{j=1}^{m}\bigcap_{k=1}^{n}\bigcap_{r<l} \mathcal{N}\left(L;\mathbf{w}_{jk}\otimes\mathbf{w}_{jk}^{*}\,1_{A_{j}};(\mathbf{v}_{k}\otimes\mathbf{v}_{m}^{*}+\mathbf{v}_{m}\otimes\mathbf{v}_{k}^{*})\, 1_{B};\frac{\varepsilon}{n^{4}m}\right)\]
\[+\bigcap_{j=1}^{m}\bigcap_{k<l}\bigcap_{q<r}\mathcal{N}\left(L; (\mathbf{w}_{jk}\otimes\mathbf{w}_{jl}^{*}+\mathbf{w}_{jl}\otimes\mathbf{w}_{jk}^{*})\,1_{A_{j}};(\mathbf{v}_{q}\otimes\mathbf{v}_{r}^{*}+\mathbf{v}_{r}\otimes\mathbf{v}_{q}^{*})\, 1_{B};\frac{2\varepsilon}{n^{4}m}\right) \]
will also fail to yield a positive tensor product with $ I_{\mathcal{M}_{n}} $. $ \square $
\paragraph{Corollary B5.12} The cone of completely positive maps is weakly closed in\\ $\mathcal{B}\left(L^{1}(\mathcal{X};\mu;\mathcal{D}(\mathsf{H})),L^{1}(\mathcal{Y};\nu;\mathcal{D}(\mathsf{J})) \right)$.
\paragraph{Comment} The preceding result also follows from showing that the spaces are norm-closed, then using the separating theorem to argue that the weak and norm topologies have the same closed, convex subsets. The approach followed here is preferable since, as has already been noted, the separating theorem depends on the axiom of choice through the Hahn-Banach theorem. For the case where the measures are trivial, the space is a dual space, $ \mathcal{D}(\mathsf{H})=\mathcal{K}(\mathsf{H})^{*} $, so it is possible to do better.
\paragraph{Proposition B5.13} The subset of $ \mathcal{B}\left(\mathcal{D}(\mathsf{H}),\mathcal{D}(\mathsf{J}) \right) $ for which the tensor product with $ I_{\mathcal{M}_{n}} $ is positive for some fixed $ n\in\lbrace 1,2,\ldots\rbrace $ is closed in the weak* topology.
\paragraph*{Proof} Following the proof of \textbf{B5.9}, if $ L\otimes I_{\mathcal{M}_{n}} $ is not positive, then there are some finite collections of vectors $ \lbrace\mathbf{w}_{k} \rbrace\subset \mathsf{H}$ and $ \lbrace\mathbf{v}_{k} \rbrace\subset \mathsf{J}$, and $ \varepsilon>0 $ such that all the maps in the weak* neighborhood
\begin{equation}
\bigcap_{k,l=1}^{n}\mathcal{N}\left( L;\mathbf{w}_{k}\mathbf{w}_{k}^{*}\,1_{A_{j}};\mathbf{v}_{l}\otimes\mathbf{v}_{l}^{*}\, 1_{B};\frac{\varepsilon}{2n^{4}}\right) 
\end{equation}
\[+\bigcap_{k=1}^{n}\bigcap_{r<l} \mathcal{N}\left( L;(\mathbf{w}_{r}\otimes\mathbf{w}_{l}^{*}+\mathbf{w}_{l}\otimes\mathbf{w}_{r}^{*})\,1_{A_{j}};\mathbf{v}_{k}\otimes\mathbf{v}_{k}^{*}\, 1_{B};\frac{\varepsilon}{n^{4}}\right)\]
\[+\bigcap_{k=1}^{n}\bigcap_{r<l} \mathcal{N}\left(L;\mathbf{w}_{k}\otimes\mathbf{w}_{k}^{*}\,1_{A_{j}};(\mathbf{v}_{k}\otimes\mathbf{v}_{m}^{*}+\mathbf{v}_{m}\otimes\mathbf{v}_{k}^{*})\, 1_{B};\frac{\varepsilon}{n^{4}}\right)\]
\[+\bigcap_{k<l}\bigcap_{q<r}\mathcal{N}\left(L; (\mathbf{w}_{k}\otimes\mathbf{w}_{l}^{*}+\mathbf{w}_{l}\otimes\mathbf{w}_{k}^{*})\,1_{A_{j}};(\mathbf{v}_{q}\otimes\mathbf{v}_{r}^{*}+\mathbf{v}_{r}\otimes\mathbf{v}_{q}^{*})\, 1_{B};\frac{2\varepsilon}{n^{4}}\right) \]
will also fail to yield a positive tensor product with $ I_{\mathcal{M}_{n}} $. $ \square $
\paragraph{Corollary B5.14} The cone of completely positive maps in $\mathcal{B}\left(\mathcal{D}(\mathsf{H}),\mathcal{D}(\mathsf{J}) \right)$ is closed in the weak* topology.
\paragraph{Proposition B5.15} Any bounded, positively-linear map \textit{L} to $ L^{1}(\mathcal{Y};\nu;\mathcal{D}(\mathsf{J})) $ that is given on the positive cone of $ L^{1}(\mathcal{X};\mu;\mathcal{D}(\mathsf{H})) $ extends uniquely by linearity to a map\\$ L\in \mathcal{B}\left(L^{1}(\mathcal{X};\mu;\mathcal{D}(\mathsf{H})),L^{1}(\mathcal{Y};\nu;\mathcal{D}(\mathsf{J})) \right)^{+} $.
\paragraph{Proof} Let $ L $ be any such map. Extend \textit{L} to $ \mathcal{B}(L^{1}(\mathcal{X};\mu;\mathcal{D}(\mathsf{H})),L^{1}(\mathcal{Y};\nu;\mathcal{D}(\mathsf{J})) )^{+} $ by
\begin{equation}
L(\rho)=L\left(\frac{\vert\rho\vert+\rho}{2}\right)- L\left(\frac{\vert\rho\vert-\rho}{2}\right)\text{ for any } \rho\in L^{1}(\mathcal{X};\mu;\mathcal{D}(\mathsf{H}))
\end{equation} 
where $  \vert \cdot\vert $ is applied pointwise with $ \vert \rho\vert(x)=\vert \rho(x)\vert=\sqrt{\rho(x)^{2}} $. It is readily seen that positive linearity implies the extended \textit{L} is now linear over $ \mathbb{R} $. Furthermore, by \textbf{B5.2}, this extension does not increase the operator norm.
\paragraph{Proposition B5.16} Any map $ L\in \mathcal{B}\left(L^{1}(\mathcal{X};\mu;\mathcal{D}(\mathsf{H})),L^{1}(\mathcal{Y};\nu;\mathcal{D}(\mathsf{J})) \right)^{+}$ extends uniquely by linearity to a map $ L\in \mathcal{B}\left(L^{1}(\mathcal{X};\mu;\mathcal{S}_{1}(\mathsf{H})),L^{1}(\mathcal{Y};\nu;\mathcal{S}_{1}(\mathsf{J})) \right)^{+}$. This extension has operator norm less than twice that of the restricted map.
\paragraph{Proof} Let $ L $ be any such map. Extend \textit{L} to $ L\in \mathcal{B}(L^{1}(\mathcal{X};\mu;\mathcal{S}_{1}(\mathsf{H})),L^{1}(\mathcal{Y};\nu;\mathcal{S}_{1}(\mathsf{J}))^{+} )$ by using Cartesian decomposition,
\begin{equation}
L(\rho)=L\left(\frac{\rho+\rho^{*}}{2}\right)+\imath L\left(\frac{\rho-\rho^{*}}{2\imath}\right)\text{ for any } \rho\in L^{1}(\mathcal{X};\mu;\mathcal{S}_{1}(\mathsf{H}))
\end{equation}
where * is applied pointwise with $ \rho^{*}(x)=\rho(x)^{*}$. It is readily seen that real linearity implies the extended \textit{L} is now linear over $ \mathbb{C} $. By definition, this extended map is a positive one. The triangle inequality implies the extension has operator norm less than twice that of the restricted map. $ \square $
\paragraph{Comment} Using
\begin{equation}
\mathcal{S}_{1}\left(L^{2}(\mathcal{X};\mu)\right)\subset\mathcal{S}_{2}\left(L^{2}(\mathcal{X};\mu)\right)\cong L^{2}\left(\mathcal{X}\times\mathcal{X};\mu\times \mu\right)
\end{equation}
(where $ \mathcal{S}_{2} $ are the Hilbert-Schmidt operators) one might imagine a map diag that extracts the diagonal of an operator $ \rho\in \mathcal{S}_{1}\left(L^{2}\left(\mathcal{X};\mu\right)\right) $ by $ \text{diag }\rho(x)=\rho(x,x) $. This does not work since the actual values of $ \rho $ are completely arbitrary on the diagonal. However, there is a useful map which does correspond to this naive notion. By singular-value decomposition, any $ \rho\in\mathcal{S}_{1}\left(L^{2}\left(\mathcal{X};\mu\right)\right) $ can be written as the countable sum $\sum_{j}\sigma_{j}f_{j}\otimes g_{j}^{*} $ for $ \langle\sigma_{j}\rangle\in \ell^{\infty+} $ and some unit length $ \lbrace f_{j},g_{j}\rbrace\subset L^{2}\left(\mathcal{X};\mu\right) $. There is a measure $ \nu $ associated to $ \rho  $, clearly absolutely continuous with respect to $ \mu $, given by  for any $ \mu $-measurable subset $ B\subset\mathcal{X} $,
\[\nu(B)=\text{tr }P_{B}\rho=\int_{B}\sum_{j}f_{j}\overline{g_{j}}\;d\mu\]
where $ P_{B} $ is the orthogonal projector associated to the subspace of $ L^{2}\left(\mathcal{X};\mu\right) $ that is zero (almost everywhere with respect to $ \mu $) outside \textit{B}. Then we have:
\paragraph{Definition B5.17} Let the map diag: $ \mathcal{S}_{1}\left(L^{2}\left(\mathcal{X};\mu\right)\right)\to L^{1}\left(\mathcal{X};\mu\right) $ be defined by the Radon-Nikod\'{y}m derivative
\[\text{diag }\rho=\frac{d\nu}{d\mu}\]
\medskip\\
The diagonal extraction map is clearly linear and positive. We also have the following properties:
\paragraph{Proposition B5.18} The diagonal extraction map has operator norm one. Furthermore, on the equivalence classes $ \mathcal{S}_{1}\left(L^{2}\left(\mathcal{X};\mu\right)\right)/\text{ker diag} $, it is an isometric bijection.
\paragraph*{Proof} Using the singular-value-decomposition form for the trace-class operator $ \rho $ as above,
\begin{align}
\Vert\text{diag }\rho\Vert&=\int_{\mathcal{X}}\left\vert\sum_{j}\sigma_{j}f_{j}\overline{g_{j}}\right\vert\;d\mu\nonumber\\&\leq \sum_{j}\sigma_{j}\int_{\mathcal{X}}\left\vert f_{j}\overline{g_{j}}\right\vert\;d\mu
\end{align}
using the triangle inequality and the Dominated Convergence theorem. Using the Cauchy-Schwartz inequality, this is less than or equal to
\begin{equation}
\sum_{j}\sigma_{j}\Vert f_{j}\Vert\Vert g_{j}\Vert=\sum_{j}\sigma_{j}=\Vert \rho\Vert_{\text{trace}}
\end{equation} 
Since $ \rho  $ was arbitrary, $ \Vert\text{diag} \Vert_{\text{op}}\leq 1 $. It is equal to one since diag is clearly an isometry on the positive cone $ \mathcal{D}\left(L^{2}(\mathcal{X};\mu)\right)^{+} $.

If the $ \sigma $-algebra for $ \mathcal{X} $ were generated by countably many atomic subsets, it is easy to see that diag is an isometric bijection--one simply notes the diagonal operator whose diagonal agrees with the diagonal of $ \rho $ is in the same equivalence class and that for this particular operator, the map clearly preserves the norm. However, in general, there are no diagonal, trace-class operators on $ L^{2}\left(\mathcal{X};\mu\right) $, so we need to work a little harder. Let $ s\in L^{\infty}\left(\mathcal{X};(\text{diag }\rho)\mu\right)$ be the phase valued ``function" given by
\begin{equation}
s=\frac{\text{diag }\rho}{\vert\text{diag }\rho\vert}
\end{equation}
(it can be left undefined where $ \text{diag }\rho $ is zero). Then the trace-class operator $ \xi=s\sqrt{\text{diag }\rho}\otimes\sqrt{\text{diag }\rho} $ has the same trace as $ \rho $; hence, it is in the same equivalence class. Therefore,
\begin{equation}
\Vert[\rho]\Vert=\inf_{\hat{\rho}\in [\rho]}\Vert\hat{\rho}\Vert_{\text{trace}}\leq \Vert\xi\Vert_{\text{trace}}=\Vert\text{diag }\rho\Vert
\end{equation}
which by the above result is less than or equal to
\begin{equation}
\inf_{\hat{\rho}\in [\rho]}\Vert\hat{\rho}\Vert_{\text{trace}}
\end{equation}
Hence, $ \Vert[\rho]\Vert=\Vert\text{diag }\rho\Vert $. Since $ \rho $ was arbitrary, diag is an isometry. It is surjective since for any ``function" $ f\in L^{1}\left(\mathcal{X};\mu\right) $, defining \textit{s} similarly to above,
\begin{equation}
s=\frac{f}{\vert f\vert}
\end{equation}
(it can be left undefined where \textit{f} is zero), then the trace-class operator $ \xi=s\sqrt{f}\otimes\sqrt{f} $ has $ \text{diag }\xi=f $. $ \square $
\paragraph{Proposition B5.19} Both the diagonal extraction map diag and its inverse $ \text{diag}^{-1}:L^{1}\left(\mathcal{X};\mu\right)\to \mathcal{S}_{1}\left(L^{2}\left(\mathcal{X};\mu\right)\right)/\text{ker diag} $ are completely-positive.
\paragraph*{Proof} By the spectral theorem for compact operators, it is enough to demonstrate positivity for $ \text{diag}\otimes I_{\mathcal{M}_{n}} $ acting on rank-one density matrices. Furthermore, since finite-tensor rank vectors are dense, it is enough show that
\begin{equation}
\int_{E}\left\langle\left(\text{diag}\otimes I_{\mathcal{M}_{n}}\right)\left(\sum_{j=1}^{m}f_{j}\otimes \mathbf{a}_{j}\sum_{k=1}^{m}f_{k}^{*}\otimes \mathbf{a}_{j}^{*}\right)\mathbf{b},\mathbf{b}\right\rangle\;d\mu\geq 0
\end{equation} 
for any $ m,n\in\mathbb{Z}^{+} $, any $ \lbrace\mathbf{a}_{j}\rbrace\subset\mathbb{C}^{n} $, any $ \mathbf{b}\in\mathbb{C}^{n}  $, any $ \lbrace f_{j}\rbrace\subset L^{2}\left(\mathcal{X};\mu\right) $, and any $ \mu $-measurable subset $ E\subset\mathcal{X} $. However, the left-hand side is equal to $ \int_{E} \vert g\vert^{2}\;d\mu $, where $ g=\sum_{j=1}^{m}\langle\mathbf{a}_{j}, \mathbf{b}\rangle f_{j} $, which is clearly greater than or equal to zero.

Firstly, note $ \text{diag}^{-1}\otimes I_{\mathcal{M}_{n}}$ is well-defined since
\begin{equation}
\mathcal{S}_{1}\left(L^{2}\left(\mathcal{X};\mu\right)\otimes \mathbb{C}^{n}\right)/\text{ker }(\text{diag}\otimes I_{\mathcal{M}_{n}})=\left(\mathcal{S}_{1}\left(L^{2}\left(\mathcal{X};\mu\right)\right)/\text{ker diag})\right)\otimes \mathcal{M}_{n}
\end{equation}
Furthermore, by \textbf{A1.3}, $ \text{diag}^{-1}\otimes I_{L^{1}(\mathcal{Y};\nu;\mathcal{S}_{1}(\mathsf{H}))}$ is well-defined for any $ \mathcal{Y},\nu,\mathsf{H} $ since
\begin{equation}
\left(\mathcal{S}_{1}\left(L^{2}\left(\mathcal{X};\mu\right)\right)/\text{ker diag})\right)\otimes L^{1}(\mathcal{Y};\nu;\mathcal{S}_{1}(\mathsf{H}))
\end{equation}
is norm-dense in
\begin{equation}
\mathcal{S}_{1}\left(L^{2}\left(\mathcal{X}\times \mathcal{Y};\mu\times \nu\right)\otimes \mathsf{H}\right)/\text{ker }\left(\text{diag}\otimes I_{ L^{1}(\mathcal{Y};\nu;\mathcal{S}_{1}(\mathsf{H}))}\right)
\end{equation}
(This will be needed in further propositions below.) A notion of positivity is provided on $ \left(\mathcal{S}_{1}\left(L^{2}\left(\mathcal{X};\mu\right)\right)/\text{ker diag})\right)\otimes \mathcal{M}_{n} $ by taking equivalence classes with positive members to be positive; this is equivalent to $ \alpha $ being positive if $ \text{tr }\beta\alpha\geq 0 $ for any positive $ \beta\in \mathcal{W}(\mathcal{X};\mu;\mathbb{C}^{n})^{+}=\text{diag }^{*}L^{\infty}(\mathcal{X};\mu)^{+}\otimes\mathcal{D}(\mathbb{C}^{n})^{+}$ ($ \mathcal{W} $ is defined below in \textbf{B5.21}, $ \text{diag }^{*} $ is defined following this proof). Any such positive $ \beta $ can be written as $ \gamma^{*}\gamma $ for some $ \gamma\in \mathcal{W}(\mathcal{X};\mu;\mathbb{C}^{n})=\text{diag }^{*}L^{\infty}(\mathcal{X};\mu)^{+}\otimes\mathcal{B}(\mathbb{C}^{n})$. 

Using the preceding, together with the definition of Bochner integrability for $ L^{1}(\mathcal{X};\mu;\mathcal{D}(\mathbb{C}^{n})) $ and the (finite dimensional) spectral theorem, to show $ \text{diag}^{-1} $ is completely positive we need to show
\begin{equation}
\text{tr }\left(\gamma^{*}\gamma(\text{diag}^{-1}\otimes I_{\mathcal{M}_{n}})1_{B}\mathbf{a}\mathbf{a}^{*}\right)\geq 0
\end{equation}
for any $ \mu $-measurable subset $ B\subset\mathcal{X} $, $ \mathbf{a}\in\mathbb{C}^{n} $, and $ \gamma\in \text{diag }^{*}L^{\infty}(\mathcal{X};\mu)^{+}\otimes\mathcal{B}(\mathbb{C}^{n}) $. Writing $ \gamma $ as $ \sum_{i,j=1}^{n}f_{ij}\mathbf{e}_{i}\mathbf{e}_{j}^{*} $ for $ \lbrace\mathbf{e}_{j}\rbrace $ an orthonormal basis for $ \mathbb{C}^{n} $ and $ f_{ij}\in L^{\infty}(\mathcal{X};\mu) $, the left-hand side is
\begin{equation}
\sum_{i,j,k=1}^{n}a_{k}\overline{a_{i}}\int_{B}\overline{f_{ji}}f_{jk}\;d\mu=\sum_{j=1}^{n}a_{k}\overline{a_{i}}\int_{B}\overline{g_{j}}g_{j}\;d\mu
\end{equation}
where $ a_{k}=\langle\mathbf{a},\mathbf{e}_{k}\rangle $ and $ g_{j}=\sum_{k=1}^{n}a_{k}f_{jk} $. This is clearly positive. $ \square $
\paragraph{Comment} Using \textbf{B5.18} with standard results for Banach spaces then gives the following bijection:
\begin{equation}
\text{diag}^{*}:\;L^{\infty}\left(\mathcal{X};\mu\right)\to \text{image }\text{diag}^{*}\subset\mathcal{B}\left(L^{2}\left(\mathcal{X};\mu\right)\right) 
\end{equation}
Note $ \text{diag}^{*}(f)g $ is the pointwise multiplication $ fg $ for $ f\in L^{\infty}\left(\mathcal{X};\mu\right) $ and $ g\in L^{2}\left(\mathcal{X};\mu\right) $. Then diag* is clearly isometric and is a representation (clearly weak* continuous) of the von Neumann algebra\footnote{A \textit{Banach algebra} is a Banach space equipped with an associative, distributive product satisfying $ \Vert ab\Vert\leq \Vert a\Vert\Vert b\Vert $. A \textit{*-algebra} is a Banach algebra with an antilinear involution. A \textit{C*-algebra} is a *-algebra where $ \Vert a^{*}a\Vert=\Vert a\Vert^{2} $. A \textit{von Neumann algebra} (also termed a \textit{W*-algebra}) is a \textit{C}*-algebra which, as a Banach space, is a dual space.} $ L^{\infty}\left(\mathcal{X};\mu\right) $ (with product given by pointwise multiplication of ``functions") as operators in $ \mathcal{B}\left(L^{2}\left(\mathcal{X};\mu\right)\right) $.

Morally, one would hope for $ L^{1}\left(\mathcal{X};\mu;\mathcal{S}_{1}(\mathsf{H})\right)^{*} $ to be $ L^{\infty}\left(\mathcal{X};\mu;\mathcal{B}(\mathsf{H})\right) $, which is a  \textit{C}*-algebra. That is almost correct, as the following proposition indicates.
\paragraph{Lemma B5.20} For any $ \Psi\in L^{1}\left(\mathcal{X};\mu;\mathcal{S}_{1}(\mathsf{H})\right)^{*} $,
\[\Vert\Psi\Vert_{\text{op}}=\sup \left\lbrace \frac{1}{\mu(B)}\Psi(1_{B}\psi\xi^{*})\left\vert\begin{array}{l} \mu\text{-measurable } B\subset\mathcal{X}\\\text{unit length }\psi,\xi\in\mathsf{H}  \end{array}\right.\right\rbrace \]
\paragraph{Proof} By the definition of the Bochner integrability for $ L^{1}\left(\mathcal{X};\mu;\mathcal{S}_{1}(\mathsf{H})\right) $, it is enough to take the supremum over simple, $ \mathcal{S}_{1}(\mathsf{H}) $-valued functions:
\begin{equation}
\Vert\Psi\Vert_{\text{op}}=\sup \left\lbrace \frac{\Psi\left(\sum_{j}t_{j}1_{B_{j}}\varphi_{j}\right)}{\sum_{j}t_{j}\mu(B_{j})}\left\vert\begin{array}{l} \mu\text{-measurable, disjoint } \lbrace B_{j}\rbrace\\\lbrace\varphi_{j}\rbrace\subset\mathcal{S}_{1}(\mathsf{H})\\\text{each with unit trace-norm}\\\lbrace t_{j}\rbrace\subset\mathbb{R}^{+}\cup\lbrace 0\rbrace \end{array}\right.\right\rbrace
\end{equation}
For fixed $\lbrace B_{j}\rbrace  $ and $ \lbrace\varphi_{j}\rbrace $, the expression the supremum is being taken over is rational linear in $ \lbrace t_{j}\rbrace $; hence it is necessarily maximized when all but one value of $ t_{j} $ are zero. Therefore, it is only necessary to take the supremum over $ \mathcal{S}_{1}(\mathsf{H}) $-valued functions of the form $ \frac{1}{\mu(B)}1_{B}\varphi $ for $ \mu$-measurable $ B\subset\mathcal{X}$ and unit-trace-norm $ \varphi\in\mathcal{S}_{1}(\mathsf{H})$. By a similar argument applied to the singular-value decomposition of $ \varphi $, it is further only necessary to take $ \varphi $ rank-one. $ \square $ 
\paragraph{Proposition B5.21} There is a well-defined, completely-positive, surjective map with operator norm one
\[\text{diag}\otimes I_{\mathcal{B}(\mathsf{H})}:\;\mathcal{S}_{1}\left(L^{2}(\mathcal{X};\mu)\otimes \mathsf{H}\right)\to L^{1}\left(\mathcal{X};\mu;\mathcal{S}_{1}(\mathsf{H})\right)\]
which is an isometry on the positive cone. This map gives rise to the isometric bijections
\begin{align}
\text{diag}\otimes I_{\mathcal{B}(\mathsf{H})}&:\;\mathcal{S}_{1}\left(L^{2}\left(\mathcal{X};\mu\right)\otimes \mathsf{H}\right)/\text{ker }(\text{diag}\otimes I_{\mathcal{B}(\mathsf{H})})\to L^{1}\left(\mathcal{X};\mu;\mathcal{S}_{1}(\mathsf{H})\right)\nonumber\\
(\text{diag}\otimes I_{\mathcal{B}(\mathsf{H})})^{*}&:\;L^{1}\left(\mathcal{X};\mu;\mathcal{S}_{1}(\mathsf{H})\right)^{*}\to \text{image }(\text{diag}\otimes I_{\mathcal{B}(\mathsf{H})})^{*}= \mathcal{W}(\mathcal{X};\mu;\mathsf{H})\subset\mathcal{B}\left(L^{2}\left(\mathcal{X};\mu\right)\otimes \mathsf{H}\right)\nonumber
\end{align}
\paragraph{Proof} Using \textbf{B4.2}, $ \mathcal{S}_{1}\left(L^{2}\left(\mathcal{X};\mu\right)\right)\otimes\mathcal{S}_{1}\left(\mathsf{H}\right) $ is trace-norm dense in $ \mathcal{S}_{1}\left(L^{2}\left(\mathcal{X};\mu\right)\otimes \mathsf{H}\right) $. By the definition of Bochner integrability, $ L^{1}\left(\mathcal{X};\mu;\mathcal{S}_{1}(\mathsf{H})\right) $ is the norm completion of $ L^{1}\left(\mathcal{X};\mu\right)\otimes\mathcal{S}_{1}(\mathsf{H}) $. Hence, $ \text{diag}\otimes I_{\mathcal{B}(\mathsf{H})} $ is a well-defined map for the given domain and range by \textbf{A1.3}. It has operator norm one by \textbf{A1.3}, \textbf{B5.19}, \textbf{B5.18}, and \textbf{B5.5}. It is completely positive by \textbf{B5.18} and \textbf{B5.8}. By the definition of Bochner integrability, the simple functions of the form $ \sum_{j}1_{B_{j}}\varphi_{j} $ for disjoint, $ \mu $-measurable subsets $ \lbrace B_{j}\rbrace $ and $ \lbrace\varphi_{j}\rbrace\subset \mathcal{S}_{1}(\mathsf{H}) $ are norm-dense in $ L^{1}\left(\mathcal{X};\mu;\mathcal{S}_{1}(\mathsf{H})\right)$. These have preimages in  $ \mathcal{S}_{1}\left(L^{2}\left(\mathcal{X};\mu\right)\otimes \mathsf{H}\right) $ given by $ \sum_{j}\left(1_{B_{j}}\otimes 1_{B_{j}}\right)\varphi_{j} $. Since $ \text{diag}\otimes I_{\mathcal{B}(\mathsf{H})} $ is bounded (hence, continuous), it is surjective. To see that it is isometric on the positive cone $ \mathcal{D}\left(L^{2}\left(\mathcal{X};\mu\right)\otimes \mathsf{H}\right)^{+} $, by the spectral theorem for compact operators, it is only necessary to check the rank-one case. By continuity and the definition of $ L^{2}\left(\mathcal{X};\mu\right)\otimes \mathsf{H} $, it is only necessary to check the case where the vectors are of finite-tensor-rank, $ \sum_{j=1}^{m}f_{j}\otimes\xi_{j} $ for $ \lbrace f_{j}\rbrace\subset L^{2}\left(\mathcal{X};\mu\right) $ and $ \lbrace\xi_{j} \rbrace\subset\mathsf{H}$. Then,
\begin{align}
\left\Vert \sum_{j=1}^{m}f_{j}\otimes\xi_{j}\sum_{k=1}^{m}f_{k}^{*}\otimes\xi_{k}^{*}\right\Vert_{\text{trace}}&=\text{tr }\sum_{j=1}^{m}f_{j}\otimes\xi_{j}\sum_{k=1}^{m}f_{k}^{*}\otimes\xi_{k}^{*}=\sum_{j,k=1}^{m}\int_{\mathcal{X}}f_{j}\overline{f_{k}}\;d\mu\langle\xi_{j},\xi_{k}\rangle\nonumber\\&=\left\Vert(\text{diag}\otimes I_{\mathcal{B}(\mathsf{H})})\left(\sum_{j=1}^{m}f_{j}\otimes\xi_{j}\sum_{k=1}^{m}f_{k}^{*}\otimes\xi_{k}^{*}\right)\right\Vert
\end{align}
so $  \text{diag}\otimes I_{\mathcal{B}(\mathsf{H})} $ is indeed isometric on the positive cone.

By the preceding results, the map
\begin{equation}
\text{diag}\otimes I_{\mathcal{B}(\mathsf{H})}:\;\mathcal{S}_{1}\left(L^{2}\left(\mathcal{X};\mu\right)\otimes \mathsf{H}\right)/\text{ker }(\text{diag}\otimes I_{\mathcal{B}(\mathsf{H})})\to L^{1}\left(\mathcal{X};\mu;\mathcal{S}_{1}(\mathsf{H})\right)
\end{equation}
is certainly a bijection. From the dense set of values considered above to establish surjectivity, it is isometric. By standard properties of Banach spaces, the adjoint map  
\begin{equation}
(\text{diag}\otimes I_{\mathcal{B}(\mathsf{H})})^{*}:\;L^{1}\left(\mathcal{X};\mu;\mathcal{S}_{1}(\mathsf{H})\right)^{*}\to \text{image }(\text{diag}\otimes I_{\mathcal{B}(\mathsf{H})})^{*}= \mathcal{W}(\mathcal{X};\mu;\mathsf{H})\subset\mathcal{B}\left(L^{2}\left(\mathcal{X};\mu\right)\otimes \mathsf{H}\right)
\end{equation}
is a bijection. To see it is an isometry, take any $ \Psi\in L^{1}\left(\mathcal{X};\mu;\mathcal{S}_{1}(\mathsf{H})\right)^{*} $. By the definition of operator norm and the Cauchy-Schwartz inequality, there is the well-known formula for the operator norm in $ \mathcal{B}(\mathsf{J}) $ for any Hilbert space \textsf{J},
\begin{equation}
\Vert \beta\Vert_{\text{op}}=\sup_{\begin{scriptsize}\begin{array}{c}\mathbf{x},\mathbf{y}\in \mathsf{J}\\\Vert\mathbf{x}\Vert\leq 1,\Vert\mathbf{y}\Vert\leq 1 \end{array}\end{scriptsize}}\langle\beta\mathbf{x},\mathbf{y}\rangle
\end{equation}
Using this, we have
\begin{align}
\Vert(\text{diag}\otimes I_{\mathcal{B}(\mathsf{H})})^{*}\Psi\Vert_{\text{op}}&=\sup_{\begin{scriptsize}\begin{array}{c}\zeta,\eta\in L^{2}\left(\mathcal{X};\mu\right)\otimes \mathsf{H}\\\Vert\zeta\Vert\leq 1,\Vert\eta\Vert\leq 1 \end{array}\end{scriptsize}}\langle(\text{diag}\otimes I_{\mathcal{B}(\mathsf{H})})^{*}\Psi\zeta,\eta\rangle\nonumber\\
&=\sup_{\begin{scriptsize}\begin{array}{c}\zeta,\zeta\in L^{2}\left(\mathcal{X};\mu\right)\otimes \mathsf{H}\\\Vert\zeta\Vert\leq 1,\Vert\eta\Vert\leq 1 \end{array}\end{scriptsize}}\Psi\left((\text{diag}\otimes I_{\mathcal{B}(\mathsf{H})})(\zeta\eta^{*})\right)
\end{align}
By the preceding lemma \textbf{B5.20}, this is certainly $ \Vert\Psi\Vert_{\text{op}} $, since we only have to consider $ \zeta $ and $ \eta $ of the form $ 1_{B}\otimes\psi $ and $ 1_{B}\otimes\xi $ respectively, for $ \mu$-measurable $ B\subset\mathcal{X}$ and unit-length $ \psi,\xi\in\mathsf{H} $, to get $ \Vert\Psi\Vert_{\text{op}} $. $ \square $
\paragraph{Proposition B5.22} The Banach space $ \mathcal{W}(\mathcal{X};\mu;\mathsf{H})\subset\mathcal{B}\left(L^{2}\left(\mathcal{X};\mu\right)\otimes \mathsf{H}\right) $ defined in the preceding proposition is a von Neumann algebra.
\paragraph{Proof} By standard properties of Banach spaces, $ \mathcal{W}(\mathcal{X};\mu;\mathsf{H}) $ is precisely composed of those elements that annihilate the kernel of $ \text{diag}\otimes I_{\mathcal{B}(\mathsf{H})} $; in other words, those $ \phi\in \mathcal{B}\left(L^{2}(\mathcal{X};\mu)\otimes \mathsf{H}\right) $ such that $ \text{tr }\varphi\alpha=0 $ for any $ \alpha $ in the kernel. Now take any $ f\in L^{\infty}(\mathcal{X};\mu) $ and consider the operator $ \gamma=\text{diag}^{*}\,f\otimes I_{\mathsf{H}} $. By using singular-value-decomposition (as in the definition of the diag map in \textbf{B5.17}) together with approximating each resulting vector by a finite-tensor-rank approximation and taking the limit as the approximations become better, it can be shown that, operating on the left or the right, the kernel is invariant under the action of $ \gamma $. Take any $ \beta\in \mathcal{S}_{1}\left(L^{2}(\mathcal{X};\mu)\otimes \mathsf{H}\right) $. By a similar argument, it can be shown that $ \text{diag}\otimes I_{\mathcal{B}(\mathsf{H})}(\gamma\beta-\beta\gamma)=0  $; hence, 
\begin{equation}
0=\text{tr }\varphi(\gamma\beta-\beta\gamma)=\text{tr }\beta(\varphi\gamma-\gamma\varphi)
\end{equation}
using the linear and cyclic properties of the trace. Since $ \beta $ was arbitrary, it must be that $ \varphi $ commutes with $ \gamma $; since \textit{f} in the definition of $  \gamma $ was arbitrary, $ \varphi $ must be in the commutant\footnote{The \textit{commutant} of a subset of an algebra is composed of all elements of the algebra that commute with the subset.} of $ \lbrace\text{diag}^{*} L^{\infty}(\mathcal{X};\mu)\otimes I_{\mathsf{H}}\rbrace $.

Now suppose $ \varphi $ is in the commutant of $ \text{diag}^{*} L^{\infty}(\mathcal{X};\mu)\otimes I_{\mathsf{H}} $. Now we will consider a certain norm-dense subset of $ \mathcal{S}_{1}\left(L^{2}\left(\mathcal{X};\mu\right)\otimes \mathsf{H}\right) $: firstly note $ \mathcal{S}_{1}\left(L^{2}\left(\mathcal{X};\mu\right)\right)\otimes\mathcal{S}_{1}\left(\mathsf{H}\right) $ is norm-dense. Using $ \mathcal{S}_{1}\left(L^{2}\left(\mathcal{X};\mu\right)\right)\subset\mathcal{S}_{2}\left(L^{2}\left(\mathcal{X};\mu\right)\right)\cong L^{2}(\mathcal{X}\times \mathcal{X};\mu\times\mu) $, we can consider $ \mathcal{S}_{1}\left(L^{2}\left(\mathcal{X};\mu\right)\right) $ as a space of ``functions" on $ \mathcal{X}\times \mathcal{X} $. Using Cartesian decomposition and the spectral theorem for compact operators (or singular value decomposition), truncating the spectral decomposition, then approximating each of the vectors in $ L^{2}(\mathcal{X};\mu) $-norm by simple functions, and then putting everything back together with the proper intersections and relative complements of subsets of $ \mathcal{X} $, the overall effect is to get something in the following subset:
\begin{equation}
\mathcal{A}=\left\lbrace \begin{array}{l}\mathcal{S}_{1}\left(\mathsf{H}\right)\text{-valued, simple functions with rectangular}\\\text{subsets } A\times B\text{ for }A,B\text{ in the finite, disjoint}\\\text{collection } \lbrace C_{1},\ldots,C_{m}\rbrace \text{ of }\mu\text{-measurable subsets of }\mathcal{X}\end{array}\right\rbrace 
\end{equation}
Therefore, the preceding subset is norm-dense.

Now let us consider the subset $ \mathcal{A}_{0}\subset\mathcal{A} $ which is composed of those $ \mathcal{S}_{1}\left(\mathsf{H}\right) $-valued, simple functions that are zero on all the rectangular subsets that include the diagonal; in other words, those that assign zero to each subset of the form $ A\times A $. Then $ \mathcal{A}_{0} $ is norm-dense within the kernel of $ \text{diag}\otimes I_{\mathcal{B}(\mathsf{H})} $. To see this, take any $ \alpha $ in the kernel and let $ \langle a_{j}\rangle $ be a sequence of elements of $ \mathcal{A} $ converging in norm to it. Each $ a_{j} $ can be written as $ b_{j}+c_{j} $, where $ b_{j}\in \mathcal{A}_{0} $ and $ c_{j} $ is zero everywhere except for the subsets of the form $ A\times A $. It is easy to see that $ \Vert c_{j}\Vert_{\text{trace}}=\Vert\text{diag}\otimes I_{\mathcal{B}(\mathsf{H})}c_{j}\Vert $. Since $ \alpha $ is in the kernel, it must be that $ \lim_{j\to \infty}\text{diag}\otimes I_{\mathcal{B}(\mathsf{H})}c_{j}= 0 $; hence, $ \lim_{j\to \infty}\Vert c_{j}\Vert_{\text{trace}}=0 $, so $ \langle b_{j}\rangle $ converges in norm to $ \alpha $. Since $ \alpha $ was arbitrary, $ \mathcal{A}_{0} $ is indeed norm-dense in the kernel.

Now again take any $ \alpha $ in the kernel and let $ \langle b_{j}\rangle $  be a sequence of elements of $ \mathcal{A}_{0} $ converging in norm to it. For each $ j $, by the definition of $ \mathcal{A} $, there is some associated finite, disjoint collection $ \lbrace C^{j}_{1},\ldots,C^{j}_{m_{j}}\rbrace $ of $ \mu $-measurable subsets of $ \mathcal{X} $; take $ f_{j} $ to be any function that takes distinct values on each subset: say $ f_{j}=\sum_{k=1}^{m_{j}}k\,1_{ C^{j}_{k}} $ for instance. Then $ b_{j} $ can be written as 
\begin{equation}
\left(\text{diag*}\,f_{j}\otimes I_{\mathcal{B}(\mathsf{H})}\right)d_{j}-d_{j}\left(\text{diag*}\,f_{j}\otimes I_{\mathcal{B}(\mathsf{H})}\right)
\end{equation} 
for a certain, unique $ d_{j}\in\mathcal{A}_{0}  $. Hence, $ \text{tr }\varphi b_{j}=0 $ for every \textit{j}, so, by continuity, $  \text{tr }\varphi \alpha=0 $. Since $ \alpha $ was arbitrary, $ \varphi $ must be in  the annihilator of the kernel; hence, it is in $ \mathcal{W}(\mathcal{X};\mu;\mathsf{H}) $.

The subset $ \text{diag}^{*} L^{\infty}(\mathcal{X};\mu)\otimes I_{\mathsf{H}} $ is itself the commutant of the $ C^{*}$-algebra given by the closure in norm of $ \text{diag}^{*} L^{\infty}(\mathcal{X};\mu)\otimes \mathcal{B}(\mathsf{H}) $, which is (since simple ``functions" are norm-dense in $L^{\infty}(\mathcal{X};\mu)$) given by $ (\text{diag}\otimes I_{\mathcal{B}(\mathsf{H})})^{*}L^{\infty}\left(\mathcal{X};\mu;\mathcal{B}(\mathsf{H})\right) $. Since $ \mathcal{W}(\mathcal{X};\mu;\mathsf{H}) $ is the bicommutant\footnote{The \textit{bicommutant} is the commutant of the commutant.} of a \textit{C}*-algebra, it is is a von Neumann algebra by von Neumann's bicommutant theorem~\cite{sakaivonneumann}. $ \square $
\paragraph{Comment} Using the isometric bijections in \textbf{B5.20}, this induces an algebraic structure on $ L^{1}\left(\mathcal{X};\mu;\mathcal{S}_{1}(\mathsf{H})\right) $. The identity operator in $ \mathcal{W}(\mathcal{X};\mu;\mathsf{H}) $ corresponds to the linear functional $ \Phi $ which agrees with the norm on elements in the positive cone $  L^{1}\left(\mathcal{X};\mu;\mathcal{D}(\mathsf{H})\right)^{+} $,
\begin{equation}
\Phi\rho=\int_{\mathcal{X}}\text{tr }\rho\,d\mu
\end{equation}
where the trace acts pointwise.
\paragraph{Proposition B5.23} The subspace $\text{ker }(\text{diag}\otimes I_{\mathcal{B}(\mathsf{H})})$ is invariant under either left or right product with elements of $ \mathcal{W}(\mathcal{X};\mu;\mathsf{H}) $. 
\paragraph{Proof} Take any $ \alpha $ in the kernel, $ \varphi $ in $ \mathcal{W}(\mathcal{X};\mu;\mathsf{H}) $, $ \mu $-measurable subset $ E\subset\mathcal{X} $, and $ \psi,\chi\in \mathsf{H} $. Then, for $ P_{E}=\text{diag}^{*}\;1_{E} $ the orthogonal projector in $ \mathcal{B}( L^{2}(\mathcal{X};\mu)) $ to the subspace of ``functions" zero almost everywhere outside of \textit{E},
\begin{equation}
\int_{E}\langle\left((\text{diag}\otimes I_{\mathcal{B}(\mathsf{H})})\alpha\varphi\right) \psi,\chi\rangle\,d\mu=\text{tr }(P_{E}\otimes\psi\chi^{*})\alpha\varphi=\text{tr }\alpha'\varphi=0
\end{equation} 
where $ \alpha'=(P_{E}\otimes\psi\chi^{*})\alpha $ is still in the kernel (as is readily seen by using singular value decomposition on $ \alpha $ as in the definition of diag in \textbf{B5.17}). Since $ E, \psi,\chi $ were arbitrary, it must be that $ (\text{diag}\otimes I_{\mathcal{B}(\mathsf{H})})\alpha\varphi=0 $; hence, $ \alpha\varphi $ is still in the kernel. Similarly, using the cyclic property of the trace,
\begin{equation}
\int_{E}\langle\left((\text{diag}\otimes I_{\mathcal{B}(\mathsf{H})})\varphi\alpha\right) \psi,\chi\rangle\,d\mu=\text{tr }(P_{E}\otimes\psi\chi^{*})\varphi\alpha=\text{tr }\alpha''\varphi=0
\end{equation}
where $ \alpha''=\alpha(P_{E}\otimes\psi\chi^{*}) $ is still in the kernel. Since $ E, \psi,\chi $ were arbitrary, it must be that $ (\text{diag}\otimes I_{\mathcal{B}(\mathsf{H})})\varphi\alpha=0 $; hence, $ \varphi\alpha $ is still in the kernel. $ \square $
\paragraph{Alternate proof} Note $ \mathcal{W}(\mathcal{X};\mu;\mathsf{H}) $ separates the elements of $\mathcal{S}_{1}\left(L^{2}\left(\mathcal{X};\mu\right)\otimes \mathsf{H}\right)/\text{ker }(\text{diag}\otimes I_{\mathcal{B}(\mathsf{H})})$. To see this, take any $ \alpha\in \mathcal{S}_{1}\left(L^{2}\left(\mathcal{X};\mu\right)\otimes \mathsf{H}\right) $. Then, by the definition of Bochner integrability, $ (\text{diag}\otimes I_{\mathcal{B}(\mathsf{H})})\alpha $ can be arbitrarily well approximated in $ L^{1}(\mathcal{X};\mu;\mathcal{S}^{1}(\mathsf{H})) $-norm by simple, $ \mathcal{S}^{1}(\mathsf{H}) $-valued functions, $ \sum_{j}1_{B_{j}}\rho_{j} $, where $ \lbrace B_{j}\rbrace $ is a finite, disjoint collection of $ \mu $-measurable subsets of $ \mathcal{X} $. Use singular value decomposition to write each $ \rho_{j} $ as $ U_{j}\Sigma_{j}V_{j}^{*} $, where $ \Sigma_{j} $ is a trace-class, diagonal operator in $ \mathcal{B}(\ell^{2}) $ (or $ \mathbb{C}^{\dim\mathsf{H}} $ if $ \dim\mathsf{H} $ is finite) and $ U_{j},V_{j} $ are partial isometries from $ \ell^{2} $ (or $ \mathbb{C}^{\dim\mathsf{H}} $ if $ \dim\mathsf{H} $ is finite) to $ \mathsf{H} $. Then, $\beta= \sum_{j}(\text{diag}^{*}1_{B_{j}})\otimes V_{j}U_{j}^{*}\in\mathcal{W}(\mathcal{X};\mu;\mathsf{H})$ has the property that
\begin{equation}
\text{tr }\beta (\text{diag}\otimes I_{\mathcal{B}(\mathsf{H})})^{-1}\sum_{j}1_{B_{j}}\rho_{j}=\left\Vert\sum_{j}1_{B_{j}}\rho_{j}\right\Vert
\end{equation} 
Hence, by taking the approximation sufficiently close, there is some such $ \beta $ such that $ \text{tr }\beta \alpha$ is arbitrarily close to $ \Vert\alpha\Vert $. (Note the Hahn-Banach theorem gives this result more easily, but it relies on the Axiom of Choice.)

Using this, $ \text{ker }(\text{diag}\otimes I_{\mathcal{B}(\mathsf{H})}) $ can be characterized as
\begin{equation}
\left\lbrace \alpha\in\mathcal{S}_{1}\left(L^{2}\left(\mathcal{X};\mu\right)\otimes \mathsf{H}\right)\left\vert\text{tr }\beta\alpha=0\text{ for all }\beta\in \mathcal{W}(\mathcal{X};\mu;\mathsf{H}) \right.\right\rbrace 
\end{equation}
Take any $ \alpha\in \text{ker }(\text{diag}\otimes I_{\mathcal{B}(\mathsf{H})}) $. For any fixed $ \beta\in\mathcal{W}(\mathcal{X};\mu;\mathsf{H})  $, we then have $ \text{tr }\gamma\beta\alpha=0 $ for all $ \gamma\in \mathcal{W}(\mathcal{X};\mu;\mathsf{H}) $; hence, $ \beta\alpha\in \text{ker }(\text{diag}\otimes I_{\mathcal{B}(\mathsf{H})}) $. Also, $ 0=\text{tr }\beta\gamma\alpha=\text{tr }\gamma\alpha\beta $ for all $ \gamma\in \mathcal{W}(\mathcal{X};\mu;\mathsf{H}) $; hence, $ \alpha\beta\in \text{ker }(\text{diag}\otimes I_{\mathcal{B}(\mathsf{H})}) $. $ \square $
\paragraph{Comment} Now given any bounded, linear map $ L:L^{1}(\mathcal{X};\mu;\mathcal{D}(\mathsf{H}))^{+} \to L^{1}(\mathcal{Y};\nu;\mathcal{D}(\mathsf{J}))$, we can extend it to $ L:L^{1}(\mathcal{X};\mu;\mathcal{S}_{1}(\mathsf{H})) \to L^{1}(\mathcal{Y};\nu;\mathcal{S}_{1}(\mathsf{J}))$ using \textbf{B5.15}. We then have the following two linear maps:
\begin{align}
\tilde{L}&:\mathcal{S}_{1}\left(L^{2}\left(\mathcal{X};\mu\right)\otimes \mathsf{H}\right)/\text{ker }(\text{diag}\otimes I_{\mathcal{B}(\mathsf{H})})\to\mathcal{S}_{1}\left(L^{2}\left(\mathcal{Y};\nu\right)\otimes \mathsf{J}\right)/\text{ker }(\text{diag}\otimes I_{\mathcal{B}(\mathsf{J})})\\
\tilde{L}^{*}&:\mathcal{W}(\mathcal{Y};\nu;\mathsf{J})\to\mathcal{W}(\mathcal{X};\mu;\mathsf{H})
\end{align}
given by
\begin{align}
\tilde{L}&=(\text{diag}_{L^{2}\left(\mathcal{X};\mu\right)}\otimes I_{\mathcal{B}(\mathsf{J})})^{-1}\circ L\circ(\text{diag}_{L^{2}\left(\mathcal{X};\mu\right)}\otimes I_{\mathcal{B}(\mathsf{H})})\\
\tilde{L}^{*}&=(\text{diag}_{L^{2}\left(\mathcal{X};\mu\right)}\otimes I_{\mathcal{B}(\mathsf{H})})^{*}\circ L^{*}\circ(\text{diag}_{L^{2}\left(\mathcal{X};\mu\right)}\otimes I_{\mathcal{B}(\mathsf{J})})^{-1*}
\end{align}
\paragraph{Proposition B5.24} Given a bounded, completely-positive, linear map $ L:L^{1}(\mathcal{X};\mu;\mathcal{D}(\mathsf{H}))^{+} \to L^{1}(\mathcal{Y};\nu;\mathcal{D}(\mathsf{J}))^{+}$, the resulting $ \tilde{L},\tilde{L}^{*} $ are completely-positive in the sense of Stinespring's theorem~\cite{stinespring}:
\[ \sum_{i,k} \text{tr }\alpha_{k}^{*}\alpha_{i}\tilde{L}(\psi_{i}\psi_{k}^{*})\geq 0\Longleftrightarrow\sum_{i,k}\left\langle\tilde{L}^{*}(\alpha_{k}^{*}\alpha_{i})\psi_{j},\psi_{i}\right\rangle\geq 0\] 
for all finite collections $ \lbrace \alpha_{k}\rbrace\subset\mathcal{W}(\mathcal{Y};\nu;\mathsf{J})  $ and $  \lbrace \psi_{k}\rbrace\subset  L^{2}(\mathcal{X};\mu)\otimes \mathsf{H}$.
\paragraph*{Proof} The two expressions are related by duality, do it is only necessary to prove the left-hand one. The map $ \tilde{L} $ is the composition of completely-positive maps; hence, it is completely-positive. Therefore, for any finite whole number \textit{n}, $ \tilde{L}\otimes I_{\mathcal{M}_{n}} $ is completely positive, so for any $ \beta\in\mathcal{W}(\mathcal{Y};\nu;\mathsf{J}\otimes \mathbb{C}^{n})  $ and $  \varphi\in  L^{2}(\mathcal{X};\mu)\otimes \mathsf{H}\otimes \mathbb{C}^{n}$,
\begin{equation}
\text{tr }\beta^{*}\beta(\tilde{L}\otimes I_{\mathcal{M}_{n}})(\varphi\varphi^{*})\geq 0
\end{equation}
For $ \lbrace\mathbf{e}_{j}\rbrace $ an orthonormal basis for $ \mathbb{C}^{n} $, write $ \beta $ as $ \sum_{i,j=1}^{n}\alpha_{ij}\mathbf{e}_{i}\mathbf{e}_{j}^{*} $ for $ \alpha_{ij}\in \mathcal{W}(\mathcal{Y};\nu;\mathsf{J}) $ and $ \varphi $  as $ \sum_{j=1}^{n}\psi_{j}\mathbf{e}_{j} $ for $ \psi_{j}\in L^{2}(\mathcal{X};\mu)\otimes \mathsf{H} $. Then we have
\begin{equation}
\sum_{i,j,k=1}^{n}\text{tr }^{*}\alpha_{jk}^{*}\alpha_{ji}\tilde{L}(\psi_{i}\psi_{k}^{*})\geq 0
\end{equation}
If the $ \alpha_{ij} $ are chosen to be independent of \textit{j}, this gives the desired result. $ \square $
\paragraph{Comment} Following the proof of Stinespring's theorem, define the positive, bilinear form $ \langle\cdot,\cdot\rangle_{\text{Stinespring}} $ on $ \mathcal{W}(\mathcal{Y};\nu;\mathsf{J})\otimes L^{2}(\mathcal{X};\mu)\otimes \mathsf{H} $ by
\begin{equation}
\left\langle \sum_{j}\alpha_{j}\otimes \psi_{j}, \sum_{k}\beta_{k}\otimes \varphi_{k}\right\rangle_{\text{Stinespring}}=\sum_{j,k}\left\langle\tilde{L}^{*}(\beta_{k}^{*}\alpha_{j})\psi_{j} ,\varphi_{k}\right\rangle=\sum_{j,k}\text{tr }\beta_{k}^{*}\alpha_{j}\tilde{L}(\psi_{j}\varphi_{k}^{*}) 
\end{equation} 
Let $ \mathcal{N} $ be the null subspace $ \lbrace\alpha\in \mathcal{W}(\mathcal{Y};\nu;\mathsf{J})\otimes L^{2}(\mathcal{X};\mu)\otimes \mathsf{H}\vert\langle\alpha,\alpha\rangle_{\text{Stinespring}}=0\rbrace $. Take any $ \alpha\in \mathcal{N} $ and $ \beta\in \mathcal{W}(\mathcal{Y};\nu;\mathsf{J})\otimes L^{2}(\mathcal{X};\mu)\otimes \mathsf{H}$; by considering $ \langle\alpha+t\beta,\alpha+t\beta\rangle_{\text{Stinespring}} $ and $ \langle\alpha+\imath t\beta,\alpha+\imath t\beta\rangle_{\text{Stinespring}} $ for small, real \textit{t}, from the positivity of $ \langle\cdot,\cdot\rangle_{\text{Stinespring}} $, it is evident $ \langle\alpha,\beta\rangle_{\text{Stinespring}}=0 $. Then $ \langle\cdot,\cdot\rangle_{\text{Stinespring}} $ is an inner product on $ \mathcal{W}(\mathcal{Y};\nu;\mathsf{J})\otimes L^{2}(\mathcal{X};\mu)\otimes \mathsf{H}/\mathcal{N} $; this space can be completed to give a Hilbert space \textsf{K}.

From the proof of Stinespring's theorem, $ \mathcal{N} $ is invariant under the left action of $ \beta\otimes I_{ L^{2}(\mathcal{X};\mu)\otimes \mathsf{H}} $ for any $ \beta\in \mathcal{W}(\mathcal{Y};\nu;\mathsf{J}) $. This then induces a representation of the von Neumann algebra, $ \zeta:\mathcal{W}(\mathcal{Y};\nu;\mathsf{J})\to \mathcal{B}(\mathsf{K}) $. The representation is readily seen to be a *-representation\footnote{A representation $ \zeta $ of a \textit{C}*-algebra is a *\textit{-representation} if $ \zeta(\alpha^{*})=\zeta(\alpha)^{*} $ for any $ \alpha $.}. 
\paragraph{Proposition B5.25} The representation $ \zeta $ has operator norm less than or equal to one and is weak* continuous with pre-adjoint $ \zeta_{*}:\mathcal{S}_{1}(\mathsf{K})\to \mathcal{S}_{1}\left(L^{2}\left(\mathcal{X};\mu\right)\otimes \mathsf{H}\right)/\text{ker }(\text{diag}\otimes I_{\mathcal{B}(\mathsf{H})}) $ being isometric on the positive cone.
\paragraph*{Proof} By the definition of operator norm and the Cauchy-Schwartz inequality, there is the well-known formula for the operator norm in $ \mathcal{B}(\mathsf{J}) $ for any Hilbert space \textsf{M},
\begin{equation}
\Vert \beta\Vert_{\text{op}}=\sup_{\begin{scriptsize}\begin{array}{c}\mathbf{x},\mathbf{y}\in \mathsf{M}\\\Vert\mathbf{x}\Vert\leq 1,\Vert\mathbf{y}\Vert\leq 1 \end{array}\end{scriptsize}}\langle\beta\mathbf{x},\mathbf{y}\rangle
\end{equation}
Using this and the norm-density of $  \mathcal{W}(\mathcal{Y};\nu;\mathsf{J})\otimes L^{2}(\mathcal{X};\mu)\otimes \mathsf{H}/\mathcal{N} $ in \textsf{K}, we have, for any $ \alpha\in \mathcal{W}(\mathcal{Y};\nu;\mathsf{J}) $, $ \Vert\zeta(\alpha)\Vert_{\text{op}} $ is given by
\begin{equation}
\sup\left\lbrace \left\langle\sum_{j}\alpha\beta_{j}\otimes\psi_{j},\sum_{k}\gamma_{k}\otimes\xi_{k} \right\rangle_{\text{Stinespring}}\left\vert\begin{array}{l}\sum_{j}\beta_{j}\otimes\psi_{j},\sum_{k}\gamma_{k}\otimes\xi_{k}\in \mathcal{W}(\mathcal{Y};\nu;\mathsf{J})\otimes L^{2}(\mathcal{X};\mu)\otimes \mathsf{H}\\\text{ with unit length using the Stinespring-norm}\end{array}\right.\right\rbrace  
\end{equation}
Absorbing $ \alpha $ into a redefinition of each $ \beta_{j} $, by the Cauchy-Schwartz inequality, this is less than or equal to $ \Vert\alpha\Vert_{\text{op}} $. Hence, $ \zeta $ has operator-norm less than or equal to one.

To see $ \zeta $ is weak* continuous, take any $ \beta\in \mathcal{W}(\mathcal{Y};\nu;\mathsf{J}) $ and rank-one $\psi\varphi^{*}\in \mathcal{S}_{1}(\mathsf{K}) $, where both $ \psi $ and $ \varphi $ are in the norm-dense subset $ \mathcal{W}(\mathcal{Y};\nu;\mathsf{J})\otimes L^{2}(\mathcal{X};\mu)\otimes \mathsf{H}/\mathcal{N} $. Let $ \hat{\psi}\in\psi $ be written $ \sum_{j}\alpha_{j}\otimes\xi_{j} $ and $ \hat{\varphi}\in\varphi $ be written $ \sum_{j}\gamma_{j}\otimes\eta_{j} $. Then 
\begin{align}
\text{tr }\zeta(\beta)\psi\varphi^{*}&=\left\langle \beta\otimes I_{ L^{2}(\mathcal{X};\mu)\otimes \mathsf{H}}\hat{\psi},\hat{\varphi}\right\rangle_{\text{Stinespring}}=\sum_{j,k}\left\langle \tilde{L}^{*}(\gamma_{k}^{*}\beta\alpha_{j})\xi_{j},\eta_{k} \right\rangle \nonumber\\
&=\sum_{j,k}\text{tr } \gamma_{k}^{*}\beta\alpha_{j}\tilde{L}(\xi_{j}\eta_{k}^{*})=\text{tr }\beta\zeta_{*}(\psi\varphi^{*})
\end{align}
where $ \zeta_{*}(\psi\varphi^{*})=\sum_{j,k}\alpha_{j}\tilde{L}(\xi_{j}\eta_{k}^{*})\gamma_{k}^{*} $. Note $ \zeta_{*}(\psi\psi^{*})$ is positive since, for any $ \beta\in \mathcal{W}(\mathcal{Y};\nu;\mathsf{J}) $,
\begin{equation}
\sum_{j,k}\text{tr }\beta^{*}\beta\alpha_{j}\tilde{L}(\xi_{j}\xi_{k}^{*})\alpha_{k}^{*}\geq 0
\end{equation}
using the cyclic property of the trace and \textbf{B5.23}. Then,
\begin{equation}
\Vert\zeta_{*}(\psi\psi^{*})\Vert=\text{tr }\zeta_{*}(\psi\psi^{*})=\Vert\psi\Vert^{2}
\end{equation}
Hence, $ \zeta_{*} $ is bounded with operator norm less than or equal to two on rank-one operators with vectors in $ \mathcal{W}(\mathcal{Y};\nu;\mathsf{J})\otimes L^{2}(\mathcal{X};\mu)\otimes \mathsf{H}/\mathcal{N}  $. Using \textbf{A1.3}, by continuity it extends to rank-one vectors with the same bound. By linearity, it extends to finite rank operators. Using the spectral theorem, it is still isometric on the positive cone; hence, it has operator norm less than or equal to two. By \textbf{A1.3}, it extends to all operators in $ \mathcal{S}_{1}(\mathsf{K}) $ while still being isometric on the positive cone and with operator norm less than or equal to two. Since $ \zeta=\zeta_{*}^{*} $, $ \zeta_{*} $ actually has operator norm less than or equal to one. $ \square $
\paragraph{Proposition B5.26} There is a Hilbert space \textsf{M} and a partial isometry $ E:\mathsf{K}\to L^{2}\left(\mathcal{X};\mu\right)\otimes \mathsf{H}\otimes\mathsf{M} $ such that the representation $ \zeta $ is given by $ \zeta(\alpha)=E^{*}(\alpha\otimes I_{\mathsf{M}})E $ for any $ \alpha\in \mathcal{W}(\mathcal{Y};\nu;\mathsf{J}) $. The partial isometry \textit{E} is such that the orthogonal projector $ EE^{*} $ commutes with all $ \lbrace\alpha\otimes I_{\mathsf{M}}\rbrace $. If all the Hilbert spaces $ L^{2}\left(\mathcal{X};\mu\right) $, \textsf{H}, $ L^{2}\left(\mathcal{Y};\nu\right) $, and \textsf{J} have finite dimension, then \textsf{M} can be taken to be $ \mathsf{J}\otimes\dim L^{2}\left(\mathcal{X};\mu\right)\otimes \mathsf{H}$.
\paragraph*{Proof} From a result by Sakai~\cite{sakai}, since $ \zeta $ is a weak*-continuous, *-representation\footnote{Termed a $\mathcal{W}^{*}$\textit{-representation} by Sakai.} by the preceding proposition, $ \zeta $ is given by the composition of an \textit{amplification} with an \textit{induction}: there exists a Hilbert space \textsf{M} and a bounded, linear map $ E:\mathsf{K}\to L^{2}\left(\mathcal{Y};\nu\right)\otimes \mathsf{J}\mathsf{M} $ such that $ \zeta(\alpha)=E^{*}(\alpha\otimes I_{\mathsf{M}})E $, $ EE^{*} $ is an orthogonal projector commuting with all $ \lbrace\alpha\otimes I_{\mathsf{M}}\rbrace $, $ EE^{*}E=E $, and $ E^{*}EE^{*}=E^{*} $. Then $ \zeta_{*} $ is given by, for any $ \beta\in\mathcal{S}_{1}(\mathsf{K}) $, $ \zeta_{*}\beta=\text{tr }_{\mathsf{M}}E\beta E^{*}$, where $ \text{tr }_{\mathsf{M}} $ is the partial trace over \textsf{M}. Since, by the preceding proposition, $ \zeta_{*} $ is an isometry on the positive cone, it must be that $ E^{*}E=I_{\mathsf{K}} $, so \textit{E} is a partial isometry. Unfortunately, the proof given by Sakai is non-constructive; the only limitation on the dimension of \textsf{M} is that it need not be larger than $ \dim \mathsf{K} $.

If all the Hilbert spaces have finite dimension, we can be far more concrete in the construction. Let $ \mathsf{M}=\mathsf{J}\otimes\dim L^{2}\left(\mathcal{X};\mu\right)\otimes \mathsf{H} $. Since the Hilbert spaces have finite dimension, $ \mathcal{W}(\mathcal{Y};\nu;\mathsf{J}) $ is isomorphic to $ L^{\infty}\left(\mathcal{Y};\nu\right)\otimes\mathcal{B}(\mathsf{J}) $, which is isomorphic to the Hilbert space $ L^{2}\left(\mathcal{Y};\nu\right)\otimes \mathsf{J}\otimes \mathsf{J}$ using $ L^{\infty}\left(\mathcal{Y};\nu\right)\cong L^{2}\left(\mathcal{Y};\nu\right) $ and\footnote{Using $ \mathcal{M}_{n}\cong \mathbb{C}^{n}\otimes\mathbb{C}^{n}$, say by using the Vec operation that stacks columns.} $\mathcal{B}(\mathsf{J})\cong \mathsf{J}\otimes \mathsf{J} $. Then, using this isomorphism, $ \mathcal{N} $ maps to a subspace $ \mathcal{N}' $ of the Hilbert space $  L^{2}\left(\mathcal{Y};\nu\right)\otimes \mathsf{J}\otimes \mathsf{M} $. Also, then \textsf{K} is isomorphic to $ L^{2}\left(\mathcal{Y};\nu\right)\otimes \mathsf{J}\otimes\mathsf{M}/\mathcal{N}'\cong {\mathcal{N}'}^{\perp}$ with no need for completion. Let $ E:\mathsf{K}\to\mathcal{N}' $ be this isomorphism. Then the operators $ \alpha\otimes I_{L^{2}\left(\mathcal{X};\mu\right)\otimes \mathsf{H}} $ acting on $ \mathcal{W}(\mathcal{Y};\nu;\mathsf{J})\otimes L^{2}\left(\mathcal{X};\mu\right)\otimes \mathsf{H} $ give rise to a representation $ \alpha\otimes I_{\mathsf{M}} $ on $ L^{2}\left(\mathcal{Y};\nu\right)\otimes \mathsf{J}\otimes \mathsf{M} $. From the proof to Stinespring's theorem, $ \mathcal{N}' $ is invariant under all the $ \lbrace\alpha\otimes I_{\mathsf{M}}\rbrace $, so the orthogonal projector $ P_{{\mathcal{N}'}^{\perp}}  $ commutes with all the $ \lbrace\alpha\otimes I_{\mathsf{M}}\rbrace $.

Any point in the space $  L^{2}\left(\mathcal{Y};\nu\right)\otimes \mathsf{J}\otimes \mathsf{M} $ can be uniquely written as $ n+m $ for $ n\in\mathcal{N}' $ and $ m\in {\mathcal{N}'}^{\perp} $. Now put a new inner-product on the space, $ \langle\cdot,\cdot\rangle_{\text{new}} $, defined by
\begin{equation}
\langle m+n,m'+n'\rangle_{\text{new}}=\langle E^{-1}(m),E^{-1}(m')\rangle_{\text{Stinespring}}+\langle n,n'\rangle
\end{equation}
(This is the step that requires all the spaces to be finite--the use of finiteness in the preceding steps is not essential.) With respect to this new inner product, \textit{E} is a partial isometry. Let $ E^{*} $ be the adjoint map with respect to the new inner product. Then the representation $ \zeta $ on \textsf{K} is provided by, for any $ k\in \mathsf{K} $, 
\begin{equation}
\zeta(\alpha)(k)=\left(E^{-1}\circ P_{{\mathcal{N}'}^{\perp}}\right)(\alpha\otimes I_{\mathsf{M}})E(k)=(E^{*}(\alpha\otimes I_{\mathsf{M}})E)(k)
\end{equation}
Hence, $ \zeta(\alpha)= E^{*}(\alpha\otimes I_{\mathsf{M}})E $. $ \square $
\paragraph{Proposition B5.27} Any completely-positive map $ L\in\mathcal{CP}(L^{1}(\mathcal{X};\mu;\mathcal{D}(\mathsf{H})), L^{1}(\mathcal{Y};\nu;\mathcal{D}(\mathsf{J}))) $ can be expressed as
\[ L\rho=(\text{diag}_{L^{2}\left(\mathcal{Y};\nu\right)}\otimes I_{\mathcal{B}(\mathsf{J})}\otimes\text{tr }_{\mathsf{M}})V((\text{diag}_{L^{2}\left(\mathcal{X};\mu\right)}\otimes I_{\mathcal{B}(\mathsf{H})})^{-1}\rho) V^{*}\]
for some Hilbert space \textsf{M} and some linear operator $ V\in \mathcal{B}(L^{2}\left(\mathcal{X};\mu\right)\otimes\mathsf{H},L^{2}\left(\mathcal{Y};\nu\right)\otimes\mathsf{J}\otimes\mathsf{M}) $. If \textit{L} is an isometry on the positive cone, then \textit{V} is a partial isometry. If all the Hilbert spaces $ L^{2}\left(\mathcal{X};\mu\right) $, \textsf{H}, $ L^{2}\left(\mathcal{Y};\nu\right) $, and \textsf{J} have finite dimension, then \textsf{M} can be taken to be $ \mathsf{J}\otimes\dim L^{2}\left(\mathcal{X};\mu\right)\otimes \mathsf{H}$.
\paragraph*{Proof} By \textbf{B.16}, \textit{L} extends uniquely to a map in $ \in\mathcal{CP}(L^{1}(\mathcal{X};\mu;\mathcal{S}_{1}(\mathsf{H})), L^{1}(\mathcal{Y};\nu;\mathcal{S}_{1}(\mathsf{J})) $. By \textbf{B.24}, the resulting $ \tilde{L}^{*} $ (defined in the comment preceding \textbf{B.24}) is completely-positive in the sense of Stinespring's theorem. Also, the von Neumann algebra $ \mathcal{W}(\mathcal{Y};\nu;\mathsf{J}) $ (see \textbf{B5.21}, \textbf{B5.22}) of course contains the identity. Hence, by Stinespring's theorem, $ \tilde{L}^{*} $ can be expressed as, for any $ \alpha\in  \mathcal{W}(\mathcal{Y};\nu;\mathsf{J}) $, $ \tilde{L}^{*}\alpha=W^{*}\zeta(\alpha)W $ for some $ W\in \mathcal{B}(L^{2}\left(\mathcal{X};\mu\right)\otimes H,\mathsf{K}) $ (where the Hilbert space \textsf{K} and the representation $ \zeta $ are defined in the comment preceding \textbf{B.25}). By the preceding proposition, $ \zeta(\alpha)=E^{*}(\alpha\otimes I_{\mathsf{M}})E $ for some Hilbert space \textsf{M} and some partial isometry $ E:\mathsf{K},L^{2}\left(\mathcal{Y};\nu\right)\otimes J) $. Also, if all the Hilbert spaces $ L^{2}\left(\mathcal{X};\mu\right) $, \textsf{H}, $ L^{2}\left(\mathcal{Y};\nu\right) $, and \textsf{J} have finite dimension, then \textsf{M} can be taken to be $ \mathsf{J}\otimes\dim L^{2}\left(\mathcal{X};\mu\right)\otimes \mathsf{H}$.

Let $ V=EW $; then $ \tilde{L}^{*}\alpha=V^{*}(\alpha\otimes I_{\mathsf{M}})V $, so $ \tilde{L}\beta=\text{tr}_{\mathsf{M}}(V\beta V^{*}) $ for any $ \beta\in \mathcal{S}_{1}(L^{2}\left(\mathcal{X};\mu\right)\otimes \mathsf{H})/\text{ker }(\text{diag}_{L^{2}\left(\mathcal{X};\mu\right)}\otimes I_{\mathcal{B}(\mathsf{H})}) $. (As a check, if $ \gamma\in\text{ker }(\text{diag}_{L^{2}\left(\mathcal{X};\mu\right)}\otimes I_{\mathcal{B}(\mathsf{H})})  $, then $ \text{tr}_{\mathsf{M}}(V \gamma V^{*}) $ is indeed in $ \text{ker }(\text{diag}_{L^{2}\left(\mathcal{Y};\nu\right)}\otimes I_{\mathcal{B}(\mathsf{J})}) $ by the alternate proof to \textbf{B5.23} since $ \text{tr }(\alpha\otimes I_{\mathsf{M}}) V \gamma V^{*}=0 $ for every $ \alpha\in \mathcal{W}(\mathcal{Y};\nu;\mathsf{J}) $.) Writing $ \tilde{L} $ in terms of \textit{L} then gives the desired form for \textit{L}.

If \textit{L} is isometric on the positive cone, then $ \tilde{L} $ is isometric on the positive cone; hence, for any $ \beta\in \mathcal{D}(L^{2}\left(\mathcal{X};\mu\right)\otimes \mathsf{H})^{+} $,
\begin{equation}
\text{tr }\beta=\text{tr }\tilde{L}\beta=\text{tr }V\beta V^{*}=\text{tr }\beta V^{*}V
\end{equation}
Therefore, it must be that $  V^{*}V=I_{\mathcal{B}(L^{2}(\mathcal{X};\mu)\otimes \mathsf{H})} $; hence, it is a partial isometry. $ \square $
\paragraph{Proposition B5.28} Any linear, positive map $ L\in \mathcal{B}(L^{1}(\mathcal{X};\mu),\mathcal{D}(L^{2}(\mathcal{X};\mu)))$ satisfying $ diag\circ L=I_{\mathcal{B}(L^{1}(\mathcal{X};\mu))} $ has image in the diagonal operators (see \textbf{B4.3}). 
\paragraph*{Proof} By the definition of the diagonal extraction map, diag, we have
\begin{equation}
1_{B}\cdot diag \rho=diag(P_{B}\rho P_{B})
\end{equation}
for any $ \mu $-measurable $ B\in \mathcal{X} $ with finite $ mu $-measure, where $ P_{B} $ is the orthogonal projector to the subspace of ``functions" zero almost everywhere outside \textit{B}. Take any ``function" $ f\in L^{2}(\mathcal{X};\mu) $. Then, 
\begin{equation}
\text{diag }P_{B}L(1_{\sim B}f)P_{B}=1_{B}\cdot diag\circ L(1_{\sim B}f=1_{B}1_{\sim B}f=0
\end{equation}
The operator $ P_{B}L(1_{\sim B}f)P_{B} $ is clearly positive; by the isometry of diag on the positive cone (see \textbf{B5.18}), it must be the zero operator. Hence, $ \langle L(1_{\sim B}f)1_{C},1_{C}\rangle=0 $ for any $ \mu $-measurable $ C\subset B $

Now take any $ \mu $-measurable, finite $ \mu $-measure $ B' $ disjoint from \textit{B}. By positivity, for any $ a\in\mathbb{R} $,
\begin{equation}
\langle L(1_{\sim B}f)(a1_{C}-1_{B'}),a1_{C}-1_{B'}\rangle\geq 0
\end{equation} 
For this to hold for all \textit{a}, no matter how large, it must be that 
\begin{equation}
\langle L(1_{\sim B}f)1_{C},1_{B'}\rangle=\langle L(1_{\sim B}f)1_{B'},1_{C}\rangle=0
\end{equation}
Similarly, since $ B'\subset\sim B $ and \textit{C} is disjoint from $ \sim B $,
\begin{equation}
\langle L(1_{B}f)1_{C},1_{B'}\rangle=0
\end{equation}
Then
\begin{equation}
0=\langle L(1_{\sim B}f)1_{C},1_{B'}\rangle+\langle L(1_{B}f)1_{C},1_{B'}\rangle=\langle L(f)1_{C},1_{B'}\rangle
\end{equation}
Taking $ C=B $, we get $ \langle L(f)1_{B},1_{B'}\rangle=0 $. Since $ B,B' $ were arbitrary, $ Lf $ is diagonal. $ \square $
\section{Vector measures}
\paragraph{Notation} The spaces of finite-norm, vector measures on the given set with the given $ \sigma $-algebra will be denoted $ \mathcal{M}(\mathcal{X};\mathcal{E};\mathsf{A})$, $ \mathcal{M}(\mathcal{Y};\mathcal{F};\mathsf{B}) $,$ \ldots $. These are Banach spaces using the total variation norm (give reference). The subset of these that is atomic will be denoted $ \mathcal{A}(\mathcal{X};\mathcal{E};\mathsf{A})$,$ \ldots $.
\paragraph{Proposition B6.1} The subset $ \mathcal{A}(\mathcal{X};\mathcal{E};\mathsf{B})$ is a closed, linear subspace of the Banach space $ \mathcal{M}(\mathcal{X};\mathcal{E};\mathsf{B})$ (hence, it is itself a Banach space).  
\paragraph*{Proof} The subset $ \mathcal{A}(\mathcal{X};\mathcal{E};\mathsf{B})$ is clearly a linear subspace. To show it is closed, take any $ \mu $ in the complement of  $ \mathcal{A}(\mathcal{X};\mathcal{E};\mathsf{B}) $. Suppose that for any $ \varepsilon>0 $ there were some countable collection of atoms $ \lbrace A^{\varepsilon}_{j}\rbrace\subset \mathcal{E}$ such that  $ \Vert \mu\left(\mathcal{X}\setminus\bigcup_{j} A^{\varepsilon}_{j}\right)\Vert<\varepsilon $. Then the countable union
\begin{equation}
B=\bigcup_{\varepsilon\in\lbrace 1,2^{-1},2^{-3},\ldots\rbrace}\bigcup_{j}A^{\varepsilon}_{j}
\end{equation} 
would satisfy $ \Vert \mu\left(\mathcal{X}\setminus B\right)\Vert=0 $, which contradicts $ \mu $ being in the complement. Hence, there is some $ \varepsilon>0 $ such that, for any countable collection of atoms $ \lbrace A_{j}\rbrace\subset \mathcal{E}$, $ \Vert \mu\left(\mathcal{X}\setminus\bigcup_{j} A_{j}\right)\Vert>\varepsilon $. Consequently, the distance from $ \mu $ to $ \mathcal{A}(\mathcal{X};\mathcal{E};\mathsf{B})$ is greater than $ \varepsilon $. Since $ \mu $ was arbitrary, the complement of $ \mathcal{A}(\mathcal{X};\mathcal{E};\mathsf{B}) $ is open; hence, $ \mathcal{A}(\mathcal{X};\mathcal{E};\mathsf{B}) $ is closed. $ \square $
\paragraph{Proposition B6.2} If the Banach spaces \textsf{A}, \textsf{B}, and \textsf{C} are such that $ \mathsf{A}\otimes\mathsf{B} $ is norm-dense in \textsf{C}, then $ \mathcal{A}(\mathcal{X};\mathcal{E};\mathsf{A})\otimes\mathcal{A}(\mathcal{Y};\mathcal{F};\mathsf{B}) $ is norm-dense in $ \mathcal{A}(\mathcal{X}\times\mathcal{Y};\mathcal{G};\mathsf{C}) $,  where $ \mathcal{G}=\sigma(\mathcal{E}\times \mathcal{F}) $ is the $ \sigma $-algebra generated by the rectangular subsets $ \mathcal{E}\times \mathcal{F} $.
\paragraph*{Proof} The atoms of $ \mathcal{G} $ are in the rectangular subsets $  \mathcal{E}\times \mathcal{F} $. Therefore, any vector measure in $ \mathcal{A}(\mathcal{X}\times\mathcal{Y};\mathcal{G};\mathsf{C}) $ can be arbitrarily well-approximated in norm by vector measures in $ \mathcal{A}(\mathcal{X}\times\mathcal{Y};\mathcal{G};\mathsf{C}) $ of the form $ \sum_{j}\mathsf{c}_{j}\delta_{E_{j}}\times\delta_{F_{j}} $ for finite collections of atoms $ \lbrace E_{j}\rbrace\subset \mathcal{E}$ and $ \lbrace F_{j}\rbrace\subset \mathcal{F}$ and a finite collection $ \lbrace \mathsf{c}_{j}\rbrace\subset\mathsf{C} $. By assumption, each $ \mathsf{c}\in\mathsf{C} $ can be arbitrarily well-approximated in norm by elements in $\mathsf{A}\otimes\mathsf{B} $. Hence, $ \mathcal{A}(\mathcal{X};\mathcal{E};\mathsf{A})\otimes\mathcal{A}(\mathcal{Y};\mathcal{F};\mathsf{B}) $ is norm-dense in $ \mathcal{A}(\mathcal{X}\times\mathcal{Y};\mathcal{G};\mathsf{C}) $. $ \square $

%% file: appendix3ver2.tex
\chapter{Propositions for option II}
\paragraph{Notation} All topological spaces $ \mathcal{X},\mathcal{Y},\ldots $ are compact and Hausdorff. $ \mathcal{C}(\mathcal{X})$, $\mathcal{C}(\mathcal{Y}),\ldots $ are the spaces of real-valued, continuous functions, which are Banach space employing the maximum norm. All partitions of unity will be assumed composed of continuous functions. Hilbert spaces, denoted \textsf{H}, \textsf{J},$ \ldots $  are complete, sesquilinear inner-product spaces, with no restriction as to their dimension or separability. $ \mathcal{K}(\mathsf{H} )$, $\mathcal{K}(\mathsf{J} ),\ldots$ denote the spaces of compact\footnote{An operator is \textit{compact} if the image of a bounded sequence necessarily contains a convergent subsequence.}, self-adjoint operators on the specified Hilbert space; these are Banach spaces using the operator norm. $ \mathcal{C}(\mathcal{X};\mathcal{K}(\mathsf{H}))$, $\mathcal{C}(\mathcal{Y};\mathcal{K}(\mathsf{J}))),\ldots $ are the spaces of compact-operator-valued, continuous functions, which are Banach spaces employing the norm given by first applying the operator norm on the operators pointwise, then the maximum norm over the space. 
\section{Real-valued, continuous functions} 
\paragraph{Comment} The following proposition strengthens the well-known result, which is a special case of a result by Grothendieck~\cite{grothendieck}, that $ \mathcal{C}(\mathcal{X}\times\mathcal{Y} )=\mathcal{C}(\mathcal{X})\check{\otimes}\mathcal{C}(\mathcal{Y} ) $, where $ \check{\otimes} $ indicates completion in the injective norm\footnote{The \textit{injective norm} on $ \mathsf{A}\otimes\mathsf{B} $ is the norm induced by its being a subspace of \text{Bilinear}($ \mathsf{A}^{*},\mathsf{B}^{*} $), $ \Vert\mathsf{c}\Vert_{\vee}=\sup \mathsf{c}(\varphi,\psi)$, where the supremum is taken over $ \varphi $ in the unit ball of $ \mathsf{A}^{*} $ and $ \psi $ in the unit ball of $ \mathsf{B}^{*} $.}.
\paragraph{Proposition C1.1} The finite-nonnegative-tensor-rank\footnote{Using only positive real scalars.} continuous functions are dense in the maximum norm topology for positive, continuous functions on $ \mathcal{X}\times \mathcal{Y} $.
\paragraph*{Proof} Let $ f $ be any positive, continuous function on $ \mathcal{X}\times \mathcal{Y} $ and take any $ \varepsilon>0 $. Since $ f $ is continuous, for each $ (x,y)\in \mathcal{X}$ there are open subsets $ U_{(x,y)}\subset\mathcal{X} $ and $ V_{(x,y)}\subset\mathcal{Y} $ such that $ f(U_{(x,y)}\times V_{(x,y)})\subset \left(f(x,y)-\frac{\varepsilon}{2},f(x,y)+\frac{\varepsilon}{2} \right) $. Since $ \mathcal{X}\times\mathcal{Y} $ is necessarily compact, there is a finite subcover, $ \left\lbrace U_{(x,y)_{j}}\times V_{(x,y)_{j}}\right\rbrace_{j=1}^{n} $. For each $ x\in\mathcal{X} $, define $ U_{x} $ by the intersection over all $ U_{(x,y)_{j}} $ containing $ x $. Similarly, for each $ y\in\mathcal{Y} $, define $ V_{y} $ by the intersection over all $ V_{(x,y)_{j}} $ containing $ y $. Since $ \mathcal{X} $ and $ \mathcal{Y} $ are compact, there are finite subcovers, $ \left\lbrace U_{x_{j}}\right\rbrace_{j=1}^{l} $ and $ \left\lbrace V_{y_{j}}\right\rbrace_{j=1}^{m} $, where each $ U_{x_{i}}\times V_{y_{k}}  $ is a subset of one of the $ U_{(x,y)_{j}}\times V_{(x,y)_{j}} $. Since compact, Hausdorff sets are normal, there are finite partitions of unity, $ \lbrace\phi_{j}\rbrace_{j=1}^{l} $ and $ \lbrace\psi_{j}\rbrace_{j=1}^{m} $, dominated by $ \left\lbrace U_{x_{j}}\right\rbrace_{j=1}^{l} $ and $ \left\lbrace V_{y_{j}}\right\rbrace_{j=1}^{m} $ respectively~\cite{munkrespartition}~\cite{roydenpartition}. Then 
\begin{equation}
\sum_{i=1}^{l}\sum_{k=1}^{m} f(x_{i},y_{l})\phi_{i}\psi_{k}
\end{equation} 
is a finite-nonnegative-tensor-rank continuous function that is everywhere on  $ \mathcal{X}\times \mathcal{Y} $ within $ \varepsilon $ of $ f $. $ \square $
\section{Maps on real-valued, continuous functions} 
\paragraph{Corollary C2.1} Positive maps $ L\in\mathcal{B}(\mathcal{C}(\mathcal{Y}),\mathcal{C}(\mathcal{X}))^{+}$ are completely-positive\footnote{A map \textit{L} is \textit{completely-positive} if $ L\otimes I_{\mathcal{B}(\mathcal{C}(\mathcal{Z}))} $ is positive for every $ \mathcal{Z} $.}.
\paragraph*{Proof} Use \textbf{C1.1} and the positivity of both maps \textit{L} and $ I_{\mathcal{B}(\mathcal{C}(\mathcal{Z}))} $. $ \square $
\paragraph{Corollary C2.2} Maps $ L\in\mathcal{B}(\mathcal{C}(\mathcal{Z}),\mathcal{C}(\mathcal{X}))$ and $ K\in\mathcal{B}(\mathcal{C}(\mathcal{W}),\mathcal{C}(\mathcal{Y}))$ satisfy $ (K\otimes L)^{*}(\mu\times\nu)=K^{*}\mu\times L^{*}\nu $ for any Radon measures $ \mu $ on $ \mathcal{X} $ and $ \nu $ on $ \mathcal{Y} $.
\paragraph*{Proof} By the preceding proposition, it is sufficient to demonstrate that for any finite collections $ \lbrace f_{j}\rbrace\subset\mathcal{C}(\mathcal{Z}) $ and $ \lbrace g_{j}\rbrace\subset\mathcal{C}(\mathcal{W}) $,
\begin{equation}
\sum_{j}\int_{\mathcal{Z}\times \mathcal{W}}f_{j}\otimes g_{j}\,d(K\otimes L)^{*}(\mu\times\nu)=\sum_{j}\int_{\mathcal{Z}\times \mathcal{W}}f_{j}\otimes g_{j}\,d(K^{*}\mu\times L^{*}\nu)
\end{equation}
but this follows from Tonelli's theorem~\cite{roydentonelli}, which is applicable since $ \mu $, $ \nu $ are necessarily finite measures,
\begin{align}
&\sum_{j}\int_{\mathcal{Z}\times \mathcal{W}}f_{j}\otimes g_{j}\,d(K\otimes L)^{*}(\mu\times\nu)=\sum_{j}\int_{\mathcal{X}\times \mathcal{Y}}Kf_{j}\otimes Lg_{j}\,d(\mu\times\nu)=\sum_{j}\int_{\mathcal{X}}Kf_{j}\,d\mu\int_{\mathcal{Y}} Lg_{j}\,d\nu\\&=\sum_{j}\int_{\mathcal{Z}}f_{j}\,dK^{*}\mu\int_{\mathcal{W}} g_{j}\,dL^{*}\nu=\sum_{j}\int_{\mathcal{Z}\times \mathcal{W}}f_{j}\otimes g_{j}\,d(K^{*}\mu\times L^{*}\nu)\hspace{.4 in}\square\nonumber
\end{align}
\paragraph{Comment} For any positive map $  L\in\mathcal{B}(\mathcal{C}(\mathcal{Y}),\mathcal{C}(\mathcal{X}))^{+}$, clearly $ \Vert L\Vert_{\text{op}}= \Vert L(1_{\mathcal{Y}})\Vert_{\max}$. Hence, the cone of positive maps is clearly normal.
\paragraph{Proposition C2.3} For any $ L\in\mathcal{B}(\mathcal{C}(\mathcal{Y}),\mathcal{C}(\mathcal{X}))$,
 $ \Vert L\Vert_{\text{op}}=\sup_{x\in\mathcal{X}}\Vert  L^{*}(\delta_{x})\Vert $.
\paragraph*{Proof} By definition of the operator norm,
\begin{align}
 \Vert L\Vert_{\text{op}}&=\sup_{f\in\mathcal{C}(\mathcal{Y}),\Vert f\Vert_{\max}\leq 1}\Vert  Lf\Vert_{\max}=\sup\left\lbrace\vert  (Lf)(x)\vert\left\vert\begin{array}{l}f\in\mathcal{C}(\mathcal{Y})\\\Vert f\Vert_{\max}\leq 1\\x\in\mathcal{X}\end{array}\right. \right\rbrace\\&=\sup\left\lbrace\left\vert  \int_{\mathcal{X}}Lf \,d\delta_{x}\right\vert\left\vert\begin{array}{l}f\in\mathcal{C}(\mathcal{Y})\\\Vert f\Vert_{\max}\leq 1\\x\in\mathcal{X}\end{array}\right. \right\rbrace=\sup\left\lbrace\left\vert  \int_{\mathcal{X}}f \,dL^{*}(\delta_{x})\right\vert\left\vert\begin{array}{l}f\in\mathcal{C}(\mathcal{Y})\\\Vert f\Vert_{\max}\leq 1\\x\in\mathcal{X}\end{array}\right. \right\rbrace\nonumber\\&=\sup_{x\in\mathcal{X}}\Vert  L^{*}(\delta_{x})\Vert_{\text{total variation}}\hspace{.4 in}\square\nonumber
\end{align} 
\paragraph{Corollary C2.4} Maps $ L\in\mathcal{B}(\mathcal{C}(\mathcal{Z}),\mathcal{C}(\mathcal{X}))$ and $ K\in\mathcal{B}(\mathcal{C}(\mathcal{W}),\mathcal{C}(\mathcal{Y}))$ satisfy $ \Vert K\otimes L\Vert_{\text{op}}=\Vert K\Vert_{\text{op}}\Vert L\Vert_{\text{op}} $.
\paragraph*{Proof} Using the preceding proposition,
\begin{equation}
\Vert K\otimes L\Vert_{\text{op}}\geq  \sup_{(x,y)\in\mathcal{X}\times\mathcal{Y}}\Vert(K\otimes L)^{*}(\delta_{(x,y)})\Vert_{\text{total variation}}
\end{equation}
Since $ \delta_{(x,y)}=\delta_{x}\times\delta_{y} $, from \textbf{C2.3}, this is equal to
\begin{equation}
\sup_{(x,y)\in\mathcal{X}\times\mathcal{Y}}\Vert K^{*}(\delta_{x})\times L^{*}(\delta_{y}) \Vert_{\text{total variation}}
\end{equation}
By \textbf{B1.5}, this is equal to
\begin{equation}
\sup_{x\in\mathcal{X}}\Vert K^{*}(\delta_{x})\Vert_{\!\!\!\begin{scriptsize}\begin{array}{l}\text{total}\\\text{variation}\end{array}\end{scriptsize}} \sup_{y\in\mathcal{Y}}\Vert L^{*}(\delta_{y}) \Vert_{\!\!\!\begin{scriptsize}\begin{array}{l}\text{total}\\\text{variation}\end{array}\end{scriptsize}}
\end{equation}
which, using the preceding proposition again, is equal to $ \Vert K\Vert_{\text{op}}\Vert L\Vert_{\text{op}} $. $ \square $
\paragraph{Proposition C2.5} $ \mathcal{B}(\mathcal{C}(\mathcal{Y}),\mathcal{C}(\mathcal{X})) $ is \underline{not} in general a vector lattice.
\paragraph*{Counterexample} Take $ \mathcal{X}=\mathcal{Y}=[-1,1] $ with the usual topology. Take $ L\in \mathcal{B}(\mathcal{C}([-1,1])) $ to be
\begin{equation}
Lf(x)=\begin{cases}0&\text{if }x\leq 0\\f(x)-f(-x)&\text{if }x> 0\end{cases}
\end{equation}
Then $ L\vee 0 $ should be
\begin{equation}
(L\vee 0)f(x)=\begin{cases}0&\text{if }x\leq 0\\f(x)&\text{if }x> 0\end{cases}
\end{equation}
but this sends some continuous functions to discontinuous ones. $ \square $
\paragraph{Proposition C2.6} $ \mathcal{B}(\mathcal{C}(\mathcal{Y}),\mathcal{C}(\mathcal{X})) $ is \underline{not} in general directed-complete.
\paragraph*{Counterexample} Take $ \mathcal{X}=\mathcal{Y}=[-1,1] $ with the usual topology. Take $ L_{j}\in \mathcal{B}(\mathcal{C}([0,1])) $ to be
\[L_{j}f(x)=\begin{cases}0&\text{if }x\leq 0\\\sqrt[j]{x}\, f(x)&\text{if }x> 0\end{cases}\]
Then $ \sup_{j}L_{j} $ should be
\[(L\vee 0)f(x)=\begin{cases}0&\text{if }x\leq 0\\f(x)&\text{if }x> 0\end{cases}\]
but this  sends some continuous functions to discontinuous ones. $ \square $
\paragraph{Proposition C2.7} For any positive $ L\in\mathcal{B}(\mathcal{C}(\mathcal{Y}),\mathcal{C}(\mathcal{X}))^{+}$, there is a transition function $ \tau(\cdot\vert\cdot):\text{Open}_{\mathcal{Y}}\times\mathcal{X} $, which is: \textit{(i)} an additive set function on the open subsets of $ \mathcal{Y} $; \textit{(ii)} pointwise countably additive on the open subsets of $ \mathcal{Y} $; and \textit{(iii)} lower-semi-continuous for any fixed open subset $ O\subset\mathcal{Y} $--such that
\[(L^{*}\mu)(O)=\int_{x\in\mathcal{X}}\tau( O\vert x)\,d\mu(x)\]
for any open $ O\subset\mathcal{Y}$ and Radon measure $ \mu $ on $ \mathcal{X} $.
\paragraph*{Proof} Let \textit{O} be any open subset of $ \mathcal{Y} $. Then, by the definition of Radon measures for open subsets in the proof of the Riesz-Markov theorem~\cite{roydenriesz},
\begin{align}
(L^{*}\mu)(O)&=\sup_{\begin{scriptsize}\begin{array}{c}f\in\mathcal{C}(\mathcal{Y})\\0\leq f\leq 1_{O}\end{array}\end{scriptsize}}\int_{\mathcal{Y}} f\;dL^{*}\mu=\sup_{\begin{scriptsize}\begin{array}{c}f\in\mathcal{C}(\mathcal{Y})\\0\leq f\leq 1_{O}\end{array}\end{scriptsize}}\int_{\mathcal{X}} Lf\;d\mu\nonumber\\
&=\sup_{\begin{scriptsize}\begin{array}{c}f\in\mathcal{C}(\mathcal{Y})\\0\leq f\leq 1_{O}\end{array}\end{scriptsize}}\int_{x\in\mathcal{X}}\left(\int_{x'\in\mathcal{X}} Lf\;d\delta_{x}(x')\right)\;d\mu(x)=\sup_{\begin{scriptsize}\begin{array}{c}f\in\mathcal{C}(\mathcal{Y})\\0\leq f\leq 1_{O}\end{array}\end{scriptsize}}\int_{x\in\mathcal{X}}\left(\int_{\mathcal{Y}} f\;d(L^{*}\delta_{x})\right)\;d\mu(x)
\end{align} 
Then for any $ \varepsilon>0 $, there is some continuous \textit{f} between zero and $ 1_{O} $ such that 
\begin{equation}
(L^{*}\mu)(O)>\int_{x\in\mathcal{X}}\left(\int_{\mathcal{Y}} f\;d(L^{*}\delta_{x})\right)\;d\mu(x)-\frac{\varepsilon}{2}
\end{equation}

Now consider the sequence of continuous functions $ \langle g_{j}\rangle $, each between zero and $ 1_{O} $, given by
\begin{equation}
g_{j}(y)=\begin{cases}1&\text{if }\text{dist}(y,\sim O)>2^{-j}\\2^{j}\cdot\text{dist}(y,\sim O)&\text{if }\text{dist}(y,\sim O)\leq 2^{-j}\end{cases}
\end{equation}
for $ j\in\mathbb{Z}^{+} $ and $ y\in\mathcal{Y} $. Since $ \mathcal{Y} $ is compact and metric, \textit{f} is uniformly continuous~\cite{roydenuniformcont}~\cite{munkresuniformcont}, so there is some $ \delta $ such that
\begin{equation}
f\left(\text{Ball}(y,\delta)\right)\subset\left(f(y)-\frac{\varepsilon}{2\Vert L\Vert_{\text{op}}\mu(\mathcal{X})},f(y)+\frac{\varepsilon}{2\Vert L\Vert_{\text{op}}\mu(\mathcal{X})}\right) 
\end{equation}
for any $ y\in\mathcal{Y}$.For \textit{j} sufficiently large that $ 2^{-j}<\delta $, then $ g_{j} $ is greater than or equal to \textit{f} everywhere except for the set $ \left\lbrace y\in\mathcal{Y}\vert \text{dist}(y,\sim O)\leq 2^{-j} \right\rbrace $, on which it cannot be smaller by more than $ \frac{\varepsilon}{22\Vert L\Vert_{\text{op}}\mu(\mathcal{X}} $; hence,
\begin{equation}
\int_{x\in\mathcal{X}}\left(\int_{\mathcal{Y}} f\;d(L^{*}\delta_{x})\right)\;d\mu(x)-\int_{x\in\mathcal{X}}\left(\int_{\mathcal{Y}} f\;d(L^{*}\delta_{x})\right)\;d\mu(x)<\int_{x\in\mathcal{X}}\frac{\varepsilon}{2\Vert L\Vert_{\text{op}}\mu(\mathcal{X}}(L^{*}\delta_{x})(\mathcal{Y})\;d\mu(x)\leq \frac{\varepsilon}{2}
\end{equation}
Therefore,
\begin{equation}
(L^{*}\mu)(O)>\int_{x\in\mathcal{X}}\left(\int_{\mathcal{Y}} g_{j}\;d(L^{*}\delta_{x})\right)\;d\mu(x)-\varepsilon
\end{equation}
Since $ \varepsilon $ was arbitrary, it must be that
\begin{equation}
(L^{*}\mu)(O)=\lim_{j\to \infty}\int_{x\in\mathcal{X}}\left(\lim_{j\to \infty}\int_{\mathcal{Y}} g_{j}\;d(L^{*}\delta_{x})\right)\;d\mu(x)
\end{equation}
which is equal to
\begin{equation}
\int_{x\in\mathcal{X}}\left(\lim_{j\to \infty}\int_{\mathcal{Y}} g_{j}\;d(L^{*}\delta_{x})\right)\;d\mu(x)
\end{equation}
by the Dominated Convergence theorem~\cite{roydendominated}. Using the Dominated Convergence theorem again, we have
\begin{equation}
(L^{*}\mu)(O)=\int_{x\in\mathcal{X}}\left(\int_{\mathcal{Y}} 1_{O}\;d(L^{*}\delta_{x})\right)\;d\mu(x)=\int_{x\in\mathcal{X}}(L^{*}\delta_{x})(O)\;d\mu(x)
\end{equation} 
Also,
\begin{equation}
(L^{*}\delta_{x})(O) =\lim_{j\to \infty}\int_{\mathcal{Y}} g_{j}\;d(L^{*}\delta_{x})=\lim_{j\to \infty} (Lg_{j})(x)
\end{equation}
is measurable by~\cite{roydenmeasurable} (as the supremum over a sequence of increasing, continuous functions, it is actually clearly lower-semi-continuous). Hence, $ \tau(O\vert x)=(L^{*}\delta_{x})(O) $ has the desired properties. $ \square $
\paragraph*{Alternate proof} Let $ O\subset\mathcal{Y} $ be open. Define $ \tau(O\vert\cdot)$ by
\begin{equation}
\tau( O\vert x)=\left(\bigvee_{\begin{scriptsize}\begin{array}{c}f\in\mathcal{C}(\mathcal{Y})\\0\leq f\leq 1_{O}\end{array}\end{scriptsize}}Lf\right)(x)
\end{equation}
Then $ \tau(O\vert\cdot) $ is lower-semi-continuous, since, for any $ a\in\mathbb{R} $,
\begin{equation}
\tau(O\vert\cdot)^{-1}((a,\infty))=\bigcup_{\begin{scriptsize}\begin{array}{c}f\in\mathcal{C}(\mathcal{Y})\\0\leq f\leq 1_{O}\end{array}\end{scriptsize}}(Lf)^{-1}((a,\infty))
\end{equation}
which is the union of open sets; hence, it is open. Also, for any Radon measure $ \mu $, by inner regularity,  
\begin{equation}
(L^{*}\mu)(O)=\sup_{\text{compact }K\subset O}(L^{*}\mu)(K)
\end{equation}
By Urysohn's lemma~\cite{roydenurysohn}, this is equal to 
\begin{equation}
\sup_{\begin{scriptsize}\begin{array}{c}f\in\mathcal{C}(\mathcal{Y})\\0\leq f\leq 1_{O}\end{array}\end{scriptsize}}\int_{y\in\mathcal{Y}}f(y)\,d(L^{*}\mu)(y)=\sup_{\begin{scriptsize}\begin{array}{c}f\in\mathcal{C}(\mathcal{Y})\\0\leq f\leq 1_{O}\end{array}\end{scriptsize}}\int_{x\in\mathcal{X}}(Lf)(x)\,d\mu(x)
\end{equation}
which is less than or equal to $ \int_{x\in\mathcal{X}}\tau( O\vert x)\,d\mu(x) $, which is less than or equal to
\begin{equation}
\left(\bigvee_{\begin{scriptsize}\begin{array}{c}f\in\mathcal{C}(\mathcal{Y})\\0\leq f\leq 1_{O}\end{array}\end{scriptsize}}L^{**}f\right)(\mu)
\end{equation}
which exists since dual spaces are necessarily directed-complete. This is in turn less than or equal to $ (L^{*}\mu)(O) $; hence, $ (L^{*}\mu)(O)=\int_{x\in\mathcal{X}}\tau( O\vert x)\,d\mu(x) $. $ \square $
\medskip\\
The bounded, Borel functions on $ \mathcal{Y} $, $ \mathcal{B}(\mathcal{Y}) $, form a Banach space using the supremum norm. The space $ \mathcal{B}(\mathcal{Y}) $, acting on Radon measures via integration, can be considered a subspace of the linear functionals $ \mathcal{C}(\mathcal{Y})^{**} $. Note the preceding shows $ \tau(O\vert x)=(L^{*}\delta_{x})(O) $ and that $ L^{**}1_{O}=\tau(O\vert \cdot) $ is a function in $ \mathcal{B}(\mathcal{X}) $ rather than just a general pseudo-function in $ \mathcal{C}(\mathcal{X})^{**} $~\cite{semadenipseudofunction}.
\paragraph{Proposition C2.8} For any positive $ L\in\mathcal{B}(\mathcal{C}(\mathcal{Y}),\mathcal{C}(\mathcal{X}))^{+}$, if $ \mathcal{Y} $ is metric there is a transition function (see \textbf{A3.1}) $ \tau(\cdot\vert\cdot):\text{Borel}_{\mathcal{Y}}\times\mathcal{X} $ such that
\[(L^{*}\mu)(B)=\int_{x\in\mathcal{X}}\tau( B\vert x)\,d\mu(x)\]
for any Borel subset $ B\subset\mathcal{Y} $ and finite Radon measure $ \mu $ on $ \mathcal{X} $.
\paragraph*{Proof} Note we cannot just immediately begin with $ \tau(B\vert x)=L\delta_{x}(B) $ for a general Borel subset $ B\subset\mathcal{Y} $ because it is not evident that such a construction is measurable in \textit{x}. Instead, we must proceed step-by-step in an inductive argument. Following Hausdorff~\cite{hausdorffborel}, let $ G_{0} $ be the collection of the open subsets of $ \mathcal{Y} $. For each ordinal $ \alpha $, let $ G_{\alpha} $ be the collection of subsets of $ \mathcal{Y} $ that are the countable intersection of subsets from the various collections $ G_{\beta} $ for ordinals $ \beta<\alpha $ if $ \alpha $ is odd and that are the countable union of subsets from the various collections $ G_{\beta} $ for ordinals $ \beta<\alpha $ if $ \alpha $ is even (where all limit ordinals--those without a predecessor--are taken even). For clarification, using the standard notation~\cite{roydengdelta}, $ G_{1}=G_{\delta} $, $ G_{2}=G_{\delta\sigma} $, and so on for the finite ordinals.

Fix any ordinal $ \alpha $ and suppose that for all subsets $ B $ in all collections $ G_{\beta} $ for $ \beta<\alpha $ we have the following property: there is a Borel measurable $ \tau(B\vert\cdot) $ such that $ (L^{*}\mu)(B)=\int_{x\in\mathcal{X}}\tau( B\vert x)\,d\mu(x) $ for any Radon measure $ \mu $ on $ \mathcal{X} $. Take any $ C\in G_{\alpha} $. If $ \alpha $ is even, we have a sequence $ \langle B_{j}\rangle $ of subsets from the various $ G_{\beta} $ with $ \beta<\alpha $ such that $ C=\bigcup_{j}B_{j} $. Let $ \tau(C\vert\cdot)=\bigvee_{j}\tau(B_{j}\vert\cdot) $. Then $ \tau(C\vert\cdot) $ is measurable (see~\cite{roydenmeasurable}) and, by the Dominated Convergence theorem~\cite{roydendominated}, $ (L^{*}\mu)(C)=\int_{x\in\mathcal{X}}\tau( C\vert x)\,d\mu(x) $ for any Radon measure $ \mu $ on $ \mathcal{X} $; hence, since \textit{C} was arbitrary, $ G_{\alpha} $ has the property. Similarly, if $ \alpha $ is odd, we have a sequence $ \langle B_{j}\rangle $ of subsets from the various $ G_{\beta} $ with $ \beta<\alpha $ such that $ C=\bigcap_{j}B_{j} $. Let $ \tau(C\vert\cdot)=\bigwedge_{j}\tau(B_{j}\vert\cdot) $. Then $ \tau(C\vert\cdot) $ is measurable and, by the dominated convergence theorem, $ (L^{*}\mu)(C)=\int_{x\in\mathcal{X}}\tau( C\vert x)\,d\mu(x) $ for any Radon measure $ \mu $ on $ \mathcal{X} $; hence, $ G_{\alpha} $ also has the property. However, by the above proposition \textbf{C2.7}, $ G_{0} $ has the property, so by transfinite induction~\cite{hausdorfftransfinite}~\cite{sierpinskitransfinite} all the $ G_{\alpha} $ have the property.

Since $ \mathcal{Y} $ is metric, its closed subsets are also in $ G_{1}=G_{\delta} $~\cite{kuratowskiclosed}; hence, following Kuratowski~\cite{kuratowskiborel}, the Borel subsets of $ \mathcal{Y} $ are in the union $ \bigcup_{\alpha}G_{\alpha} $. Therefore, the Borel subsets also have the property. $ \square $ 
\paragraph*{Comment} The preceding proof does not depend on the axiom of choice since the union only needs to be taken up to the ordinal number for the minimal uncountable well-ordered set~\cite{hausdorffborel}~\cite{kuratowskiborel}, whose existence does not depend on the axiom of choice~\cite{munkrestransfinite}.
\paragraph{Corollary C2.9} For any positive $ L\in\mathcal{B}(\mathcal{C}(\mathcal{Y}),\mathcal{C}(\mathcal{X}))^{+}$, if $ \mathcal{Y} $ is metric then $ L^{**} $ acting on $ \mathcal{B}(\mathcal{Y})\subset\mathcal{C}(\mathcal{Y})^{**} $ has image in $ \mathcal{B}(\mathcal{X})\subset\mathcal{C}(\mathcal{X})^{**} $.
\paragraph*{Proof} Use the density of simple functions in $ \mathcal{B}(\mathcal{Y}) $ together with the above propositions \textbf{C2.8} and \textbf{A1.3}.
\section{Compact-operator-valued, continuous functions}
\paragraph{Proposition C3.1} $ \mathcal{C}(\mathcal{X};\mathcal{K}(\mathsf{H}))\otimes\mathcal{C}(\mathcal{Y};\mathcal{K}(\mathsf{J}))$ is norm dense within $ \mathcal{C}(\mathcal{X}\times \mathcal{Y};\mathcal{K}(\mathsf{H}\otimes \mathsf{J})) $.
\paragraph*{Proof} Finite-rank operators are norm-dense among compact operators, so proceed as in the first part of the proof for \textbf{B4.1}. Then following the argument in the proof of \textbf{C1.1} involving partitions-of-unity gives the desired result. $ \square $
\section{Operator inequalities}
\paragraph{Proposition C4.1} For any operators $ \varphi,\xi\in\mathcal{K}(\mathsf{H}) $,
\[\Vert \vert \varphi\vert-\vert \xi\vert\Vert_{\text{op}}\leq \sqrt{\Vert  \varphi- \xi\Vert_{\text{op}}\Vert \varphi+ \xi\Vert_{\text{op}}} \]
where $ \vert \chi\vert=\sqrt{\chi^{2}} $.
\paragraph*{Proof} By the properties of compact operators, $ \vert \varphi\vert-\vert \xi\vert $ is compact. By the spectral theorem for compact operators, there is a unit-length $ \psi\in\mathsf{H} $ which is an eigenvector $ \psi $ of $ \vert \varphi\vert-\vert \xi\vert $ and with corresponding eigenvalue $ \lambda $ equal to $ \Vert \vert \varphi\vert-\vert \xi\vert\Vert_{\text{op}} $ in magnitude. Then, using the triangle inequality,
\begin{align}
\Vert \vert \varphi\vert-\vert \xi\vert\Vert_{\text{op}}^{2}&=\left\vert\lambda\left\langle\left(\vert \varphi\vert-\vert \xi\vert\right) \psi,\psi\right\rangle\right\vert \leq \left\vert\lambda\left\langle\left(\vert \varphi\vert+\vert \xi\vert\right) \psi,\psi\right\rangle\right\vert\\&=\left\vert\frac{1}{2}\left\langle\left(\vert \varphi\vert+\vert \xi\vert\right)\left(\vert \varphi\vert-\vert \xi\vert\right) \psi,\psi\right\rangle+\frac{1}{2}\left\langle\left(\vert \varphi\vert-\vert \xi\vert\right)\left(\vert \varphi\vert+\vert \xi\vert\right) \psi,\psi\right\rangle\right\vert\nonumber\\&=\left\vert\left\langle\left( \varphi^{2}-\xi^{2}\right)\psi,\psi\right\rangle\right\vert\leq \Vert\varphi^{2}-\xi^{2}\Vert_{\text{op}}=\left\Vert\frac{1}{2}( \varphi+ \xi)( \varphi- \xi)+\frac{1}{2}( \varphi- \xi)( \varphi+ \xi)\right\Vert_{\text{op}}\nonumber\\&\leq\frac{1}{2}\left\Vert( \varphi+ \xi)( \varphi- \xi)\right\Vert_{\text{op}}+\frac{1}{2}\left\Vert( \varphi- \xi)( \varphi+ \xi)\right\Vert_{\text{op}}\leq \Vert  \varphi- \xi\Vert_{\text{op}}\Vert \varphi+ \xi\Vert_{\text{op}}\hspace{.4 in}\square\nonumber
\end{align}
\medskip\\

The second needed inequality, which involves subsets of direct products of spaces of operators, is lengthy to state, although it has a very short proof. For real numbers $ a=(a_{1},\ldots, a_{n}) $ in the simplex $ \lbrace x\in [0,1]^{\times n}\vert x_{1}+\cdots+x_{n}=1\rbrace $, let $ A_{n}(a,\mathsf{H})\subset\mathcal{K}(\mathsf{H})^{\times n}$ be the set
\begin{equation}
\left\lbrace(\phi_{1},\ldots, \phi_{n})\in \mathcal{K}(\mathsf{H})^{\times n}\left\vert \left\Vert\sum_{j=1}^{n}a_{j}\phi_{j}\right\Vert_{\text{op}}\leq 1\right.\right\rbrace
\end{equation}
Let $ \mathbb{B}(\mathcal{B}(\mathsf{H}))$ denote the closed, unit ball (using the operator norm) in $ \mathcal{B}(\mathsf{H}) $.
By the triangle inequality, $ \mathbb{B}(\mathcal{B}(\mathsf{H}))^{\times n}\subset A_{n}(a,\mathsf{H})$ for any allowed choice of $ a $. Now consider the set $ C_{n}(\varepsilon,\mathsf{H}) \subset\mathcal{K}(\mathsf{H})^{\times n} $ given by
\begin{equation}
\left\lbrace(\phi_{1},\ldots, \phi_{n})\in\lbrace B\in \mathcal{K}(\mathsf{H})^{\times n}\left\vert \max_{j<k}\left\Vert \phi_{j}-\phi_{k}\right\Vert_{\text{op}}\leq \varepsilon\right.\right\rbrace
\end{equation}
for $ \varepsilon>0 $.

Define a distance from $ A\in \mathcal{K}(\mathsf{H})^{\times n}  $ to a subset $ F\subset\mathcal{K}(\mathsf{H})^{\times n}$ by
\begin{equation}
\text{dist }(A,F)=\inf_{E\in F}\max_{j}\Vert A_{j}-E_{j}\Vert_{\text{op}}
\end{equation}
Using this to define a Hausdorff distance between subsets then gives:
\paragraph{Proposition C4.2} The Hausdorff distance between $ A_{n}(a,\mathsf{H})\cap C_{n}(\varepsilon,\mathsf{H}) $ and $ \mathbb{B}(\mathcal{B}(\mathsf{H}))^{\times n}\cap C_{n}(\varepsilon,\mathsf{H}) $ is bounded by $ \varepsilon $.
\paragraph*{Proof} Take any  allowed $ a $ and any $ \phi\in A_{n}(a,\mathsf{H})\cap C_{n}(\varepsilon,\mathsf{H}) $. Then, by the triangle inequality, for any $ j\in\lbrace 1,2,\ldots, n\rbrace $,
\begin{equation}
\left\Vert \phi_{j}-\sum_{k=1}^{n}a_{k}\phi_{k}\right\Vert_{\text{op}}\leq \sum_{k=1}^{n}a_{k}\left\Vert \phi_{j}-\phi_{k}\right\Vert_{\text{op}}\leq \varepsilon
\end{equation} 
However, by the definition of $ A_{n}(a,\mathsf{H}) $, $ \sum_{k=1}^{n}a_{k}\phi_{k} $ is in $ \mathbb{B}(\mathcal{B}(\mathsf{H})) $. $ \square $
\medskip\\
Note the compactness of the operators was not used in the preceding proof so the proposition holds for the corresponding subsets of self-adjoint operators in $ \mathcal{B}(\mathsf{H}) $, although we will not require that generalization.

For the following proposition, let $ \lbrace A_{j}\rbrace $ be any finite collection of positive, Hermitian $ n\times n $-matrices and let \textit{P} be any orthogonal projector. 
\paragraph{Proposition C4.3} There is a constant $ c_{0} $, independent of \textit{P}, \textit{n}, and \textit{r}, such that
\[\sum_{j=1}^{r}\Vert A_{j}- PA_{j}P\Vert_{\text{trace}}\leq c_{0}\sqrt{\left(\sum_{j=1}^{r} \text{tr }(A_{j}- PA_{j}P)\right)\left(\sum_{j=1}^{r} \text{tr } A_{j}\right)} \]
\paragraph*{Proof} Observe that for any $ a_{1},\ldots,a_{m}\in\mathbb{R}\setminus\lbrace 0\rbrace $, the quantity $ \sum_{j=1}^{m}\frac{a_{j}^{2}}{x_{j}} $ is minimized over $ x_{1},\ldots,x_{m}\in\mathbb{R}^{+} $ for fixed $ \sum_{j=1}^{m}x_{j} $ when for each \textit{j}, $ x_{j}=\vert a_{j}\vert $. Now consider the problem of minimizing the product $ \text{tr }D \text{ tr }C D^{-1}C^{*}$ over strictly-positive, Hermitian, $ m\times m $-matrices \textit{D}, given any $ n\times m $-matrix \textit{C}. First take the case of $ n\geq m $ and $ C^{*}C $ strictly positive. The minimization over \textit{D} is the same as minimizing over its eigenvalues, $ x_{1},\ldots,x_{m}\in\mathbb{R}^{+} $, and eigenvectors, \textit{W}, with \textit{W} unitary. For fixed value of \textit{W} and the trace of \textit{D}, $ x_{1}+\cdots+x_{m} $, using the preceding observation, this will occur for
\begin{equation}
x_{j}=\dfrac{(x_{1}+\cdots+x_{m})e_{j}}{e_{1}+\cdots+e_{m}}
\end{equation}
where $ e_{j} $ is the square root of the \textit{j}th diagonal entry of $ W^{*}C^{*}CW $. Then, $ \text{tr }D \text{ tr }C D^{-1}C^{*}=(e_{1}+\cdots+e_{m})^{2} $. Since squaring is monotonic for positive reals and the square root function is concave, this is minimized over all \textit{W} when $ W^{*}C^{*}CW $ is diagonal, so $ e_{1}+\cdots+e_{m}=\text{tr }\sqrt{C^{*}C} $. Therefore, $ \text{tr }D \text{ tr }C D^{-1}C^{*}\geq \left(\text{tr }\sqrt{C^{*}C}\right)^{2} $. By the ability to embed matrices into larger matrices and the density of invertible matrices, this inequality actually holds for all $ n\times m $-matrices \textit{C}.

By the triangle inequality,
\begin{equation}\label{eq:triangleinequality3}
\left\Vert\left[\begin{array}{cc}0&C\\C^{*}&D\end{array}\right]\right\Vert_{\text{trace}}^{2}\leq \left(\left\Vert\left[\begin{array}{cc}0&C\\C^{*}&0\end{array}\right]\right\Vert_{\text{trace}}+\left\Vert D\right\Vert_{\text{trace}}\right)^{2}=\left(2\text{tr }\sqrt{C^{*}C}+\text{tr }D\right)^{2}
\end{equation}
By the arithmetic-geometric mean inequality, this is less than or equal to
\begin{equation}
8\left(\text{tr }\sqrt{C^{*}C}\right)^{2}+2\left(\text{tr }D\right)^{2}\leq 8\left(\left(\text{tr }\sqrt{C^{*}C}\right)^{2}+\left(\text{tr }D\right)^{2}\right)
\end{equation}
which, by the above result, is less than or equal to 
\begin{equation}\label{eq:boundinequality3}
8\left(\left(\text{tr }C D^{-1}C^{*}+\text{tr }D\right)\text{tr }D\right)
\end{equation}

For any Hermitian matrix of the form $ \left[\begin{array}{cc}B&C\\C^{*}&D\end{array}\right] $ to be positive, it is necessary (and sufficient) that $ B-C D^{-1}C^{*} $ is positive; otherwise, if there were some vector \textbf{v} such that $ \langle( B-C D^{-1}C^{*})\mathbf{v} , \mathbf{v}\rangle<0 $, then the vector $ \mathbf{w}=\left[\begin{array}{c}\mathbf{v}\\-D^{-1}C^{*}\mathbf{v}\end{array}\right] $ would give
\begin{align}
\left\langle \left[\begin{array}{cc}B&C\\C^{*}&D\end{array}\right]\mathbf{w},\mathbf{w}\right\rangle&=\langle( B-C D^{-1}C^{*})\mathbf{v} , \mathbf{v}\rangle+\left\langle D\left[\begin{array}{cc}D^{-1}C^{*} &I_{m}\end{array}\right]\mathbf{w},\left[\begin{array}{cc}D^{-1}C^{*} &I_{m}\end{array}\right]\mathbf{w}\right\rangle\\&=\langle( B-C D^{-1}C^{*})\mathbf{v} , \mathbf{v}\rangle<0\nonumber
\end{align} 
which is a contradiction. Therefore, for any positive, Hermitian matrix of the form $ \left[\begin{array}{cc}B&C\\C^{*}&D\end{array}\right] $ with \textit{D} strictly positive,
\begin{equation}
\left\Vert\left[\begin{array}{cc}0&C\\C^{*}&D\end{array}\right]\right\Vert_{\text{trace}}^{2}\leq 8\left(\left(\text{tr }B+\text{tr }D\right)\text{tr }D\right)
\end{equation}
By the density of strictly-positive matrices among positive ones, this holds even if \textit{D} is not strictly positive. Since any orthogonal projector can be brought into the form where there are ones along the upper-left diagonal and zeros everywhere else, this proves the case of $ r=1 $,
\begin{equation}
\Vert A- PAP\Vert_{\text{trace}}\leq c_{0}\sqrt{\text{tr }(A- PAP) \text{tr } A}
\end{equation}
with $ c_{0}=2\sqrt{2} $.

The case of $ r>1 $ then follows. Take any row vectors $ \textbf{a},\textbf{b},\textbf{c}\subset\mathbb{R}^{+} $ that are independent and whose projections to any of the subspace of dimension greater than two using diagonal projectors are still independent. Then, using Lagrange multipliers, the maximum value of $ \sum_{j=1}^{r}c_{j}x_{j} $ over $ x_{1},\ldots,x_{r}\geq 0 $ for fixed value of the product $ \left(\sum_{j=1}^{r}a_{j}x_{j}\right)\left(\sum_{k=1}^{r}b_{k}x_{k}\right) $ can only occur if all but two of the $ x_{j} $'s are zero. Even then, the maximum value can only be attained for both $ x_{j} $'s  nonzero if it is also attained for one of them zero. Since the given conditions on $ \textbf{a},\textbf{b},\textbf{c} $ describe a dense subset, then it always the case that the maximizing value is attained when all the $ x_{j} $'s are zero except for one. Then, since
\begin{equation}
\sup\dfrac{\sum_{j=1}^{r}\Vert A_{j}- PA_{j}P\Vert_{\text{trace}}}{\sqrt{\left(\sum_{k=1}^{r} \left( \text{tr }(A_{k}- PA_{k}P)\right)\right)\left(\sum_{l=1}^{r} \text{tr } A_{l}\right)}}
\end{equation}
over all \textit{P}, \textit{n}, \textit{r}, and $ \lbrace A_{j}\rbrace $, is the same as 
\begin{equation}
\sup\dfrac{\sum_{j=1}^{r}x_{j}\Vert A_{j}- PA_{j}P\Vert_{\text{trace}}}{\sqrt{\left(\sum_{k=1}^{r} x_{k}\left( \text{tr }(A_{k}- PA_{k}P)\right)\right)\left(\sum_{l=1}^{r} x_{l}\text{tr } A_{l}\right)}}
\end{equation}
over all \textit{P}, \textit{n}, \textit{r}, $ \lbrace x_{j}\rbrace\subset\mathbb{R}^{+}\cup\lbrace 0\rbrace $, and $ \lbrace A_{j}\rbrace $, it is the same as
\begin{equation}
\sup\dfrac{\Vert A- PAP\Vert_{\text{trace}}}{\sqrt{ \text{tr }(A- PAP) \text{tr } A}}
\end{equation} 
over all \textit{P}, \textit{n}, and \textit{A}, which is bounded by $ 2\sqrt{2} $ by the above. $ \square $
\paragraph*{Remark} Apparently $ c_{0}=2 $ is sufficient, since, by numerical calculation, (\ref{eq:boundinequality3}) can be replaced by 
\begin{equation}
\left\Vert\left[\begin{array}{cc}0&C\\C^{*}&D\end{array}\right]\right\Vert_{\text{trace}}^{2}\leq 4\left(\left(\text{tr }C D^{-1}C^{*}+\text{tr }D\right)\text{tr }D\right)
\end{equation}
where the bound is tight since equality is approached as $ \varepsilon\to 0^{+} $ for $ D=\varepsilon\,\sqrt{C^{*}C} $. However, a proof of this tight bound is lacking. 
\section{Maps on compact-operator-valued, continuous functions}
\paragraph{Proposition C5.1} Any positive map $ L\in\mathcal{B}\left(\mathcal{C}(\mathcal{Y};\mathcal{K}(\mathsf{J})),\mathcal{C}(\mathcal{X};\mathcal{K}(\mathsf{H})) \right)^{+} $ satisfies
\begin{equation}
\Vert L\Vert_{\text{op}}=\sup_{\begin{scriptsize}\begin{array}{c}
f\in \mathcal{C}(\mathcal{Y};\mathcal{K}(\mathsf{J}))^{+}\\\Vert f\Vert\leq 1\end{array}\end{scriptsize}}\Vert L f\Vert=\sup_{\text{finite-dimensional }\mathsf{K}\subset\mathsf{J}}\Vert L (I_{\mathsf{K}} 1_{\mathcal{Y}})\Vert
\end{equation}
\paragraph*{Proof} From the proof of \textbf{C3.1}, functions of the form $ \sum_{j}g_{j}\phi_{j} $ for finite collections $ \lbrace \phi_{j}\rbrace\subset\mathcal{K}(\mathsf{J}) $ and partition-of-unity $ \lbrace g_{j}\rbrace $ are norm-dense in $ \mathcal{C}(\mathcal{Y};\mathcal{K}(\mathsf{J})) $. Furthermore, for any $ \varepsilon>0 $, it is possible to enforce the constraint that $ \left\Vert\phi_{j}-\phi_{k}\right\Vert_{\text{op}}<\varepsilon $ if the support of $ g_{j} $ intersects the support of $g_{k} $ while maintaining the density property. 

Then, for any value of $ \varepsilon>0 $, $ \Vert L\Vert_{\text{op}} $ is equal to
\begin{equation}
\sup\left\lbrace \left\Vert\sum_{j=1}^{n}L(\phi_{j} g_{j})\right\Vert\left\vert\begin{array}{l}\text{partition-of-unity } \lbrace g_{1},\ldots,g_{n}\rbrace, \\ \lbrace\phi_{j}\rbrace\subset\mathcal{K}(\mathsf{J}), \left\Vert\phi_{j}-\phi_{k}\right\Vert_{\text{op}}<\varepsilon \\\text{if }\text{support } g_{j}\cap\text{support } g_{k}\neq \varnothing,\\\left\Vert \sum_{j=1}^{n}\phi_{j} g_{j}\right\Vert\leq 1 \end{array}\right.\right\rbrace 
\end{equation}
Now consider $ b(\varepsilon) $ given by
\begin{equation}\label{eq:defbepsilon}
\sup\left\lbrace \left\Vert\sum_{j=1}^{n}L(\phi_{j} g_{j})\right\Vert\left\vert\begin{array}{l}\text{partition-of-unity } \lbrace g_{1},\ldots,g_{n}\rbrace, \\ \lbrace\phi_{j}\rbrace\subset\mathcal{K}(\mathsf{J}), \left\Vert\phi_{j}-\phi_{k}\right\Vert_{\text{op}}<\varepsilon \\\text{if }\text{support }g_{j}\cap\text{support } g_{k}\neq \varnothing,\\\left\Vert\phi_{j}\right\Vert_{\text{op}}\leq 1 \end{array}\right.\right\rbrace 
\end{equation}
Note $ b(\varepsilon) $ is a decreasing function of $ \varepsilon $, so its limit as $ \varepsilon\to 0^{+} $ certainly exists. By \textbf{C4.2},
\begin{equation}
b(\varepsilon)\leq  \Vert L\Vert_{\text{op}} \leq b(\varepsilon)+\varepsilon\Vert L\Vert_{\text{op}}
\end{equation}
so
\begin{equation}
b(\varepsilon)\leq\Vert L\Vert_{\text{op}} \leq \frac{b(\varepsilon)}{1-\varepsilon}
\end{equation}
Hence, $ \Vert L\Vert_{\text{op}} $ is equal to $ \lim_{\varepsilon \to 0^{+}}b(\varepsilon) $.

The additional constraint that all the $ \lbrace \phi_{j}\rbrace $ are positive can only reduce the supremum or leave it unchanged. On the other hand, since \textit{L} is positive  
\begin{equation}
\left\Vert\sum_{j}L(g_{j}\phi_{j})\right\Vert=\sup_{x\in\mathcal{X}}\sup_{\mathbf{v}\in\mathsf{H},\Vert\mathbf{v}\Vert\leq 1}\sum_{j}\left\langle L(g_{j}\phi_{j})(x)\mathbf{v},\mathbf{v}\right\rangle 
\end{equation}
is certainly less than
\begin{equation}
\left\Vert\sum_{j}L(g_{j}\vert\phi_{j}\vert)\right\Vert=\sup_{x\in\mathcal{X}}\sup_{\mathbf{v}\in\mathsf{H},\Vert\mathbf{v}\Vert\leq 1}\sum_{j}\left\langle L(g_{j}\vert\phi_{j}\vert)(x)\mathbf{v},\mathbf{v}\right\rangle 
\end{equation}
In addition, examining the constraints on $ \lbrace\phi_{j}\rbrace $ in (\ref{eq:defbepsilon}), $ \left\Vert\,\vert\phi_{j}\vert\,\right\Vert_{\text{op}}=\left\Vert\phi_{j}\right\Vert_{\text{op}} $ and, by \textbf{C4.1},
\begin{equation}
\left\Vert\vert\phi_{j}\vert-\vert\phi_{k}\vert\right\Vert_{\text{op}}\leq \sqrt{\left\Vert\phi_{j}-\phi_{k}\right\Vert_{\text{op}}\left\Vert\phi_{j}+\phi_{k}\right\Vert_{\text{op}}}\leq \sqrt{\varepsilon}\sqrt{\left\Vert\phi_{j}+\phi_{k}\right\Vert_{\text{op}}}
\end{equation} 
By the triangle inequality and the other constraints, this is less than or equal to $ \sqrt{2\varepsilon} $ (so this new condition gives rise to a subset that necessarily includes the previous one), which can simply be replaced by $ \varepsilon $ since $ \varepsilon\to 0 $. Hence, the additional constraint that all the $ \lbrace \phi_{j}\rbrace $ are positive can only reduce the supremum or leave it unchanged. Therefore, it must leave it unchanged.

However, now starting with $\sup_{\begin{scriptsize}\begin{array}{c}
f\in \mathcal{C}(\mathcal{Y};\mathcal{K}(\mathsf{J}))^{+}\\\Vert f\Vert\leq 1\end{array}\end{scriptsize}}\Vert L f\Vert$ and repeating the process would lead to the same result, so it is only necessary to take the supremum over the positive cone. Finite-rank operators are norm-dense among compact operators, so it is only necessary to take the supremum over them. Any collection of finite-rank operators live collectively on some finite-dimensional subspace $ \mathsf{K}\subset\mathsf{J} $. Since \textit{L} is positive, the supremum for that particular \textsf{K} then occurs for the constant function with value $ I_{\mathsf{K}} $. $ \square $
\paragraph{Proposition C5.2} If $ L\in\mathcal{B}\left(\mathcal{C}(\mathcal{Y};\mathcal{K}(\mathsf{J})),\mathcal{C}(\mathcal{X};\mathcal{K}(\mathsf{H})) \right) $ is completely bounded, then, for any compact space $ \mathcal{Z} $ and any Hilbert space \textsf{K}, $ \Vert L\otimes I\Vert_{\text{op}}\leq \Vert L\Vert_{\text{matrix}} $ with $ I $ the identity map in $ \mathcal{B}\left(\mathcal{C}(\mathcal{Z};\mathcal{K}(\mathsf{K}))\right) $.
\paragraph*{Proof} By the definition of operator norm and the definition of the tensor product of maps, $ \Vert L\otimes I\Vert_{\text{op}} $ is equal to
\begin{equation}
\sup\left\lbrace\left\Vert (L\otimes I)f\right\Vert\left\vert\begin{array}{l}\text{finite-tensor-rank }f\in \mathcal{C}(\mathcal{Y}\times\mathcal{Z};\mathcal{K}(\mathsf{J}\otimes\mathsf{K}) \\\text{with }\Vert f\Vert\leq 1\end{array}\right.\right\rbrace
\end{equation}
By the preceding proof, this is the same as
\begin{equation}
\lim_{\varepsilon\to 0}\sup\left\lbrace\left\Vert \sum_{j=1}^{n}\sum_{k=1}^{p}\sum_{l=1}^{m}L(\phi_{jkl}g_{j})\otimes \tau_{jkl}h_{k}\right\Vert\right.
\end{equation}
\[\left.\left\vert \begin{array}{l}m\in\lbrace 1,2,\ldots\rbrace,\\\text{partition-of-unity }\lbrace g_{1},\ldots,g_{n}\rbrace,\text{partition-of-unity } \lbrace h_{1},\ldots,h_{p}\rbrace, \\\lbrace \phi_{jkl}\rbrace\subset\mathcal{K}(\mathsf{J}), \lbrace \tau_{jkl}\rbrace\subset\mathcal{K}(\mathsf{K}),\\\left\Vert\sum_{l=1}^{m}(\phi_{jkl}\otimes\tau_{jkl}-\phi_{j'k'l}\otimes\tau_{j'k'l})\right\Vert_{\text{op}}<\varepsilon \\\text{if }\text{support } g_{j}\cap\text{support } g_{j'}\neq \varnothing\text{ and }\text{support } h_{k}\cap\text{support } h_{k'}\neq \varnothing\\\left\Vert\sum_{l=1}^{m}\phi_{jkl}\otimes \tau_{jkl} \right\Vert_{\text{op}}\leq 1\end{array}\right.\right\rbrace\]

Writing 
\begin{equation}
\left\Vert \sum_{j=1}^{n}\sum_{k=1}^{p}\sum_{l=1}^{m}L(\phi_{jkl}g_{j})\otimes \tau_{jkl}h_{k}\right\Vert
\end{equation}
out as
\begin{equation}
\sup_{x\in\mathcal{X}}\sup_{z\in\mathcal{Z}}\left\Vert \sum_{j=1}^{n}\sum_{k=1}^{p}\sum_{l=1}^{m}L(\phi_{jkl}g_{j})(x)\otimes \tau_{jkl}h_{k}(z)\right\Vert_{\text{op}}
\end{equation}
then, since $ \mathcal{Z} $ is compact, the maximum value is achieved for a certain $ z_{*} $ (which depends on all the other quantities the supremum is taken over). Incorporating the value of $ h_{k}(z_{*}) $ into $ \tau_{jkl} $, then the constraint:
\begin{equation}
\left\Vert\sum_{l=1}^{m}\phi_{jkl}\otimes \tau_{jkl} \right\Vert_{\text{op}}\leq 1 
\end{equation} 
becomes
\begin{equation}
\left\Vert\sum_{l=1}^{m}\phi_{jkl}\otimes \tau_{jkl} \right\Vert_{\text{op}}\leq h_{k}(z_{*})
\end{equation} 
which implies 
\begin{equation}
 \sum_{k=1}^{p}\left\Vert\sum_{l=1}^{m}\phi_{jkl}\otimes \tau_{jkl} \right\Vert_{\text{op}}\leq 1
\end{equation}
which by the triangle inequality implies
\begin{equation}
\left\Vert\sum_{k=1}^{p}\sum_{l=1}^{m}\phi_{jkl}\otimes \tau_{jkl} \right\Vert_{\text{op}}\leq 1
\end{equation}

Similarly, the constraint: 
\begin{equation}
\left\Vert\sum_{l=1}^{m}(\phi_{jkl}\otimes\tau_{jkl}-\phi_{j'k'l}\otimes\tau_{j'k'l})\right\Vert_{\text{op}}<\varepsilon 
\end{equation}
\[\text{ if support }g_{j}\cap\text{support }g_{j'}\neq \varnothing \text{ and support } h_{k}\cap\text{support }h_{k'}\neq \varnothing \]
becomes
\begin{equation}
\left\Vert\sum_{l=1}^{m}(\phi_{jkl}\otimes\tau_{jkl}-\phi_{j'kl}\otimes\tau_{j'kl})\right\Vert_{\text{op}}<\varepsilon h_{k}(z_{*}) \end{equation}
\[\text{ if support } g_{j}\cap\text{support } g_{j'}\neq \varnothing\]
which implies
\begin{equation}
\sum_{k=1}^{p}\left\Vert\sum_{l=1}^{m}(\phi_{jkl}\otimes\tau_{jkl}-\phi_{j'kl}\otimes\tau_{j'kl})\right\Vert_{\text{op}}<\varepsilon   
\end{equation} 
\[\text{ if support } g_ {j}\cap\text{support } g_{j'}\neq \varnothing\]
which, by the triangle inequality, implies
\begin{equation}
\left\Vert\sum_{k=1}^{p}\sum_{l=1}^{m}(\phi_{jkl}\otimes\tau_{jkl}-\phi_{j'kl}\otimes\tau_{j'kl})\right\Vert_{\text{op}}<\varepsilon  
\end{equation} 
\[\text{ if support } g_{j}\cap\text{support } g_{j'}\neq \varnothing\]

Using the new $ \tau_{jkl}$ together with the new, weaker constraints can only increase the value of the supremum. Furthermore, the sums over \textit{k} and \textit{l} can now be combined, yielding a bound to $ \Vert L\otimes I\Vert_{\text{op}} $ given by
\begin{equation}
\lim_{\varepsilon\to 0}\sup\left\lbrace\left\Vert \sum_{j=1}^{n}\sum_{k=1}^{m}L(\phi_{jk}g_{j})\otimes \tau_{jk}\right\Vert\left\vert \begin{array}{l}m\in\lbrace 1,2,\ldots\rbrace,\\\text{partition-of-unity }\lbrace g_{1},\ldots,g_{n}\rbrace, \\\lbrace \phi_{jk}\rbrace\subset\mathcal{K}(\mathsf{J}), \lbrace \tau_{jk}\rbrace\subset\mathcal{K}(\mathsf{K}),\\\left\Vert\sum_{k=1}^{m}(\phi_{jk}\otimes\tau_{jk}-\phi_{j'k}\otimes\tau_{j'kl})\right\Vert_{\text{op}}<\varepsilon \\\text{if }\text{support } g_{j}\cap\text{support } g_{j'}\neq \varnothing\\\left\Vert\sum_{k=1}^{m}\phi_{jk}\otimes \tau_{jk} \right\Vert_{\text{op}}\leq 1\end{array}\right.\right\rbrace
\end{equation}
Since the finite-rank operators are norm-dense among the compact operators, this bound is unchanged by requiring the $ \tau_{jk} $ to be finite-rank. However, the finite-rank operators collectively live on a finite-dimensional subspace of \textsf{K} that is isomorphic to $ \mathbb{C}^{N} $ for some integer $ N\leq \dim\mathsf{K} $. Incorporating this into the preceding expression for the bound, we have
\begin{equation}
\Vert L\otimes I\Vert_{\text{op}}\leq\sup_{N\leq \dim\mathsf{K}}\Vert L\otimes I_{\mathcal{M}_{N}}\Vert_{\text{op}}\hspace{.4 in}\square
\end{equation} 
\paragraph{Proposition C5.3} If the positive map $L\in\mathcal{B}\left(\mathcal{C}(\mathcal{Y};\mathcal{K}(\mathsf{J})),\mathcal{C}(\mathcal{X};\mathcal{K}(\mathsf{H})) \right)^{+} $, compact space $ \mathcal{Z} $,  and Hilbert space \textsf{K} are such that $ L\otimes I $ is positive, for $ I $ the identity map in $ \mathcal{B}\left(\mathcal{C}(\mathcal{Z};\mathcal{K}(\mathsf{K}))\right) $, then $ \Vert L\otimes I\Vert_{\text{op}}=\Vert L\Vert_{\text{op}} $.
\paragraph*{Proof} If $ L\otimes I $ is positive, then clearly $ L\otimes I_{\mathcal{M}_{n}} $ is positive for every positive integer $ n $ less than or equal to $ \dim\mathsf{K} $. By the preceding proposition, it is therefore only necessary to show that if $ L\otimes I_{\mathcal{M}_{n}} $ is positive, then $ \Vert L\otimes I_{\mathcal{M}_{n}}\Vert_{\text{op}}=\Vert L\Vert_{\text{op}} $. However, if $ L\otimes I_{\mathcal{M}_{n}} $ is positive, then by \textbf{C5.1},
\begin{align}
\Vert L\otimes I_{\mathcal{M}_{n}}\Vert_{\text{op}}&=\sup_{\text{finite-dimensional }\mathsf{L}\subset\mathsf{J}}\Vert (L\otimes I_{\mathcal{M}_{n}}) (I_{\mathsf{L}\otimes\mathbb{C}^{n}} 1_{\mathcal{Y}})\Vert\\&=\sup_{\text{finite-dimensional }\mathsf{L}\subset\mathsf{J}}\;\;\sup_{x\in\mathcal{X}}\Vert L(I_{\mathsf{L}}1_{\mathcal{Y}})(x)\otimes I_{n}\Vert_{\text{op}}\nonumber\\&=\sup_{\text{finite-dimensional }\mathsf{L}\subset\mathsf{J}}\;\;\sup_{x\in\mathcal{X}}\Vert L(I_{\mathsf{L}}1_{\mathcal{Y}})(x)\Vert_{\text{op}}\nonumber
\end{align}
which is $ \Vert L\Vert_{\text{op}} $ by \textbf{C5.1}. $ \square $
\paragraph{Corollary C5.4} The completely-positive maps
\[\mathcal{B}\left(\mathcal{C}(\mathcal{Y};\mathcal{K}(\mathsf{J})),\mathcal{C}(\mathcal{X};\mathcal{K}(\mathsf{H})) \right)^{\text{cp}}\]
are completely bounded.
\paragraph{Corollary C5.5} The cone of completely-positive maps
\[\mathcal{B}\left(\mathcal{C}(\mathcal{Y};\mathcal{K}(\mathsf{J})),\mathcal{C}(\mathcal{X};\mathcal{K}(\mathsf{H})) \right)^{\text{cp}}\]
is normal in either the induced operator norm or the matrix norm.
\paragraph{Proposition C5.6} If $ L\in\mathcal{B}\left(\mathcal{C}(\mathcal{Y};\mathcal{K}(\mathsf{J})),\mathcal{C}(\mathcal{X};\mathcal{K}(\mathsf{H})) \right)^{\text{cp}} $, then, for any compact space $ \mathcal{Z} $ and any Hilbert space \textsf{K}, $  L\otimes I $ is positive, with $ I $ the identity map in $ \mathcal{B}\left(\mathcal{C}(\mathcal{Z};\mathcal{K}(\mathsf{K}))\right) $. 
\paragraph*{Proof} Since \textit{L} is completely positive, by \textbf{C5.4} it is completely bounded. Hence, $ L\otimes I $ exists by \textbf{C5.2}. Furthermore, by \textbf{C3.1} and \textbf{A1.3}, $ L\otimes I $ is unique, so it is meaningful to speak of it being positive.

Now suppose there were some compact space $ \mathcal{Z} $ and some Hilbert space \textsf{K} such that $  L\otimes I $ were not positive. Then there would be some positive $ f\in \mathcal{C}(\mathcal{Y}\otimes\mathcal{Z};\mathcal{K}(\mathsf{J}\otimes\mathsf{K}))^{+} $ such that $ (L\otimes I)f $ is not positive. Since the cone $  \mathcal{C}(\mathcal{X}\otimes\mathcal{Z};\mathcal{K}(\mathsf{H}\otimes\mathsf{K}))^{+} $ is norm-closed and $ L\otimes I $ is continuous, that implies there is a relatively open neighborhood of $ f $ in the cone $  \mathcal{C}(\mathcal{Y}\otimes\mathcal{Z};\mathcal{K}(\mathsf{J}\otimes\mathsf{K}))^{+} $ whose image under $ L\otimes I $ does not intersect $  \mathcal{C}(\mathcal{X}\otimes\mathcal{Z};\mathcal{K}(\mathsf{H}\otimes\mathsf{K}))^{+} $. 

Now approximating $ f $ as in the proof of \textbf{C5.2} as $ \sum_{jkl}\varphi_{jkl}\otimes \tau_{jkl}\;g_{j}\otimes h_{k} $ (but without the need for any $ \varepsilon $-constraints), one finds that for this to occur there must be some $ n\in\lbrace 1,2,\ldots\rbrace $ for which $ L\otimes I_{\mathcal{M}_{n}} $ is not positive; however, that is a contradiction. $ \square $
\paragraph{Proposition C5.7} If either $ \dim\mathsf{H} $ or $ \dim\mathsf{J} $ is finite and a positive map\\$ L\in\mathcal{B}\left(\mathcal{C}(\mathcal{Y};\mathcal{K}(\mathsf{J})),\mathcal{C}(\mathcal{X};\mathcal{K}(\mathsf{H})) \right)^{+}$ is such that $ L\otimes I_{\mathcal{M}_{m}} $ is positive for $ m=\min\lbrace\dim\mathsf{H}, \dim\mathsf{J}\rbrace $, then \textit{L} is completely positive.
\paragraph*{Proof} Clearly, since $ L\otimes I_{\mathcal{M}_{m}} $ is positive, so is $ L\otimes I_{\mathcal{M}_{n}} $ for all $ n<m $. Now take $ n>m $. $ L\otimes I_{\mathcal{M}_{n}} $ will be positive if for every $ f\in  \mathcal{C}(\mathcal{Y};\mathcal{K}(\mathsf{J}\otimes\mathbb{C}^{n}))^{+} $, $ \mathbf{y}\in\mathsf{H}\otimes\mathbb{C}^{n} $, and $ x\in\mathcal{X} $,
\begin{equation}
\langle((L\otimes I_{\mathcal{M}_{n}})f)(x)\mathbf{y},\mathbf{y}\rangle\geq 0
\end{equation}
It is enough to show this for \textit{f} in a dense subset, so \textit{f} can be restricted to the form $ \sum_{j}\varphi_{j}g_{j} $ for $ \lbrace g_{j}\rbrace $ a partition-of-unity and $ \lbrace \varphi_{j}\rbrace $ a collection of compact operators. By the spectral theorem for compact operators, it is enough to show this for the $ \varphi_{j} $'s all rank one. Following the argument of \textbf{B5.9} then gives the desired result. $ \square $
\paragraph{Proposition C5.8} The space of completely bounded maps,
\[\mathcal{CB}\left(\mathcal{C}(\mathcal{Y};\mathcal{K}(\mathsf{J})),\mathcal{C}(\mathcal{X};\mathcal{K}(\mathsf{H})) \right)\]
is a Banach space with respect to the matrix norm.
\paragraph*{Proof} Same argument as for \textbf{B5.10} $ \square $   
\paragraph{Proposition C5.9} The subset of $\mathcal{B}\left(\mathcal{C}(\mathcal{Y};\mathcal{K}(\mathsf{J})),\mathcal{C}(\mathcal{X};\mathcal{K}(\mathsf{H})) \right)$ for which the tensor product with $ I_{\mathcal{M}_{n}} $ is positive for some fixed $ n\in\lbrace 1,2,\ldots\rbrace $ is closed in the weak topology.
\paragraph*{Proof} We will show the complement is open. Take such a map \textit{L} that is not in the subset. By the argument in the proof for \textbf{C5.7} and \textbf{B5.9}, that implies there are some $x\in \mathcal{X}$, partition-of-unity $ \lbrace g_{j}\rbrace $, finite collections of vectors $ \lbrace\mathbf{w}_{jk} \rbrace\subset \mathsf{J}$ and $ \lbrace\mathbf{v}_{k} \rbrace\subset \mathsf{H}$, and $ \varepsilon>0 $ such that
\begin{equation}
\sum_{j=1}^{m}\sum_{k,l=1}^{n} \langle L(\mathbf{w}_{jk}\otimes\mathbf{w}^{*}_{jl}\; g_{j})(x)\mathbf{v}_{l},\mathbf{v}_{k}\rangle<-\varepsilon 
\end{equation}
Then, by the triangle inequality, all the maps in the weak neighborhood
\begin{equation}
\bigcap_{j=1}^{m}\bigcap_{k,l=1}^{n}\mathcal{N}\left( L;\mathbf{w}_{jk}\mathbf{w}_{jk}^{*}\,g_{j};\mathbf{v}_{l}\otimes\mathbf{v}_{l}^{*}\, \delta_{x};\frac{\varepsilon}{2n^{4}m}\right) 
\end{equation}
\[+\bigcap_{j=1}^{m}\bigcap_{k=1}^{n}\bigcap_{r<l} \mathcal{N}\left( L;(\mathbf{w}_{jr}\otimes\mathbf{w}_{jl}^{*}+\mathbf{w}_{jl}\otimes\mathbf{w}_{jr}^{*})\,g_{j};\mathbf{v}_{k}\otimes\mathbf{v}_{k}^{*}\,  \delta_{x};\frac{\varepsilon}{n^{4}m}\right)\]
\[+\bigcap_{j=1}^{m}\bigcap_{k=1}^{n}\bigcap_{r<l} \mathcal{N}\left(L;\mathbf{w}_{jk}\otimes\mathbf{w}_{jk}^{*}\,g_{j};(\mathbf{v}_{k}\otimes\mathbf{v}_{m}^{*}+\mathbf{v}_{m}\otimes\mathbf{v}_{k}^{*})\,  \delta_{x};\frac{\varepsilon}{n^{4}m}\right)\]
\[+\bigcap_{j=1}^{m}\bigcap_{k<l}\bigcap_{q<r}\mathcal{N}\left(L; (\mathbf{w}_{jk}\otimes\mathbf{w}_{jl}^{*}+\mathbf{w}_{jl}\otimes\mathbf{w}_{jk}^{*})\,g_{j};(\mathbf{v}_{q}\otimes\mathbf{v}_{r}^{*}+\mathbf{v}_{r}\otimes\mathbf{v}_{q}^{*})\,  \delta_{x};\frac{2\varepsilon}{n^{4}m}\right) \]
will also fail to yield a positive tensor product with $ I_{\mathcal{M}_{n}} $. $ \square $
\paragraph{Corollary C5.10} The cone of completely positive maps is weakly closed in\\ $\mathcal{B}\left(\mathcal{C}(\mathcal{Y};\mathcal{K}(\mathsf{J})),\mathcal{C}(\mathcal{X};\mathcal{K}(\mathsf{H})) \right)$.
\paragraph{Comment} See the comment following \textbf{B5.12} concerning the use of the axiom of choice.
\paragraph{Proposition C5.11} For any positive functional $ \Phi\in\mathcal{C}(\mathcal{X};\mathcal{K}(\mathsf{H}))^{*} $, there is some $ \mathcal{D}(\mathsf{H})^{+} $-valued, Radon vector measure $ \mu $ such that 
\[\Phi f=\int_{\mathcal{X}} f  d\mu\]
\paragraph*{Proof} Since $ \mathcal{K}(\mathsf{H})^{*}\cong\mathcal{D}(\mathsf{H}) $, there is some $ \rho\in \mathcal{D}(\mathsf{H})^{+} $ such that $ \Phi(\phi 1_{\mathcal{X}})=\text{tr }\rho\phi $ for any $ \phi\in \mathcal{K}(\mathsf{H})$. Since finite rank operators are dense in compact ones, $ \rho $ lives on a separable subspace of \textsf{H}; let $ \lbrace\mathbf{e}_{j}\rbrace $ be an orthonormal basis for this subspace and $ \langle P_{j}\rangle $ an increasing sequence of projectors onto the subspaces spanned by the first $ j $ basis vectors. Applying the Riesz-Markov theorem entry-wise, there is a sequence of Radon vector measures, $ \langle \mu_{j}\rangle $, with each $ \mu_{j} $ taking values in $ \mathcal{D}(P_{j}\mathsf{H})^{+} $. Take $ j>k $; then $ \Vert\mu_{j}-\mu_{k}\Vert $ is given by
\begin{equation}
\sup\left\lbrace\sum_{l=1}^{r}\Vert\mu_{j}(E_{l})-\mu_{k}(E_{l})\Vert_{\text{trace}}\left\vert\begin{array}{l}r\in\lbrace 1,2,\ldots\rbrace,\text{disjoint, Borel}\\\text{subsets }\lbrace E_{1},\ldots,E_{n}\rbrace\\\text{with }\bigcup_{l=1}^{r}E_{l}=\mathcal{X}\end{array}\right. \right\rbrace 
\end{equation} 
Applying \textbf{C4.3}, $ \sum_{l=1}^{r}\Vert\mu_{j}(E_{l})-\mu_{k}(E_{l})\Vert_{\text{trace}} $ is less than or equal to
\begin{equation}
 c_{0}\;\sqrt{\left(\sum_{l=1}^{r}\text{tr }\left(\mu_{j}(E_{l})-\mu_{k}(E_{l})\right)\right)\left(\sum_{l=1}^{r}\text{tr }\mu_{j}(E_{l})\right) }
\end{equation}
\[=c_{0}\;\sqrt{\text{tr }\left(\mu_{j}(\mathcal{X})-\mu_{k}(\mathcal{X})\right)\text{tr }\mu_{j}(\mathcal{X}) } \]
which is less than or equal to $ c_{0}\;\sqrt{\text{tr }\left(\rho-\mu_{k}(\mathcal{X})\right)\text{tr }\rho } $. This goes to zero as $ k\to \infty $ since the $ \mu_{k}(\mathcal{X}) $'s are truncations of $ \rho $, which converge to $ \rho $ in trace norm, which can be seen either by using \textbf{C4.3} again or by first demonstrating that the truncations converge in norm for rank-one $ \rho $ (which is readily shown); by the spectral theorem for compact operators, the truncations converge in trace norm for every $ \rho $. Therefore, $ \langle \mu_{j}\rangle $ is a Cauchy sequence. Since $ \mathcal{D}(\mathsf{H})^{+} $-valued vector measures are complete with the given norm~\cite{ryancompletevectormeasure}, the limit is the desired $ \mu $. $ \square $  
\medskip\\
For the following proposition, restrict the Hilbert space \textsf{H} to be separable. Then there is some $ \rho\in\mathcal{D}(\mathsf{H})^{+} $ that is strictly positive. Let $ \lbrace\mathbf{e}_{j}\rbrace $ be an orthonormal basis for \textsf{H} composed of eigenvectors of $ \rho $, with eigenvalues $ \lambda_{j} $ in decreasing order. For any Radon measure $ \mu $ on $ \mathcal{X} $, by the preceding proposition there is the induced variation measure $ \nu=\left\vert L^{*}(\rho\, 1_{\mathcal{X}} \mu)\right\vert $ on $ \mathcal{Y} $, which is also Radon~\cite{ryancompletevectormeasure}. Then we have the following result:
\paragraph{Proposition C5.12} For any positive $L\in \mathcal{B}\left(\mathcal{C}(\mathcal{Y};\mathcal{K}(\mathsf{J})),\mathcal{C}(\mathcal{X};\mathcal{K}(\mathsf{H})) \right)^{+}$, the adjoint map $ L^{*} $ induces a map $ K $ that sends $ L^{1}(\mathcal{X};\mu;\mathcal{D}(\mathsf{H}))^{+} $ into $ L^{1}(\mathcal{Y};\nu;\mathcal{D}(\mathsf{J}))^{+} $.
\paragraph*{Proof} Let $ C\subset\mathcal{Y} $ be any closed, $ \nu$-null subset. By outer-regularity, there are open sets containing $ C $ with arbitrarily small $ \nu $-measure. Therefore, by Urysohn's lemma~\cite{roydenurysohn},
\begin{equation}
0=\nu(A)=\inf_{f\in\mathcal{C}(\mathcal{Y}),1_{\mathcal{X}}\geq f\geq 1_{C}}\int_{\mathcal{Y}} f\,d\nu
\end{equation}
Hence, by the definition of $ \nu $, for any $ \phi\in \mathcal{K}(\mathsf{J})^{+}$,
\begin{equation}
0=\inf_{f\in\mathcal{C}(\mathcal{Y}),1_{\mathcal{Y}}\geq f\geq 1_{C}}\int_{x\in\mathcal{X}}\text{tr }\rho L(\phi f)(x)\,d\mu(x)
\end{equation}
For \textit{A} any Borel subset of $ \mathcal{X} $ and $ \tau $ any positive, self-adjoint operator on the span of finitely many of the $ \lbrace\mathbf{e}_{j}\rbrace $, any single-term simple function, $ \tau 1_{A} $, can be scaled to be less than $ \rho\, 1_{\mathcal{X}} $. Then
\begin{equation}
0=\inf_{f\in\mathcal{C}(\mathcal{Y}),1_{\mathcal{Y}}\geq f\geq 1_{C}}\int_{x\in\mathcal{X}}1_{A}(x)\text{tr }\tau L(\phi f)(x)\,d\mu(x)
\end{equation}
Now take $ \lbrace\tau_{j}\rbrace $ and $ \lbrace A_{j}\rbrace$ to be finite collections of such operators and subsets. For allowed $ f_{1},\ldots,f_{n} $, the pointwise product $ f_{1}\cdots f_{n} $ is also allowed and is less than or equal to each of the $ f_{j} $; hence,
\begin{align}
&\inf_{f\in\mathcal{C}(\mathcal{Y}),1_{\mathcal{Y}}\geq f\geq 1_{C}}\sum_{j=1}^{n}\int_{x\in\mathcal{X}}1_{A_{j}}(x)\text{tr }\tau_{j} L(\phi f)(x)\,d\mu(x)\\&=\sum_{j=1}^{n}\inf_{f\in\mathcal{C}(\mathcal{Y}),1_{\mathcal{X}}\geq f\geq 1_{C}}\int_{x\in\mathcal{X}}1_{A_{j}}(x)\text{tr }\tau_{j} L(\phi f)(x)\,d\mu(x)=0\nonumber
\end{align}

Since simple functions of the form $ \sum_{j=1}^{n}\tau_{j}\,1_{A_{j}} $ are dense in $L^{1}(\mathcal{X};\mu;\mathcal{D}(\mathsf{H}))^{+}  $, then for any element $ \xi\in L^{1}(\mathcal{X};\mu;\mathcal{D}(\mathsf{H}))^{+} $,
\begin{equation}
0=\inf_{f\in\mathcal{C}(\mathcal{Y}),1_{\mathcal{Y}}\geq f\geq 1_{C}}\int_{x\in\mathcal{X}}\text{tr }\xi(x) L(\phi f)(x)\,d\mu(x)=\inf_{f\in\mathcal{C}(\mathcal{Y}),1_{\mathcal{Y}}\geq f\geq 1_{C}}\int_{\mathcal{Y}} f\,d(\phi L^{*}(\xi\mu))
\end{equation}
where $ \phi L^{*}(\xi\mu) $ is a Radon measure. Therefore, it must be that $ (\phi L^{*}(\xi\mu))(C)=0 $. Since $ \phi $ was arbitrary, $ L^{*}(\xi\mu)(C) $ must be zero. 

By inner regularity, any Borel set can have its measure approximated arbitrarily well by closed sets it contains. Therefore, for any $ \nu $-null, Borel subset \textit{A}, $ L^{*}(\xi\mu)(A) $ is also zero. Therefore, $ L^{*}(\xi\mu)$ is absolutely continuous with respect to $ \nu $. Since $ \mathcal{D}(\mathsf{H}) $ has the Radon-Nikod\'{y}m property~\cite{uhl}, there is a $ \psi\in L^{1}(\mathcal{Y};\nu;\mathcal{D}(\mathsf{J}))^{+} $ such that $ L^{*}\mu=\psi\nu $, which gives the desired map \textit{K} by $ K\xi= \psi$. $ \square $